# Application-level Fault-Tolerance Protocols

Vincenzo De Florio
University of Antwerp,
Middelheimlaan 1, 2020 Antwerp, Belgium

# Preface

## 1 INTRODUCTION

The central topic of this book is application-level fault-tolerance, that is the methods, architectures, and tools that allow to express a fault-tolerant system in the application software of our computers. Application-level fault-tolerance is a sub-class of software fault-tolerance that focuses on the problems of expressing the problems and solutions of fault-tolerance in the top layer of the hierarchy of virtual machines that constitutes our computers. This book shows that application-level fault-tolerance is a key ingredient to craft truly dependable computer systems—other approaches, such as hardware fault-tolerance, operating system fault-tolerance, or fault-tolerant middleware, are also important ingredients to achieve resiliency, but they are not enough. Failing to address the application layer means leaving a backdoor open to problems such as design faults, interaction faults, or malicious attacks, whose consequences on the quality of service could be as unfortunate as, e.g., a physical fault affecting the system platform. In other words, in most cases it is simply not possible to achieve complete coverage against a given set of faults or erroneous conditions without embedding fault-tolerance provisions also in the application layer. In what follows the provisions for application-level fault-tolerance are called application-level fault-tolerance protocols.

As a lecturer in this area, I wrote this book as my ideal textbook for a possible course on resilient computing and for my doctoral students in software dependability at the University of Antwerp. Despite this, the main goal of this book is not—only—education. The main mission of this book is first of all spreading the awareness of the necessity of application-level fault-tolerance. Another critical goal is highlighting the role of several important concepts that are often neglected or misunderstood: The fault and the system models, i.e., the assumptions on top of which our computer services are designed and constructed. Last but not the least of our goals, this book aims to provide a clear view to the state-of-the-art of application-level fault-tolerance, also highlighting in the process a number of lessons learned through hands-on experiences gathered in more than 10 years of work in the area of resilient computing.

It is our belief that any person who wants to include dependability among the design goals of their intended software services should have a clear understanding of concepts such as dependability, system models, failure semantics, and fault models and of their influence on their final product's quality of experience. Such information is often scattered among research papers while it is presented here in a unitary framework and from the viewpoint of the application-level dependable software engineer.
Another aspect that makes this book unique from all others in the field is the fact that concepts are described with examples that, in some cases, reach a deep level of detail. This is not the case in all chapters, as it reflects the spectrum of working experiences that the author had during more than a decade of research in this area. Any such spectrum is inherently not uniformly distributed. As a consequence some chapters provide the reader with in-depth knowledge, down to the level of source code examples, while others just introduce the reader into the subject, explain the main concepts, and place the topic in the wider context of methods treated in this book. To increase readability, we isolated some of the most technical texts into quoted sections typed in blue.
Furthermore, this book has a privileged viewpoint, which is the one of real-time, concurrent, and embedded system design. This book does not focuses in particular on the design of fault-tolerance provisions for service-oriented applications, such as web services, and does not cover fault-tolerance in the middleware layer.
In what follows the background top-level information and the structure of this book are introduced.

| |
|---|
| Application-level fault-tolerance is defined in what follows as the sub-class of software fault-tolerance that focuses on how to express the problems and solutions of fault-tolerance in the top layer of the hierarchy of virtual machines that constitutes our computers. Traditionally research in this sub-class was initiated by Brian Randell with his now classical article on which system structure to give our programs in order to be tolerant to faults (Randell, 1975). The key problem expressed in that paper was that of a cost-effective solution to embed fault-tolerance in the application software. Recovery blocks (treated in Chapter 4) was the proposed solution. Randell was also the first to state the insufficiency of fault-tolerance solutions based exclusively on hardware designs and the need of appropriate structuring techniques such that the incorporation of a set of fault-tolerance provisions in the application software could be performed in a simple, coherent, and well structured way. A first proposal for the embedding recovery blocks in a programming language was proposed shortly afterwards (Shrivastava, 1978). Leaving the safe path of hardware fault-tolerance brought about new problems and challenges: Hardware redundancy guarantees random component failures, while software replication does not guarantee statistical independence of failures. In other words, a single cause may produce many (undesirable) effects. This means that "in software the redundancy required is not simple replication of programs but *redundancy of design*" (Randell, 1975). An answer to this problem and another important milestone was the conception of $N$-version programming by Algirdas Avižienis (Avižienis, 1985), which combines hardware and information redundancy in the attempt to reduce the chance of correlated failures in the software components. At the same time, the very meaning of computing and programming was evolving, again bringing new possibilities but also opening up new problems and challenges: The spread of distributed systems meant also the end of the purely synchronous model for computing and communication (see for instance (Jalote, 1994) and (Lamport, Shostak, & Pease, 1982) and Chapter 2); object orientation made it possible to easily reuse third-party software components, but turned our applications into a chain of links of unknown strength and trustworthiness (Green, 1997). The logics for assembling the links together is in our applications, hence it is clear that the logics to prevent the break of those links to lead to disaster must also involve the application layer (Saltzer, Reed, & Clark, 1984). Luckily from the object model there began to stem several variants, such as composition filters, distributed objects, or fragmented objects, that would provide the programmer with powerful tools for fault-tolerance programming in the application layer (see Chapter 6 for a few examples). Other approaches are also being devised, e.g. aspect-oriented programming—though their potential as fault-tolerance language is yet to be confirmed (see Chapter 8 for a brief introduction). Still other approaches are also discussed in this book. A special accent is given to those approaches where the author had first-hand experience with. In one case—the Ariel recovery language—the reader is provided with enough details to even understand how the approach has been crafted. We are now at the verge of yet another change, with ubiquitous computing, service orientation and the novel Web technology promising to serve us as even more powerful solutions to accompany us in the transition towards the Information Society of tomorrow. Such topics would require a book on their own and have not been treated here. Still the problems of application-level fault-tolerance are with us, while to date no ultimate and general-purpose solution has been found out. This book is about this possibly unique case in computer science and engineering of a problem yet unsolved though being formulated more than 30 years ago. |

Table 1: A short introduction to application-level fault-tolerance

## 2 BACKGROUND AND STRUCTURE

No man conceived tool in human history has ever permeated so many aspects of human life as the computer has been doing for the last 60 years. An outstanding aspect of this success story is certainly given by an overwhelming increase in computer performance. Another one, also very evident, is the continuous decrease of costs of computer devices—A 1000$ PC today provides its user with more performance, memory, and disk space of a million dollar mainframe of the Sixties. Clearly performance and costs are “foreground figures”—society at large is well aware of their evolution and of the societal consequences of the corresponding spread of computers. On the other hand this process is also characterized by “background figures”, that is, properties that are often overlooked despite their great relevance. Among such properties it is worth mentioning the growth in complexity and the crucial character of the roles nowadays assigned to computers: Human society more and more expects and relies on good quality of complex services supplied by computers. More and more these services become vital, in the sense that lack of timely delivery ever more often can have immediate consequences on capitals, the environment, and even human lives. Strangely though it may appear, the common man is well aware that computers get ever more powerful and less expensive, but doesn’t seem to be aware or even care about computers being safe and up to their ever more challenging tasks. The turn of the century brought about this problem for the first time—the Millennium Bug, also known as Y2K, reached the masses with striking force, as a tsunami of sudden awareness that “yes, computers are powerful, but even computers can fail.”
Y2K ultimately did not show up, and the dreaded scenarios of a society simultaneously stripped by its computer services ended up in a few minor accidents.

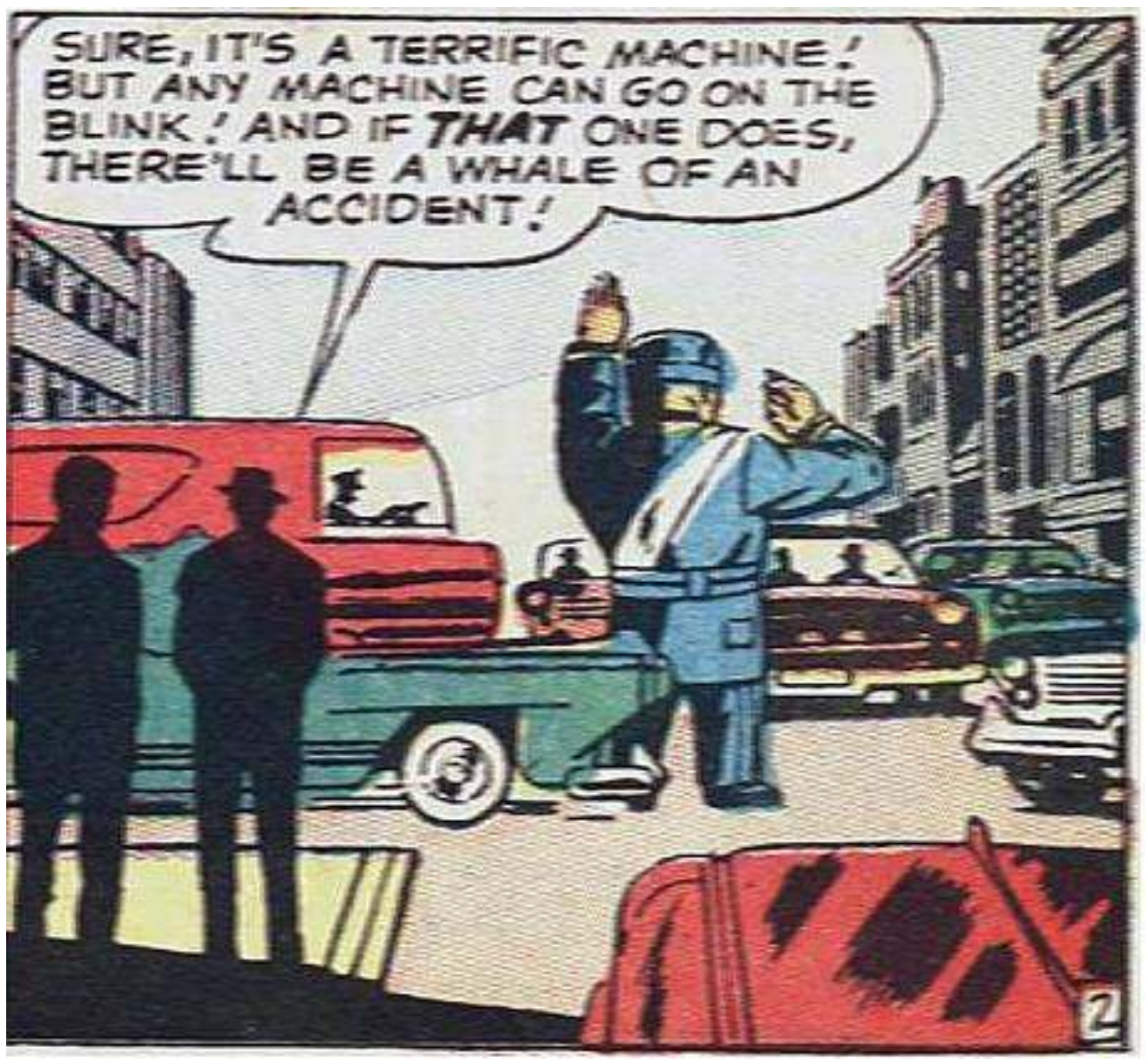

Figure 1: Perception of dependability in the Sixties (TM & © 2008 Marvel Characters, Inc. All Rights Reserved.)

But society had had a glimpse to some of the problems that are central to this book: Why do we trust computer services? Are there modeling, design, development practices, conceptual tools, and concrete methods, to convince me that when I take a computer service, that service will be reliable, safe, secure, available? In other words, is there a science of computer dependability, such that reliance of computer systems can be measured, hence quantitatively justified? And, is there an engineering of computer dependability, such that trustworthy computer services can be effectively achieved? Dependability—the discipline that studies those problems—is introduced in Chapter 1.

This book in particular focuses on **fault-tolerance**, which is described in **Chapter 1** as one of the “means” for dependability: Fault-tolerance is one of the four classes of methods and techniques enabling one to provide the ability to deliver a service on which reliance can be placed, and to reach confidence in this ability (together with fault prevention, fault removal, and fault forecasting). Its core objective is “preserving the delivery of expected services despite the presence of fault-caused errors within the system itself” (Avižienis, 1985). The exact meaning of faults and errors is also given in the cited chapter, together with an introduction to fault-tolerance mainly derived from the works of Jean-Claude Laprie (Laprie, 1992, 1995, 1998, 1985). What is important to remark here is that fault-tolerance acts after faults have manifested themselves in the system: Its main assumption is that faults are inevitable, but they must be tolerated, which is fundamentally different from other approaches where, e.g., faults are sought to be avoided in the first place. Why focusing on fault-tolerance, why is it so important? For the same reason referred above as a background figure in the history of the relationship between human society and computers: The growth in complexity. Systems get more and more complex, and there are no effective methods that can provide us with a zero-fault guarantee. The bulk of the research of computer scientists and engineers concentrated on methods to pack conveniently ever more complexity in computer systems. **Software** in particular has become a point of accumulation of complexity, and the main focus so far has been on how to express and compose complex software modules so as to tackle ever new challenging problems rather than dealing with the inevitable faults introduced by that complexity. Layered design is a classical method to deal with complexity.

Software, software fault-tolerance, and application-level software fault-tolerance, are the topics of **Chapter 2**. It is explained what does it mean that a program is fault-tolerant and what are the properties expected from a fault-tolerant program. The main objective of Chapter 2 is introducing two sets of design assumptions that shape the way people structure their fault-tolerant software—the system and the fault models. Often misunderstood or underestimated, those models describe

- what is expected from the execution environment in order to let our software system function correctly,
- and what are the faults that our system is going to consider. Note that a fault-tolerant program shall (try to) tolerate only those faults stated in the fault model, and will be as defenseless against all other faults as any non fault-tolerant program.

Together with the system specification, the fault and system models represent the foundation on top of which our computer services are built. Not surprisingly enough, weak foundations often result in fragile constructions. To provide evidence to this, the chapter introduces three well-known accidents—the Ariane 5 flight 501 and Mariner-1 disasters and the Therac-25 accidents (Leveson, 1995). In each case it has been stressed out what went wrong, what were the biggest mistakes, and how a careful understanding of fault models and system models would have helped highlighting the path to avoid catastrophic failures that cost considerable amounts of money and even the lives of innocent people.

After this, the chapter focuses on the core topic of this book, application-level software fault-tolerance. Main questions addressed here are: How to express and achieve fault-tolerance in the mission layer? And, why is application-level software fault-tolerance so important? The main reason for this is that a computer service is the result of the concurrent execution of several "virtual" and physical machines (see Fig. 2). Some of these machines run a predefined, special-purpose service, meant to serve—unmodified—many different applications. The hardware, the operating system, the network layers, the middleware, a programming language's run-time executive, and so forth, are common names of those machines. A key message in this book is that *tolerating the faults in one machine does not protect from faults originating in another one*. This includes the application layer. Now, while the machines "below" the application provide architectural (special-purpose) complexity, the mission layer contributes to computer services with general-purpose complexity, which is intrinsically less reliable. This and other reasons justifying the need for application-level software fault-tolerance are given in that chapter. The main references here are (Randell, 1975; Lyu, 1998a, 1998b).
Chapter 2 also introduces what the author considers to be the three main properties of application-level software fault-tolerance: Separation of design concerns, adaptability, and syntactical adequacy (De Florio & Blondia, 2008b). In this context the key questions are: Given a certain fault-tolerance provision, is it able to guarantee an adequate separation of the functional and non-functional design concerns? Does it tolerate a fixed, predefined set of faulty scenarios, or does it dynamically change that set? And, is it flexible enough as to host a large number of different strategies, or is it a "hardwired" solution tackling a limited set of strategies?
Finally, this chapter defines a few fundamental fault-tolerance services, namely watchdog timers, exception handling, transactions, and checkpointing-and-rollback.
After having described the context and the "rules of the game", this book discusses the state of the art in application-level fault-tolerance protocols. First, in **Chapter 3**, the focus is on so-called single-version and multiple-version software fault-tolerance (Avižienis, 1985).

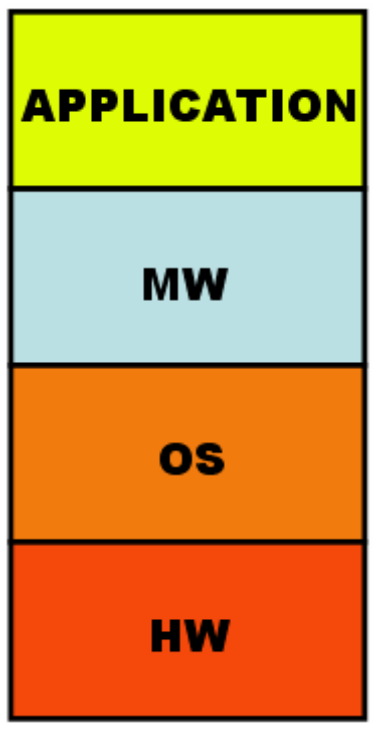

Figure 2: Computer Services are the result of layered designs. The higher you go, the more specialized is the layer.

- Single-version protocols are methods that use a non-distributed, single task provision, running side by side with the functional software, often available in the form of a library and a run-time executive.
- Multiple-version protocols are methods that use actively a form of redundancy, as explained in what follows. In particular the chapter discusses recovery blocks and N-version programming.

Chapter 3 also features several in-depth case studies deriving from the author's research experiences in the field of resilient computing. In particular the EFTOS fault-tolerance library (Deconinck, De Florio, Lauwereins, & Varvarigou, 1997; Deconinck, Varvarigou, et al., 1997) is introduced as an example of application-level single-version software fault-tolerance approach. In that general framework, the EFTOS tools for exception handling, distributed voting, watchdog timers, fault-tolerant communication, atomic transactions, and data stabilization, are discussed. The reader is also given a detailed description of RAFTNET (Raftnet, n.d.), a fault-tolerance library for data parallel applications.
A second large class of application-level fault-tolerance protocols is the focus of **Chapter 4**, namely the one that works "around" the programming language, that is to say either embedded in the compiler or via language transformations driven by translators. In that chapter it is also discussed the design of a translator supporting language-independent extensions called reflective and refractive variables and linguistic support for adaptively redundant data structures.

- Reflective and refractive variables (De Florio & Blondia, 2007a) are syntactical structures to express adaptive feedback loops in the application layer. This is useful to resilient computing because a feedback loop can attach error recovery strategies to error detection events.

- Redundant variables (De Florio & Blondia, 2008a) are a tool that allows designers to make use of adaptively redundant data structures with commodity programming languages such as C or Java. Designers using such tool can define redundant data structures in which the degree of redundancy is not fixed once and for all at design time, but rather it changes dynamically with respect to the disturbances experienced during the run time.

The chapter shows that by a simple translation approach it is possible to provide sophisticated features such as adaptive fault-tolerance to programs written in any programming language.
In **Chapter 5** te reader gets in touch with methods that work at the level of the language itself: Custom fault-tolerance programming languages. In this approach fault-tolerance is not embedded in the program, nor around the programming language, but provided through the syntactical structures and the run-time executives of fault-tolerance programming languages. Also in this case application-level complexity is enucleated from the source code and shifted to the architecture, where it is much easier and cost-effective to tame. Three classes of approaches are treated—object-oriented languages, functional languages, and hybrid languages. In the latter class special emphasys is given to Oz (Müller, Müller, & Van Roy, 1995), a multi-paradigm programming language that achieves both transparent distribution and translucent failure handling.
A separate chapter is devoted to a large case study in fault-tolerant languages: The so-called recovery language approach (De Florio, 2000; De Florio, Deconinck, & Lauwereins, 2001). In **Chapter 6** the concept of recovery language is first introduced in general terms and then proposed through an implementation: the Ariel recovery language and a supporting architecture. That architecture is an evolution of the EFTOS system described in Chapter 3, and targets distributed applications with non-strict real-time requirements, written in a procedural language such as C, to be executed on distributed or parallel computers consisting of a predefined set of processing nodes. Ariel and its run-time system provide the user with a fault-tolerance linguistic structure that appears to the user as a sort of second application-level especially conceived and devoted to address the error recovery concerns. This separation is very useful at design time, as it allows to bound design complexity. In Ariel this separation holds also at run-time, because even the executable code for error recovery is separated from the functional code. This means that, in principle, the error recovery code could change dynamically so as to match a different set of internal and environmental conditions. This can be used to avoid "hardwiring" a fault model into the application—an important property especially when, e.g., the service is embedded in a mobile terminal (De Florio & Blondia, 2005).

**Chapter 7** discusses fault-tolerance protocols based on aspect-oriented programming (Kiczales et al., 1997), a relatively novel structuring technique with the ambition to become the reference solution for system development, the way object-orientation did starting with the Eighties. We must remark how aspects and their currently available implementations have not yet reached a maturity comparable with that of the other techniques discussed in this book. For instance, the chapter remarks how no aspect-oriented fault-tolerance language has been proposed to date and, at least in same cases, the adequacy of aspects as a syntactical structure to host fault-tolerance provisions has been questioned. On the other hand, aspects allow regarding the source code as a flexible web of syntactic fragments that the designer can rearrange with great ease, deriving modified source codes matching particular goals, e.g. performance and, hopefully in the near future, dependability. The chapter explains how aspects allow to separate design concerns, which bounds complexity and enhances maintainability, and presents three programming languages: AspectJ (Kiczales, 2000), AspectC++ (Spinczyk, Lohmann, & Urban, 2005) and GluonJ (GluonJ, n.d.).

The following chapter, **Chapter 8**, deals with failure detection protocols in the application layer. First the concept of failure detection (Chandra & Toueg, 1996), a fundamental building block to develop fault-tolerant distributed systems, is introduced. Then the relationship between failure detection and system models is highlighted—the key assumptions on which our dependable services are built, which were introduced in Chapter 2. Then it is introduced a tool for the expression of this class of protocols (De Florio & Blondia, 2007b), based on a library of objects called time-outs (V. De Florio, 2006). Finally a case study is described in detail: The failure detection protocol employed by the so-called EFTOS DIR net (De Florio, Deconinck, & Lauwereins, 2000), a distributed "backbone" for fault-tolerance management which was introduced in Chapter 3 and that later evolved into the so-called Backbone discussed in Chapter 6.

Hybrid approaches are the focus of **Chapter 9**, that is, fault-tolerance protocols that blend two or more methods among those reported in previous chapters. In more detail $\mathcal{REL}$inda is introduced—a system coupling the recovery language approach of Chapter 6 and generative communication, one of the models introduced in Chapter 4 (De Florio & Deconinck, 2001). After this the recovery language-empowered extensions of two single-version mechanisms previously introduced in Chapter 3 are described, namely a distributed voting mechanism and a watchdog timer (De Florio, Donatelli, & Dondossola, 2002). The main lessons learned in this case are that the recovery language approach allows to fast-prototype complex strategies by composing a set of building blocks together and by building system-wide, recovery-time coordination strategies with the Ariel language. This allows set up sophisticated fault-tolerance systems while keeping the management of their complexity outside of the user application. Other useful properties achieved in this way are transparency of replication and transparency of location.

**Chapter 10** provides three examples of approaches used to assess the dependability of application-level provisions. In the first case reliability analysis is used to quantify the benefits of coupling an approach such as recovery languages to a distributed voting mechanism (De Florio, Deconinck, & Lauwereins, 1998). Then a tool is used to systematically inject faults onto the adaptively redundant data structure discussed in Chapter 4 (De Florio & Blondia, 2008a). Monitoring and fault-injection are the topic of the third case, where a hypermedia application to watch and control a dependable service is introduced (De Florio, Deconinck, Truyens, Rosseel, & Lauwereins, 1998).

**Chapter 11** concludes the book by summarizing the main lessons learned. It also offers a view to the internals of the application-level fault-tolerance provision described in Chapter 6—the Ariel recovery language.

# 3 SUMMARY OF CONTRIBUTIONS

Application software development is not an easy task; writing truly dependable fault-tolerant applications is even more difficult, not only in itself for the additional complexity required by fault-tolerance but often also because of the lack of awareness which is necessary in order to master the complexity of this tricky task.

The first and foremost contribution of this book is increasing the awareness of the role and significance of application-level fault-tolerance. This has been reached by highlighting important concepts that are often neglected or misunderstood, as well as introducing the available tools and approaches that can be used to craft high-quality dependable services by working also in the application layer.

Secondly, this book summarizes the most widely known approaches to application-level software fault-tolerance. A base of properties in which those approaches can be compared and assessed is defined.

Finally, this book features a collection of several research experiences the author had in the field of resilient computing through his participation to several research projects funded by the European Community. This large first-hand experience is reflected into the deep level of detail that is reached in some cases.

We hope that the above contributions will prove useful to the readers and intrigue them into entering the interesting arena of resilient computing research and development. Also, too many times the lack of awareness and know-how in resilient computing has brought the designers to supposedly robust systems whose failures had in some cases dreadful consequences on capitals, the environment, and even human lives—as a joke we call them sometimes “endangeneers”. We hope that this book may contribute to the spread of that awareness and know-how that should always be part of the education of dependable software engineers. This important requirement is witnessed by several organizations such as the European Workshop on Industrial Computer Systems Reliability, Safety and Security, technical committee 7), whose mission is “To promote the economical and efficient realization of programmable industrial systems through education, information exchange, and the elaboration of standards and guidelines" (EWICS, n.d.), and the ReSIST network of excellence (ReSIST, n.d.), which is developing a resilient computing curriculum recommended to all people involved in teaching dependability-related subjects.

# DEPENDABILITY AND FAULT-TOLERANCE: BASIC CONCEPTS AND TERMINOLOGY

## Contents

# 1 INTRODUCTION

The general objective of this chapter is to introduce the basic concepts and the terminology of the domain of dependability. Concepts such as reliability, safety, or security, have been used inconsistently by different communities of researchers: The real-time system community, the secure computing community, and so forth, each had its own "lingo" and was referring to concepts such faults, errors and failures without the required formal foundation. This changed in the early Nineties, when Jean-Claude Laprie finally introduced a tentative model for dependable computing. To date, the Laprie model of dependability is the most widespread and accepted formal definition for the terms that play a key role in this book. As a consequence, the rest of this chapter introduces that model.

# 2 DEPENDABILITY, RESILIENT COMPUTING, AND FAULT-TOLERANCE

As just mentioned the central topic of this chapter is **dependability**, defined in (Laprie, 1985) as the trustworthiness of a computer system such that *reliance* can *justifiably* be placed on the service it delivers. In this context,

**service** means the manifestations of a set of external events perceived by the user as the behavior of the system (Avižienis, Laprie, & Randell, 2004)

**user** means another system, e.g., a human being, or a physical device, or a computer application, interacting with the former one.

The concept of dependability as described herein was first introduced by Jean-Claude Laprie (Laprie, 1985) as a contribution to an effort by IFIP Working Group 10.4 (Dependability and Fault-Tolerance) aiming at the establishment of a standard framework and terminology for discussing reliable and fault-tolerant systems. The cited paper and other works by Laprie are the main sources for this chapter—in particular (Laprie, 1992), later revised as (Laprie, 1995) and (Laprie, 1998). A more recent work in this framework is (Avižienis, Laprie, Randell, & Landwehr, 2004). Professor Laprie is continuously revising his model, also with the contributions of various teams of researchers in Europe and abroad—let me just cite here the EWICS TC7 (European Workshop on Industrial Computer Systems Reliability, Safety and Security, technical committee 7), whose mission is "To promote the economical and efficient realization of programmable industrial systems through education, information exchange, and the elaboration of standards and guidelines" (EWICS, n.d.), and the ReSIST network of excellence (ReSIST, n.d.), boasting a 50-million items resilience knowledge base (Anderson, Andrews, & Fitzgerald, 2007), which developed a resilient computing curriculum recommended to all involved in teaching dependability-related subjects.

Laprie's is the most famous and accepted definition of dependability, but it is certainly not the only one. Not surprisingly, due to the societal relevance of such a topic, dependability has also slightly different definitions (Motet & Geffroy, 2003). According to Sommervilla (Sommervilla, 2006), for instance, dependability is "The extent to which a critical system is trusted by its users". This is clearly a definition that focuses more on how the user perceives the system than on how the system actually *is* trustworthy. It reflects the extent of the user's confidence that the system will operate as users expect and in particular without failures.

In other words, dependability is considered by Sommervilla and others as a measure of the quality of experience of a given user and a given service. From this descends that the objective of dependability engineers is *not* to make services failure-proof, but to let its users believe so! Paraphrasing Patterson and Hennessy (Patterson & Hennessy, 1996), if a particular hazard does not occur very frequently, it may not be worth the cost to avoid it. This means that residual faults are not only inevitable, but sometimes even expected. It's the notion of "dependability economics": Because of the very high costs of dependability achievement, in some cases it may be more cost effective to accept untrustworthy systems and pay for failure costs. This is especially relevant when time-to-market is critical to a product's commercial success. Reaching the market sooner with a sub-optimal product may bring more revenues than doing so with a perfectly reliable product surrounded by early bird competitors that have already captured the interest and trust of the public.

In what follows the book shall stick to Laprie's model of dependability. Following such model, a precise and complete characterization of dependability is given

1. by enumerating its basic properties or *attributes*,
2. by explaining which phenomena constitute potential *impairments* to it, and
3. by reviewing the scientific disciplines and the techniques that can be adopted as *means* for improving dependability.

Attributes, impairments, and means can be globally represented into one picture as a tree, traditionally called the dependability tree (Laprie, 1995) (see Fig. 1).

## 2.1 The Attributes of Dependability

As just mentioned, dependability is a general concept that embraces a number of different properties (Walkerdine, Melville, & Sommerville, 2002). These properties correspond to different viewpoints from which the user perceives the quality of the offered service—in other words, for different users there will be in general different key properties corresponding to a positive assessment of the service:

- **Availability** is the name of the property that addresses the readiness for usage.
- **Reliability** is the property that measures the continuity of service delivery.

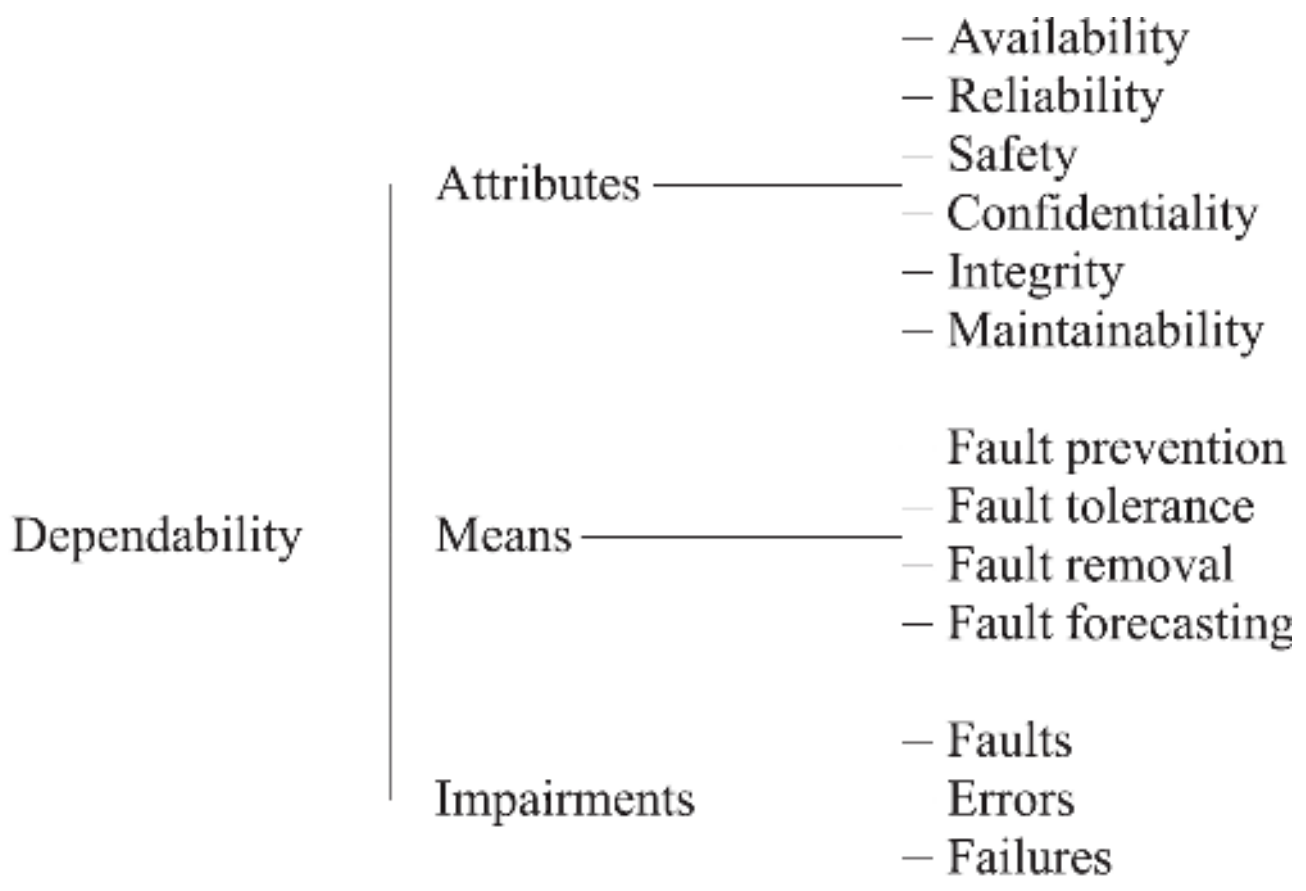

Figure 1: The dependability tree

- The property expressing the reliance on the non-occurrence of events with catastrophic consequences on the environment is known as **safety**.
- The property that measures the reliance on the non-occurrence of unauthorized disclosure of information has been called **confidentiality**.
- The property that measures the reliance on the non-occurrence of improper alterations of information has been called **integrity**.
- The property that expresses the ability to undergo repairs and upgrades has been called **maintainability**.

These properties qualify dependability, and therefore are known as its **attributes** (Laprie, 1995). Certain combinations of these attributes received a special name—**security** (Jonsson, 1998; Bodeaum, 1992), for instance, is defined as the conjoint requirement for integrity, availability, and confidentiality.
This section defines a number of important measures of a system's quality of service, including some of the attributes presented in Sect. 2.1 that are most relevant in what follows.

### 2.1.1 Reliability

When we take a plane, or even just a lift, the key property we expect from the computer system behind the service is that it proceed without flaws for the entire duration of the service. Any disruption of the service in the middle of the run would be disastrous. **Reliability** is the property that measures the continuity of service delivery. In other words, one expects—and hopes!—that airborne systems be reliable throughout their flights!

> More formally, reliability is defined as the conditional probability that the system will perform correctly throughout interval $[t_0, t]$, given that the system was performing correctly at time $t_0$ (Johnson, 1989). Time $t_0$ is usually omitted and taken as the current time. The general notation for reliability is therefore $R(t)$.

The negative counterpart of reliability, **unreliability**, is defined as $Q(t) = 1 - R(t)$, and represents the conditional probability that the system will perform incorrectly during the interval $[t_0, t]$, given that the system was performing *correctly* at time $t_0$. Unreliability is also known as the **probability of failure**.

### 2.1.2 Mean Time To Failure, Mean Time To Repair, and Mean Time Between Failures

If a system is known to fail, it makes sense to ask how long the system can be expected to run without problems. Such figure is called Mean Time to Failure (MTTF). MTTF is defined as the expected time that a system will operate before the occurrence of its first failure.
Another important property is Mean Time to Repair (MTTR). MTTR is defined as the average time required for repairing a system. It is often specified by means of a repair rate $\mu$, namely the average number of repairs that occur per time unit.
Mean Time Between Failures (MTBF) is the average time between any two consecutive failures of a system. This is slightly different from MTTF which regards on a system's very first failure. The following relation holds:

$$\text{MTBF} = \text{MTTF} + \text{MTTR}.$$

As it is usually true that MTTR is a small fraction of MTTF, it is usually allowed to assume that MTBF $\approx$ MTTF.

### 2.1.3 Availability

When we need to perform a banking transaction, or when we press the brake pedal while driving our car, or when we take an elevator and press the key corresponding to the floor we need to reach, the key property we expect from the system is that it serve us—it allow us to complete our transaction, or to slow down our car, or simply that the elevator works. What really matters is not how long the system worked so far, but that it works the moment we need it. This property is called availability.
Availability is defined as a function of time representing the probability that a service is operating correctly and is available to perform its functions at the instant of time $t$ (Johnson, 1989). It is usually represented as function $A(t)$.
Availability represents a property at a given *point* in time, whereas reliability concerns time *intervals*. These two properties are not to be mistaken with each other—a system might exhibit a good degree of availability and yet be rather unreliable, e.g., when inoperability is pointwise or rather short.

Availability can be approximated as the total time that a system has been capable of supplying its intended service divided by the elapsed time that system has been in operation, i.e., the percentage of time that the system is available to perform its expected functions. The steady-state availability can be proven (Johnson, 1989) to be

$$A_{\mathrm{ss}} = \frac{\mathrm{MTTF}}{\mathrm{MTTF} + \mathrm{MTTR}}.$$

### 2.1.4 Safety

Safety is the attribute of dependability that measures a system's ability to operate, *normally or not*, without damaging that system's environment (Sommervilla, 2006). So though it might seem a little strange at first, safety does not take into account the correctness of the system. In other words, while e.g. reliability is a functional attribute, this is not the case for safety. With reliability, quality is related to conformance to the functional specifications. With safety, quality is non-functional. A crashed system would exhibit minimal reliability and maximal safety. As remarked by Sommervilla, because of the increasing number of software-based control systems, software safety is being recognized as an important aspect of overall system safety. Systems where the issue of safety is particularly important, to the point that failures may lead to loss of lives or severe environmental damages are called life-critical or mission-critical systems (Storey, 1996). An example architecture for safety-critical systems is Cardamon (Corsaro, 2005).

### 2.1.5 Maintainability

Maintainability is a function of time representing the probability that a failed system will be repaired in a time less than or equal to $t$.

Maintainability can be estimated as

$$M(t) = 1 - \exp^{-\mu t},$$

$\mu$ being the repair rate, assumed to be constant (see Sect. 2.1.2).

## 2.2 Impairments to Dependability

Hardware and software systems must conform to certain specifications, i.e., agreed upon descriptions of the expected system response corresponding to any initial system state and input, as well as the time interval within which the response should occur. This includes a description of the functional behavior of the system—basically, what the system is supposed to do, or in other words, a description of its service—and possibly a description of other, non-functional requirements. Some of these requirements may concern the dependability of the service.

In real life, any system is subject to internal or external events that can affect in different ways the quality of its service. These events have been partitioned into three classes by their cause-effect relationship: depending on this, an impairment can be classified as a **fault**, an **error**, or a **failure**. When the delivered service of a system deviates from its specification, the user of the system experiences a **failure**. Such failure is due to a deviation from the correct state of the system, known as an **error**. That deviation is due to a given cause, for instance related to the physical state of the system, or to bad system design. This cause is called a **fault**. A failure of a system could give rise to an event that is perceived as a fault by the user of that system, bringing to a concatenation of cause-and-effects events known as the "fundamental chain" (Laprie, 1985):

$$\ldots \text{fault} \Rightarrow \text{error} \Rightarrow \text{failure} \Rightarrow \text{fault} \Rightarrow \text{error} \Rightarrow \text{failure} \Rightarrow \ldots$$

(symbol "$\Rightarrow$" can be read as "brings to"). Attributes defined in Sect. 2.1 can be negatively affected by faults, errors, and failures. For this reason, failures, errors, and faults have been collectively termed as the "impairments" of dependability. They are characterized in the following three paragraphs.

### 2.2.1 Failures

Failures System failures occur when the system does not behave as agreed in the system specifications or when the system specification did not properly describe its function. This can happen in many different ways (Cristian, 1991):

**omission** failures occur when an agreed reply to a well defined request is missing. The request appears to be ignored;

**timing** failures occur when the service is supplied, though outside the real-time interval agreed upon in the specifications. This may occur when the service is supplied *too soon* (early timing failure), or *too late* (late timing failure, also known as **performance** failure);

**response** failures happen either when the system supplies an incorrect output (in which case the failure is said to be a **value** failure), or when the system executes an incorrect state transition (**state transition** failure);

**crash** failure is when a system continuously exhibits omission failures until that system is restarted. In particular, a **pause-crash** failure occurs when the system restarts in the state it had right before its crash, while a **halting-crash** occurs when the system simply never restarts. When a restarted system re-initialises itself wiping out the state it had before its crash, that system is said to have experienced an **amnesia crash**. It may also be possible that some part of a system's state is re-initialized while the rest is restored to its value before the occurrence of the crash—this is called a **partial-amnesia crash**.

Defining the above failure classes allows extending a system's specification—that is, the set of its failure-free behaviors—with failure semantics, i.e., with the failure behavior that system is likely to exhibit upon failures. This is important when programming strategies for recovery after failure (Cristian, 1991). For instance, if the service supplied by a communication system may delay transmitted messages but never lose or corrupt them, then that system is said to have *performance* failure semantics. If that system can delay and also lose them, then it is said to have *omission/performance* failure semantics.
In general, if the failure semantics of a system $s$ allows it to exhibit a behavior in the union of two failure classes $F$ and $G$, then $s$ is said to have $F/G$ failure semantics. In other words, the "slash" symbol can be read as the union operator among sets. For any given $s$ it is possible to count the possible failure behaviors in a failure class. Let us call $b$ this function from the set of failure classes to integers. Then, given failure classes $F$ and $G$,

$$b(F/G) = b(F \mathbf{U} G) = b(F) + b(G).$$

> Failure semantics can be partially ordered by means of function $b$: Given any two failure semantics $F$ and $G$, then $F$ is said to exhibit a weaker (less restrictive) failure semantics than $G$:
>
> $$F < G \quad \equiv \quad b(F) > b(G).$$
>
> In particular, it is true that $F/G < F$. Therefore, the union of all possible failure classes represents the weakest failure semantics possible. If system $s$ exhibits such semantics, $s$ is said to have **arbitrary failure semantics**, i.e., $s$ can exhibit *any* failure behavior, without any restriction. By its definition, arbitrary failure semantics is also weaker than arbitrary *value* failure semantics. This latter is also known as **Byzantine failure semantics** (Lamport, Shostak, & Pease, 1982).

In the case of stateless systems, pause-crash and halting-crash behaviors are subsets of omission failure behaviors (Cristian, 1991), so omission failure semantics is in this case weaker than pause-crash and halting-crash failure semantics.
As clearly stated in (Cristian, 1991), it is the responsibility of a system designer to ensure that it properly implements a specified failure semantics. For instance, in order to implement a processing service with crash failure semantics, one can use duplication with comparison: Two physically independent processors executing in parallel the same sequence of instructions and comparing their results after the execution of each instruction. As soon as a disagreement occurs, the system is shut down (Powell, 1997). Another possibility is to use self-checking capabilities. Anyway, given any failure semantics $F$, it is up to the system designer to decide how to implement it, also depending on the designer's other requirements, e.g., those concerning

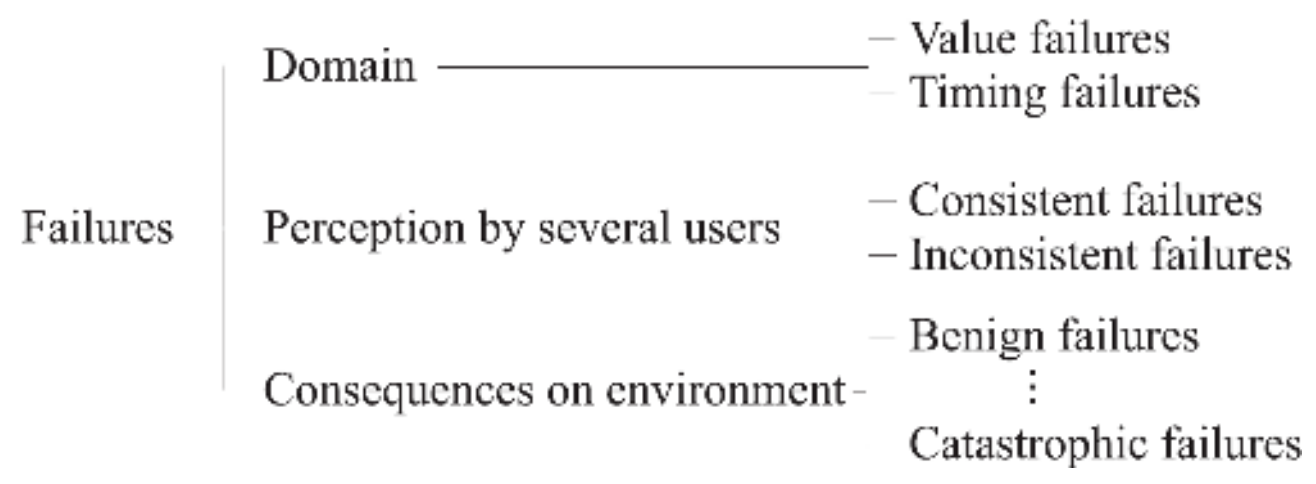

Figure 2: Failure classes.

costs and expected performance. In general, *the weaker the failure semantics, the more expensive and complex it is its implementation.* Moreover, a weak failure semantics imply *higher costs in terms of redundancy exhaustion* (see Sect. 3) and, often, higher performance penalties. For this reason, the designer may leave the ultimate choice to the user—for instance, the designer of the Motorola C compiler for the PowerPC allows the user to choose between two different modes of compilation—the fastest mode does not guarantee that the state of the system pipeline be restored on return from interrupts (Sun, 1996). This translates into behaviors belonging to the partial-amnesia crash semantics. The other mode guarantees the non-occurrence of these behaviors at the price of a lower performance for the service supplied by that system—programs compiled with this mode run slower.
Failures can also be characterized according to the classification in Fig. 2 (Laprie, 1995), corresponding to the different viewpoints of

- failure **domain** (i.e., whether the failure manifests itself in the time or value domain),
- failure **perception** (i.e., whether any two users perceive the failure in the same way, in which case the failure is said to be *consistent*, or differently, in which the failure is said to be *inconsistent*),
- and **consequences on the environment**. In particular a failure is said to be *benign* when consequences are of the same order as the benefits provided by normal system operation, while it is said *catastrophic* when consequences are incommensurably more relevant than the benefits of normal operation (Laprie, 1995).

Systems that provide a given failure semantics are often said to exhibit a "failure mode". For instance, systems having arbitrary failure semantics (in both time and value domains) are called **fail-uncontrolled** systems, while those only affected by benign failures are said to be **fail-safe** systems; likewise, systems with halt-failure semantics are referred to as **fail-halt** systems. These terms are also used to express the behavior a system should have when dealing with multiple failures—for instance, a "fail-op, fail-op, fail-safe" system is one such that is able to withstand two failures and then behaves as a fail-safe system (Rushby, 1994) (fail-op stands for "after failure, the system goes back to operational state"). Finally, it is worth mentioning the **fail-time-bounded** failure mode, introduced in (Cuyvers, 1995), which assumes that all errors are detected within a pre-defined, bounded period after the fault has occurred.

### 2.2.2 Errors

An error is the manifestation of a fault (Johnson, 1989) in terms of a deviation from accuracy or correctness of the system state. An error can be either **latent**, i.e., when its presence in the system has not been yet perceived, or **detected**, otherwise. Error latency is the length of time between the occurrence of an error and the appearance of the corresponding failure or its detection.

### 2.2.3 Faults

A fault (I. Lee & Iyer, 1993) is a defect, or an imperfection, or a lack in a system's hardware or software component. It is generically defined as the adjudged or hypothesised cause of an error. Faults can have their origin within the system boundaries (*internal faults*) or outside, i.e., in the environment (*external faults*). In particular, an internal fault is said to be *active* when it produces an error, and *dormant* (or *latent*) when it does not. A dormant fault becomes an active fault when it is *activated* by the computation process or the environment. Fault latency is defined as either the length of time between the occurrence of a fault and the appearance of the corresponding error, or the length of time between the occurrence of a fault and its removal.
Faults can be classified according to five viewpoints (Laprie, 1992, 1995, 1998)—phenomenological cause, nature, phase of creation or occurrence, situation with respect to system boundaries, persistence. Not all combinations can give rise to a fault class—this process only defines 17 *fault classes*, summarized in Fig. 3. These classes have been further partitioned into three "groups", known as combined fault classes.
The combined fault classes that are more relevant in the rest of the book are now briefly characterized:

**Physical faults:**

- Permanent, internal, physical faults. This class concerns those faults that have their origin within hardware components and are continuously active. A typical example is given by the fault corresponding to a worn out component.
- Temporary, internal, physical faults (also known as *intermittent faults*) (Bondavalli, Chiaradonna, Di Giandomenico, & Grandoni, 1997). These are typically internal, physical defects that become active depending on a particular pointwise condition.
- Permanent, external, physical faults. These are faults induced on the system by the physical environment.

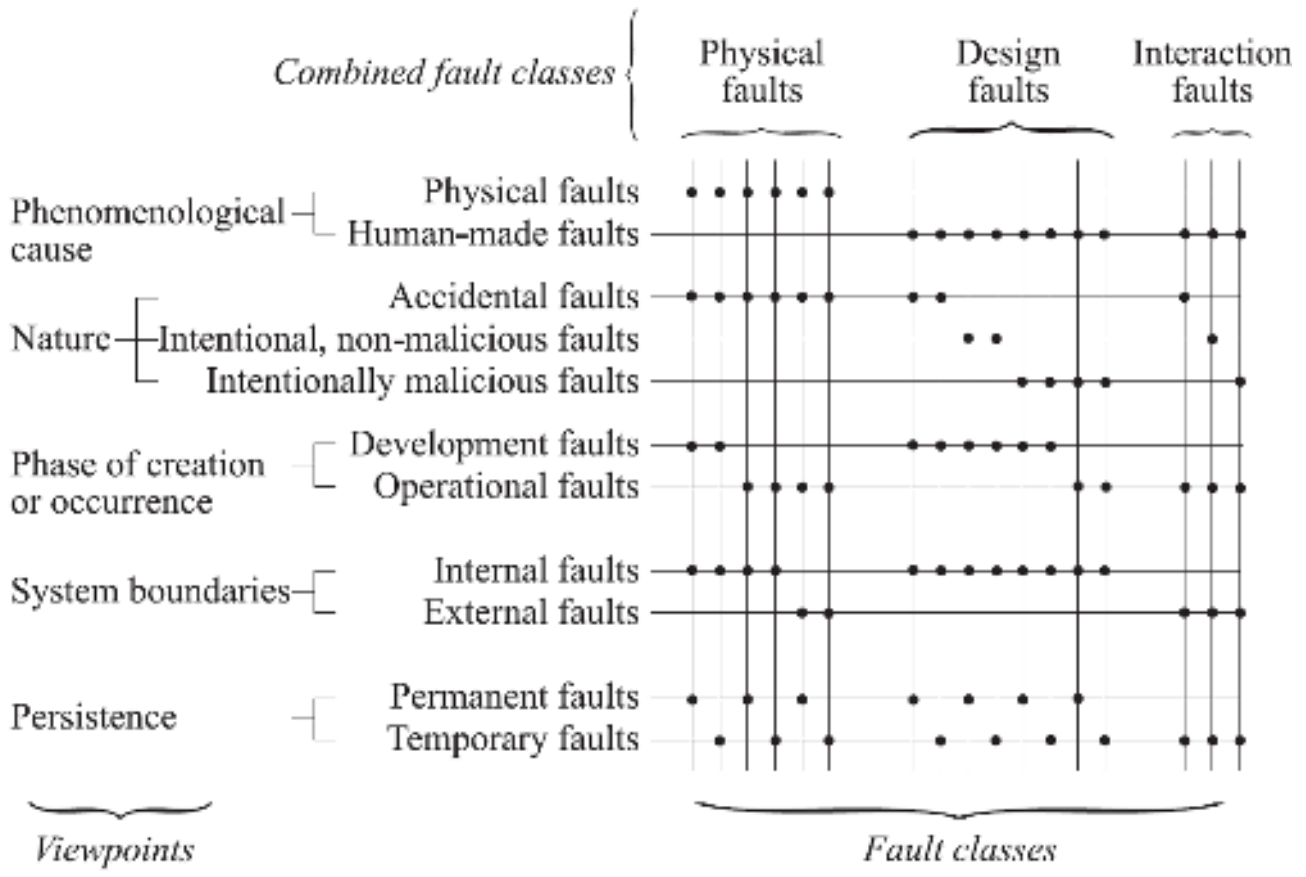

Figure 3: Laprie's fault classification scheme.

- Temporary, external, physical faults (also known as *transient faults*) (Bondavalli et al., 1997). These are faults induced by environmental phenomena, e.g., EMI.

**Design faults:**

- Intentional, though not malicious, permanent / temporary design faults. These are basically trade-offs introduced at design time. A typical example is insufficient dimensioning (underestimations of the size of a given field in a communication protocol[1], and so forth).
- Accidental, permanent, design faults (also called systematic faults, or Bohrbugs): flawed algorithms that systematically turn into the same errors in the presence of the same input conditions and initial states—for instance, an unchecked divisor that can result in a division-by-zero error.
- Accidental, temporary design faults (known as Heisenbugs, for "bugs of Heisenberg", after their elusive character): while systematic faults have an evident, deterministic behavior, these bugs depend on subtle combinations of the system state and environment.

**Interaction faults:**

- Temporary, external, operational, human-made, accidental faults. These include operator faults, in which an operator does not correctly perform his or her role in system operation.
- Temporary, external, operational, human-made, non-malicious faults: "neglect, interaction, or incorrect use problems" (Sibley,

1998). Examples include poorly chosen passwords and bad system parameter setting.

- Temporary, external, operational, human-made, malicious faults. This class includes the so-called malicious replication faults, i.e., faults that occur when replicated information in a system becomes inconsistent, either because replicates that are supposed to provide identical results no longer do so, or because the aggregate of the data from the various replicates is no longer consistent with system specifications.

## 2.3 Means for Dependability

Developing a dependable service, i.e., a service on which reliance can be placed justifiably, calls for the combined utilisation of a set of methods and techniques globally referred to as the "means for dependability" (Laprie, 1998):

**fault prevention** aims at preventing the occurrence or introduction of faults. Techniques in this category include, e.g., quality assurance and design methodologies;

**fault-tolerance** groups methods and techniques to set up services capable of fulfilling their function in spite of faults;

**fault removal** aims at reducing the number, incidence, and consequences of faults. Fault removal is composed of three steps: verification, diagnosis and correction. Verification checks whether the system adheres to certain properties—the verification conditions—during the design, development, production or operation phase; if it does not, the fault(s) preventing these conditions to be fulfilled must be diagnosed, and the necessary corrections (corrective maintenance) must be made;

**fault forecasting** investigates how to estimate the present number, the future incidence and the consequences of faults. Fault forecasting is conducted by evaluating the system behavior with respect to fault occurrence or activation. Qualitatively, it aims at identifying, classifying and ordering failure modes or at identifying event combinations leading to undesired effects. Quantitatively, it aims at evaluating (in terms of probabilities) some of the attributes of dependability.

Of the above mentioned methods, fault-tolerance represents the core tool for the techniques and tools presented in this book. Because of this, it is discussed in more detail in Sect. 2.3.1.

### 2.3.1 Fault-Tolerance

Fault-tolerance methods come into play the moment a fault enters the system boundaries. Its core objective is "preserving the delivery of expected services despite the presence of fault-caused errors within the system itself" (Avižienis, 1985). Fault-tolerance has its roots in hardware systems, where the assumption of *random* component failures is substantiated by the physical characteristics of the adopted devices (Rushby, 1994).
According to (Anderson & Lee, 1981), fault-tolerance can be decomposed into two sub-techniques—error processing and fault treatment.

**Error processing** aims at removing errors from the computational state (if possible, before failure occurrence). It can be based on the following primitives (Laprie, 1995):

- **Error detection** , which focuses on detecting the presence in the system of latent errors before they are activated. This can be done, e.g., by means of built-in self-tests or by comparison with redundant computations (Rushby, 1994).
- **Error diagnosis** i.e., assessing the damages caused by the detected errors or by errors propagated before detection.
- **Error recovery** consists of methods to replace an erroneous state with an error-free state. This replacement takes one of the following forms:
    1. Compensation, which means reverting the erroneous state into an error-free one exploiting information redundancy available in the erroneous state, predisposed, e.g., through the adoption of error correcting codes (Johnson, 1989).
    2. Forward recovery, which finds a new state from which the system can operate (frequently in degraded mode). This method only allows recovering from errors of which the damage can be *anticipated*[2]—therefore, this method is system dependent (P. Lee & Anderson, 1990). The main tool for forward error recovery, according to (Cristian, 1995), is exception handling.
    3. Backward recovery, which substitutes an erroneous state by an error-free state prior to the error occurrence. As a consequence, the method requires that, at different points in time (known as *recovery points*), the current state of the system be saved in some stable storage means. If a system state saved in a recovery point is error-free, it can be used to restore the system to that state, thus wiping out the effects of transient faults. For the same reason, this technique allows also to recover from errors of which the damage cannot or has not been anticipated. The need for backward error recovery tools and techniques stems from their ability to prevent the occurrence of failures originated by transient faults, which are many times more frequent than permanent faults (Rushby, 1994). The main tools for backward error recovery are based on checkpoint-and-rollback (Deconinck, 1996) and recovery blocks (Randell, 1975) (see Chapter 3).

According to (Rushby, 1994), an alternative technique with respect to error recovery is **fault masking**, classically achieved by modular redundancy (Johnson, 1989): a redundant set of components perform independently on the same input value. Results are voted upon. The basic assumption of this method is again that of random component failures—in other words, to be effective, modular redundancy requires statistical independence, because correlated failures translate in contemporary exhaustion of the available redundancy. Unfortunately a number of experiments (Eckhardt et al., 1991) and theoretical studies (Eckhardt & Lee, 1985) have shown that often this assumption is incorrect, to the point that even independent faults are able to produce correlated failures. In this context, the concept of design diversity (or $N$-version programming) came up (Avižienis, 1985). It is discussed in Chapter 3.

**Fault treatment** aims at preventing faults from being re-activated. It can be based on the following primitives (Laprie, 1995):

**Fault diagnosis** i.e., identifying the cause(s) of the error(s), in location and nature, i.e. determining the fault classes to which the faults belong. This is different from error diagnosis; besides, different faults can lead to the same error.

**Fault passivation** i.e., preventing the re-activation of the fault. This step is not necessary if the error recovery step removes the fault, or if the likelihood of re-activation of the fault is low enough.

**Reconfiguration** updates the structure of the system so that non-failed components fulfill the system function, possibly at a degraded level, even though some other components have failed.

# 3 FAULT-TOLERANCE, REDUNDANCY, AND COMPLEXITY

A well-known result by Shannon (Shannon, Winer, & Sloane, 1993) tells us that, from any unreliable channel, it is possible to set up a more reliable channel by increasing the degree of information redundancy. This means that *it is possible to trade off reliability and redundancy* of a channel. The author of this book observes that the same can be said for a fault-tolerant system, because fault-tolerance is in general the result of some strategy effectively exploiting some form of redundancy—time, information, and/or hardware redundancy (Johnson, 1989). This redundancy has a cost penalty attached, though. Addressing a weak failure semantics, able to span many failure behaviors, effectively translates in higher reliability—nevertheless,

1. it **requires** large amounts of extra resources, and therefore implies a high cost penalty, and

2. it **consumes** large amounts of extra resources, which translates into the rapid exhaustion of the extra resources.

For instance, a well-known result by Lamport *et al.* (Lamport et al., 1982) sets the minimum level of redundancy required for tolerating Byzantine failures to a value that is greater than the one required for tolerating, e.g., value failures. Using the simplest of the algorithms described in the cited paper, a 4-modular-redundant (4-MR) system can only withstand any *single Byzantine failure*, while the same system may exploit its redundancy to withstand up to *three crash faults—though no other kind of fault* (Powell, 1997). In other words:

> After the occurrence of a crash fault, a 4-MR system with strict Byzantine failure semantics has exhausted its redundancy and is no more dependable than a non-redundant system supplying the same service, while the crash failure semantics system is able to survive to the occurrence of that and two other crash faults. On the other hand, the latter system, subject to just one Byzantine fault, would fail regardless its redundancy.

Therefore, for any given level of redundancy, *trading complexity of failure mode against number and type of faults tolerated* may be considered as an important capability for an effective fault-tolerant structure. Dynamic adaptability to different environmental conditions[3] may provide a satisfactory answer to this need, especially when the additional complexity does not burden (and jeopardize) the application. Ideally, such complexity should be part of a custom architecture and not of the application. On the contrary, the embedding in the application of complex failure semantics, covering many failure modes, implicitly promotes complexity, as it may require the implementation of many recovery mechanisms. This complexity is detrimental to the dependability of the system, as it is in itself a significant source of design faults. Furthermore, the isolation of that complexity outside the user application may allow cost-effective verification, validation and testing. These processes may be unfeasible at application level.
The author of this book conjectures that a satisfactory solution to the design problem of the management of the fault-tolerance code (presented in Chapter 2) may translate in an optimal management of the failure semantics (with respect to the involved penalties). The fault-tolerance linguistic structure proposed in Chapter 6 allows solving the above problems by means of its *adaptability*.

# 4 CONCLUSION

This chapter has introduced the reader to Laprie's model of dependability describing its attributes, impairments, and means. The central topic of this book, fault-tolerance, has also been briefly discussed. The complex relation between the management of fault-tolerance, of redundancy, and of complexity, has been pointed out. In particular, a link has been conjectured between attribute adaptability and the dynamic ability to trade off the complexity of failure mode against number and type of faults being tolerated.

# Notes

[1]A noteworthy example is given by the bad dimensioning of IP addresses. Currently, an IP address consists of four sections separated by periods. Each section contains an 8-bit value, for a total of 32 bits per address. Normally this would allow for more than 4 billion possible IP addresses—a rather acceptable value. Unfortunately, due to a lavish method for assigning IP address space, IP addresses are rapidly running out. A new protocol, IPv6 (Hinden & Deering, 1995), is going to fix this problem through larger data fields (128-bit addresses) and a more flexible allocation algorithm.

[2]In general, program specifications are not *complete*: there exist input states for which the behavior of the corresponding program $P$ has been left unspecified. No forward recovery technique can be applied to deal with errors resulting from executing $P$ on these input states. On the contrary, if a given specification is complete, that is, if each input state is covered in the set $G$ of all the standard and exceptional specifications for $P$, and if $P$ is *totally correct*, i.e. fully consistent with what prescribed in $G$, then $P$ is said to be *robust* (Cristian, 1995). In this case forward recovery can be used as an effective tool for error recovery.

[3]The following quote by J. Horning (Horning, 1998) captures very well how relevant may be the role of the environment with respect to achieving the required quality of service: "What is the most often overlooked risk in software engineering? That the environment will do something the designer never anticipated".

page

# FAULT-TOLERANT SOFTWARE: BASIC CONCEPTS AND TERMINOLOGY

## Contents

# 1 INTRODUCTION AND OBJECTIVES

After having described the main characteristics of dependability and fault-tolerance, it is analyzed here in more detail what does it mean that a program is fault-tolerant and what are the properties expected from a fault-tolerant program. The main objective of this chapter is introducing two sets of design assumptions that shape the way our fault-tolerant software is structured—the system and the fault models. Often misunderstood or underestimated, those models describe

- what is expected from the execution environment in order to let our software system function correctly,
- and what are the faults that our system is going to consider. Note that a fault-tolerant program shall (try to) tolerate only those faults stated in the fault model, and will be as defenseless against all other faults as any non fault-tolerant program.

Together with the system specification, the fault and system models represent the foundation on top of which our computer services are built. It is not surprising that weak foundations often result in falling constructions. What is really surprising is that in so many cases little or no attention had been given to those important factors in fault-tolerant software engineering. To give an idea of this, three well-known accidents are described—the Ariane 5 flight 501 and Mariner-1 disasters and the Therac-25 accidents. In each case it is stressed what went wrong, what were the biggest mistakes, and how a careful understanding of fault models and system models would have helped highlighting the path to avoid catastrophic failures that cost considerable amounts of money and even the lives of innocent people.
The other important objective of this chapter is introducing the core subject of this book: Software fault-tolerance situated at the level of the application layer. First of all, it is explained why targeting (also) the application layer is not an open option but a mandatory design choice for effective fault-tolerant software engineering. Secondly, given the peculiarities of the application layer, three properties to measure the quality of the methods to achieve fault-tolerant application software are introduced:

1. Separation of design concerns, that is, how good the method is in keeping the functional aspects and the fault-tolerance aspects separated from each other.
2. Syntactical adequacy, namely how versatile the employed method is in including the wider spectrum of fault-tolerance strategies.
3. Adaptability: How good the employed fault-tolerance method is in dealing with the inevitable changes characterizing the system and its run-time environment, including the dynamics of faults that manifest themselves at service time.

Finally, this chapter also defines a few fundamental fault-tolerance services, namely watchdog timers, exception handling, transactions, and checkpointing-and-rollback.

# 2 WHAT IS A FAULT-TOLERANT PROGRAM?

So what makes a program fault-tolerant? In order to answer this key question, let us further detail what a service is: In the following a *service* is considered as a set of manifestations of external events that, if compliant to what agreed upon in a formal specification, can be considered by a watcher as being "correct". This said, a *program* can be defined as a physical entity, stored for instance as voltage values in a set of memory cells, which is supposed to drive the production of a service. One of the main goals of software engineering is being able to set up of a robust link (in mathematical terms, a emphhomomorphism/) between a service's high-level specification and a low-level computer design (the program).

> More formally, for some functions $f$ and $g$ it is true that
>
> $$\text{Service} = f(\text{program}), \text{program} = g(\text{specification}).$$
>
> A first (obvious) conclusion is the hard link between the service and its specification:
>
> $$\text{Service} = g \cdot f(\text{specification}).$$
>
> Building robust versions of $f$ and $g$ is well known as being a difficult, non trivial job.
>
> Now let us concentrate on the range of $g$ (the software set).

For any two systems $a$ and $b$, if $a$ relies on $b$ to provide its service, then the expression "$a$ depends on $b$" will be used. We shall represent this through the following notation:

$$a \Rightarrow b.$$

This relation is called the "dependence" among the two systems. Clearly it is true that, for instance, Service $\Rightarrow$ program, program $\Rightarrow$ CPU, and CPU $\Rightarrow$ memory. Trying to develop an exhaustive list of dependent systems may be a long-lasting exercise, and most likely it would end up with an incomplete categorization. Figure 1 provides an arbitrary incomplete expansion of the dependence relation.

As evident from that picture, dependences call for *Dependability*, i.e., the fundamental property to achieve *Dependable services* which has been characterized in Chapter 1. A dependable service is then one that persists even when, for instance, its corresponding program experiences faults—to some agreed upon extent.

When designing a fault-tolerant program, two important steps are:

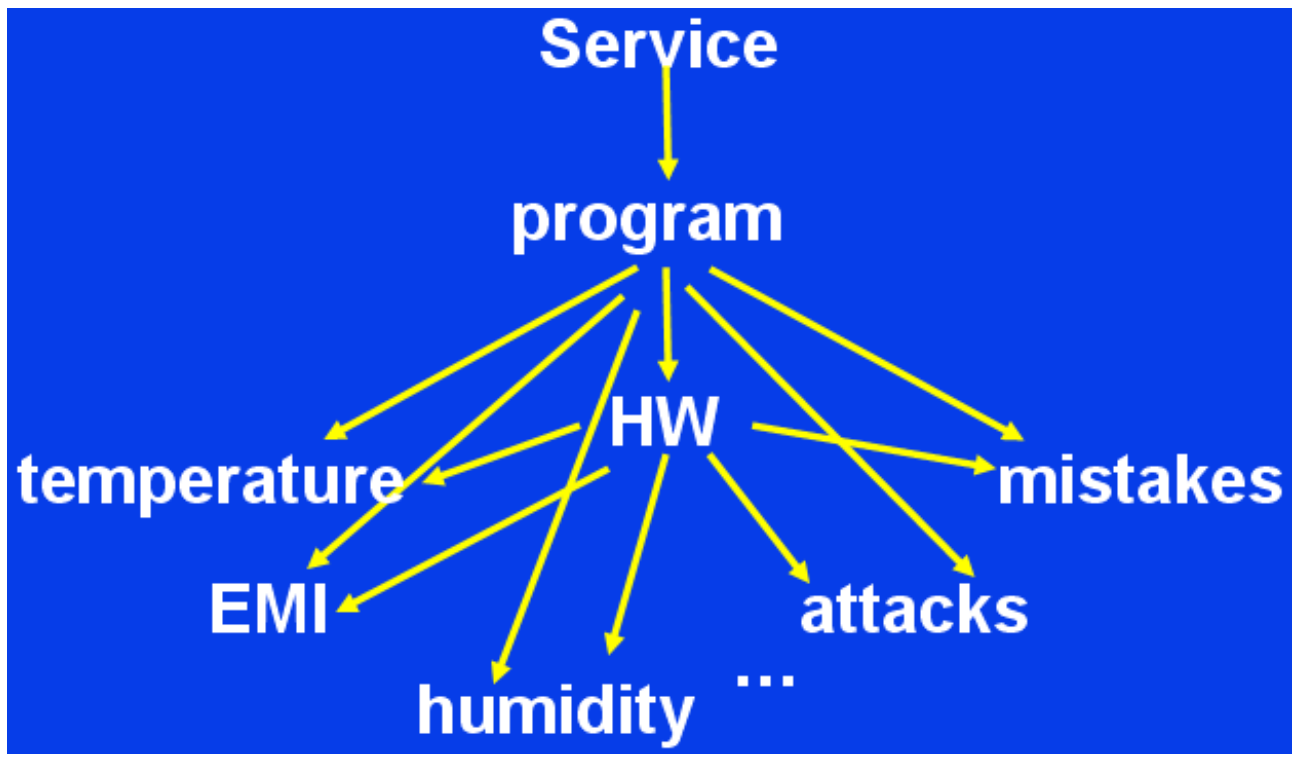

Figure 1: An expansion of the dependence relation.

1. Defining the **System model**, which declares the characteristics we expect from the run-time environment—the main system features our program will depend upon at run-time.

2. Defining the **Fault model**, which enumerates the erroneous cases that one considers and aims to tolerate.

Summarizing, the main three variables that the designer of a fault-tolerant service needs to take into account in order to preserve its functions are: The specification, the system model, and the fault model:

$$\text{Fault-tolerant Service} \Rightarrow (\text{specification}, \text{system model}, \text{fault model}).$$

In the following two sections the system and fault models are characterized.

## 2.1 Dependable Services: The System Model

System Model The system model characterizes the properties of the system components our program depends upon. It could be represented as a tree like the one in Fig. 1 whose leaves are computation, communication and clock sub-systems. These leaves are annotated with statements representing some expected properties of the corresponding sub-systems.

### 2.1.1 Synchronous System Model.

A well-known system model is the synchronous model: In such a system the service depends on "perfect" (ideal) computation, communication and clock sub-systems. In particular, that model dictates that it is possible to know precisely how long will it take for any task to be fully executed by the available CPUs and for any message to be sent and eventually received through the available communication networks. Moreover, the hardware clocks available on different nodes are perfect—no drift is possible.

The main benefit of such a model is that it facilitates considerably the task of the designer and the developer: The system is assumed to be perfectly stable, which means that no disruptions are deemed as likely to occur. This paves the way to the adoption of simple software structures, such as connection-oriented communication: Any two tasks willing to communicate with each other first establish a connection and then synchronously exchange messages through it. This structure is very simple and much more effective than, e.g., datagram-based communication—where messages are sent asynchronously, and each of them must be routed separately to destination.
Clearly opting for the synchronous system model is an optimistic approach, though not always a very realistic one. Writing a program with these assumptions means basically shifting problems to deployment time. This is because whatever violation of the system assumptions *becomes a fault*. Possible events, such as a late message arrival or a missed deadline, violate the model assumptions and can lead to a failure. Even momentary disruptions e.g. a node becoming unavailable for a small fraction of a second and then back on-line are not compatible with the synchronous system assumption—for instance, they break all the connections between the tasks on that node and those residing elsewhere in the system. This single event becomes a fault that triggers potentially many errors. Tolerating that single fault requires a non trivial error treatment, e.g. re-establishing all the broken connections through some distributed protocol. Of course in some cases it can be possible to build up a system that strictly obeys the synchronous system model. But such a system would require custom, non-standard hardware/software components: For instance, synchronous Ethernet could be used for communication instead of the inherently non-deterministic CSMA/CD Ethernet. These choices clearly have the consequence to strengthen the dependence between service and target platform. Embedded systems are exactly this—a combination of custom hardware and software assembled so as to produce a well defined, special purpose service. In some other cases—for instance, hard real-time systems—the synchronous system model is the only option, as the service specification dictates strict deterministic execution for all processing and communication tasks.

### 2.1.2 Asynchronous System Model.

At the other extreme in the spectrum of possible system models is the asynchronous system model. Its main assumptions are:

- No bounds on the relative speed of process execution.
- No bounds on message transmission delays.
- No hardware clocks are available or otherwise there are no bounds to clock drift.

As can be clearly understood this model is quite simple, does not impose special constraint on the hardware platform and (in a sense) is more close to reality: It recognizes that non-determinism and asynchrony are common and does not try to deny or fight this matter of fact. This matches many common life execution environments such as the Internet. Unfortunately as Einstein would say this system model is *too simple*: It was proven that given these assumptions one cannot come up with effective solutions to services such as time-based coordination and failure detection (Fischer, Lynch, & Paterson, 1985).

### 2.1.3 Partially Synchronous System Model.

Given the disadvantages of these two main system models, designers have been trying to devise new models combining the best of both aspects. Partial synchrony models belong to this category. Such models consider that for some systems and some environments there are long period of times where the system is obeying the synchronous hypotheses and physical time bounds are respected. Such periods are followed by brief periods where delays are experienced on processing and communication tasks. One such model is the so-called *timed asynchronous system model* (Cristian & Fetzer, 1999), which is characterized by the following assumptions: Timed asynchronous system model

- All tasks communicate through the network via a datagram service with omission/performance failure semantics (see Chapter 1).
- All services are timed: specifications prescribe not only the outputs and state transitions that should occur in response to inputs, but also the time intervals within which a client task can expect these outputs and transitions to occur.
- All tasks (including those related to the OS and the network) have crash/performance failure semantics (again, see Chapter 1).
- All tasks have access to a local hardware clock. If more than one node is present, clocks on different nodes have a bounded drift rate.
- A "time-out" service is available at application-level: Tasks can schedule the execution of events so that they occur at a given future point in time, as measured by their local clock[1].

In particular, this model allows a straightforward modeling of system partitioning: As a result of sufficiently many omission or performance communication failures, correct nodes may be temporarily disconnected from the rest of the system during so-called periods of instability (Cristian & Fetzer, 1999). Moreover it is assumed that, at reset, tasks or nodes restart from a well-defined, initial state—partial-amnesia crashes (defined in Chapter 1) are not considered.

As clearly explained in the cited paper, the above hypotheses match well current distributed systems based on networked workstations—as such, they represent an effective model on which to build our fault-tolerant services.

## 2.2 Dependable Services: The Fault Model

Another important step when designing a fault-tolerant system is the choice of which erroneous conditions one wants to tackle so as to prevent them to lead to system failures. This set of conditions that our fault-tolerant system is to tolerate is the fault model, $F$.

What is $F$ exactly? It is a set of events that

- may hinder the service distribution, and that
- are considered as likely to occur, and that
- one aims to tolerate (that is, prevent them from turning into failures).

Clearly $F$ is a very important property for any fault-tolerant program, because even the most sophisticated fault-tolerant program $p$ will be defenseless when any other condition than the ones in its fault model takes place. To highlight this fact, programs shall be referred to as functions of $F$, e.g. one shall write $p(F)$. A special case is given by non fault-tolerant programs, that is, programs with an empty fault model. In this case one shall write $p(\emptyset)$. The same applies for the service produced by program $p(F)$. In what follows such a service will be referred to as an $F$-dependable service. In other words an $F$-dependable service is one that persists despite the occurrence of faults as described in its fault model $F$.

An important property of $F$ is that, in turn, it is a function of an environment $E$ where the service (or better, its corresponding program) is operating. Clearly an $F$-dependable service may tolerate faults in $E'$ and may not do so for those in $E''$: An airborne service may well experience different events than, e.g., one meant in an electrical energy primary substation[2] (Unipede, 1995). Obviously the choice of $F$ is an important aspect towards a successful development of a dependable service. Imagine for instance what may happen if our fault model $F$ matches the wrong environment, or if the target environment changes its characteristics (e.g. a rising of temperature due to a firing). One may argue that all the above cases are exceptional, and that most of the time they do not take place. This was maybe the case in the past, when services were *stable*. Our services now run in a very fluid environment, where the occurrence of changes is the rule, not the exception. As a consequence, software engineering for fault-tolerant systems should allow to consider the nature of faults as a dynamic system, i.e., a system evolving in time, and by modeling faults as a function $F(t)$. The author is convinced that any current fault-tolerance provision should adopt such structure for its fault model. Failing to do so leaves the designer with two choices:

1. Overshooting, i.e., over-dimensioning the fault-tolerance provisions with respect to the actual threats being experienced, or
2. undershooting, namely underestimating the threat in view of an economy of resources.

Note how those two risks turn into a crucial dilemma to the designer: Wrong choices here can lead to either unpractical, too costly designs, or to cheap but vulnerable provisions: Fault-tolerant codes that are not dependable enough to face successfully the threats actually experienced.
In Chapter 4 it is introduced and discussed an example of fault-tolerant software whose fault model dynamically changes tracking the environment.
Next section focuses on a few cases where static fault models and wrong system models led to catastrophic consequences.

# 3 (IN)FAMOUS ACCIDENTS

## 3.1 Faulty Fault Models: The Ariane 5 Flight 501

On June 4, 1996, the maiden flight of the unmanned Ariane 5 rocket ended in a failure just forty seconds after its lift-off from Kourou, French Guiana. At an altitude of about 3700 meters, the launcher veered off its flight path, broke up and exploded. The rocket was on its first voyage, and it took the European Space Agency (ESA) more than a decade of intense development costing $7 billion.
Designed as a successor to the successful Ariane 4 series, the Ariane 5 was designed to be capable of hurling a cargo of several tons—four identical scientific satellites that were designed to establish precisely how the Earth's magnetic field interacts with solar winds—into orbit each launch, and was intended to give Europe a leading edge in the commercial space business.
After the failure, the ESA set up an independent Inquiry Board to identify the causes of the failure. It was their task to determine the causes of the launch failure, investigate whether the qualification tests and acceptance tests were appropriate in relation to the problem encountered and recommend corrective actions. The recommendations of the Board concern mainly around software engineering practices like testing, reviewing and the construction of specifications and requirements. The case of the Ariane 5 is particularly meaningful to what discussed so far, because it provides us with an example of a fault-tolerant design that did not consider the right fault model. This was the ultimate cause of its failure. In the following we discuss what happened and which have been the main mistakes with respect to the discussion so far.
The Flight Control System of the Ariane 5 was of a standard design. The attitude of the launcher and its movements in space were measured by an Inertial Reference System (SRI). The SRI had its own internal computer, in which angles and velocities were calculated on the basis of information from a strap-down inertial platform, with laser gyros and accelerometers. The data from the SRI were transmitted through the data-bus to an On-Board Computer (OBC), which executed the flight program and controlled the nozzles of the solid boosters and the so-called Vulcain cryogenic engine, via servovalves and hydraulic actuators. As already mentioned, this system was fault-tolerant: In order to improve its reliability two SRI's were operating in parallel, with identical hardware and software. For the time being the consequences of this particular design choice will not be considered—Chapter 3 will go back to this issue.

One SRI was active and one was in hot stand-by—as soon as the OBC detected that the active SRI had failed it immediately switched to the other one, provided that this unit was functioning properly. Likewise the system was equipped with two OBC's, and a number of other units in the Flight Control System were also duplicated.

The software used in the Ariane 5 SRI was mostly reused from that of the Ariane 4 SRI. The launcher started to disintegrate about 39 seconds after take-off because of high aerodynamic loads due to an angle of attack of more than 20 degrees that led to separation of the boosters from the main stage, in turn triggering the self-destruct system of the launcher. This angle of attack was caused by full nozzle deflections of the solid boosters and the so-called Vulcain main engine. These nozzle deflections were commanded by the OBC software on the basis of data transmitted by the active Inertial Reference System (SRI 2). Part of these data at that time did not contain proper flight data, but showed a diagnostic bit pattern of the computer of the SRI 2, which was interpreted as flight data. The reason why the active SRI 2 did not send correct attitude data was that the unit had declared a failure due to a software exception. The OBC could not switch to the backup SRI 1 because that unit had already ceased to function during the previous data cycle (72 millisecond period) for the same reason as SRI 2. The internal SRI software exception was caused during execution of a data conversion from 64-bit floating point to 16-bit signed integer value. The floating point number which was converted had a value greater than what could be represented by a 16-bit signed integer. This resulted in an Operand Error. The data conversion instructions (in Ada code) were not protected from causing an Operand Error, although other conversions of comparable variables in the same place in the code were protected.

No reference to justification of this decision was found directly in the source code. Given the large amount of documentation associated with any industrial application, the assumption, although agreed, was essentially obscured, though not deliberately, from any external review. The reason for the three remaining variables, including BH, the one denoting horizontal bias, being unprotected was that further reasoning indicated that they were either physically limited or that there was a large margin of safety, a reasoning which in the case of the variable BH turned out to be faulty.

The main reason behind the failure was a software reuse error in the Inertial Reference System (ISR). Specifically, the conversion from horizontal velocity of the rocket (represented as a 64-bit floating-point number) with respect to the platform to a 16-bit signed integer resulted in an overflow, as the number was larger than the largest integer storable in a 16-bit unsigned integer, resulting in an overflow exception being thrown. This failure caused complete loss of guidance and attitude information approximately 37 seconds after the start of the main engine ignition sequence. Ariane 5 had been deprived of its basic faculties: Its perception of where it was in the sky and which direction it had to proceed. This loss of information was due to specification and design errors in the ISR, upon which depends the Flight Control Computer (FCC). This software was originally developed and successfully used in the Ariane 4 but was not altered to support the new flight trajectory and increase in horizontal acceleration resulting from the new Vulcain engines. Because of this, the ISR memory banks were quickly overloaded with information that could not be processed and fed to the onboard computers fast enough. The FCC could thus no longer ensure the correct guidance and control and from that instant the launcher was lost.
Several are the reasons behind the Ariane 5 failure—in what follows the focus shall go on the one more pertaining to this chapter: Several faults resulting in Operand Errors were included in the Ariane 4 fault model, $F$. Treating these faults introduced some run-time overhead. To minimize this overhead, some of these faults were not included in the fault model, reduced to a smaller $F'$. One of the faults in $F$ but not in $F'$ triggered the chain of events that ultimately led to the Ariane failure.

## 3.2 Faulty Specifications: The Mariner 1.

The Mariner Program, a series of ten spacecrafts, was started by NASA on July 22, 1962 with the launch of Mariner 1 and ended on November 3, 1973, with the launch of Mariner 10. Other spacecraft, based on these, were continued with different names, like Voyager and Viking.
Mariner 1, a 202.8 kg unmanned spacecraft, was sent to Venus for a flyby with several scientific instruments on board, such as a microwave radiometer, an infrared radiometer and a cosmic dust detector. These should investigate Venus and its orbits.
The Mariner was made by the Jet Propulsion Laboratory (JPL) and to be used by NASA. The total costs for this spacecraft were close to $ 14 million. For getting into space, the Mariner 1 was attached to an Atlas-Agena rocket. Such type of rocket had been already used for launching missiles. It used different antennas to be controllable by a ground control unit, but it had its own on-board control system in case of a failing communication.
The launch was rescheduled to July 20, 1962. That day at Cape Canaveral launch platform the first countdown started, after which a fey delays occurred because of problems in the range safety command system. The countdown was stopped and restarted once before being canceled because of a blown fuse in the range safety circuits. At 23:08 local time on July 21, the countdown began again. Another three holds gave the technicians time to fix minor issues such as power fluctuations in the radio guidance system. At 09:21:23 UTC, the countdown ended and the spaceship started its launch. Let us call this time as time X. A few minutes later, the range safety officer noticed that the spacecraft was going out of course, and at X + 4 minutes, it was clear that manual correction was necessary. However, the spacecraft did not react as hoped and went more and more off course. A strict deadline at this stage was time X + 300, corresponding to the separation of the Atlas/Agena rocket. After this time destroying the Mariner1 would not be possible anymore. To prevent serious damage, the range safety officer decided to enforce destroyment at X + 293 seconds. The Mariner 1 radio transponder kept sending signals until X + 357 seconds.
The investigation on what went wrong includes many factors. A brief overview is given in what follows.
It is sometimes stated that a misspelling in a Fortran program was responsible for the crash of the Mariner 1. However, this is not true—such bug existed in another system in the Mercury project, which was fixed before being able to do any harm. In fact, the faulty software was used in several missions, but corrected before the resulting inaccurate data resulted in a flight failure. The bug was caused by two main factors: Fortran, which ignores spaces, and a design fault—a small typo, a period instead of a comma, resulting in a line like "DO 5 K = 1, 3" (an iterative loop) being interpreted as "DO5K = 1.3" (an assignment).
The actual cause of the crash started with a hardware malfunction of the Atlas antennas. The beacon for measuring the rate of the spacecraft failed to pass signals for four periods ranging from 1.5 to 61 seconds. During the absence of correct data, a smoothed function should guide the spacecraft into the right direction. However, *the smooth function had not been implemented*, resulting in fast changes in the course direction. To counteract these drastic changes, the course was changed over and over again, with the vehicle going more and more off its intended course.

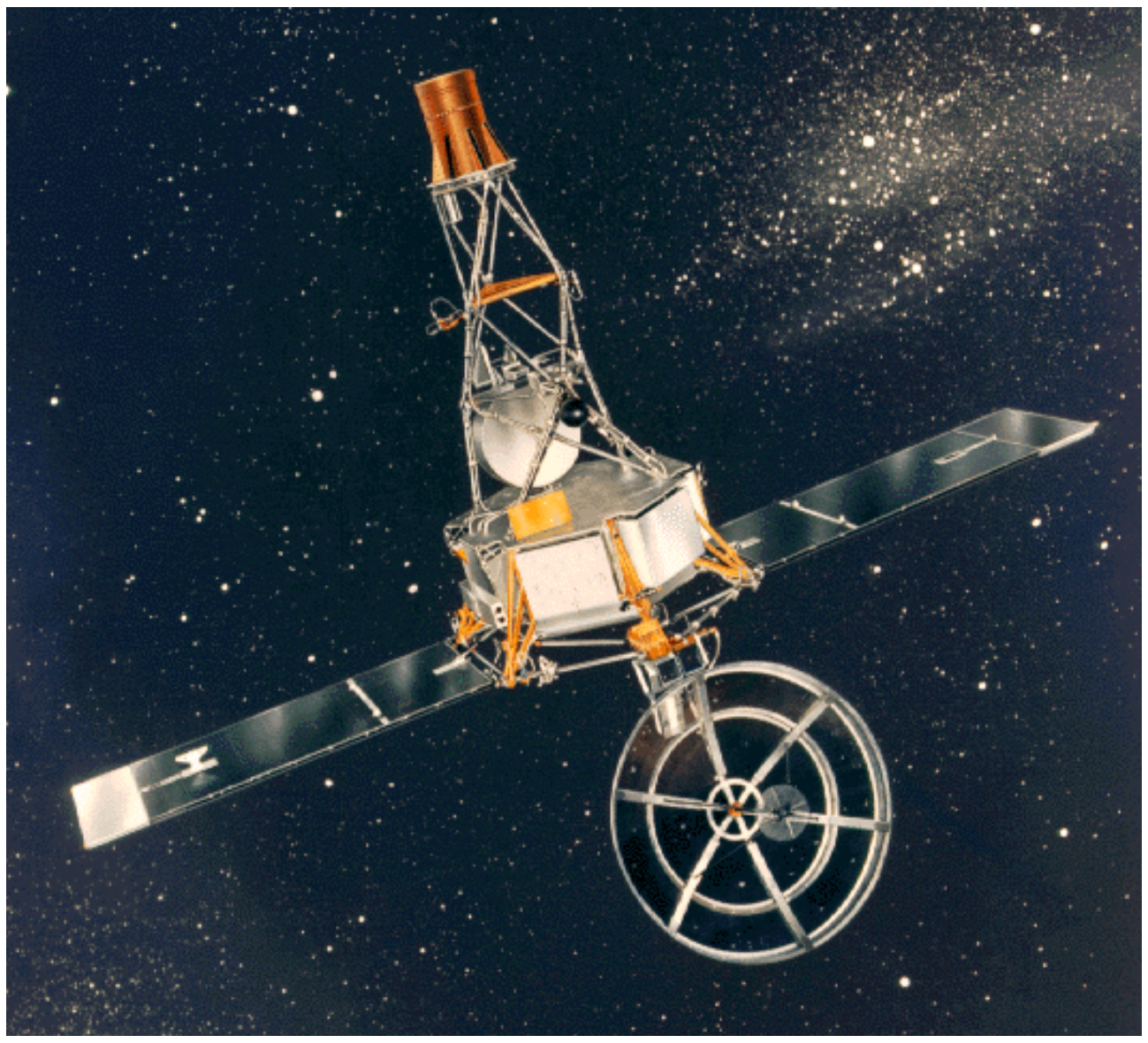

Figure 2: The Mariner 1

For each flight, a Range Safety Officer made sure that, should the spacecraft go out of a safety zone, it would be destructed before being able to do any harm to people or the environment. After the Range Safety Officer saw that the flight was uncontrollable before being out of reach he ordered to let the Mariner 1 explode, to prevent further damage. This happened only 7 seconds before the separation of the Mariner 1 and the Atlas-Agena rocket, which held the explosives.
Why the smooth function had not been implemented? The error had occurred when an equation was being transcribed by hand in the specification for the guidance program. The writer missed the superscript bar in $\overline{\dot{r}_n}$ (the $n$-th smoothed value of the time derivative of a radius). Without the smoothing function indicated by the bar, the program treated normal minor variations of velocity as if they were serious, causing spurious corrections that sent the rocket off course. Because of that the Range Safety Officer had to shut it down. As the method would be used only in case of communication failure, and such failure had not been injected during testing experiments, the simulation did not verify the consequences of the hardware failure and did neither notice the slight but catastrophic difference between the expected and the real function values.
It is possible to conclude that the Mariner 1 is a classic example of the consequences of a faulty or misinterpreted specification: As mentioned before,

$$\text{Service} = f(\text{program}), \text{program} = g(\text{specification}),$$

and a flawed specification fatally translates in a failed service.

## 3.3 Faulty Models: The Therac-25 Accidents

The Therac-25 accidents have been recognized as “the most serious computer-related accidents to date” (Leveson, 1995). Herein they are briefly discussed to give an idea of the consequences of faulty system and fault models. The Therac-25 was a *linac*, that is, a medical linear accelerator that uses accelerated electrons to create high-energy beams able to destroy tumors with minimal impact on the surrounding healthy tissue. It was the latest member of a successful family of linacs, including e.g. the Therac-6 and the Therac-20, built by Atomic Energy Commission Limited (AECL), a Canadian company. Compared to its ancestors, the Therac-25 had a revolutionary design: it was smaller, cheaper and more powerful.
In the past AECL had built several successful medical linear accelerators, including the Therac-6 and the Therac-20. Compared to its ancestors, the Therac-25 had three advantages: It was more compact, cheaper and had more features. The compactness was due to the so-called “double-pass” concept used for the Therac-25. This double-pass design of the accelerator meant that the accelerator itself was much more compact, rendering the total size of the machine much smaller. This was achieved by folding the mechanism that accelerates the electrons (a little like it is for the French horn among wind instruments).

The cheaper cost of the Therac-25 came from several factors—it was a dual-mode linacs, that is, it was able to produce both electron and photon beams, which required normally two machines; also, the Therac-6 and the Therac-20 both had hardware interlocks to ensure safe operation. With the development of the Therac-25 however, AECL decided that such interlocks were an unnecessary burden for the customer, raising the costs without bringing extra benefits. Most of the extra complexity of the Therac-25, including safety issues, was managed in software. This is a key difference between the new model and its ancestors—in the latter, software played a very limited role and "merely added convenience to the existing hardware, which was capable of standing alone" (Leveson, 1995). Such software was custom built but reused routines of the original Therac-6 and Therac-20 code. The Therac-25 software was developed over a period of several years by a single programmer using the PDP 11 assembly language. Even the system software was not standard but custom built. One could argue that when used to compose a life-critical service such as this, software should come with guarantees about its quality and its fault-tolerant features; unfortunately this was not the case at the time[3]. Not only no fault model or system model document had been produced—the safety analysis carried out by AECL was a fault tree where only hardware faults had been considered! AECL apparently considered their software as error-free. It is interesting to note the assumptions AECL drew on software, as they could be considered as the three main mistakes in fault-tolerant software development:

1. Programming errors have been significantly reduced by extensive testing on a simulator. *Any residual software error is not included in the analysis.*

2. Program software does not wear out or degrade.

3. Possible faults belong to the following two classes: Hardware physical faults and transient physical faults induced by alpha particles and electromagnetic noise.

The Therac-25 software was very complex. It consisted of four major components: Stored data, a scheduler, a set of critical and non-critical tasks, and interrupt services. It used the interrupt-driven software model, and inter-process communication among concurrent tasks was managed through shared memory access. Analysis revealed that no proper synchronization was put into place to secure accesses to shared memory. This introduced race conditions that would cause some of the later accidents.
One of the tasks of the software was to monitor the machine status. In particular the treatment unit had an interlock system designed to remove power in case of a hardware malfunction. The software monitored this interlock system and, when faults got detected, either prevented a treatment from being started or, if the treatment was in progress, it suspended or put in hold the treatment. The software had been developed relying on the availability of said interlock system—in other words, it had to be part of the system model document. Changing the system and reusing the software led to disaster. Indeed hardware interlocks had been the only reason that prevented deadly overdoses to be delivered while using the old members of the Therac family of devices. A proof of this was found later with the Therac-20. At the University of Chicago, students could exercise radiation therapy with the Therac-20. In the beginning of each academic year, there were a lot of defected machines. Most of the time, the main problem was blown fuses. After about three weeks, these failures would typically go away. After carefully studying this behavior, it was concluded that due to the random faulty configurations entered by students who did not know the machine, overdose charges took place. Fortunately fuses were in place to prevent any overdose damage. Would these fuses also have been in place in the Therac-25, many of the accidents could have been avoided[4].

# 4 SOFTWARE FAULT-TOLERANCE

Research in fault-tolerance concentrated for many years on *hardware* fault-tolerance, i.e., on devising a number of effective and ingenious hardware structures to cope with faults (Johnson, 1989). For some time this approach was considered as the only one needed in order to reach the requirements of availability and data integrity demanded by nowadays complex computer services. Probably the first researcher who realized that this was far from being true was B. Randell who in 1975 (Randell, 1975) questioned hardware fault-tolerance as the only approach to employ—in the cited paper he states:

> "Hardware component failures are only *one* source of unreliability in computing systems, decreasing in significance as component reliability improves, while software faults have become increasingly prevalent with the steadily increasing size and complexity of software systems."

Indeed most of the complexity supplied by modern computing services lies in their software rather than in the hardware layer (Lyu, 1998a, 1998b; Huang & Kintala, 1995; Wiener, 1993; Randell, 1975). This state of things could only be reached by exploiting a powerful conceptual tool for managing complexity in a flexible and effective way, i.e., devising hierarchies of sophisticated abstract machines (Tanenbaum, 1990). This translates in implementing software with high-level computer languages lying on top of other software strata—middleware, the device drivers layers, the basic services kernel, the operating system, the run-time support of the involved programming languages, and so forth.

Partitioning the complexity into stacks of software layers allowed the implementor to focus exclusively on the high-level aspects of their problems, and hence it allowed managing a larger and larger degree of complexity. But *though made transparent, still this complexity is part of the overall system* being developed. A number of complex algorithms are concurrently executed by the hardware, resulting in the simultaneous progress of many system states—under the hypothesis that no involved abstract machine, nor the actual hardware, be affected by faults. Unfortunately, as in real life faults do occur, the corresponding deviations are likely to jeopardize the system's function, also propagating from one layer to the other, unless appropriate means are taken to avoid in the first place, or to remove, or to tolerate these faults.
In particular, faults may also occur in the **application layer**, that is, in the abstract machine on top of the software hierarchy[5]. These faults, possibly having their origin at design time, or during operation, or while interacting with the environment, *are not different in the extent of their consequences from those faults originating, e.g., in the hardware or the operating system.*
An efficacious argument to bring evidence to the above statement is the case of the so-called "millennium bug", i.e., the most popular class of design faults that ever emerged in the history of computing technologies, also known as "the year 2000 problem", or as "Y2K". The source of this problem is simple: Most of the software still in use today was developed using a standard where dates are coded in a 6-digit format. According to this standard, two digits were considered as enough to represent the year. Unfortunately this translates into the impossibility to distinguish, e.g., year 2000 from year 1900, which by the en of last century was recognized as the possible cause of an unpredictably large number of failures when calculating time elapsed between two calendar dates, as for instance year 1900 was not a leap year while year 2000 is. Choosing the above mentioned standard to represent dates resulted in a hidden, almost forgotten design fault, never considered nor tested by application programmers. As society got closer and closer to the year 2000, the possible presence of this design fault in our software became a nightmare that seemed to jeopardize all those crucial functions of our society today appointed to programs manipulating calendar dates, such us utilities, transportation, health care, communication, public administration, and so forth. Luckily the expected many and possibly crucial system failures due to this one application-level fault (Hansen, LaSala, Keene, & Coppola, 1999) were not so many and not that crucial, though probably for the first time the whole society became aware of the extent of the relevance of dependability in software.
These facts and the above reasoning suggest that, the higher the level of abstraction, the higher the complexity of the algorithms into play and the consequent error proneness of the involved (real or abstract) machines.
As a conclusion, full tolerance of faults and the complete fulfillment of the dependability design goals of a complex software application *call for the adoption of protocols to avoid, remove, or tolerate faults working at all levels, including the application layer.*

# 5 SOFTWARE FAULT-TOLERANCE IN THE APPLICATION LAYER

The need of software fault-tolerance provisions located in the application layer is supported by studies that showed that the majority of failures experienced by nowadays computer systems are due to *software faults*, including those located in the application layer (Lyu, 1998a, 1998b; Laprie, 1998); for instance, NRC reported that 81% of the total number of outages of US switching systems in 1992 were due to software faults (NRC, 1993). Moreover, nowadays application software systems are increasingly networked and distributed. Such systems, e.g., client-server applications, are often characterized by a loosely coupled architecture whose global structure is in general more prone to failures[6]. Due to the complex and temporal nature of interleaving of messages and computations in distributed software systems, no amount of verification, validation and testing can eliminate all faults in an application and give complete confidence in the availability and data consistency of applications of this kind (Huang & Kintala, 1995). Under these assumptions, *the only alternative (and effective) means for increasing software reliability is that of incorporating in the application software provisions of software fault-tolerance* (Randell, 1975).

Another argument that justifies the addition of software fault-tolerance means in the application layer is given by the widespread adoption of object orientation, components and service orientation. Structuring one's software into a web of objects, components and services is a wonderful conceptual tool which allows to quickly compose a service out of *reusable components.* This has wonderful relapses on many aspects including development and maintenance times and costs, but has also a little drawback: it promotes the composition of software systems from third-party objects the sources of which are unknown to the application developers.

In other words, the object, component and service abstractions fostered the capability to deal with higher and higher levels of complexity in software and at the same time eased and therefore encouraged software reuse. As just mentioned, this has very positive impacts though it translates the application in a sort of collection of reused, pre-existing components or objects made by third parties. The reliability of these software entities and hence their impact on the overall reliability of the user application is often unknown, to the point that Grey refers as an "art" to the ability to create reliable applications using off-the-shelf software components (Green, 1997). The case of the Ariane 501 flight and that of the Therac-25 linear accelerator (see Chapter 2) are well-known examples that show how improper reuse of software may produce severe consequences (Inquiry, 1996).

But probably the most convincing reasoning for not excluding the application layer from a fault-tolerance strategy is the so-called "end-to-end argument"—a system design principle introduced in (Saltzer, Reed, & Clark, 1984). Such principle states that, rather often, functions such as reliable file transfer, can be *completely* and *correctly* implemented only with the knowledge and help of the application standing at the endpoints of the underlying system (for instance, the communication network).

This does not mean that everything should be done at the application level—fault-tolerance strategies in the underlying hardware and operating system can have a strong impact on the system's performance. However, an extraordinarily reliable communication system that guarantees that no packet is mistreated (lost, duplicated, or corrupted, or delivered to the wrong addressee) does not reduce the burden of the application programmer to ensure reliability: For instance, for reliable file transfer, the application programs that perform the transfer must still supply a file-transfer-specific, end-to-end reliability guarantee.

The main message of this chapter can be summarized as follows:

> Pure hardware-based or operating system-based solutions to fault-tolerance, though often characterized by a higher degree of transparency, are not *fully* capable of providing complete end-to-end tolerance to faults in the user application. Relying solely on the hardware, the middleware, or the operating system is a mistake:
>
> - It develops only partially satisfying solutions.
> - It requires a large amount of extra resources and costs.
> - And often it is characterized by poor service portability (Saltzer et al., 1984; Siewiorek & Swarz, 1992).

# 6 STRATEGIES, PROBLEMS, AND KEY PROPERTIES

The above conclusions justify the strong need for application-level fault-tolerance. As a consequence of this need, several approaches to application-level fault-tolerance have been devised during the last three decades (see chapters 5–8 for an extensive survey). Such a long research period hints at the complexity of the design problems underlying Application-level fault-tolerance engineering, which include:

1. How to incorporate fault-tolerance in the application layer of a computer program.
2. Which fault-tolerance provisions to support.
3. How to manage the fault-tolerance code.

Problem 1 is also known as the problem of the **system structure to software fault-tolerance**, which was first proposed by B. Randell in 1975 (Randell, 1975). It states the need of appropriate structuring techniques such that the incorporation of a set of fault-tolerance provisions in the application software might be performed in a simple, coherent, and well structured way. Indeed, poor solutions to this problem result in a huge degree of **code intrusion**: in such cases, the application code that addresses the functional requirements and the application code that addresses the fault-tolerance requirements are mixed up into one large and complex application software.

- This greatly complicates the task of the developer and demands expertise in both the application and the fault-tolerance domains. Negative repercussions on the development times and costs are to be expected.
- The maintenance of the resulting code, *both for the functional part and for the fault-tolerance provisions*, is more complex, costly, and error prone.
- Furthermore, the overall complexity of the software product is increased—which, as mentioned in Chapter 1, is in itself a source of faults.

One can conclude that, with respect to the first problem, an ideal system structure should guarantee an adequate **Separation between the functional and the fault-tolerance Concerns** (in what follows this property will be referred to as "SC").

Moreover, the design choice of *which fault-tolerance provisions to support* can be conditioned by the adequacy of the syntactical structure at "hosting" the various provisions. The well-known quotation by B. L. Whorf efficaciously captures this concept:

> "Language shapes the way we think, and determines what we can think about":

Indeed, as explained in Chapter 1, a non-optimal answer to Problem 2 may

- require a high degree of redundancy, and
- rapidly consume large amounts of the available redundancy,

which at the same time would increase the costs and reduce the reliability. One can conclude that, devising a syntactical structure offering *straightforward support* to *a large set of fault-tolerance provisions*, can be an important aspect of an ideal system structure for application-level fault-tolerance. In the following this property will be called **Syntactical Adequacy** (or more briefly "SA").

Finally, one can observe that another important aspect of any application-level fault-tolerance architecture is *the way the fault-tolerance code is managed*, at compile time as well as at execution time. If one wants to realize $F$-dependable systems where the fault model $F$ can change over time, as mentioned in Chapter 2, then our architecture must allow the fault-tolerance code to be changed as well so as to track the changing fault model.
A possible way to do so is for instance to have an architectural component to monitor the observed faults and check whether the current fault model is still valid or not. When this is not the case, the component should extend the fault model and change the fault-tolerance code accordingly, either loading some pre-existing code or synthesizing a new one matching the current threat. In both cases, the architecture must allow disabling the old code and enabling the new one.
**Adaptability** (or A for brevity) is defined herein as the ability of an application-level fault-tolerant architecture such that it allows on-line (dynamic) or at least off-line management of the fault-tolerance provisions and of their parameters. This would allow letting the fault-tolerance code *adapt* itself to the current environment or at least allow service portability. Clearly an important requirement for any such solution is that it does not overly increase the complexity of the resulting application—which would be detrimental to dependability.
The three properties SC, SA and A will be referred to in what follows as the *structural attributes* of application-level fault-tolerance.

# 7 SOME WIDELY USED SOFTWARE FAULT-TOLERANCE PROVISIONS

In this section the key ideas behind some widely used software fault-tolerance building blocks will be introduced: the watchdog timer, exception handling, transactions, and checkpointing and rollback. Such building blocks will be studied in depth in the rest of the book.

## 7.1 Watchdog Timers

Clov: Wait! Yes Yes! I have it! I set the alarm.
Hamm: This is perhaps not one of my bright days, but frankly
Clov: You whistle me. I don't come. The alarm rings. I'm gone. It doesn't ring. I'm dead. [*Pause.*]
Hamm: Is it working? [*Pause. Impatiently.*] The alarm, is it working?
Clov: Why would'nt it be working?
Hamm: Because it's worked too much.
Clov: But it's hardly worked at all.
Hamm: [*Angrily.*] Then because it's worked too little!
(Samuel Beckett, *Endgame.*)

Watchdogs are versatile and effective tools in detecting processing errors. The idea is very simple: Let us suppose there is a process $p$ that needs to perform cyclically a critical operation and then releases the locks that keep in a waiting state several concurrent processes. Clearly the pending processes are dependant on $p$: A single fault affecting $p$ and preventing it to continue would result in blocking indefinitely all the pending processes. Obviously it is very important to make sure that a fault stopping $p$ be timely detected. (Such first service would then lead to proper error recovery steps, e.g. releasing the locks). A watchdog timer is an additional process $w$ that monitors the execution of $p$ by requiring the latter to send $w$ periodically a "sign of life"—clearing a shared memory flag or sending a heartbeat message to $w$. By checking whether the flag has been cleared or the heartbeat has arrived, process $w$ can assess that $p$, at least in the last period, had been at least able to timely send the sign of life.

> In more formal terms, using the vocabulary of Chapter 2, one could say that watchdog timers protect $p$ against crash/performance failures.

If $w$ does not receive the expected sign of life, then it is said to "fire." Despite its simplicity, the watchdog calls for important choices at design and configuration time. In particular,

- How often should $p$ send the sign of life implies a trade off of performance and failure detection latency. Moreover, quite often the dependency chain between $p$ and $w$ is not simple: For instance, $p$ may rely on a communication system $C$ to deliver a heartbeat message, which complicates the matter at hand considerably ($p \Rightarrow C$, so is it $C$ or $p$ that failed when $w$ fired?) Of course also dependency chains such as these are an arbitrary simplification, and $C$ could be more precisely identified as a long cascade of dependent sub-services, whose emergent behavior is that of a communication system, each component of which could be the actual source of a problem ultimately resulting in the firing of $w$.

- How often should $w$ check for the arrival of a sign of life from $p$ implies the adherence to a system model, explicitly defined or otherwise. A synchronous system model corresponds to a hard real-time system assumptions, which would allow for a very "tight" watchdog cycle. The farther one goes from that assumption (the more asynchronous the system model, so to say), the larger the chance to introduce unexpected latencies in the execution of both $p$ and $w$. A consequence of this is that, if one wants to reduce the probability that $w$ erroneously declares $p$ as faulty, then he shall need to compensate for late processing and late messages by widening the watchdog cycle. Of course this implies a larger failure detection latency. Bringing this to the limit, in a fully asynchronous system, the compensation time grows without bound, at the cost of not being able to accomplish any sensible task anymore! This result was proven in the famous article (Fischer et al., 1985), which I usually refer to as "the FLoP paper" (a little kidding with the names of its authors). Scientists have found a way to deal with this ostensible conundrum, and the idea is based on using a web of "extended watchdogs"—so called failure detectors (see Chapter 8 .)

- What to do if the watchdog timer fails. Everything has a coverage, and this includes watchdogs, so it is unwise assuming that a watchdog cannot fail. Furthermore, the failure can be as simple to deal with as a crash, or as tricky as a Byzantine (arbitrary) failure. Failures could be for instance the result of either
  - a design fault in the algorithm of the watchdog, or
  - a physical fault, e.g. causing the processing node of the watchdog to get disconnected from the network, or
  - an attack, e.g. a man-in-the-middle attack or an identity spoofing attack (Harshini, Sridhar, & Sridhar, 2004).

Watchdog timers are often used in software fault-tolerance designs. The problem this book focuses on, as mentioned already, is how to *express* watchdog timers and their configuration. And again, an important factor in measuring the effectiveness of the available solutions is for us how such solutions perform with respect to the three structural properties of application-level software fault-tolerance, and to SC in particular. The less code intrusion an approach requires, the higher our assessment. This book presents two examples: The EFTOS watchdog timer (a library and run-time executive requiring full code intrusion, see Chapter 3) and the Ariel watchdog timer (an evolution of the EFTOS watchdog timer that makes use of the Ariel language to enucleate the configuration statements from the source code—thus reducing code intrusion; see Chapter 6 and Appendix "The Ariel Internals.")

## 7.2 Exceptions and Exception Handlers

An exception is an event that is triggered at run time due to the interaction with the environment and results in a (temporary or permanent) suspension of the current application so to manage the event. Let us consider the following C fragment:

```
void main(void) {
    int a, b, c;

    a = function1();
    b = function2();

    c = a / b; /* danger here */
}
```

Clearly instruction `a / b` is unprotected against a division-by-zero exception: When `b` is zero the division is undefined and—unless the division instruction is faulty—the CPU does not know how to deal with it.
Other examples of exceptions are:

- An overflow or underflow condition.

- A not-a-number (NaN) floating point constant.
- Misalignments.
- A breakpoint is encountered.
- Access to protected or non existing memory areas.
- Power failures.
- A sub-system failure, e.g. a disk crash while accessing a file.

As the CPU has no clue about how to recover from the condition, either the program stops or it tries to deal with it with some code supplied by the programmer precisely for that, that is, to catch the exception and deal with it. The following version of the above code fragment does prevent the exception to occur:

```
void main(void) {
    int a, b, c;

    a = function1();
    b = function2();

    if (b != 0)
        c = a / b; /* no more danger here */
    else {
        fprintf(stderr, "An exception has been avoided: division-by-zero\n");
    }
}
```

As can be seen from the above mentioned examples, not all the exceptions can be avoided. Hence it is very important that a programming language hosts some mechanisms that allow the user to catch exceptions and properly deal with them. One such language is Java. Java is particularly interesting for not only it allows, but *mandates* (with some exception, if you excuse me for the pun) that the programmer supplies proper code to deal with all the exceptions that can be raised by the sub-services the application depends upon (Pelliccione, Guelfi, & Muccini, 2007). For example, if one tries to compile an instruction such as this:

```
ImageFile input = new OpenImageFile("edges.png");
```

whose purpose is to open an image file and associate its descriptor with a local variable, the Java compiler would report an error complaining the lack of proper instructions to deal with the case that the `OpenImageFile` method fails due to a `java.io.FileNotFoundException`. In more detail, the Java compiler would emit a message like "unreported exception *i*; must be caught or declared to be thrown *s*", where *i* is the exception and *s* is the lacking statement, and report an unrecoverable error. The only way to compile successfully the above Java fragment is through the following syntax:

```
try {
   ImageFile input = new OpenImageFile("edges.png");
}
catch (FileNotFoundException exception) {
   System.out.println("Exception: Couldn't open file edges.png");
   exception.printStackTrace();
}
```

whose semantics is: First try to execute the statements in the `try` block; if everything goes well, skip the catch statement and go on; otherwise if the `catch` block refers the raised exception, execute it. In this case the handling of the exception is a simple printed message and a dump of the program execution stack though method `printStackTrace`, which reports on where in the control flow graph the execution took place and how it propagated through the system and application modules. Note how a Try-Catch block is a nice syntactical construct to build mechanisms such as Recovery Blocks (discussed in Chapter 3)—that is, the Syntactical Adequacy (SA) of Java to host mechanisms such as Recovery Blocks is very high.
The general syntax for exception handling in Java is

```
try {
   ...Instructions possibly raising exceptions...
}
catch (ExceptionType1 exception1) {
   ...Instructions to deal with exception Exception1...
}
catch (ExceptionType2 exception2) {
   ...Instructions to deal with exception Exception2...
}
...
finally {
   ...Instructions to be executed in any case at the end of the try block...
}
```

An example follows:

```
    try {
      x = new BufferedReader(new
          FileReader(argv[0])); // this instruction
                                // throws FileNotFoundException

      String s = x.readLine();  // this one throws IOException
      while(s != null) {
        System.out.println(s);
```

```
      s = x.readLine();       // this one throws IOException
    }
  }
  catch(FileNotFoundException e1) {
    System.out.println("I can't open a file.");
    e1.printStackTrace();
  }
  catch(IOException e2) {
    System.out.println("I can't read from a file");
    ioe.printStackTrace();
  }
  finally {
    x.close();                // this one throws IOException
                              // and NullPointerException
  }
```

Java defines a large number of exceptions, divided into two classes: Checked and unchecked exceptions. Checked exceptions are basically recoverable exceptions, which include e.g. those due to input/output failures or network failures. Checked exceptions mandatorily call for a corresponding `try`...`catch` block. Unchecked exceptions are unrecoverable conditions corresponding to the exhaustion of the system assets (e.g. an out of memory error or a segment violation).
Java also offers a mechanism to propagate an exception from an invoked module to the invoking one—this is known as "throwing" an exception.
Java and other systems offer so-called Automated Exception Handling or Error Interception tools, which continuously monitor the execution of programs recording debugging information about exceptions and other conditions. Such tools allow tracking the cause of exceptions taking place in Java programs that run in production, testing or development environments. An example of an exception handling mechanism is given in Chapter 3.

## 7.3 Checkpointing and Rollback

Checkpointing and Rollback (CR) is a widely used fault-tolerance mechanism. The idea is simple: Someone (the user, or the system, or the programmer) takes a periodic snapshot of the system state and, if the system fails afterwards, the snapshot is reloaded so as to restore the system to a working and (hopefully) correct situation. The fault model of most of the available CR tools is transient (design and physical) faults, i.e., faults that might not show up again when the system re-executes. Checkpointing is also a basic building blocks for more complex fault-tolerance mechanisms, such as Recovery Blocks (described in Chapter 3), where after rollback a new software version is tried out, or task migration (supported by language Ariel, see Chapter 6), where the snapshot is loaded on a different processing node of the system. Clearly in the latter case the fault model may be somewhat extended so as to consider permanent faults having their origin in the originating machine (for instance in its local run-time executives, or compilers, or shared libraries, and so forth). CR is also a key requirement to achieve atomic actions (see Sect. 7.4).
CR packages can be divided into three categories:

- application-level libraries, such as psncLibCkpt (Meyer, 2003),
- user commands, e.g. Dynamite (Iskra et al., 2000) or ckpt (Zandy, n.d.),
- operating system mechanisms and patches, e.g. psncC/R (Meyer, 2003).

Another classification is given by the logics for initiating checkpointing, which can be:

- Time-based ("every $t$ time units do checkpointing"). This is supported e.g. by ckpt and libckpt (Plank, Beck, Kingsley, & Li, 1995). The latter in particular supports *incremental* checkpointing (only the data that changed from last checkpointing needs to be stored.)
- Event based (e.g., when the user generates a signal, e.g. with UNIX command "kill"). An example of this is psncLibCkpt. A special case is (Shankar, 2005), where the signal can actually terminate the process and create a dump file that can be "revived" afterwards).
- Algorithmic (that is, when the algorithm enters a given phase, e.g., the top of a loop; obviously application-level libraries allow this).

Also in the case of checkpointing there are several important design and configuration issues: In particular,

- How often should the checkpointing occur? Suppose one has executed a series of checkpointings, $c_1, c_2, \dots c_n$, and after the last one and before the next one the system experiences a failure. The normal practice in CR is to reload $c_n$ and retry. Are we sure that the corresponding fault occurred between $c_{n-1}$ and $c_n$? In other words, are we sure that the period of checkpointing is large enough to compensate for fault latency and error latency (see Chapter 1 for their definitions)?
- Are we sure that the checkpointed state includes the whole of the system state? The state of the system may include e.g. descriptors of open TCP connections, the state of low-level system variables, the contents of files distributed throughout the network, and so forth. Failing to restore the whole system state may well result in a system failure.
- Are we sure the the checkpointed state resides in a safe part of the system? Are we sure that we will be able to access it, unmodified, when rollback is needed? In other words, are we making use of a reliable *stable storage* for checkpointed states? Recall that everything has a coverage, and this includes stable storage; so how stable is our stable storage? See further on for a section on stable storage.

CR has been specialized in several different contexts, such as distributed systems, parallel computers, clusters and grid systems (Schneider, Kohmann, & Bugge, n.d.)
Our focus is on how to *express* checkpointing and rollback, so mainly in CR libraries and their configuration. As usual the less code intrusion an approach requires, the better its SC.
Chapter 3 briefly discusses two CR libraries, PsncLibCkpt and Libckpt.

## 7.4 Transactions

An important building block to fault-tolerance is transactions. A transaction bundles an arbitrary number of instructions of a common programming language together and makes them "atomic", that is indivisible: It is not possible to execute one such bundle partially, it either executes completely or not at all. More formally a transaction must obey the so-called ACID properties:

Atomicity: In a transaction involving two or more blocks of instructions, either all of the blocks are committed or none are.

Consistency: A transaction either brings the system into a new valid processing state, or, if any failure occurs, returns it to exact state the system was before the transaction was started.

Isolation: Running, not yet completed transactions must remain isolated from any other transaction.

Durability: Data produced by completed transactions is saved in a stable storage that can survive a system failure or a system restart.

A common protocol to guarantee the ACID properties is so-called two-phase commit, described e.g. in (Moss, 1985). Two important services required by transactions are stable storage and checkpointing and rollback.
As mentioned in (Kienzle & Guerraou, 2002), transactions act like a sort of firewall for failures and may be considered as effective building blocks for the design of dependable distributed services. Another important feature of transactions is that they mask concurrency, which makes transaction-based systems eligible for being executed on a parallel machine.
As it is always the case in fault-tolerant computing, the hypotheses behind transaction processing are characterized by their coverage, that is, a probability of being effectively achieved. A so-called transaction monitor is a sort of Watchdog controlling and checking the execution of transactions.
Transactions are common in database management systems, where operations such as database updating must be either fully completed or not at all in order to avoid inconsistencies possibly leading to financial disasters. This explains why transactions are supported by SQL, the Structured Query Language that is the standard database user and programming interface.

Transactions require considerable run-time support. One system supporting transactions is OPTIMA (Kienzle & Guerraou, 2002), a highly configurable, object-oriented framework that offers support for open multithreaded transactions and guarantees the ACID properties for transactional objects. Written in Java, it provides its users with a procedural interface that allows an application programmer to start, join, commit, and abort transactions.
Argus and Arjuna (discussed in Chapter 5) are examples of transactional languages. The C programming language does not provide any support for transactions in its standard library; for this reason, a custom tool was developed for that within the EFTOS project. Such tool is described in Chapter 3.

# 8 CONCLUSION

Together with system specifications, two important ingredients to craft correct fault-tolerant systems are the system model and the fault model. After describing those models, it has been shown how relevant their choice can be on the dependability of important services. Configurable communication protocols and services are collections of modules that can be combined into different configurations. This allows designing system that can be customized with respect to the requirements of the system and fault models. This allows to put those models in the foreground and to fine-tune the system towards the application requirements (Hiltunen, Taïani, & Schlichting, 2006). As a side effect of this, one would obtain a system characterized by less overhead and higher performance.
This chapter also reviewed a few famous accidents. What is surprising is that, quite often, the reports summarizing the “things that went wrong” all lead to the same conclusions, which have been nicely summarized in (Torres-Pomales, 2000):

> ”In a system with relaxed control over allowable capabilities, a damaged capability can result in the execution of undesirable actions and unexpected interference between components.”

The various approaches to application-level fault-tolerance surveyed in this book provide different system structures to solve the above mentioned problems. Three “structural attributes” are used in the next chapters in order to provide a qualitative assessment of those approaches with respect to various application requirements. The structural attributes constitute, in a sense, a *base* with which to perform this assessment. One of the outcomes of this assessment is that regrettably *none* of the approaches surveyed in this book is capable to provide the best combination of values of the three structural attributes in *every* application domain. For specific domains, such as object-oriented distributed applications, satisfactory solutions have been devised at least for SC and SA, while only partial solutions exist, for instance, when dealing with the class of distributed or parallel applications *not based on the object model*.
The above matter of facts has been efficaciously captured by Lyu, who calls this situation "*the software bottleneck*" of system development (Lyu, 1998b): in other words, there is evidence of an urgent need for *systematic approaches to assure software reliability within a system* (Lyu, 1998b) while effectively addressing the above problems. In the cited paper, Lyu remarks how "developing the required techniques for software reliability engineering is a major challenge to computer engineers, software engineers and engineers of related disciplines".
This chapter concludes our preliminary discussion on dependability, fault-tolerance and the general properties of application-level provisions for fault-tolerance. From next chapter onward various families of methods for the inclusion of fault-tolerance in our programs will be discussed.

## Notes

[1]In Chapter 8 we describe in detail a time-out service.

[2]See Chapter 3 for a characterization of the faults typical of a primary substation, as well as for a case study of a fault-tolerant service for primary substations.

[3]Quoting Frank Houston of the US Food and Drug Administration (FDA), "A significant amount of software for life-critical systems comes from small firms, especially in the medical device industry; firms that fit the profile of those resistant to or uninformed of the principles of either system safety or software engineering."

[4]A full report about the Therac-25 accidents is out of the scope of this book; the reader may refer e.g. to (Leveson, 1995) for that.

[5]In what follows, the application layer is to be intended as the programming and execution context in which a complete, self-contained program that performs a specific function directly for the user is expressed or is running.

[6]As Leslie Lamport efficaciously synthesised in his quotation, "a distributed system is one in which I cannot get something done because a machine I've never heard of is down".

page

# FAULT-TOLERANT PROTOCOLS USING SINGLE- AND MULTIPLE-VERSION SOFTWARE FAULT-TOLERANCE

## 1 INTRODUCTION AND OBJECTIVES

This chapter discusses two large classes of fault-tolerance protocols:

- Single-version protocols, that is, methods that use a non-distributed, single task provision, running side by side with the functional software, often available in the form of a library and a run-time executive.
- Multiple-version protocols, which are methods that use actively a form of redundancy, as explained in what follows. In particular recovery blocks and N-version programming will be discussed.

The two families have been grouped together in this chapter because of the several similarities they share.
The chapter also introduces two important structures for software fault-tolerance, namely exception handling and transactions, and proposes several examples of single-version and multiple version tools.

## 2 FAULT-TOLERANT PROTOCOLS USING SINGLE- AND MULTIPLE-VERSION SOFTWARE FAULT-TOLERANCE

A key requirement for the development of fault-tolerant systems is the availability of **replicated resources**, in hardware or software. A fundamental method employed to attain fault-tolerance is **multiple computation**, i.e., $N$-fold ($N > 1$) replications in three domains:

**Time** That is, repetition of computations.

**Space** I.e., the adoption of multiple hardware channels (also called “lanes”).

**Information** That is, the adoption of multiple versions of software.

Following Avižienis (Avižienis, 1985), it is possible to characterize at least some of the approaches towards fault-tolerance by means of a notation resembling the one used to classify queuing systems models (Kleinrock, 1975):

$$n\mathrm{T}/m\mathrm{H}/p\mathrm{S},$$

the meaning of which is "$n$ executions, on $m$ hardware channels, of $p$ programs". The non-fault-tolerant system, or 1T/1H/1S, is called *simplex* in the cited paper.

## 2.1 Single-version Software Fault-Tolerance: Libraries of Tools

Single-version software fault-tolerance (SV) is basically the embedding into the user application of a simplex system of error detection or recovery features, e.g., atomic actions (Jalote & Campbell, 1985), checkpoint-and-rollback (Deconinck, 1996), or exception handling (Cristian, 1995). The adoption of SV in the application layer requires the designer to concentrate in one physical location, namely, the source code of the application, both the specification of what to do in order to carry on some user computation and the strategy such that faults are tolerated when they occur. As a result, the size of the problem addressed is increased. A fortiori, this translates into increasing the size of the user application. This induces loss of transparency, maintainability, and portability while increasing development times and costs.
A partial solution to this loss in portability and these higher costs is given by the development of libraries and frameworks created under strict software engineering processes. In the following, three examples of this approach are presented—the EFTOS library and the SwIFT system. Special emphasis is reserved in particular to the first system, for which the author of this book designed a number of contributions.

### 2.1.1 The EFTOS library.

EFTOS (Deconinck, De Florio, Lauwereins, & Varvarigou, 1997; Deconinck, Varvarigou, et al., 1997) (the acronym stands for "embedded, fault-tolerant supercomputing") is the name of ESPRIT-IV project 21012. The aims of this project were to integrate fault-tolerance into embedded distributed high-performance applications in a flexible and effective way. The EFTOS library has been first implemented on a Parsytec CC system (Parsytec, 1996b), a distributed-memory MIMD supercomputer consisting of processing nodes based on PowerPC 604 microprocessors at 133MHz, dedicated high-speed links, I/O modules, and routers. As part of the project, this library has been then ported to a Microsoft Windows NT / Intel PentiumPro platform and to a TEX / DEC Alpha platform (TXT, 1997; DEC, 1997) in order to fulfill the requirements of the EFTOS application partners. The main characteristics of the CC system are the adoption of the thread processing model and of the message passing communication model: communicating threads exchange messages through a proprietary message passing library called EPX (Parsytec, 1996a). The porting of the EFTOS library was achieved by porting EPX on the various target platforms and developing suitable adaptation layers. Through the adoption of the EFTOS library, the target embedded parallel application is plugged into a hierarchical, layered system whose structure and basic components (depicted in Fig. 1) are:

- At the base level, a distributed net of "servers" whose main task is mimicking possibly missing (with respect to the POSIX standards) operating system functionalities, such as remote thread creation;

- One level upward (detection tool layer), a set of parameterizable functions managing error detection, referred to as "Dtools". These basic components are plugged into the embedded application to make it more dependable. EFTOS supplies a number of these Dtools, including:

  – A watchdog timer thread (see Sect. 4);
  – a trap-handling mechanism (described in Sect. 5);
  – in Sect. 6, a tool to manage transactions.

  and an API for incorporating user-defined EFTOS-compliant tools;

- At the third level (control layer), a distributed application called "DIR net" (its name stands for "detection, isolation, and recovery network") is used to coherently combine the Dtools, to ensure consistent fault-tolerance strategies throughout the system, and to play the role of a backbone handling information to and from the fault-tolerance elements (Deconinck et al., 1999). The DIR net can be regarded as a fault-tolerant network of crash-failure detectors, connected to other peripheral error detectors. Each node of the DIR net is "guarded" by an <I'm Alive> thread that requires the local component to send periodically "heartbeats" (signs of life). For this reason the algorithm of the DIR net shall be described (in Chapter 8, devoted to failure detection protocols.)

  A special component of the DIR net, called RINT, manages error recovery by interpreting a custom language called RL—the latter being a sort of ancestor of the programming language described in this book in Chapter 6;

- At the fourth level (application layer), the Dtools and the components of the DIR net are combined into dependable mechanisms, among which will be described:

– In Sect. 3, a distributed voting mechanism called “voting farm” (De Florio, 1997; De Florio, Deconinck, & Lauwereins, 1998a, 1998c).
– In Sect. 7, a so-called data stabilizing tool.

Other tools not described in what follows include e.g. a virtual Stable Memory (Deconinck, Botti, Cassinari, De Florio, & Lauwereins, 1998).

- The highest level (presentation layer) is given by a hypermedia distributed application based on standard World-Wide Web technology, which monitors the structure and the state of the user application (De Florio, Deconinck, Truyens, Rosseel, & Lauwereins, 1998). This application is based on a special CGI script (E. Kim, 1996), called DIR Daemon, which continuously takes its inputs from the DIR net, translates them into HTML (Berners-Lee & Connolly, 1995), and remotely controls a WWW browser (Zawinski, 1994) so that it renders these HTML data. A description of this application is in Chapter 10.

A system of communication daemons, called Server network in the EFTOS lingo, manages communication among the processing nodes in a way somewhat similar to that used in the Parallel Virtual Machine (see (Geist et al., 1994) for more details on this).
The author of this book contributed to this project designing and developing a number of basic tools, e.g., its distributed voting system (described in detail in Sect. 3), the EFTOS monitoring tool (see Chapter 10), the RL language and its run-time system (that is, the task responsible for the management of error recovery (De Florio, Deconinck, & Lauwereins, 1998b, 1998c), which will evolve into the ARIEL language discussed in Chapter 6). Furthermore, he took part in the design and development of various versions of the DIR net (De Florio, 1998).

### 2.1.2 The SwIFT System.

SwIFT (Huang, Kintala, Bernstein, & Wang, 1996), whose name stands for Software Implemented Fault-Tolerance, is a system including a set of reusable software components (`watchd`, a general-purpose UNIX daemon watchdog timer; `libft`, a library of fault-tolerance methods, including single-version implementation of recovery blocks and $N$-version programming (see Sect. 2.3); `libckp`, i.e., a user-transparent checkpoint-and-rollback library; a file replication mechanism called `REPL`; and `addrejuv`, a special “reactive” feature of `watchd` (Huang, Kintala, Kolettis, & Fulton, 1995) that allows for software rejuvenation[1]. The system derives from the HATS system (Huang & Kintala, 1995) developed at AT&T. Both have been successfully used and proved to be efficient and economical means to increase the level of fault-tolerance in a software system where residual faults are present and their toleration is less costly than their full elimination (Lyu, 1998). A relatively small overhead is introduced in most cases (Huang & Kintala, 1995).

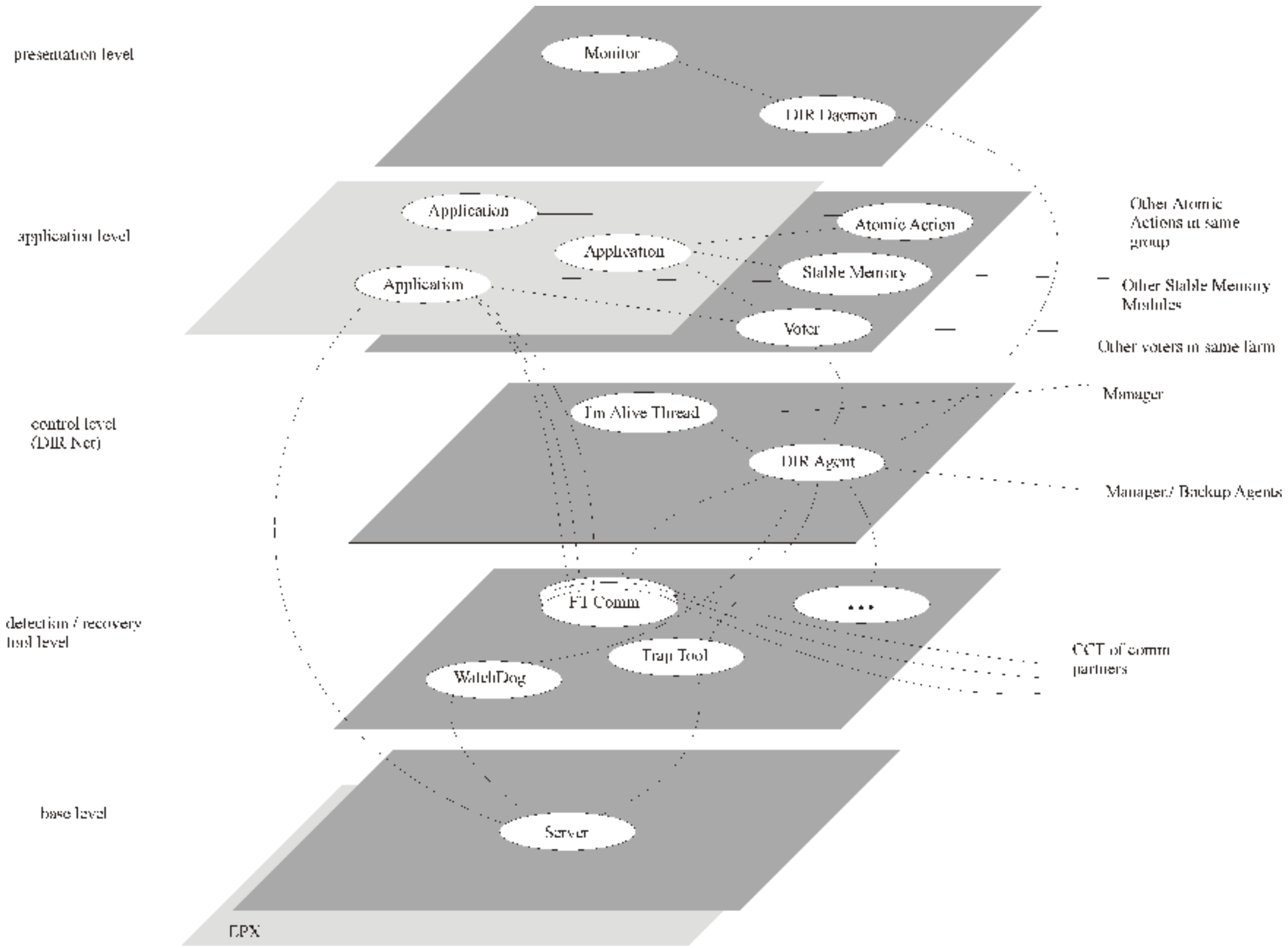

Figure 1: The structure of the EFTOS library. Light gray has been used for the operating system and the user application, while dark gray layers pertain EFTOS.

### 2.1.3 Two libraries for Checkpointing and Rollback

As mentioned in Chapter 2, checkpointing and rollback (CR) is an important mechanism to achieving software fault-tolerance. The focus here goes on two packages working in the application layer.

**Library psncLibCkpt.** PsncLibCkpt (Meyer, 2003) is a library for applications written in C. psncLibCkpt has been designed for simplicity—very few changes in the application software allow to add the CR functionality. Such changes are so simple that could be applied automatically, e.g. through the C preprocessor "#define" statement. In practice, only the main function needs to be renamed as ckpt_target, with no modification on its parameters. Once this is done, the application is ready to catch signals of type "SIGFREEZE" and to save a checkpoint as a response. Restarting the application on the last saved checkpoint is quite easy: calling the program with argument "=recovery" makes psncLibCkpt load the checkpoint. Configuration is also quite simple and can be done through a configuration file or by editing a header file. The latter case requires compiling the application.

**Library libckpt.** Libckpt (Plank, Beck, Kingsley, & Li, 1995) is another CR library for C applications. It performs several optimizations such as "main memory checkpointing" (a 2-stage pipeline overlapping application execution and flushing of the checkpointed state onto disk) and state compression. The main reason for our interest in libckpt is its support for so-called "user-directed checkpointing", which means that libckpt makes intense use of the application layer to optimize processing.
One of these optimizations is user-driven exclusion of memory blocks from the state to be checkpointed. This allows not to include, e.g., clean data (memory yet to be initialized or used). Two function calls are available,

```
exclude_bytes(address, size, usage);
include_bytes(address, size);
```

which allow to adapt the checkpointed state dynamically at run-time.
Another application-level mechanism is so-called "synchronous checkpointing": The user can specify, *in the application program*, points where checkpointing the state would have more sense from an algorithmic point of view. Function `checkpoint_here` does exactly this. There are also parameters allowing the express a minimum and a maximum amount of time between checkpointings. In the cited articles the authors of libckpt show how the adoption of user-directed checkpointing on the average brought to halving the checkpoint size.

**Conclusions.** Two libraries for checkpointing and rollback, both of them targeting the same class of applications,have been discussed. The first case only manages user commands while the second one allows more control in the application layer. Apart from performance issues, one can observe that the second case allows greater control but exhibits lower SC. Such control may be used to achieve adaptive resizing of the checkpointed state, so a slightly better A.

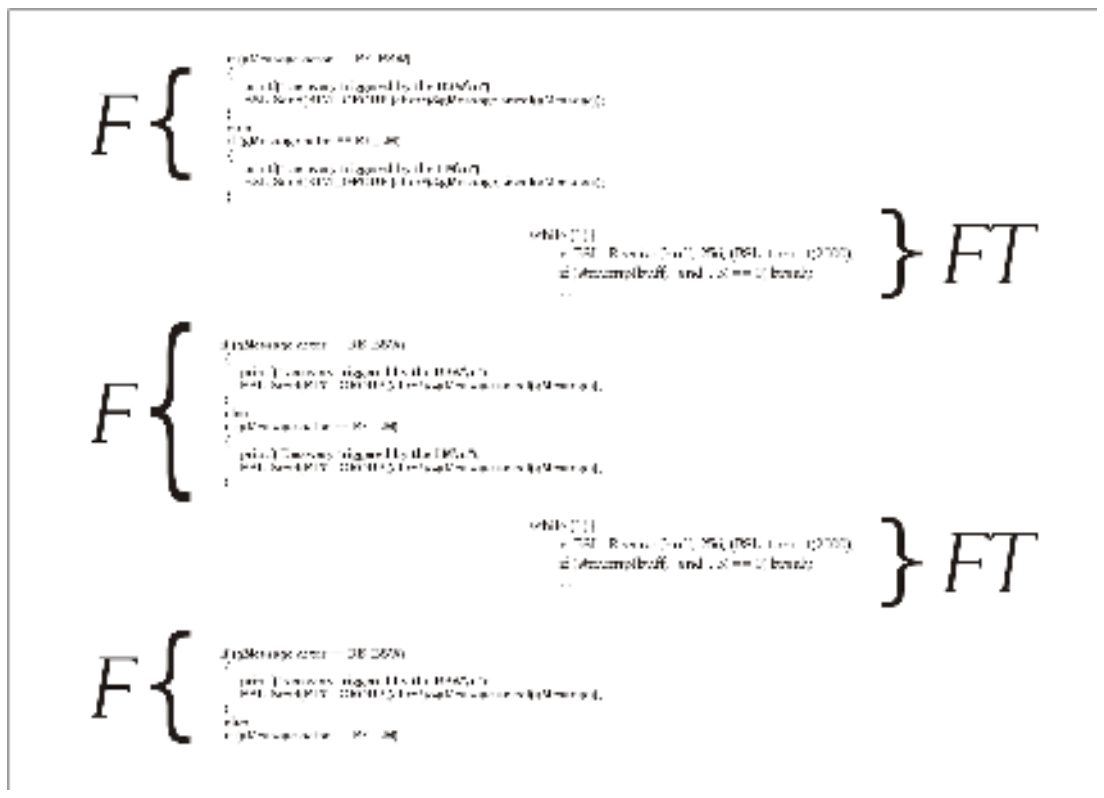

Figure 2: A fault-tolerant program according to a SV system.

## 2.2 Conclusions.

Figure 2 synthesizes the main characteristics of the SV approach: the functional and the fault-tolerant code are intertwined and the developer has to deal with the two concerns at the same time, even with the help of libraries of fault-tolerance provisions. In other words, SV requires the application developer to be an expert in fault-tolerance as well, because he (she) has to integrate in the application a number of fault-tolerance provisions among those available in a set of ready-made basic tools. His (hers) is the responsibility for doing it in a coherent, effective, and efficient way. As it has been observed in Chapter 2, the resulting code is a mixture of functional code and of custom error-management code that does not always offer an acceptable degree of portability and maintainability. The functional and non-functional design concerns are not kept apart with SV, hence one can conclude that (qualitatively) SV exhibits poor separation of concerns (SC). This in general has a bad impact on design and maintenance **costs**.
As to syntactical adequacy (SA), one can easily observe how following SV the fault-tolerance provisions are offered to the user through an interface based on a general-purpose language such as C or C++. As a consequence, very limited SA can be achieved by SV as a system structure for application-level software fault-tolerance.
Furthermore, little or no support is provided for off-line and on-line configuration and reconfiguration of the fault-tolerance provisions. Consequently the adaptability (A) of this approach is deemed as insufficient. On the other hand, tools in SV libraries and systems give the user the ability to deal with fault-tolerance "atoms" without having to worry about their actual implementation and with a good ratio of costs over improvements of the dependability attributes, sometimes introducing a relatively small overhead. Using these toolsets the designer can re-use existing, long tested, sophisticated pieces of software without having each time to "re-invent the wheel".
It is also important to remark that, in principle, SV poses no restrictions on the class of applications that may be tackled with it.
As a final remark, it is interesting to note how, at least judging from the following recent work (Liu, Meng, Zhou, & Wu, 2006), it appears that the concept of a reusable "library" of fault-tolerance services is re-emerging in the context of service-oriented architectures.

## 2.3 Multiple-version Software Fault-Tolerance: Structures for Design Diversity

This section describes multiple-version software fault-tolerance (MV), an approach that requires $N$ ($N > 1$) independently designed versions of software. MV systems are therefore $x$T/$y$H/$N$S systems. In MV, a same service or functionality is supplied by $N$ pieces of code that have been designed and developed by different, independent software teams[2]. The aim of this approach is to reduce the effects of design faults due to human mistakes committed at design time. The most used configurations are $N$T/1H/$N$S, i.e., $N$ sequentially applicable alternate programs using the same hardware channel, and 1T/$N$H/$N$S, based on the parallel execution of the alternate programs on $N$, possibly diverse, hardware channels.
Two major approaches exist: the first one is known as recovery block (Randell, 1975; Randell & Xu, 1995), and is dealt with in Sect. 2.3. The second approach is the so-called $N$-version programming (Avižienis, 1985, 1995). It is described in Sect. 2.3.

**The Recovery Block Technique.** Recovery Blocks are usually implemented as $N$T/1H/$N$S systems. The technique addresses residual software design faults. It aims at providing fault-tolerant functional components which may be nested within a sequential program. Other versions of the approach, implemented as 1T/$N$H/$N$S systems, are suited for parallel or distributed programs (Scott, Gault, & McAllister, 1985; Randell & Xu, 1995). The recovery blocks technique is similar to the hardware fault-tolerance approach known as "stand-by sparing", which is described, e.g., in (Johnson, 1989). The approach is summarized in Fig. 3: on entry to a recovery block, the current state of the system is checkpointed. A primary alternate is executed. When it ends, an acceptance test checks whether the primary alternate successfully accomplished its objectives. If not, a backward recovery step reverts the system state back to its original value and a secondary alternate takes over the task of the primary alternate. When the secondary alternate ends, the acceptance test is executed again. The strategy goes on until either an alternate fulfills its tasks or all alternates are executed without success. In such a case, an error routine is executed. Recovery blocks can be nested—in this case the error routine invokes recovery in the enclosing block (Randell & Xu, 1995). An exception triggered within an alternate is managed as a failed acceptance test. A possible syntax for recovery blocks is as follows:

```
ensure       acceptance test
by           primary alternate
else by      alternate 2
             .
             .
else by      alternate N
else error
```

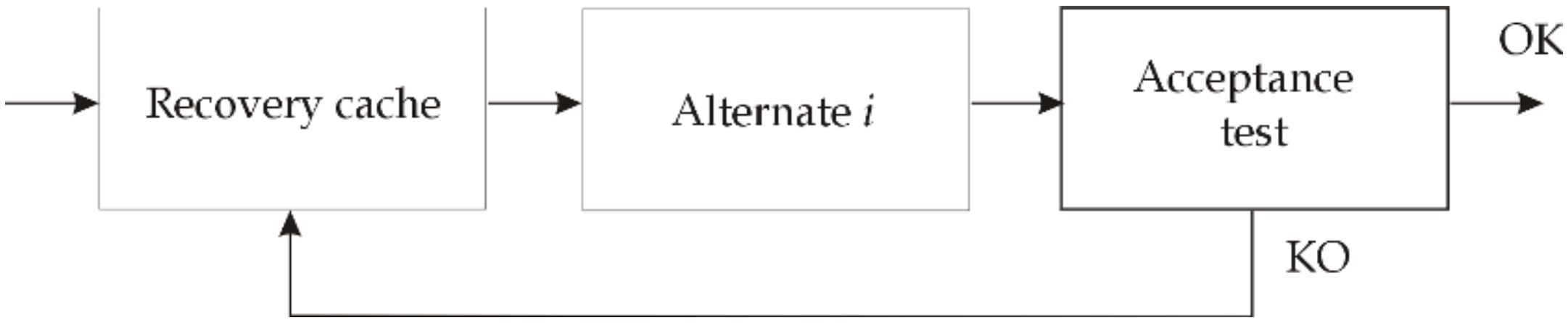

Figure 3: The recovery block model with two alternates. The execution environment is charged with the management of the recovery cache and the execution support functions (used to restore the state of the application when the acceptance test is not passed), while the user is responsible for supplying both alternates and the acceptance test.

Note how this syntax does not explicitly show the recovery step that should be carried out transparently by a run-time executive.
The effectiveness of recovery blocks rests to a great extent on the coverage of the error detection mechanisms adopted, the most crucial component of which is the acceptance test. A failure of the acceptance test is a failure of the whole recovery blocks strategy. For this reason, the acceptance test must be simple, must not introduce huge run-time overheads, must not retain data locally, and so forth. It must be regarded as the ultimate means for detecting errors, though not the exclusive one. Assertions and run-time checks, possibly supported by underlying layers, need to buttress the strategy and reduce the probability of an acceptance test failure. Another possible failure condition for the recovery blocks approach is given by an alternate failing to terminate. This may be detected by a time-out mechanism that could be added to recovery blocks. This addition, obviously, further increases the complexity.
The SwIFT library that has been described in Sect. 2.1 implements recovery blocks in the C language as follows:

```
#include <ftmacros.h>
```

```
...
ENSURE(acceptance-test) {
          primary alternate;
} ELSEBY {
          alternate 2;
} ELSEBY {
          alternate 3;
}
...
ENSURE;
```

Unfortunately this approach does not cover any of the above mentioned requirements for enhancing the error detection coverage of the acceptance test. This would clearly require a run-time executive that is not part of this strategy. Other solutions, based on enhancing the grammar of pre-existing programming languages such as Pascal (Shrivastava, 1978) and Coral (Anderson, Barrett, Halliwell, & Moulding, 1985), have some impact on portability. In both cases, code intrusion is not avoided. This translates into difficulties when trying to modify or maintain the application program without interfering "much" with the recovery structure, and vice-versa, when trying to modify or maintain the recovery structure without interfering "much" with the application program. Hence one can conclude that recovery blocks are characterized by unsatisfactory values of the structural attribute SC. Furthermore, a system structure for application-level software fault-tolerance based exclusively on recovery blocks does not satisfy attribute SA[3]. Finally, regarding attribute A, one can observe that recovery blocks are a rigid strategy that does not allow off-line configuration nor (*a fortiori*) code adaptability.

On the other hand, recovery blocks have been successfully adopted throughout 30 years in many different application fields. It has been successfully validated by a number of statistical experiments and through mathematical modeling (Randell & Xu, 1995). Its adoption as the sole fault-tolerance means, while developing complex applications, resulted in some cases (Anderson et al., 1985) in a failure coverage of over 70%, with acceptable overheads in memory space and CPU time.
A negative aspect in MV system is given by development and maintenance **costs**, that grow as a monotonic function of $x, y, z$ in any $x$T/$y$H/$z$S system. Development costs may be alleviated by using an approach such as diversity for off-the-shelf products (Gashi & Popov, 2007; Gashi, Popov, & Strigini, 2006). Other researchers have sought cost-effective diversity through the use of different computer architectures, different compilers, or different programming languages (Meulen & Revilla, 2005). A recent approach is using diversity for security concerns (Cox et al., 2006).

**N-Version Programming.** *N*-Version Programming (NVP) systems are built from generic architectures based on redundancy and consensus. Such systems usually belong to class 1T/$N$H/$N$S, less often to class $N$T/1H/$N$S.
NVP is defined by its author (Avižienis, 1985) as "the independent generation of $N > 1$ functionally equivalent programs from the same initial specification." These $N$ programs, called versions, are developed for being executed in parallel. This system constitutes a fault-tolerant software unit that depends on a generic decision algorithm to determine a consensus or majority result from the individual outputs of two or more versions of the unit.
Such a strategy (depicted in Fig. 4) has been developed under the fundamental conjecture that independent designs translate into random component failures—i.e., statistical independence. Such a result would guarantee that correlated failures do not translate into immediate exhaustion of the available redundancy, as it would happen, e.g., using $N$ copies of the same version. Replicating software would also mean replicating any dormant software fault in the source version—see, e.g., the accidents with the Therac-25 linear accelerator (Leveson, 1995) or the Ariane 5 flight 501 (Inquiry, 1996).
According to Avižienis, independent generation of the versions significantly reduces the probability of correlated failures. A number of experiments (Eckhardt et al., 1991) and theoretical studies (Eckhardt & Lee, 1985) questioned the correctness of this assumption, though a more recent study involving a large number of independently developed multiple software versions claims otherwise (Lyu, Huang, Sze, & Cai, 2003).
The main differences between recovery blocks and NVP are:

- Recovery blocks (in its original form) is a sequential strategy whereas NVP allows concurrent execution;
- Recovery blocks require the user to provide a fault-free, application-specific, effective acceptance test, while NVP adopts a generic consensus or majority voting algorithm that can be provided by the execution environment (EE);
- Recovery blocks allow different correct outputs from the alternates, while the general-purpose character of the consensus algorithm of NVP calls for a single correct output[4].

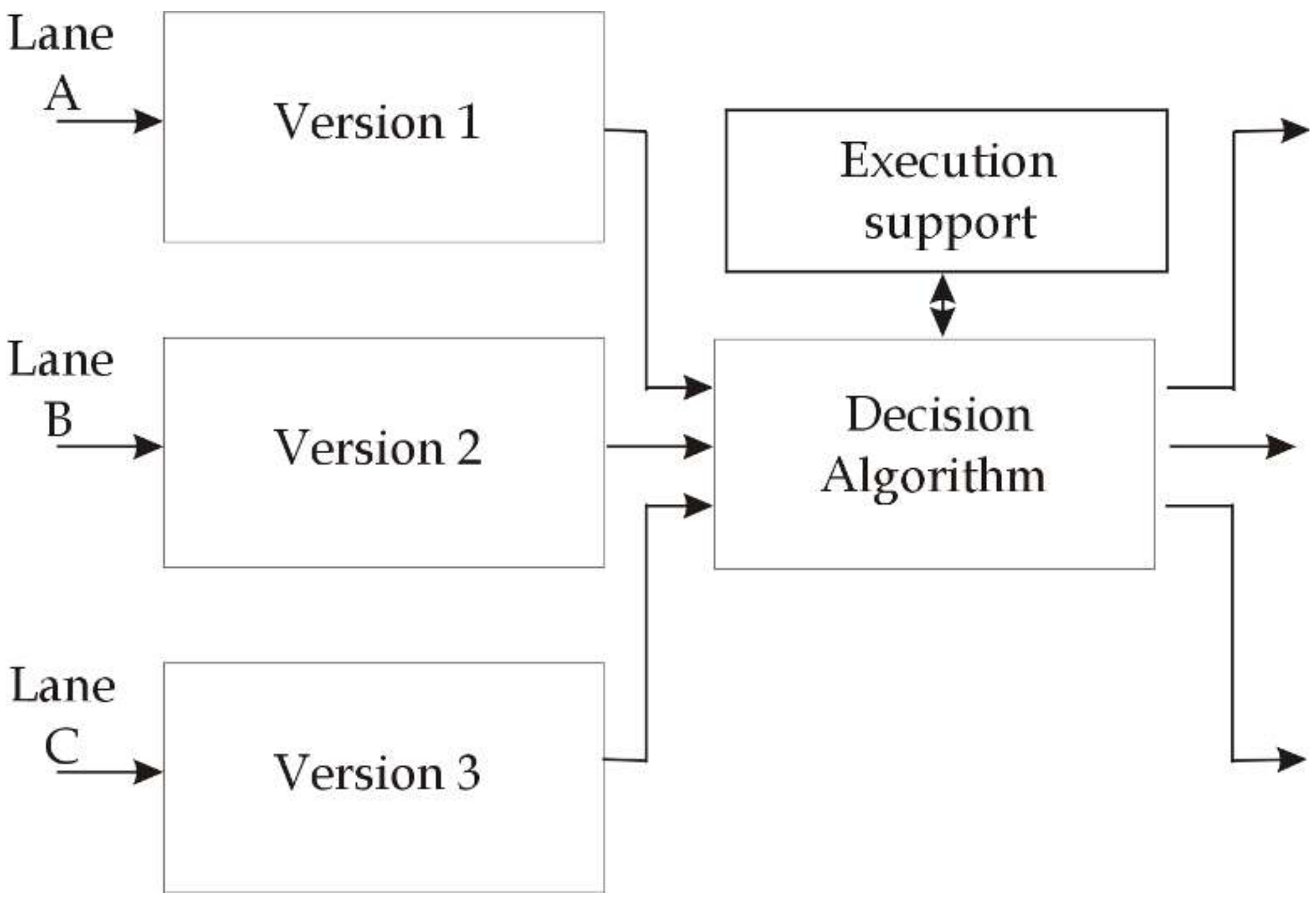

Figure 4: The $N$-Version Software model when $N = 3$. The execution environment is charged with the management of the decision algorithm and with the execution support functions. The user is responsible for supplying the $N$ versions. Note how the Decision Algorithm box takes care also of multiplexing its output onto the three hardware channels—also called "lanes".

The two models collapse when the acceptance test of recovery blocks is done as in NVP, i.e., when the acceptance test is a consensus on the basis of the outputs of the different alternates.
A few hybrid designs derived by coupling the basic ideas of recover blocks and NVP are now briefly discussed.

**Variations on the Main Theme.** $N$ Self-Checking Programming (Laprie, Arlat, Beounes, & Kanoun, 1995) couples recovery blocks with $N$-version programming: as in $N$-version programming, $N$ independently produced versions are executed, sequentially or in parallel. Each version is associated to a separate acceptance test, possibly different from the others, which tells whether the version passed the test and also produces a "rank". A selection module then chooses as overall output the one produced by the version with the highest rank. A variant of this technique organized versions in couples and performs comparison between the outputs of their versions as a general-purpose acceptance test. To the best of our knowledge, no application-level support for $N$ Self-Checking Programming has been proposed to date.
Consensus recovery blocks (Vouk, McAllister, Eckhardt, & Kim, 1993) targets the chance that the $N$-version programming scheme fail because it is not possible to find a majority vote among the output of the replicas. When this is the case, instead of declaring failure the outputs are assessed by acceptance tests (as in recovery blocks), which then have the last word in choosing the overall system output or declaring failure. Reliability analysis proves this approach to be better than $N$-version programming and recovery blocks, though the added complexity may well translate in higher chances of introducing faults in the architecture (Torres-Pomales, 2000).

Distributed recovery blocks (K. Kim & Welch, 1989) (DRB) may be considered as a parallel computing extension of recovery blocks.

> In DRB there is not a single couple of primary and alternate versions. Instead, several couples are running concurrently on different interconnected processing nodes. Each couple executes the recovery block scheme in parallel. Nodes and couples are organized hierarchically. When the execution of the top-level couple finishes, one queries the result of the acceptance test. If the test is passed by either primary or alternate, then the system declares success. If the test is not passed, instead of declaring failure as in plain recovery blocks, DRB goes on checking the acceptance test at the top-minus-one node. Global failure is only declared if no successful acceptance test can be found when orderly scanning the nodes. A time acceptance test is also used to handle performance failure of the acceptance tests.

**Again on the Ariane 5.** Chapter 2 briefly reported on the case of the Ariane 5 disaster. As it was mentioned there, the chain of events that brought to the Ariane 5 failure started within the Inertial Reference System (SRI), a component responsible for the measurement of the attitude of the launcher and its movements in space. To enhance the dependability of the system, the SRI was equipped two computers. Such computers were operating in parallel, *with identical hardware and software.* As described in the mentioned chapter, the SRI software had a number of data conversion instructions. Some of these instructions were "protected" (proper exception handling code had been associated to them), while some others were considered "safe enough" and were not protected so as to reduce the overhead on performance. One of the unprotected variables experienced an Operand Error. If the Ariane 5 designers had divided the SRI variables into two blocks, and had protected one block on the primary SRI and the other block on the backup SRI, they would have had no increased performance penalty and the failure would not have occurred.

### 2.3.1 A hybrid case: Data Diversity

A special, hybrid case is given by data diversity (Ammann & Knight, 1988). A data diversity system is a 1T/$N$H/1S (less often a $N$T/1H/1S). It can be concisely described as an NVP system in which $N$ equal replicas are used as versions, but each replica receives a different minor perturbation of the input data. Under the hypothesis that the function computed by the replicas is non chaotic, that is, it does not produce very different output values when fed with slightly different inputs, data diversity may be a cost-effective way to fault-tolerance. Clearly in this case the voting mechanism does not run a simple majority voting but some vote fusion algorithm (Lorczak, Caglayan, & Eckhardt, 1989). A typical application of data diversity is that of real time control programs, where sensor re-sampling or a minor perturbation in the sampled sensor value may be able to prevent a failure. Being substantially an NVP system, data diversity reaches the same values for the structural attributes. The greatest advantage of this technique is that of drastically decreasing design and maintenance costs, because design diversity is avoided.

**Conclusions.** As in recovery blocks, also NVP has been successfully adopted for many years in various application fields, including safety-critical airborne and spaceborne applications. The generic NVP architecture, based on redundancy and consensus, addresses parallel and distributed applications written in any programming paradigm. A generic, parameterizable architecture for real-time systems that supports the NVP strategy straightforwardly is GUARDS (Powell et al., 1999).
It is noteworthy to remark that the EE (also known as $N$-Version Executive) is a complex component that needs to manage a number of basic functions, for instance the execution of the decision algorithm, the assurance of input consistency for all versions, the inter-version communication, the version synchronization and the enforcement of timing constraints (Avižienis, 1995). On the other hand, this complexity is not part of the application software—the $N$ versions—and it does not need to be aware of the fault-tolerance strategy. An excellent degree of transparency can be reached, thus guaranteeing a good value for attribute SC. Furthermore, as mentioned in Chapter 2, costs and times required by a thorough verification, validation, and testing of this architectural complexity may be acceptable, while charging them to each application component is certainly not a cost-effective option. Regarding attribute SA, the same considerations provided when describing recovery blocks hold for NVP: also in this case a single fault-tolerance strategy is followed. For this reason NVP is assessed here as unsatisfactory regarding attribute SA.
Off-line adaptability to "bad" environments may be reached by increasing the value of $N$—though this requires developing new versions—a costly activity for both times and costs. Furthermore, the architecture does not allow any dynamic management of the fault-tolerance provisions. One concludes that attribute A is poorly addressed by NVP. In other words, the choices of the designer about the fault model are very difficult to maintain and change. Portability is restricted by the portability of the EE and of each of the $N$ versions. Maintainability actions may also be problematic, as they need to be replicated and validated $N$ times—as well as performed according to the NVP paradigm, so not to impact negatively on statistical independence of failures. Clearly the same considerations apply to recovery blocks as well. In other words, the adoption of multiple-version software fault-tolerance provisions always implies a penalty on maintainability and portability.
Limited NVP support has been developed for "conventional" programming languages such as C. For instance, `libft` (see Sect. 2.1) implements NVP as follows:

```
#include <ftmacros.h>
...
NVP
```

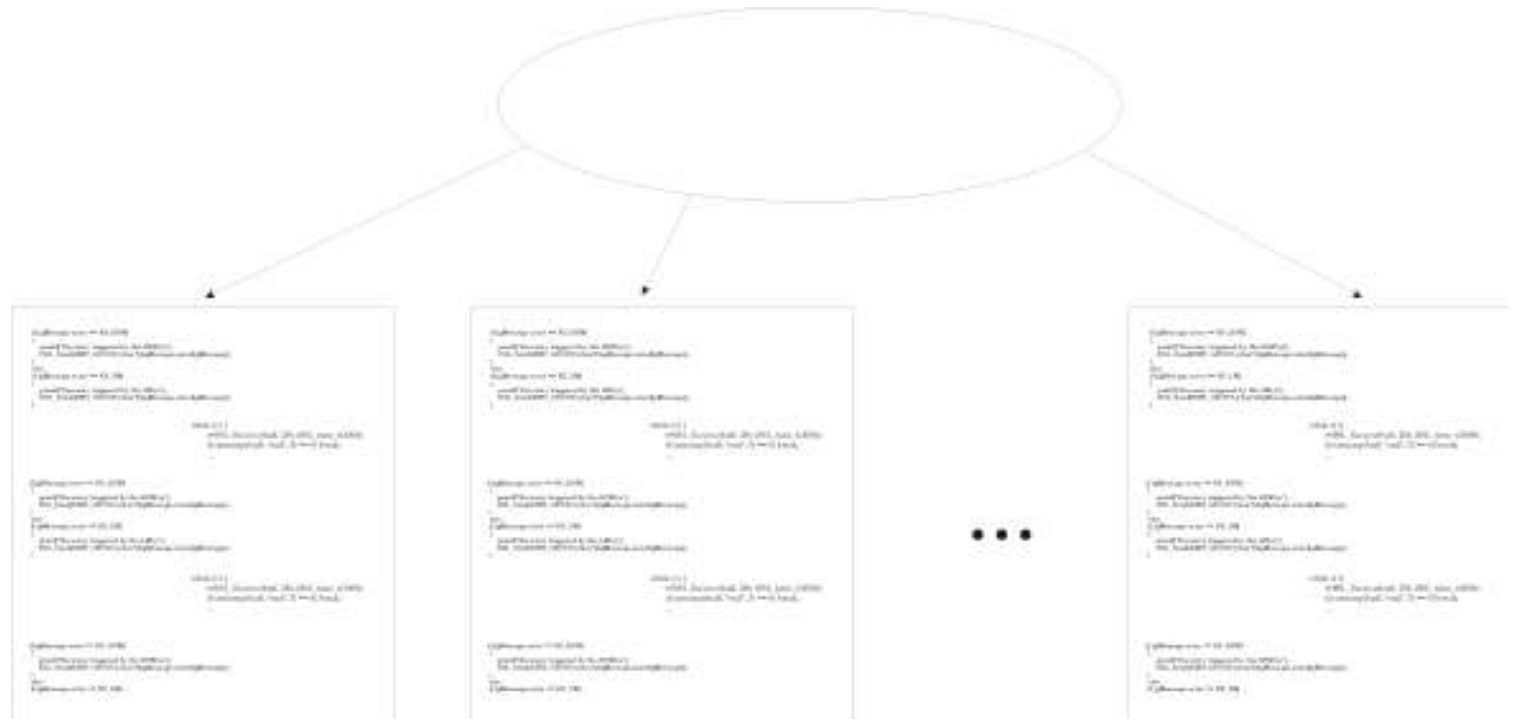

Figure 5: A fault-tolerant program according to a MV system.

```
VERSION{
          block 1;
          SENDVOTE(v_pointer, v_size);
}
VERSION{
          block 2;
          SENDVOTE(v_pointer, v_size);
}
...
ENDVERSION(timeout, v_size);
if (!agreeon(v_pointer)) error_handler;
ENDNVP;
```

Note that this particular implementation extinguishes the potential transparency that in general characterizes NVP, as it requires some non-functional code to be included. This translates into an unsatisfactory value for attribute SC. Note also that the execution of each block is in this case carried out sequentially.

It is important to remark how the adoption of NVP as a system structure for application-level software fault-tolerance requires a substantial increase in development and maintenance **costs**: both 1T/$N$H/$N$S and $N$T/1H/$N$S systems have a cost function growing with the square of $N$. The author of the NVP strategy remarks how such costs are paid back by the gain in trustworthiness. This is certainly true when dealing with systems possibly subjected to catastrophic failures—let us recall once more the case of the Ariane 5 flight 501 (Inquiry, 1996). Nevertheless, the risks related to the chances of rapid exhaustion of redundancy due to a burst of correlated failures caused by a single or few design faults (Motet & Geffroy, 2003) justify and call for the adoption of other fault-tolerance provisions within and around the NVP unit in order to deal with the case of a failed NVP unit.
Figure 5 synthesizes the main characteristics of the MV approach: several replicas of (portions of) the functional code are produced and managed by a control component. In recovery blocks this component is often coded side by side with the functional code while in NVP this is usually a custom hardware box.

# 3 The EFTOS Tools: The EFTOS Voting Farm

In this section the EFTOS voting farm— a library of functions written in the C programming language and implementing a distributed software voting mechanism—is described: This tool could be used to implement NVP systems in the application software. It has developed in the framework of project EFTOS, which was introduced in Sect. 2.1.1.
The Voting Farm was designed to be used either as a stand-alone tool for fault masking or as a basic block in a more complex fault tolerance structure set up within the EFTOS fault tolerance framework. In what follows the design and structure of the stand-alone voting farm are described as a means to orchestrate redundant resources with fault transparency as primary goal. It is also described how the user can exploit said tool to straightforwardly set up systems consisting of redundant modules and based on voters. An example of such system is given by so-called "restoring organs."

## 3.1 Basic Structure and Features of the EFTOS Voting Farm

A well-known approach to achieve fault masking and therefore to hide the occurrence of faults is the so-called $N$-modular redundancy technique (NMR), valid both on hardware and at software level. To overcome the shortcoming of having one voter, whose failure leads to the failure of the whole system even when each and every other module is still running correctly, it is possible to use $N$ replicas of the voter and to provide $N$ copies of the inputs to each replica, as described in Fig. 6. This approach exhibits among others the following properties:

1. Depending on the voting technique adopted in the voter, the occurrence of a limited number of faults in the inputs to the voters may be masked to the subsequent modules (Lorczak et al., 1989); for instance, by using majority voting, up to $ceil(N/2) - 1$ faults can be made transparent.

2. If one considers a pipeline of such systems, then a failing voter in one stage of the pipeline can be simply regarded as a corrupted input for the next stage, where it will be restored.

The resulting system is easily recognizable to be more robust than plain NMR, as it exhibits no single-point-of-failure. Dependability analysis confirms intuition. Property 2. in particular explains why such systems are also known as "restoring organs" (Johnson, 1989).

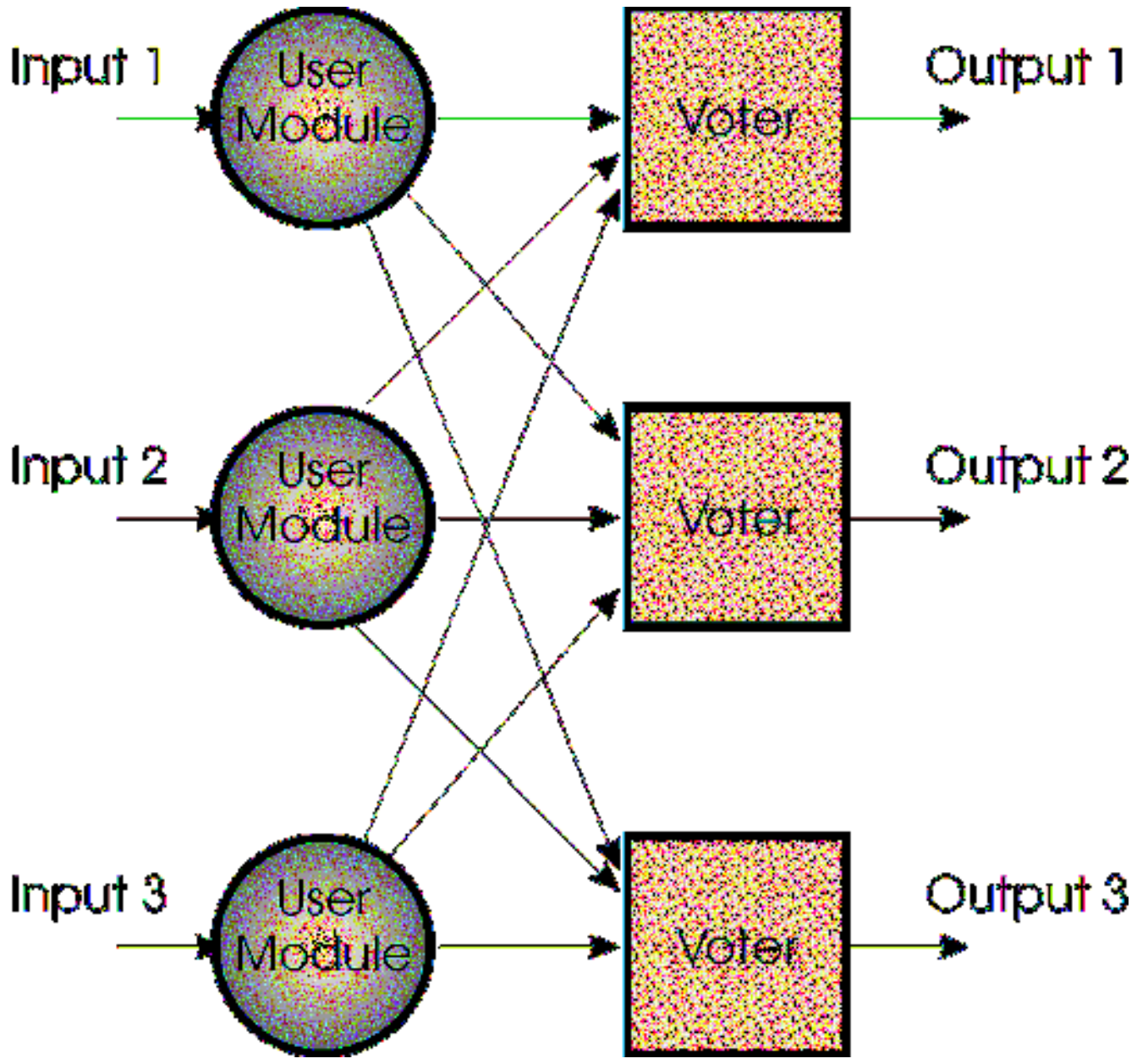

Figure 6: A restoring organ, i.e., an $N$-modular redundant system with $N$ voters, when $N = 3$.

From the point of view of software engineering, this system though has two major drawbacks:

- Each module in the NMR must be aware of and responsible for interacting with the whole set of voters;
- The complexity of these interactions, which is a function increasing with the square of $N$, the cardinality of the voting farm, burdens each module in the NMR.

Within EFTOS the two above flaws were recognized as serious impairments to our design goals, which included replication transparency, ease of use, and flexibility (De Florio, Deconinck, & Lauwereins, 1998a).
In order to reach the full set of our requirements, the design of the system was slightly modified as described in Fig. 7: In this new picture each module only has to interact with and be aware of *one* voter, regardless the value of $N$. Moreover, the complexity of such a task is fully shifted to the voter, i.e., it is transparent to the user.
The basic component of our tool is therefore the *voter* (see Fig.8) which is defined as follows:

> A voter is a local software module connected to *one* user module and to a farm of fully interconnected fellows. Attribute "local"

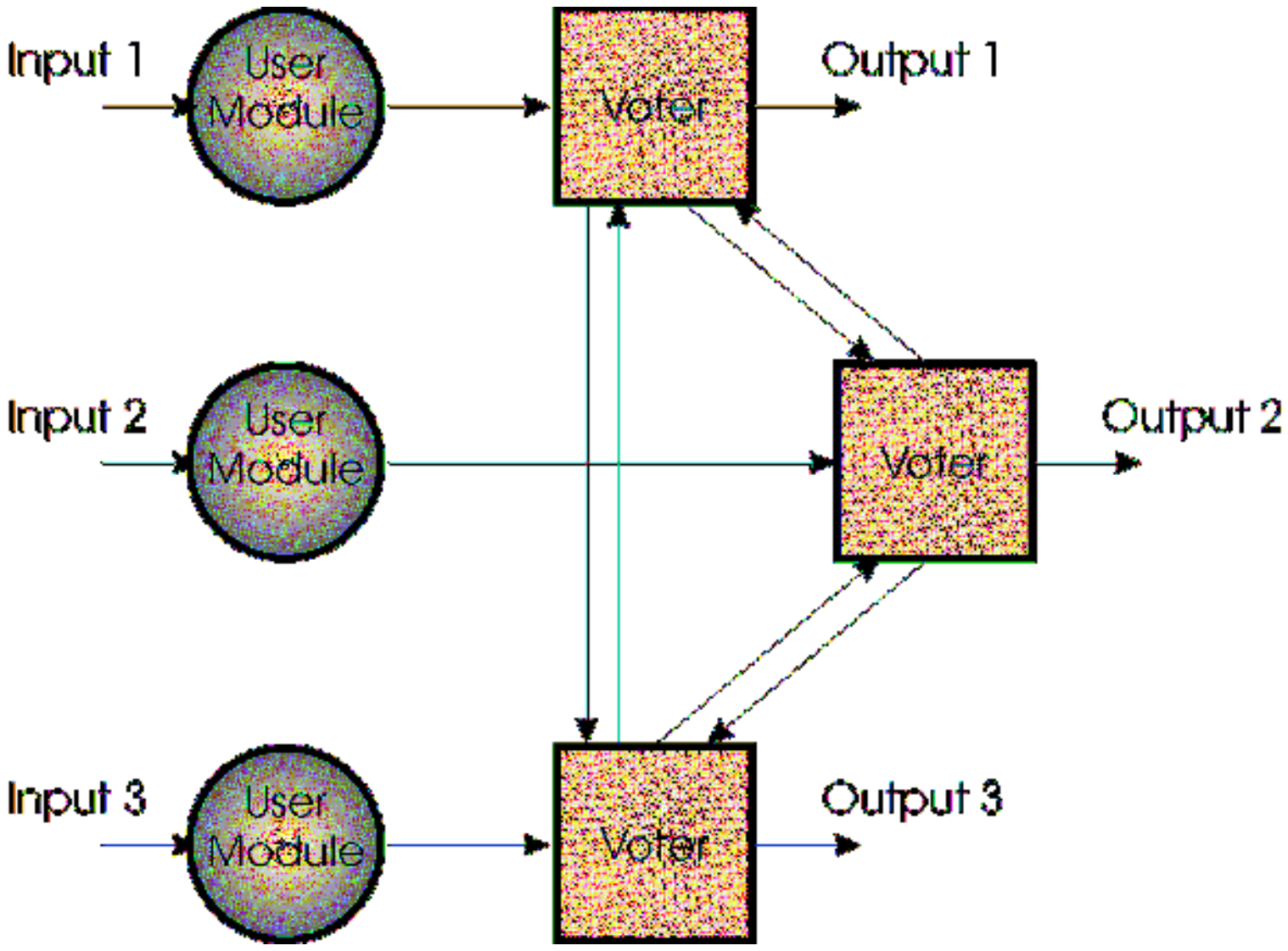

Figure 7: Structure of the EFTOS voting farm mechanism for a NMR system with $N = 3$ (the well-known triple modular redundancy system, or TMR).

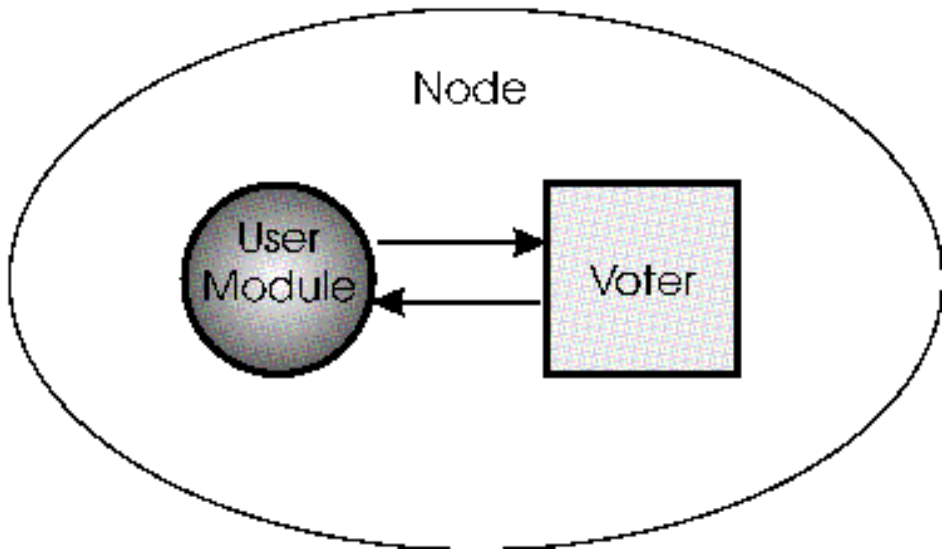

Figure 8: A user module and its voter. The latter is the only member of the farm of which the user module should be aware of: from the user point of view, messages will only flow between these two ends. This has been designed so as to minimize the burden of the user module and to keep it free to continue undisturbed as much as possible.

means that both user module and voter run on the same processing node.

As a consequence of the above definition, the user module has no other interlocutor than its voter, whose tasks are completely transparent to the user module. It is therefore possible to model the whole system as a simple client-server application: On each user module the same client protocol applies (see Sect. 3.1.1) while the same server protocol is executed on every instance of the voter (see Sect. 3.1.3).

### 3.1.1 Client-Side of the Voting Farm: the User Module

Table 3 gives an example of the client-side protocol to be executed on each processing node of the system in which a user module runs: a well-defined, ordered list of actions has to take place so that the voting farm be coherently declared and defined, described, activated, controlled, and queried: In particular, *describing* a farm stands for creating a static map of the allocation of its components; *activating* a farm substantially means spawning the local voter (Sect. 3.1.3 will shed more light on this); *controlling* a farm means requesting its service by means of control and data messages; finally, a voting farm can also be *queried* about its state, the current voted value, etc.
As already mentioned, the above steps have to be carried out in the same way on each user module: this coherency is transparently supported in Single-Process, Multiple-Data (SPMD) architectures. This is the case, for instance, of Parsytec EPX (*Embedded Parallel eXtensions to UNIX*, see, e.g., (Parsytec, 1996a, 1996b)) whose "initial load mechanism" transparently runs the same executable image of the user application on each processing node of the user partition.
This protocol is available to the user as a class-like fault-tolerant library of functions dealing with opaque objects referenced through pointers. A tight resemblance with the FILE set of functions of the standard C programming language library (Kernighan & Ritchie, 1988) has been sought so to shorten as much as possible the user's learning time—the API and usage of Voting Farm closely resemble those of FILE (see Table 1).

| **phase** | FILE **class** | VotingFarm_t **class** |
|---|---|---|
| declaration | FILE* f; | VotingFarm_t* vf; |
| opening | f = fopen(...); | vf = VF_open(...); |
| control | fwrite(f, ...); | VF_control(vf, ...); |
| closings | fclose(f); | VF_close(vf); |

Table 1: The C language standard class for managing file is compared with the VF class. The tight resemblance has been sought in order to shorten as much as possible the user's learning time.

The Voting Farm has been developed using the CWEB system of structured documentation (De Florio, 1997)—an invaluable tool both at design and at development time (Knuth, 1984).

### 3.1.2 System and Fault Models

A fault and system model document allows to bring to the foreground all the assumptions and dependencies that were used while designing a service. This is done so that when porting that service to a new platform all those underlying dependencies and assumptions do not slip the attention of the designer—see Chapter 2 for possible consequences of such a mistake.
The EFTOS target platform was a dedicated system with a custom, dedicated communication network. Accordingly, the adopted system model was that of partially synchronous systems. This assumption is in this case a realistic one, at least for parallel environments like that of the Parsytec EPX, which was equipped with a fast and dedicated communication subsystem, such that processors did not have to compete "too much" for the network. Such subsytem also offered a reliable communication means and allowed to transparently tolerate faults like, e.g., the break of a link, or a router's failure.
The internal algorithms of the Voting Farm are assumed to have fail/stop behavior. Upper bounds are known for communication delays. A means to send and to receive messages across communication links is assumed to be available. Let us call these functions Send and Receive. Furthermore, the following semantics is assumed for those functions: Send blocks the caller until the communication system has fully delivered the specified message to the specified (single) recipient, while Receive blocks the caller until the communication system has fully transported a message directed to the caller, or until a user-specified timeout has expired.
The Voting Farm can deal with the following classes of faults (Laprie, 1995):

- physical as well as human-made,
- accidental as well as intentional,
- development as well as operational,
- internal and external faults,
- permanent and temporary,

as long as the corresponding failure domain consists only of value failures. Timing errors are also considered, though the delay must not be larger than some bounded value (which is assumed to be the case in the system model). The tool is only capable of dealing with one fault at a time—the tool is ready to deal with other new faults only after having recovered from the present one. Consistent value errors are tolerated. Under this assumption, arbitrary in-code value errors may occur.
As a final remark, let us recall what mentioned in Chapter 2: software engineering for fault-tolerant systems should allow considering the nature of faults as a dynamic system, i.e., a system evolving in time, and by modeling faults as a function $F(t)$. The EFTOS Voting Farm allows to do so: If a service using the voting farm is moved to a new environment, for instance one characterized by a higher frequency of faults affecting the voters, the designer has just to choose a new value for $N$, the number of voters. Nothing changes in the application layer except that value. Of course this is an example of off-line adaptation, as it requires recompiling the service programs. In Chapter 4 an example of a tool will be described, which tracks the environment adjusting its fault model accordingly.

### 3.1.3 Server-Side of the Voting Farm: the Voter

The local voter thread represents the server-side of the voting farm. After the set up of the static description of the farm (Table 3, Step 3) in the form of an ordered list of processing node identifiers (positive integer numbers), the server-side of our application is launched by the user by means of the VF_run function. This turns the static representation of a farm into an "alive" (running) object, the voter thread.
This latter connects to its user module via inter-process communication provisions (so called "local links") and to the rest of the farm via synchronous, blocking channels ("virtual links").
Once the connection is established, and in the absence of faults, the voter reacts to the arrival of the user messages as a finite-state automaton: In particular, the arrival of input messages triggers a number of broadcasts among the voters—as shown in Fig.9—which are managed through the distributed algorithm described in Table 2.
When faults occur and affect up to $M < N$ voters, no arrival for more than $\Delta t$ time units is interpreted as an error. As a consequence, variable input_messages is incremented as if a message had arrived, and its faulty state is recorded. By doing so one can tolerate up to $M < N$ errors at the cost of $M\Delta t$ time units. Note that even though this algorithm tolerates up to $N-1$ faults, the voting algorithm may be intrinsically able to cope with much less than that: for instance, majority voting fails in the presence of faults affecting $ceil(N/2)$ or more voters. As another example, algorithms computing a weighted average of the input values consider all items whose "faulty bit" is set as zero-weight values, automatically discarding them from the average. This of course may also lead to imprecise results as the number of faults gets larger.

> Besides the input value, which represents a request for voting, the user module may send to its voter a number of other requests—some of these are used in Table 3, Step 5. In particular, the user can choose to adopt a voting algorithm among the following ones:
>
> - Formalized majority voting technique,
> - Generalized median voting technique,

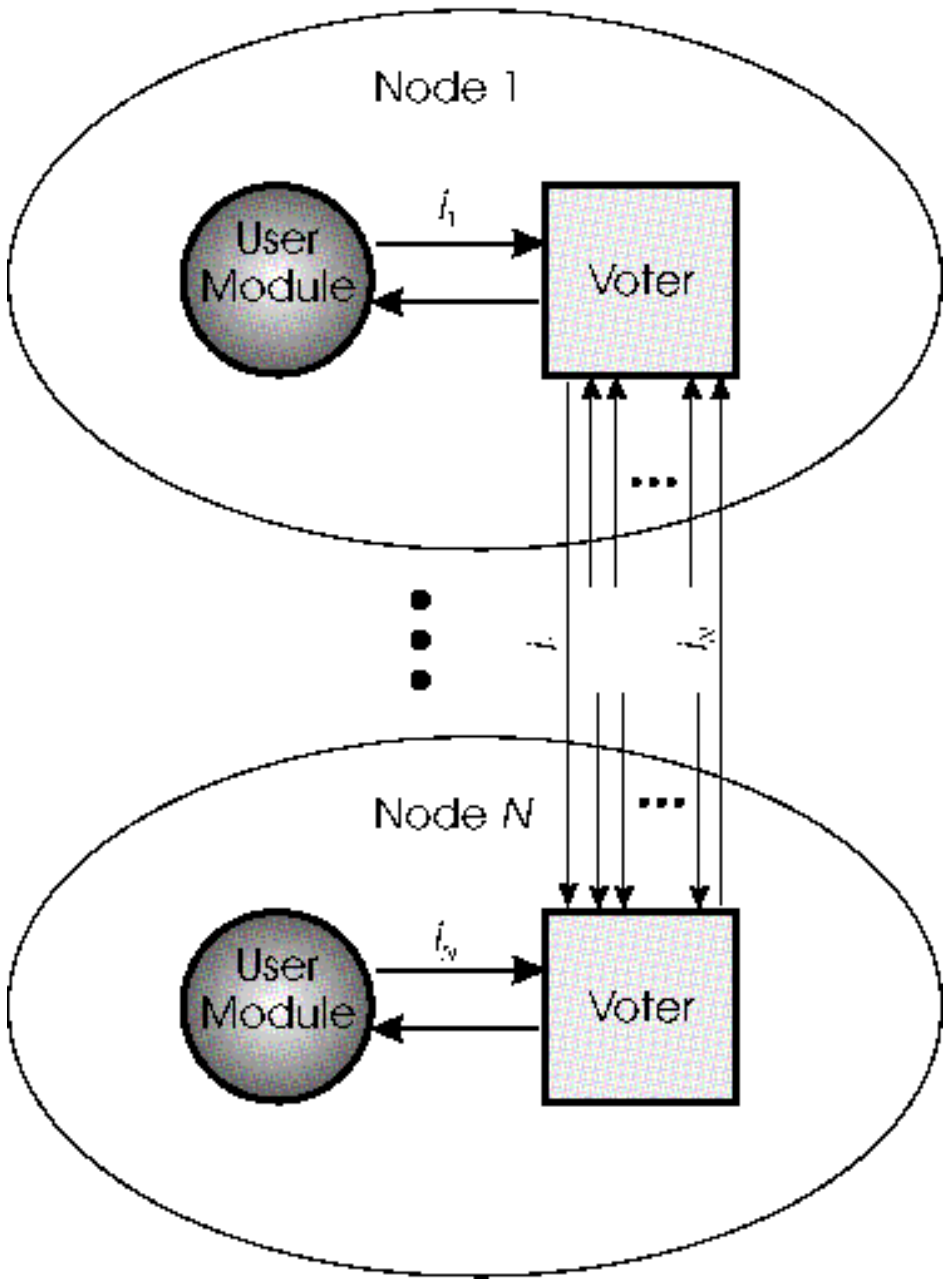

Figure 9: The "local" input value has to be broadcast to $N-1$ fellows, and $N-1$ "remote" input values have to be collected from each of the fellows. The voting algorithm takes place as soon as a complete set of values is available.

```
1  /* each voter gets a unique voter_id in {1, ..., N} */
   voter_id = who-am-i;
2  /* all messages are first supposed to be valid */
   For all i : valid_i = TRUE;
3  /* keep track of the number of received input messages */
   i = input_messages = 0;
4  do {
5      /* wait for an incoming message or a timeout */
       Wait_Msg_With_Timeout(Δt);
6      /* u points to the user module's input */
       if ( Sender == USER ) u = i;
7      /* read it */
       if ( ¬ Timeout ) msg_i = Receive;
8      /* or invalidate its entry */
       else valid_i = FALSE;
9      /* count it */
       i = input_messages = input_messages + 1;
10     if (voter_id == input_messages) Broadcast(msg_u);
11 } while (input_messages ¬ = N);
```

Table 2: The distributed algorithm needed to regulate the right to broadcast among the $N$ voters. Each voter waits for a message for a time which is at most $\Delta t$, then it assumes a fault affected either a user module or its voter. Function Broadcast sends its argument to all voters whose id is different from voter_id. It is managed via a special sending thread so to circumvent the case of a possibly deadlock-prone Send.

- Formalized plurality voting technique,
- Weighted averaging technique,
- Consensus,

the first four items being the voting techniques that were generalized in (Lorczak et al., 1989) to "arbitrary $N$-version systems with arbitrary output types using a metric space framework." To use these algorithms, a metric function can be supplied by the user when he or she "opens" the farm (Table 3, Step 2, function objcmp): this is exactly the same approach used in opaque C functions like e.g., bsearch or qsort (Kernighan & Ritchie, 1988). A default metric function is also available.

Note how the fault model assumption: "arbitrary in-code value errors may occur" is due to the fact that the adopted metric approach is not able to deal with non-code values.

The choice of the algorithm, as well as other control choices are managed via function VF_control, which takes as argument a voting farm pointer plus a variable number of control argument—in Table 3, Step 5, these arguments are an input message, a virtual link for the output vote, an algorithm identifier, plus an argument for that algorithm.
Other requests include the setting of some algorithmic parameters and the removal of the voting farm (function VF_close).
The voters' replies to the incoming requests are straightforward. In particular, a VF_DONE message is sent to the user module when a broadcast has been performed; for the sake of avoiding deadlocks, one can only close a farm after the VF_DONE message has been sent. Any failed attempt causes the voter to send a VF_REFUSED message. The same refusing message is sent when the user tries to initiate a new voting session sooner than the conclusion of the previous session.
Note how function VF_get (Table 3, Step 6) simply sets the caller in a waiting state from which it exits either on a message arrival or on the expiration of a time-out.

```
1   /* declaration */
    VotingFarm_t *vf;
2   /* definition */
    vf = VF_open(objcmp);
3   /* description */
    For all i in {1,...,N} : VF_add(vf, node_i, ident_i);
4   /* activation */
    VF_run(vf);
5   /* control */
    VF_control(vf, VF_input(obj, sizeof(VFobj_t)),
                   VF_output(link),
                   VF_algorithm (VFA_WEIGHTED_AVERAGE),
                   VF_scaling_factor(1.0) );
6   /* query */
    do {} while (VF_error==VF_NONE   and   VF_get(vf)==VF_REFUSED);
7   /* deactivation */
    VF_close(vf);
```

Table 3: An example of usage of the voting farm.

### 3.1.4 Voting Farm: An Example

This section introduces and discusses a program simulating a NMR (N modular redundant) restoring organ which makes use of the Voting Farm class. N is set to the cardinality of that list of values to vote on.

```
// An example of usage of the EFTOS voting farm
// We exploit the SPMD mode to launch the same executable on all target nodes

// First the necessary header files are loaded
#include <vf.h>
```

```
#include "tmr.h"

void main(int argc, char *argv[])
{
VotingFarm_t *vf;            // vf is the pointer to the Voting Farm descriptor
VF_msg_t *m;                 // m is a Voting Farm message object
double metrics(void*,void*); // metrics is the opaque function to compare votes
double sf = 0.5;             // sf is the scaling factor for voting algorithm
double d;                    // d is an input value to vote upon, read from the command line
int this;                    // this is the processor id (the node on which the code runs)
int i;


    // this is the id of the processor I'm running on
    this= GET_ROOT()->ProcRoot->MyProcID;

    // up to argc processors are to be used
    if (this >= argc-1) return;

    // declare a voting farm, with metrics() as metric function
    vf = VF_open(metrics);

    // add version i @ node i
    for (i=0; i<argc-1 && i<NPROCS; i++)
     VF_add(vf, i, i);

    // spawn the farm
    VF_run(vf);

    // read the value to be voted
    sscanf(argv[this+1], "%lf", &d);

    // send vf three parameters: scaling factor...
    VF_send(vf, 3, VFO_Set_Scaling_Factor(&sf)
                // ...voting algorithm...
             , VFO_Set_Algorithm(VFA_MAJORITY)
                // ...an input value
             , VFO_Set_Input_Message(&d,sizeof(d))
          );

    // wait for a message from the farm
    do {
          m = VF_get(vf);
          // keep on waiting while there's no error and return
          // code is VF_REFUSED ("refused attempt to close VF")
    } while ( VF_error == 0 && m->code == VF_REFUSED);
```

```
          // when there an error or a different message, let's check: done?
          if (m->code == VF_DONE)
          {
           // was it possible to find a majority vote?
           if (m->msglen == VF_FAILURE)
                 printf("<user %d> : no output vote is available\n", this);
           else
                 printf("<user %d> : output vote is %lf\n", this, DOUBLE(m->msg));

           // anyway, close the farm
           VF_close(vf);

           // wait for an acknowledgment or error
           do {
                 m = VF_get(vf);
           } while ( VF_error == 0 && m->code != VF_QUIT );

           return;
          }

          return;
}

// metrics reveals the nature of the two opaque input values:
// they are double precision floating point numbers, and their
// distance is abs(a-b)
double metrics(void *a,void *b) {
       double *d1, *d2;
       d1 = (double*)a, d2 = (double*)b;
       if (*d1 > *d2) { return *d1 - *d2; }
       return *d2 - *d1;
}
```

### 3.1.5 Voting Farm: Some Conclusions

The EFTOS Voting Farm is currently available for a number of message passing environments, including Parsytec EPX, Windows, and TXT TEX. A special version has been developed for the latter, which adopts the mailbox paradigm as opposed to message passing via virtual links. In this latter version, the tool has been used in a software fault tolerance implementation of a stable memory system for the high-voltage substation controller of ENEL, the main Italian electricity supplier (Deconinck et al., 1998). This stable memory system is based on a combination of temporal and spatial redundancy to tolerate both transient and permanent faults, and uses two voting farms, one with consensus and the other with majority voting.

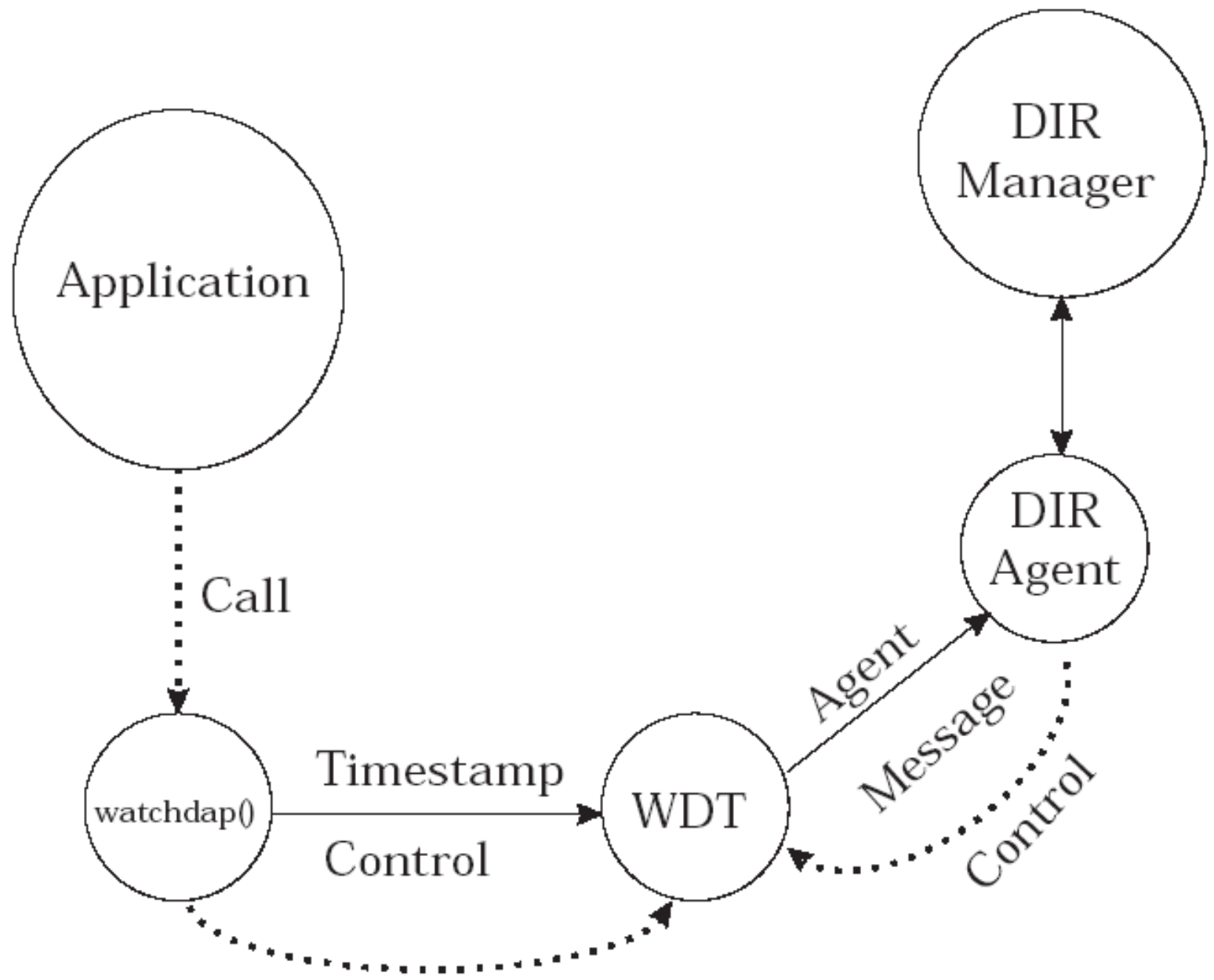

Figure 10: The interaction between Watchdog Timer, DIR net and the application. The dotted line represents control flow, the full line stands for data flow.

The Voting Farm can be used as a stand-alone tool, as seen so far; but it can also be used as a tool to compose more complex dependable mechanisms within a wider framework. Chapter 9 shall describe how to use our tool with the so-called "recovery language approach", a linguistic framework and an architecture for dependable automation services.
As a conclusion the Voting Farm is characterized by limited support for SC and bad SA (as it only targets a single provision). As for Aone may observe how, despite that tool exhibits no support for adaptability in the form described in this section, this aspect could be enhanced by using an hybrid approach such as the one described in Chapter 9.

# 4 The EFTOS Tools: The Watchdog Timer

This section describes the EFTOS watchdog timer. It consists of a single thread. This thread does the timing and checking of user-driven timestamps, and informs a DIR Agent thread if a performance failure is detected. This concept is depicted in Fig. 10.
The whole set up of Fig. 10 is built by executing the single StartWD function when the two major system component for EFTOS, the so-called DIR net and Server net, are both used and when the Watchdog thread was pre-configured through the server net (details on how to do this have been omitted). Note that after this step any future interaction with the WatchDog Timer, done via watchdap, is characterized by a satisfactory level of transparency: The user needs not to concern about low level details such as protocols and interface; he or she has just to control the process through a high-level application-program interface.
This user-transparency can no longer be sustained if neither DIR net nor Server net are used. In this case it is the responsibility of the user to deploy the watchdog through function StartWDnd and to let it start watching by issuing function WDStart. In both cases the user interfaces its watchdog through the same function, the already mentioned watchdap.
As can be seen from Fig. 10, an active watchdog connects to a so-called DIR agent and notifies it of all performance failures experienced by its watched task. When no DIR net is used, this message must nevertheless be sent to some other task.
The following short source code illustrates the usage of the EFTOS watchdog:

```
// A worker performs some work receiving input and sending output
// through a communication link called ioLink
//
// To protect the worker, a watchdog timer is started (in this case
// by the worker itself). Within the processing loop, the watcher
// sends n heartbeat signal to the watchdog through function watchdapp
//
int worker (LinkCB˙t *ioLink)
{
        // declare the communication link with the watchdog
        LinkCB˙t *AWDLink;

        // declare the communication link with the EFTOS server net
        LinkCB˙t *mylink2server;

        // input and output buffers
        char input[1024], output[1024];

        int  size, error;



        // Connects (or spawns) the EFTOS Server net
        mylink2server = ConnectServer();

        if ((AWDLink = StartWD(link2server, ...various parameters...,
                        ...cycle times..., &error)) == NULL)
                fprintf(stderr, "Failed to initialise the WD, error:%d ", error);

        // main processing loop: get input data...
```

```
        while ((size = RecvLink(ioLink, (byte *)input, sizeof(input))) != 0)
        {
                //          ...process data...
                process(input, output);

                //          ...forward output...
                SendLink(ioLink, (byte *)output, strlen(output)+1);

                //          ...and say "I'm OK"
                if (watchdap(AWDLink,TIMESTAMP,0)!= 0)
                        fprintf(stderr, "Error re-initialising the watchdog");
        }
}
```

For more details on programming and configuring the EFTOS watchdog timer the reader may refer to (Team, 1998).

The system model of the EFTOS Watchdog Timer is the same specified for the whole EFTOS framework: A fully synchronous system—an assumption allowed by the embedded character of the EFTOS target services and platforms. The fault model includes accidental, permanent or temporary design faults, and temporary, external, physical faults.

As a final statement let us remark how, as for the structural properties, what has been said for the Voting Farm also applies to the EFTOS watchdog timer: limited support for SC, bad SA due to the single design concern, and no adaptability unless coupled with other approaches and tools. One such hybrid approach is described in Chapter 9.

The EFTOS watchdog timer was developed by Wim Rosseel at the University of Leuven.

# 5 The EFTOS Tools: The EFTOS Trap Handler

Programming languages such as C constitute powerful tools to craft efficient system services, but are streamlined "by construction" for run-time efficiency. As a consequence, their run-time executive is very simple: They lack mechanisms for bound checking in arrays, are very permissive with data type conversions, and allow all type of "dirty tricks" with pointers. *A fortiori*, the C language does not provide any support for exception handling. Within project EFTOS a so-called Trap Handler was designed and developed. This tool is basically a library and a run-time executive to manage exceptions taking place in programs written in the C programming language on Parsytec supercomputers based on PowerPC processors. The library was developed Stephan Graeber at DLR (the Deutsche Zentrum für Luft- und Raumfahrt) with the Parsytec EPX message passing library. In the following this tool is described.

## 5.1 The EFTOS Trap Handling Tool

As mentioned in Chapter 2, exception (or trap) handling is an important feature to design software fault-tolerant systems. When the processor e.g. tries to access memory that is not allocated or executes illegal instructions then a trap is generated, which causes the processor to jump to a specialized routine called trap handler. As other operating systems, also EPX provides a standard trap handling function which simply stops processing and writes a core dump file.
The EFTOS framework provides two ways to alter this behavior:

1. The Trap Handling Tool connects to a third party (by default, the EFTOS DIR net) and creates a "fault notification stream": Caught exceptions are forwarded to a remote handler. A generalization of this strategy is used in Oz (see Chapter 5) and Ariel (in Chapter 6) and, in service-oriented architectures, in the system reported in (Ardissono, Furnari, Goy, Petrone, & Segnan, 2006).

2. The programmer defines which exception to catch and how to handle them with the functions of the Trap Handling library. This is semantically equivalent to, e.g., Java exceptions, but very different from the syntactical point of view. This is because the handling is done *with* the programming language, as opposed to *in*.

## 5.2 Algorithm of the Trap Handling Tool.

The first action of StartTrapTool is to give the server network the command to create remotely a thread on a specified node with the code of the TrapTool. After that, it connects to the newly spawn trap tool and exchanges some additional information with it. After this state has been set up properly, it installs a new trap handler for the current thread. The Trap handling Tool itself first gets the connection to the StartTrapTool function and receives the additional information from there. After that it connects to the appropriate DIR agent, and waits for incoming messages for the rest of its execution time. If a trap message arises from the trap handler, the DIR net is informed and the necessary information about the trap that occurred is passed to the responsible DIR agent. The DIR net is also able to send messages to the Trap Handling Tool to enact user-defined exception handling procedures. The trap handler itself is only responsible for passing the message of a fault to the Trap Tool and to set the processor in a sleeping mode. The processor will resume only when proper actions to handle the exception are scheduled for execution.

### 5.2.1 Structure of user trap handling

The concept of user-defined trap handlers is based on a stack of functions. The first element in the stack is the default EPX trap handler. New user-defined handlers are orderly pushed onto the stack. When an exception is caught, the stack is visited from top to bottom calling each visited function. When a function successfully handles the exception the Trap Handler stops this procedure, otherwise the stack reaches its bottom and EPX performs termination and memory dump.
A user defined trap handler can handle either one or more classes of traps. Traps are processor-dependent, e.g. the PowerPC defines among others the following classes:

1. DSI exception: A data memory access cannot be performed because of a memory protection violation, or because the instruction is not supported for the type of memory addressed.

2. ISI exception: An instruction fetch cannot be performed. Reasons may be that an attempt is made to fetch an instruction from a non-execute segment, or that a page fault occurred when translating the effective address, or that the fetch access violates memory protection.

3. Alignment exception: processor cannot perform a memory access because of an incorrect alignment of the requested address.

4. Program exception: This may have several reasons. E.g. the execution of an instruction is attempted with an illegal opcode.

The user-defined trap handler should be defined as a function with the following prototype:

```
int MyTrapHandler (int TrapNo)
```

With TrapNo this function gets the trap number, that is the exception code corresponding to the exception that was actually caught by the system. This corresponds to the exception code returned by Java in the `catch` statements.
The user defined trap handler function should return 1 if the trap was handled and the system has to be recovered, otherwise the function should return 0. If a trap occurs, in some cases the whole node has to be rebooted. In such cases a user defined trap handler can be used for instance to store state information on another node, so as to restart execution from there or on the same node after reboot. Obviously this procedure only covers transient faults. In other cases the fault shall represent itself and cause the occurrence of the same failure again.
With the function NewTrapHandler the user can push a trap handling function on top of the handling functions stack. When the function is called for the first time, a stack manager is installed as internal trap handler and the stack of functions is initialized.
With function ReleaseTrapHandler the user can remove the function at the top of the stack. To remove all functions and bring the stack to its initialization state with just the original EPX trap handler, the user can invoke function SetDefaultTrapHandler.

As can be clearly seen, the EFTOS Trap Handler is not as easy and intuitive as e.g. the exception handling mechanism used in Java: As mentioned already, syntactical adequacy (SA, defined in Chapter 2) has a strong link with complexity. The other side of the coin is given by efficiency: The EFTOS trap handling tool is characterized by a very limited overhead and consumes quite few system resources.
The following short source code illustrates the usage of user defined trap handlers:

```
// Function MyTraphandler returns 1 if an exception
// is caught and processed, and 0 otherwise.
//
int MyTrapHandler (int TrapNo)
        {
        switch (TrapNo) {
                // this is the equivalent of the Java catch statement.
                // NK_TRA_DFETCH means in EPX ``data access exception''
                case NK_TRAP_DFETCH:

                // what follows is the handling of the data access exception
                ...
                return 1;

        // other cases may follow here... */
                default:

                return 0;
        }
}


void main(void)
{
        // some work is done here
        ...

        // right before an operation that may result in a data exception
        NewTrapHandler(&MyTrapHandler);

        // here there is an operation that may result in a data exception
        ...

        // the default handler is finally restored
        ReleaseTrapHandler();
}
```

### 5.2.2 System and Fault Models of the EFTOS Trap Handling Tool

The system model of the EFTOS Trap Handling Tool is the same specified for the whole EFTOS framework: A fully synchronous system—an assumption allowed by the embedded character of the EFTOS target services and platforms. Target faults are clearly exceptions and system errors such as the one presented in Chapter 2. The fault model includes temporary design faults, and temporary external physical faults.

### 5.2.3 Conclusions

A single-version software fault-tolerance tool has been introduced, addressing exception handling and fault information forward. Developed in the framework of the EFTOS project, the tool is characterized by limited support for SC, bad SA due to its single design concern, and no adaptability.

# 6 The EFTOS Tools: Atomic Actions

The main goal of the functions described in what follows is to provide a mechanism for atomic transactions: The actions checked by these functions either end properly or are not executed at all. A description of transactions can be found in Chapter 2.

## 6.1 The EFTOS Atomic Action Tool

As explained in Chapter 2, an atomic action or transaction may be defined as the activity of a set of components where no information flows between that set and the rest of the system during that activity, and the activity is either fully completed or not at all (Anderson & Lee, 1981). To guarantee this property an atomic action needs to be able to checkpoint its state before the beginning of the action and roll back in case of failure.

In literature several protocols for atomic commitment have been proposed (Babaoglu, Toueg, & Mullender, 1993; Jalote & Campbell, 1985). As mentioned already, probably the best known and the simplest protocol is the two phase commit protocol (2PC) (Lampson, 1981). The 2PC protocol although very simple has as the big drawback that it may block. For example, if the coordinator fails while all the cohorts are waiting to receive a decision message, then none of these processes will be able to terminate. The cohorts need to wait until the coordinator is recovered before being able to decide on the outcome of the action. It is clear that such behavior is unacceptable. Next to the blocking aspect of several protocols, often the assumption is made that no faults can occur in the communication layer. Clearly this assumption has a coverage, which means one needs accomodate for the cases where it proves to be not valid. The tool described herein takes these aspects into account.

Let us begin by introducing our assumptions:

```
Atomic Action Algorithm:
        (save the status)
        (Synchronize)
        Check an assertion
        Set the timer t_i,
        Broadcast the result of the assertion to the other partners
                {
                while the deadline has not passed
                for all partners
                        send result within time
                if sending timed out change state to abort
                  and inform everyone hereof
                }
        Receive the result from all partners
                {
                while not received all results and deadline t_i has not passed
                        receive
                if deadline t_i passed
                abort and inform everyone
                }
        If at least one result was abort then abort
        Wait (t_2) for potential stray messages
        if result is abort
        do recovery
```

Table 4: A pseudo code sketch of the algorithm of the Atomic Action tools.

#### 6.1.1 System Model

**Assumptions.** As already remarked, any algorithm is valid under specific assumptions. In the case of the EFTOS Atomic Action Tool a partially synchronous model of computation is assumed: although not limited as in the synchronous model, an upper bound on message delays is assumed to be known.

At any time a process may be either operational or non-operational. A process is considered to be operational when it follows exactly the actions specified by the program it is executing. Any operational process may end up in a non-operational state due to a failure. In a non-operational state any information related to that process is considered to be lost, unless it was stored into some stable storage. A non-operational process may be returned to an operational state after executing some recovery protocol. During this recovery protocol the information saved in stable memory is used to restore the process. Each processor has its local clock, which does not run synchronous with the neighboring processors. Each local clock however is only used to measure time intervals, so a global time is not a necessary assumption (Lamport, 1978). The target design platforms require a bounded termination time and a low amount of communication, as communication negatively affect the communication vs. processing ratio. Therefore a reasonably simple and lightweight algorithm has been designed. It has as main constraint that all tasks should be loosely synchronized before making use of the algorithm.

**The Algorithm.**

> Figure 4 provides the reader with a pseudo code overview of the algorithm. The algorithm has a fairly simple structure as can be seen at first glance. Some pre-processing steps like saving state information and loosely synchronizing should be considered. Hereafter the algorithm will start by checking an assertion which decides on the local status. After the local status has been decided a timer $t_1$, conditional for the successful completion of the action, is set. As a next step the local status information is propagated to the other partners in the action and the algorithm starts waiting for the status information to be received from the other partners. All expected messages should be received within the time-out $t_1$ to be able to result in a successful action. Once all status messages are received, a decision is made on success or failure of the action. The decision propagation however is delayed for some more time $t_2$, so that possible failure messages can be received. This will be further elaborated in the next section where some failure cases are discussed. By means of these multiple time-outs $(t_1, t_2)$ the algorithm can guarantee successful functioning under the specified restrictions.

**Failure Mode.** Both process and communication failures are considered. For process failures, it may be clear that the time-out $(t_1)$ will trigger a transition to the ABORT state (see Fig. 11). For communication, due to the synchronous nature hereof on the design platforms, failures can be compared to process failures. This is clear in case of blocking. In such a case either one partner never joined in the communication or both partners tried to send or receive over the same communication link. Otherwise the link communication might also truly fail. This case can be considered as a failure of the two partners in communication. This assumption is valid because the only thing the processes are aware of is the fact that their communication with another process failed. In Fig. 12 some failure cases are illustrated. In the first case it is considered a process failure. The failure of this process will lead to a condition in which insufficient inputs have been received for a successful decision phase. This will lead to a time-out that will automatically trigger the ABORT behavior of the action. In the second case it is assumed that some communication fails. This will lead to the situation where some partners will decide to ABORT as they have not received sufficient inputs, while others will decide to COMMIT in the first step. The transition to ABORT however, due to insufficient input, will trigger the propagation of the ABORT message to all other partners. Upon receipt of this message all partners will still change their status to ABORT.

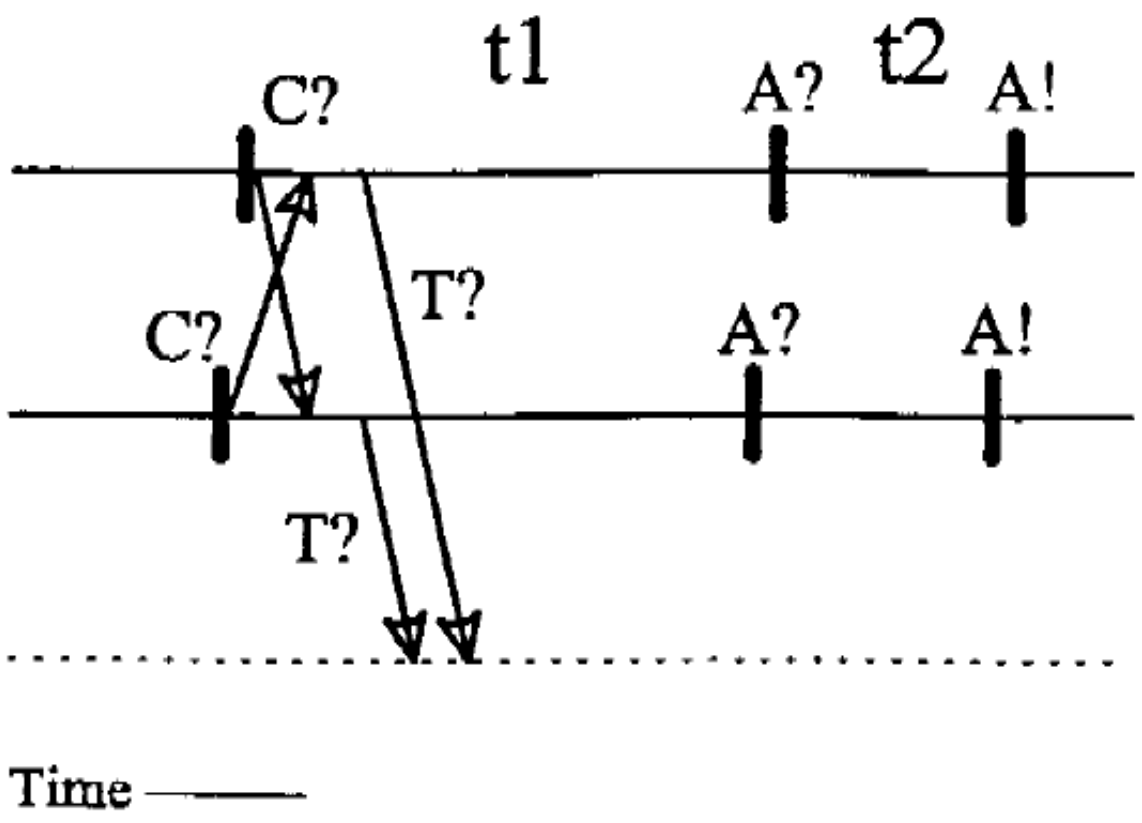

Figure 11: The behavior in case of a not responding partner. C? stands for potential commit. A? stands for potential abort, A! is an agreed abort. T? is a potential time-out in communication. $t_1$ is the primary time-out the action should respect, $t_2$ is the secondary timeout used to receive stray messages.

#### 6.1.2 The Implementation

The status of the action is decided upon by assertions provided by the user. The implementation has two working modes. In one mode only the local status is changed, unless there is a transition from COMMIT to ABORT at which point this new state is propagated to all partners in the action. In the second mode the distributed state decision is made. This is the mode illustrated in Fig. 13. Notice that the communication time-out is realized by means of a return message that should be received within time T. In Fig. 14 the state graph of the used algorithm is shown. AA_End in this graph is the intermediate state complying with the first mode of the algorithm. From this graph it is clear that any error will result in a transition to the ABORT state.

#### 6.1.3 Functionality

The whole mechanism basically is based on two levels of control (see Fig. 15. The first level embraces the local state. This is achieved by direct interaction with a local Atomic Action thread. The second level embraces a global state. This global state is maintained by the Atomic Action threads themselves, using the knowledge of the requirements to limit the communications. The communication limitation is achieved for in-block checks, where the global state will only be adapted if a request comes to change to local state to "abort". A final check always leads to a proprietary decision among all local states and the current global state. This leads to communication from every partner to every other partner, thus having a quadratic complexity.
More details about the EFTOS Atomic Action tool is available in (Team, 1998) and (Rosseel, De Florio, Deconinck, Truyens, & Lauwereins, 1998).

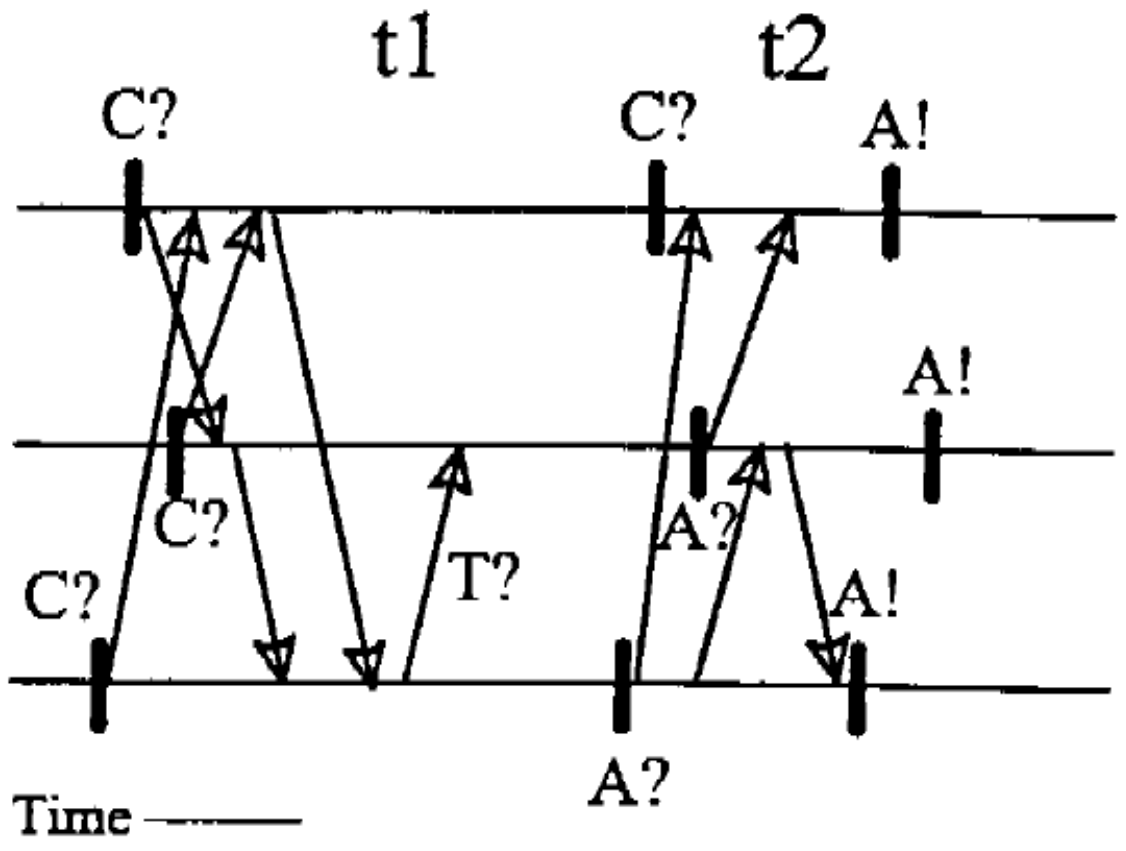

Figure 12: The behavior in case of a communication fails (time-out).

### 6.2 Conclusion

A single-version software fault-tolerance provisions for managing atomic actions has been briefly sketched. As most of the EFTOS tools, it is characterized by limited support for SC, bad SA and no adaptability.

## 7 The TIRAN Data Stabilizing Software Tool

An application-level tool is described, which implements a software system for stabilizing data values, capable of tolerating both permanent faults in memory and transient faults affecting computation, input and memory devices by means of a strategy coupling temporal and spatial redundancy. The tool maximizes data integrity allowing a new value to enter the system only after a user-parameterizable stabilization procedure has been successfully passed. Designed and developed in the framework of the ESPRIT project TIRAN (the follow-up of project EFTOS, described in more detail in Chapter 6), the tool can be used in stand-alone mode but can also be coupled with other dependable mechanisms developed within that project. Its use had been suggested by ENEL, the main Italian electricity supplier, in order to replace a hardware stable storage device adopted in their highvoltage sub-stations.

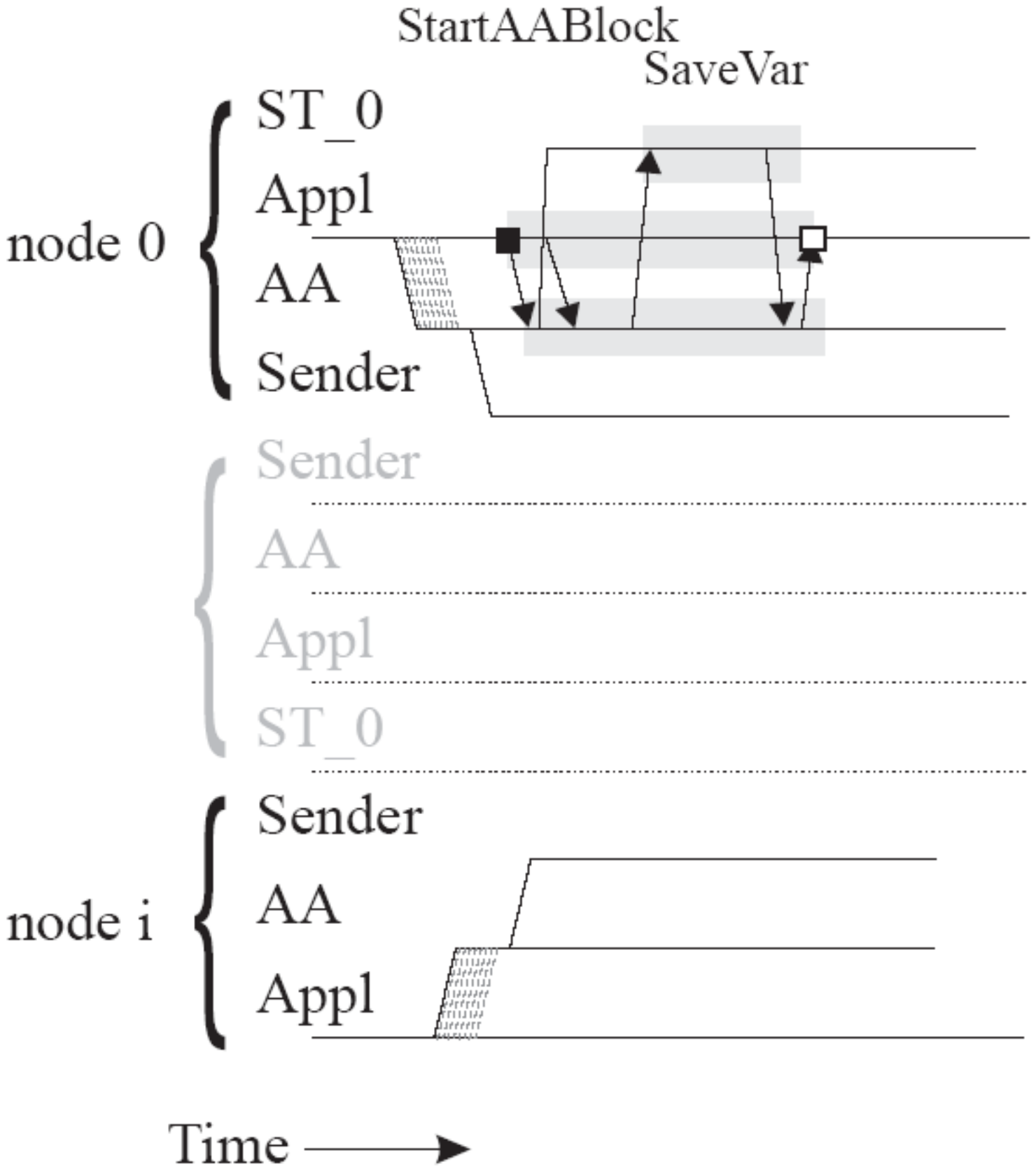

Figure 13: The message passing scheme in fault free case for the EFTOS implementation. The application starts the AA thread which will spawn a sender thread on its own. This is illustrated in the beginning of the time-scale. At this point node I also will save some status information to a stable storage entity. Once a distributed decision is to be achieved messages are exchanged according to the algorithm. Upon the agreed decision of ABORT, the first node will restore its saved status. The central zone between node 0 and node $i$ illustrates execution locked time, the black rectangle illustrates the beginning of the user-function the unfilled rectangle illustrates the returning of the user-function.

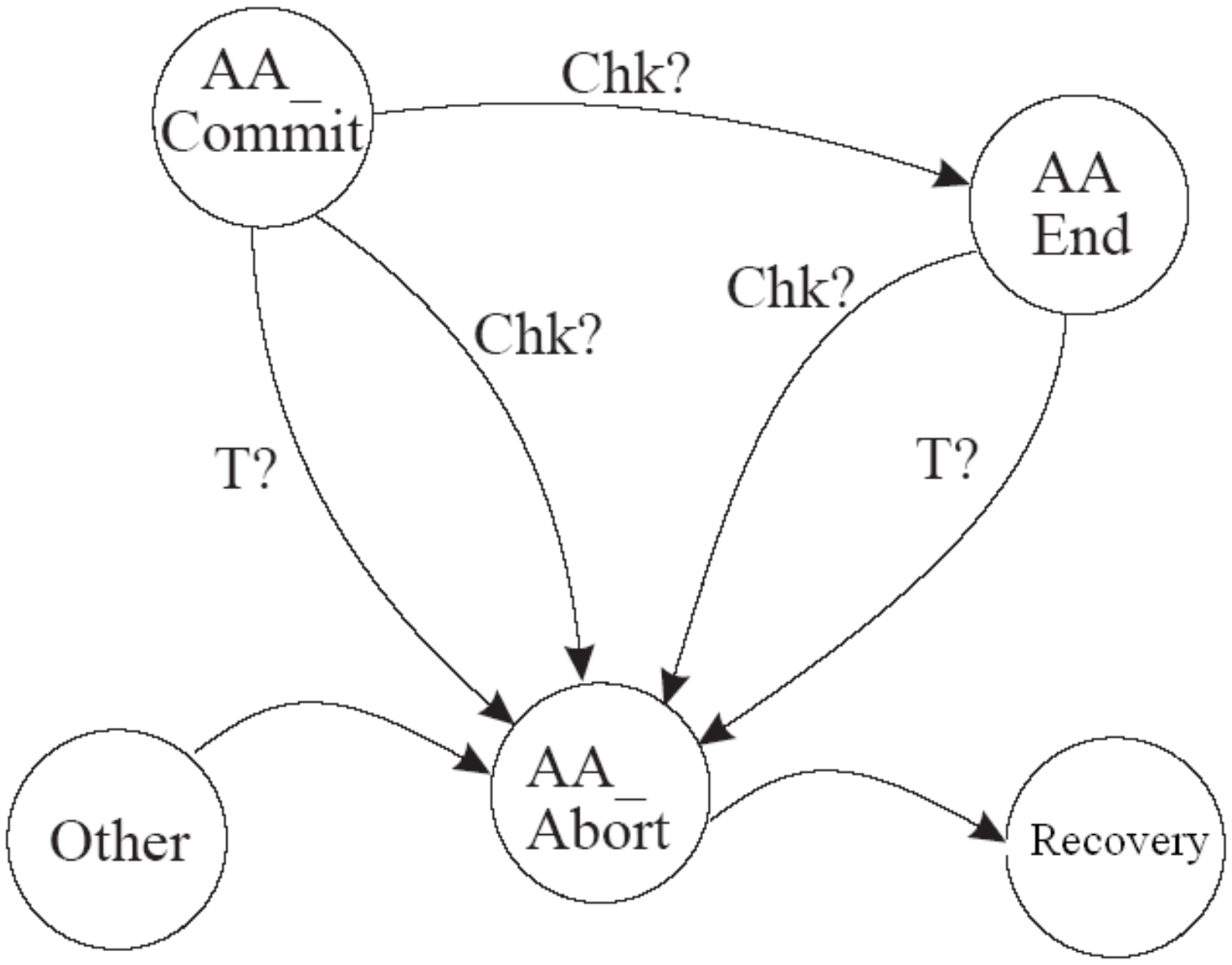

Figure 14: This is a state-graph of the implementation. AACommit is the intermediate COMMIT state, AA_END is an intermediate ABORT state and AA_Abort is the ABORT state. AA_Commit is the start state for the algorithm.

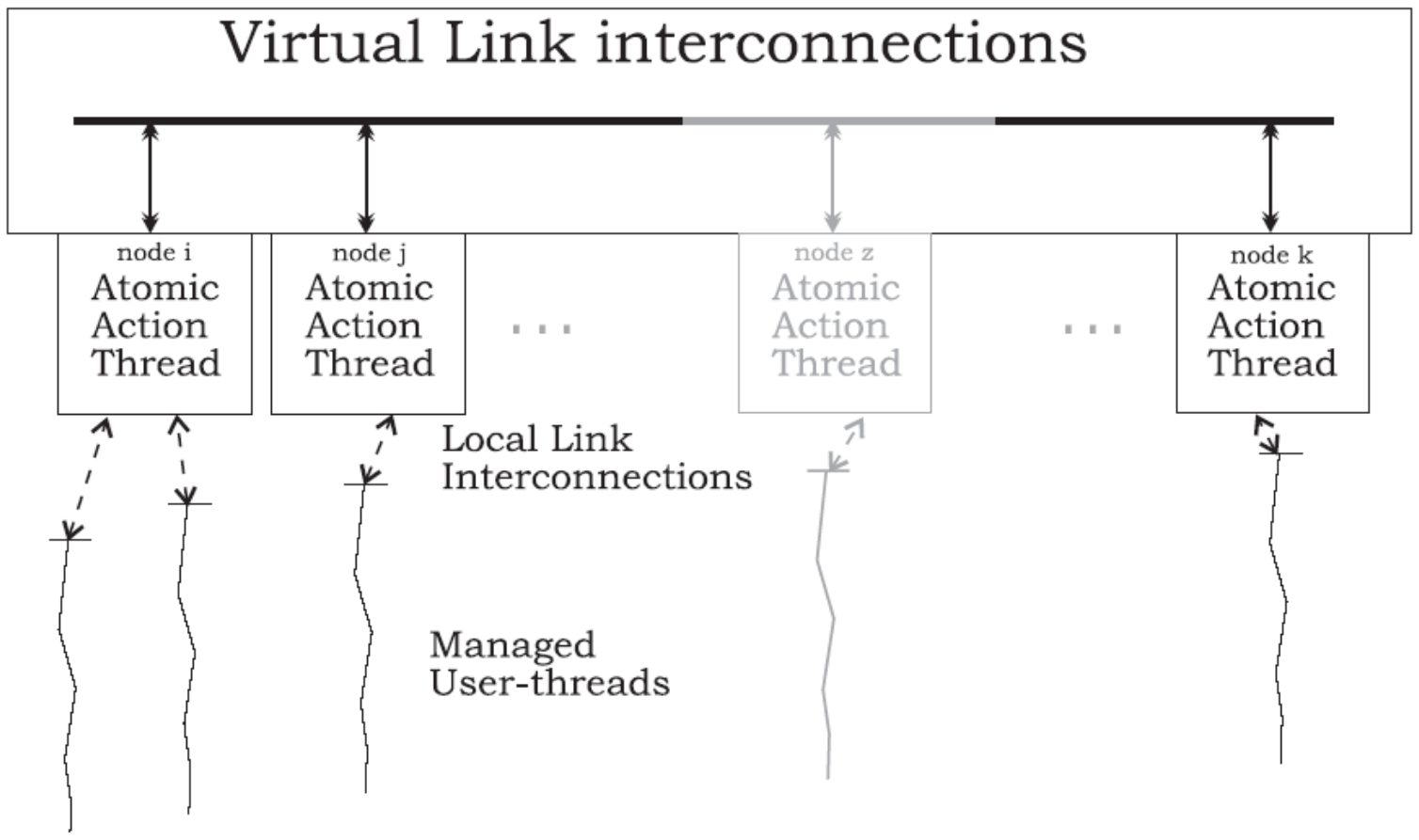

Figure 15: The Atomic Action embraces two levels of control, local within the Atomic Action thread and global in agreement with the other Atomic Action threads.

## 7.1 Introduction

In this text the design and the implementation of a data stabilizing software system are introduced. Such system is a fault-tolerant software component that allows validating input and output data by means of a stabilization procedure. Such data stabilizing system has been developed with the explicit goal of taking over a pre-existing stable storage hardware device used at ENEL S.p.A.—the main Italian electricity supplier, the third largest world-wide—-within a program for substation automation of their high voltage sub-stations. The mentioned hardware device is able to tolerate the typical faults of a highly disturbed environment subject to electro-magnetic interference: transient faults affecting memory modules and the processing devices, often resulting in bit flips or even in system crashes. This hardware component was mainly used within control applications with a cyclic behavior only dependent on their state (that is, Moore automata). Typically these applications:

- Read their current state from stable storage,
- produce with it an output that, once validated, is propagated to the field,
- then they read their input from the field and compute a tentative future state and future output.

The whole cycle is repeated a number of times in order to validate the future state. When this temporal redundancy scheme succeeds, the tentative state is declared as a valid next state and the stable storage is updated accordingly. The cyclic execution is paced by an external periodic signal which resets the CPU and re-fetches the application code from an EPROM. External memory is not affected by this step. This policy and the nature of faults (frequency, duration and so forth) allow confining possible impairment affecting the internal state of the application within one cycle.

Developed in the framework of the ESPRIT project TIRAN, a prototypic version of this data stabilizing tool has been successfully integrated in a test-bed control application at ENEL, whose cyclic behavior is regulated by a periodic restart device—the only custom, dedicated component of that architecture. Initially developed on a Parsytec CC system equipped with 4 processing nodes, the tool has been then ported to a number of runtime systems; at ENEL, the tool is currently running under the TEX nanokernel (DEC, 1997; TXT, 1997) and VxWorks on several hardware boards, each based on the DEC Alpha processor. Preliminary results on these systems show that the tool is capable to fulfill its dependability and data integrity requirements, adapting itself to a number of different simulated disturbed environments thanks to its flexibility.

In what follows an analysis of the requirements to the Data Stabilizing System tool is carried out. Basic functionalities of the tool are then summarized. The two “basic blocks” of our tool, namely a manager of redundant memories and a data stabilizer, are then introduced. Finally some conclusions are drawn, summarizing the lessons learned while developing our Data Stabilizing Software Tool.

## 7.2 Requirements for the Data Stabilizer

In an electrical power network, automation is a fundamental requirement for the subsystems concerning production, transport and distribution of energy. In many sites of such a network, remotely controlled automation systems play an important role towards reaching a more efficient and cost-effective management of the networks. While considering the option to install high performance computing nodes as controllers into such environments, the question of a software solution for a data stabilizer arose. The goal of a data stabilizer is to protect data in memory from permanent faults affecting memory devices and from transient faults affecting data of systems running in disturbed environments, as they typically arise by electro-magnetic interference, and to validate these data by means of a stabilization procedure based on the *joint exploitation of temporal and spatial redundancy.* When controlling high voltage, an important source of faults is electricity itself—because all switching actions in the field cause electrical disturbances, which enter the control computers via their I/O devices, often overcoming the filtering barriers. Furthermore, electro-magnetic interference causes disturbs in the controllers. Clearly, due to the very nature of this class of environments, such faults *cannot be avoided*; on the other hand, they should not impair the expected behavior of the computing systems that control the automation system. In order to overcome the effects of transient faults, temporal redundancy is employed. This means that all computation is repeated several times (a concept also known as "redoing" and introduced in Chapter 5), assuming that due to the nature (frequency, amplitude, and duration) of the disturbances, not all of the cyclic replications are affected. As in other redundancy schemes, a final decision is taken via a voting between the different redundant results. Clearly this calls for a memory component that be more resilient to transient faults with respect to conventional memory devices. In traditional applications often a special hardware device, called stable storage device, is used for this. The idea was to replace this special hardware with a software solution, which offers on the one hand more flexibility, while on the other hand it provides the same fault tolerance functionality. The following requirements were deduced from this:

1. The data stabilizer has to be implemented in conventional memory available on the hardware platform running the control application.

   Typical control applications show a cyclic behavior, which is represented in the following steps:

   (a) Read sensor data,

   (b) calculate control laws based on new sensor data and status variables,

(c) update status variables,

(d) and output actuator data.

The data stabilizer has to interface with such kind of applications.

2. The Data Stabilizing System has to tolerate any transient faults affecting the elaboration or the inputs. Furthermore, it has to tolerate permanent and transient faults affecting the memory itself.

3. The Data Stabilizing System has to store and to stabilize status data, i.e. if input data to the Data Stabilizing System have been confirmed a number of times, they should be considered as stable.

4. Because of this stabilization the Data Stabilizing System has to provide a high data integrity, i.e. only correct output should be released.

A few further requirements were added, namely:

1. The Data Stabilizing System has to minimize the number of custom, dedicated, hardware components in the system: In particular, the system has to work with conventional memory chips.

2. The system has to make use of the inherently available redundancy of a parallel or distributed system.

3. Its design goal must include maximizing flexibility and re-usability, so as to favor the adoption of the system in a wide application field, which, in the case of ENEL, ranges from energy production and transport to energy distribution.

4. The Data Stabilizing System has to eliminate the use of mechanisms possibly affecting the deterministic behavior, for instance by not using dynamic memory allocation during the critical phases of real-time applications.

5. A major focus of the system is on service portability (see Chapter 2) in order to have minimal dependencies with specific hardware platforms or specific operating systems.

6. The system has to be scalable, at least from 1 to 8 processors.

In the following the functionality of the Data Stabilizing System is deduced from the requirements stated in the last paragraph.
First one needs to clarify the concept of data stabilization. Let us assume one wants to develop a controller with a short cycle time with respect to the dynamics of the input and the output data. Disturbances from temporal faults can influence either the input or the output data. These temporal effects, in particular on the output data, must be eliminated. For this reason, the controller is run several times with the same input data. Let us furthermore assume that, in the absence of faults, the same output data are produced. This allows the output of the controller from several runs to be compared. If the output does not change in a number of consecutive cycles, the output is staid to be stable. The described process of repeated runs of the controller and comparison of the results is called stabilization.
The described procedure of cyclic repetition of a process is the basis of temporal redundancy (that is, redoing). In order to detect and overcome transient faults, even if their characteristics such as distribution of frequency and duration are not known, temporal redundancy can be applied. If the computation time for a process is outspoken longer than the expected duration of a transient fault, and the frequency of the disturbances is low enough, it is assumed that in several repetitive computations of the same data only one fault may show up. So if the algorithm performs the same computation several times, there will be a period of some consecutive, not impaired results.

> The number $N_{time}$ of consecutive equal data inputs to the Data Stabilizing System is the level of temporal redundancy. It is the minimum number of cycles the Data Stabilizing System has to execute until a new input can be assumed to be stable.

Another strong requirement of our design is that of maximizing data integrity. To reach this goal, the Data Stabilizing System tool adopts a strategy, to be described later on, aiming at ensuring that data are only allowed to be stored in the Data Stabilizing System the moment they have been certified as being "stable". On a read request from a given memory location, the Data Stabilizing System will then return the last stabilized data, while a write into Data Stabilizing System will actually take place only when the strategy guarantees that data that are going to be written are stable. Another important requirement is that permanent faults affecting the system should not destroy the data. A standard method for increasing the reliability of memory is replication of data in redundant memories: a "write" is then translated into writing into each of a set of redundant memories, with voting of the data when reading out. An approach like the one described in Chapter 4, that is, redundant variables, was not available yet and therefore it was not used in this case. No additional hardware is required for this, as the writings are done in the memories of the processing nodes of the target, distributed memory platform.

> Using the principles of spatial redundancy, the same data are replicated in different memory areas—let us call them banks. The spatial redundancy factor $N_{spat}$ is the number of replicas stored in the Data Stabilizing System. Changing this parameter the user is allowed to trade off dependability with performance and resource consumption.

In order to fulfill the above mentioned requirements the Data Stabilizing System implements a strategy based on two buffers, one for reading the last stabilized data, and the other for receiving the new data. These two buffers are called the *memory banks*.

- The bank used for the output of the stabilized data is called the *current bank*.
- The other bank, called *future bank*, receives the $N_{time}$ input data for the Data Stabilizing System one after the other and checks whether the results are stable.

If the results are stable the role of the banks is switched, so that the future bank becomes the new current bank and the output data are fetched from there.
During the design and implementation phases of the Data Stabilizing System tool, the idea arose to isolate the spatial redundancy from the tool to build a custom tool especially devoted to the distributed memory approach—the Distributed Memory Tool. It showed that this approach simplifies the design and the implementation of the Data Stabilizing System tool.

### 7.2.1 The Distributed Memory Tool

The Distributed Memory Tool is the sub-system responsible for the management of the spatial redundancy scheme for the Data Stabilizing System.
Let us call a *local user context* either a thread or a process, which the user application sees as one task with its own local environment.
Assume that the user application consists of several such local user contexts, which are distributed among several nodes of a multiprocessor system. The basic component of the distributed memory tool is the *local handler*, which is defined as follows:

> A local memory handler is a local software module connected to one local user context and to a set of fully interconnected fellows. The attribute "local" means that both user context and memory handler run on the same processing node and they represent the whole tool from the viewpoint of the processing node.

As a consequence of this definition, the local user context regards the local memory handler as the only interface to the distributed memory tool. The local memory handler and the attached user module are connected via an IPC mechanism based on shared memory, referred to in the following as a "local link" and borrowed from the EPX terminology[5]. Commands to the distributed memory or messages from the memory will only flow between local user context and local memory handler. The tasks of the local memory handler are completely transparent to the user module. The same design concepts used for the EFTOS Voting Farm and described in Sect. 3 have been used here.

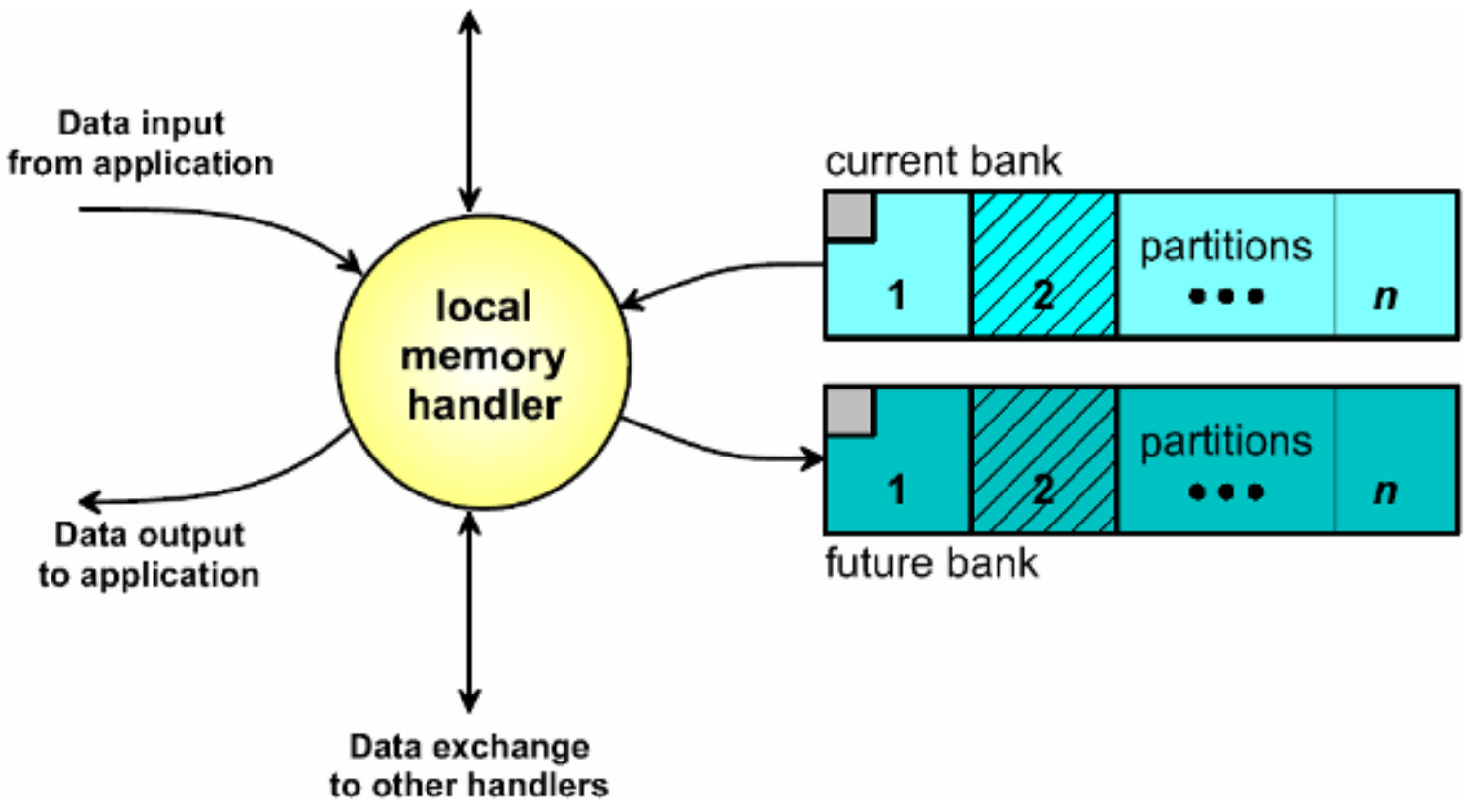

Figure 16: The local memory handler with its memory banks. The associated partitions are represented in a hatched way.

### 7.2.2 The Local Memory Handler and Its Tasks

As mentioned in the previous section, the local memory handler (see Fig. 16) is responsible for the management of two banks of memory, i.e., the current bank, that is, the bank where all read access take place, and the future bank, which is the bank for all writing actions. Each bank is cut into $N_{spat}$ partitions, where each handler is responsible for exactly one partition. This partition can be seen as the part of the redundant memory attached to the local user context, which is assigned to the local handler. This partition is referred to as the one *associated* to the user module. If a user module (i.e. a local user context) initiates to write data into its partition, this is done via a command to the local memory handler, which is sent via the connecting local link. The local handler then stores the data into the associated partition, and distributes them to the other local memory handlers residing on the other nodes. With this method the data are distributed as soon as they are received from the application.

For reading from the local memory handler there are several concurrent commands available. The common way is to request voted data from the local memory handler. In this case, one local handler receives a request for voted data from the attached user module. It then informs all other handlers and requests a voting service to vote among the replicas of the partition associated to the requesting user module. The result of the voting is provided to the calling user module as result of the read action. The kind of voting is user definable among those treated in (Lorczak et al., 1989). If a voting task fails, a time-out system allows regarding such an event as the delivery of a dummy vote. If the user application has a cyclic behavior, such that the user modules on the different nodes all execute the same cycle, and under the hypothesis of each node serving exclusively the same set of tasks, then under the hypothesis of a synchronous system model it is possible to assume that the requests for reading may be processed more or less simultaneously on all nodes. In this case, this information can be used to synchronize the nodes via the set of local memory handlers and some data transfer between the nodes can be run in an optimized way.
Clearly the Distributed Memory Tool is not aware of the logics pertaining to the stabilization mechanism, hence it is a task of the user application to inform the local memory handler about when to switch the banks (in the next section it is shown how the temporal redundancy tasks take care of this). Normally this can be done once per cycle. The switching can also be connected with a checking phase, testing whether the current banks are equal on all nodes by means of an equality voting.
Clearly, both determining the role of each bank and switching these roles are crucial points for the whole strategy. In particular, these actions need to be atomic. A fast and effective way to reach this property is the use of two binary flags—one per bank—whose contents concurrently determines the roles of the banks. These flags have been protected by storing them in the bank themselves. Table 5 shows how the coding of the flags in both banks is done. The idea is that, for changing from one state to another, *just one write action is needed in the future bank*. The current bank can therefore be regarded as being a readonly memory bank. As an example if the actual state is state 2, and one wants to switch the banks, then it suffices to change the flag in the future bank, which is bank $A$. Changing Flag A from 0 to 1 brings the system to state 4.

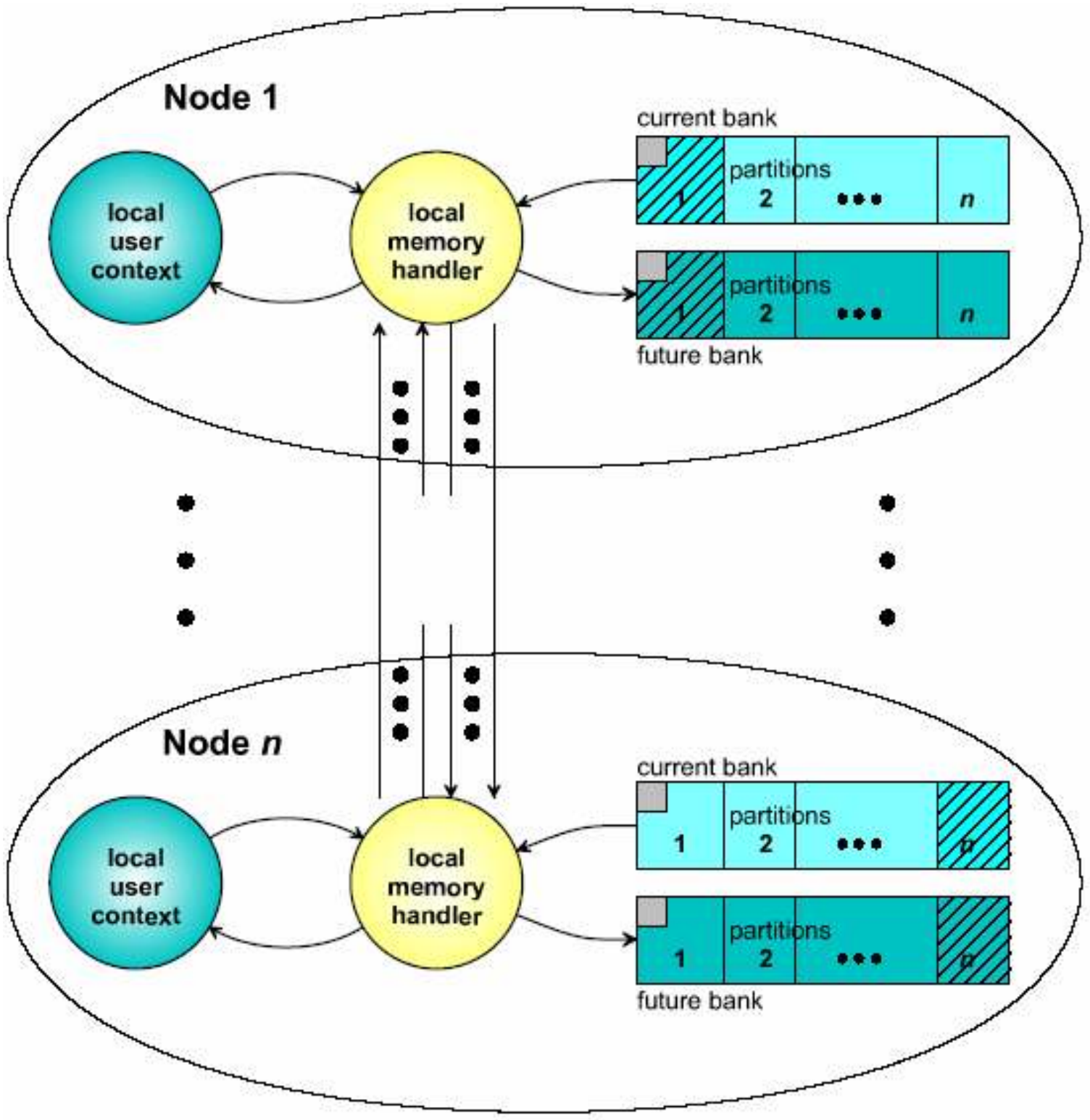

Figure 17: Structure of the Distributed Memory Tool. The associated partitions are represented hatched.

| State | Flag A | Flag B | Current | Future |
|---|---|---|---|---|
| 1 | 0 | 0 | A | B |
| 2 | 0 | 1 | B | A |
| 3 | 1 | 0 | B | A |
| 4 | 1 | 1 | A | B |

Table 5: Coding of the current and future bank flags

#### 7.2.3 Application-program Interface and Client side

Using function calls the user module is able to initialize the tool, to setup the net of local handlers, and to activate the tool. This process is done in several steps:

1. Each instance of a DMT is built up by declaring a pointer to a corresponding data structure:

   ```
   dmt_t *dmt;
   ```

   This data structure is the place to hold all information for one instance of the Distributed Memory Tool on each node. So each user module that wants to use the Distributed Memory Tool needs to declare a variable of this type.

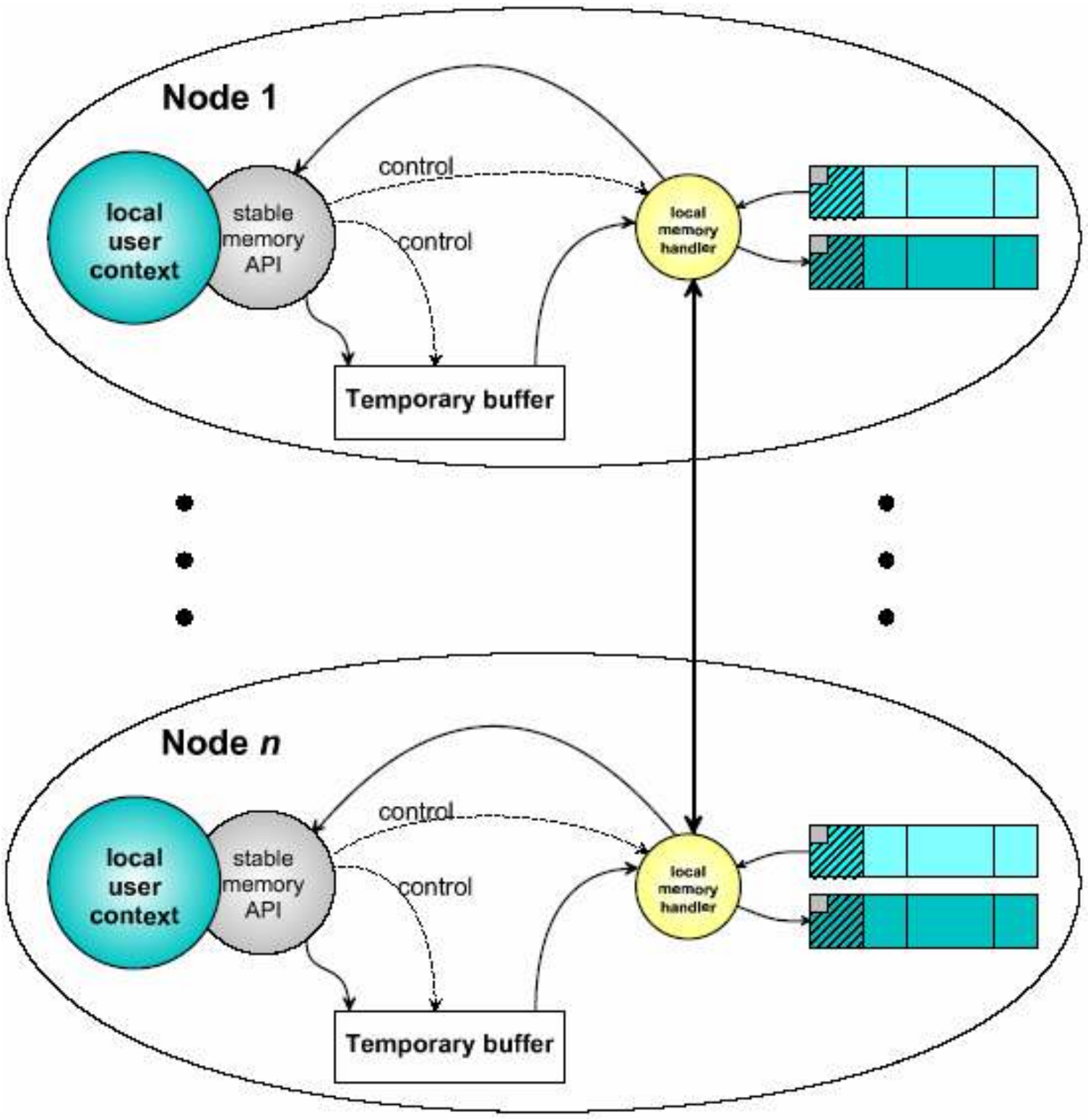

Figure 18: Structure of the Data Stabilizing System Tool.

2. In the next step, the Distributed Memory Tool is defined and described. This creates a static map which holds all necessary information to drive the tool on each node. The following code fragment:

```
dmt = DMT_open (id, VotingAlgorithm);
for (i = 0; i < NumHandlers; i++)
DMT_add (dmt, i, PartitionSize,
(i == MyNodeId));
```

   when executed on every node where the Distributed Memory Tool is intended to run, sets up a local memory handler to be used by the tool. The voting algorithm can be selected by the user via a kind of call-back function.

3. After the definition and description of the Distributed Memory Tool, the latter has to be started to set up data structures and threads. This activation is done via function

```
int DMT_run (dmt_t *dmt);
```

This function simply spawns the local memory handler thread after having checked the consistency of the structures defined in the description phase. All allocation of memory is done in the handler itself.

## 7.3 The Data Stabilizing System Tool

As already mentioned, the Data Stabilizing System tool builds on top of the Distributed Memory Tool. The latter is used for the management of the spatial redundancy (see Fig. 18), while the Data Stabilizing System takes care of the management of the temporal redundancy strategy. On each node the writing requests to the Data Stabilizing System are done into a temporal redundancy buffer, which holds a user definable number $N_{time}$ of copies of the last inputs. As the temporal buffers are only handled locally, this fits well to the concept of local memory handler. The Data Stabilizing System module performs a voting on the contents of the temporal buffers and, if this voting is successful, the result is fed into the Distributed Memory Tool. The Data Stabilizing System handles all accesses as well as the temporal voting transparently of the local user context.

### 7.3.1 Algorithms

The Data Stabilizing System Module as a whole gives each local user context a combination of temporal and spatial redundant memory buffers, which are able to keep the local state variables. The set of all local user contexts that store data in an instance of a Data Stabilizing System is called the *context family* associated to that tool. The Data Stabilizing System completely hides the memory handling and the handling of the Distributed Memory Tool, so that the user only needs to write to and read from the memory—all other handling is done transparently. This guarantees an acceptable separation of design concerns (SC), which could be further improved by using some translator as in Chapter 4.
The following steps are executed automatically:

1. The new data is written into the temporal buffer.

2. Using internal flag values the current and the future banks are determined. This is done within the Distributed Memory Tool, therefore only the flag values of the spatial redundancy buffers are used.

3. A voting on the temporal buffer takes place. If the temporal buffer is stabilized, i.e. the voting is positive, the content of the temporal buffer is stored into the spatial redundancy buffers, i.e. the respective calls to the DMT are done.

4. If the evaluation of the internal flags allows it, the Distributed Memory Tool switches the memory banks.

5. A voting is applied to evaluate the output of the current bank in the distributed memory. Such output is returned to the calling local user context.

### 7.3.2 Application-Program Interface

The Data Stabilizing System is identified via a control block structure. It must be seen in connection with the Distributed Memory Tool—in fact it is a kind of front end to that tool, which provides additional functionality. The steps to be performed to set up the Data Stabilizing System are similar to those elaborated for the Distributed Memory Tool:

1. Each instance of a Data Stabilizing System Tool is built up by declaring a pointer to a corresponding data structure:

   ```
   smCB_t * sm;
   ```

   Each local user context that is member of the associated context family needs to declare a pointer to a variable of type smCB_t. Similarly to the Distributed Memory Tool, the Data Stabilizing System Tool has to be defined and described. With the SM_open statement a local instance of the system is created. In addition to the parameters of the DMT_open statement some parameters regarding the temporal redundancy are passed to this statement.

   ```
   sm = SM_open (id, N_time,
   TemporalVotingAlgorithm, SpatialVotingAlgorithm);
   for (i = 0; i< N_spat; i++)
   SM_add(sm, i, PartitionSize, (i == MyNodeId) ) ;
   ```

   The above code sets up an instance of the Data Stabilizing System tool, provided it is called in every local user context that needs to take part in the processing of the tool.

2. Up to this point the Data Stabilizing System Tool is defined and described, that is its structure is set up, but no instance has been installed, no memory has been allocated and no handler for the spatial redundant memory has been started yet. To do this the tool must be activated. This is done via the function SM_run. This function allocates the temporal redundancy buffers, initializes the variables, and activates the attached Distributed Memory Tool using the function DMT_run. The Data Stabilizing System can only be started up if all local user contexts belonging to its context family call SM_run at the same point in their start-up phase.

At run-time, the Data Stabilizing System Tool is controlled via two functions, which read the data from the application or provide stabilized and voted data to the application:

```
int SM_write (smCB_t *MyCB, void *SM_in);
int SM_read (smCB_t*MyCB, void *SM_out);
```

The user provides data to the Data Stabilizing System via function SM_write, which is then responsible for the handling of these values. This function is used to input the local data of a local user context to the Data Stabilizing System. The parameters submitted to the function describe the Data Stabilizing System control block structure (MyCB) and the address of the data to be copied into the Data Stabilizing System (SM_in). When the function is returned, all data provided to the function using the SM_in pointer are copied out of this memory location into the temporary buffer of the Data Stabilizing System. The SM_read function writes the local data of the calling local user context back to the address submitted through the pointer SM_out. The Data Stabilizing System control block structure MyCB is used to identify the Data Stabilizing System.
Switching from the current to the future bank is achieved by means of the following procedure:

- Determine the current meaning of the banks from their internal flags.
- Write data into the specified partition(s) of the future bank of the memory.
- Output data via a voting between the distributed copies of the current bank.
- If the contents of the future banks of all nodes are equal, switch the role of the memory banks.

### 7.3.3 System and Fault Models

As already mentioned, the embedded and hard real-time character of the application makes it reasonable to assume a synchronous system model. The overall strategy implemented in the Data Stabilizing System allows to mask a number of transient faults resulting in:

- An erroneous input value.
- Errors affecting the circular buffer.
- Errors affecting the temporal redundancy modules.
- Errors affecting the flag values,

occurring during the execution cycle, or caused by an external disturbance, or a wrong flag value, and so forth. These are tolerated either through the voting sessions (temporal redundancy) or are masked via the periodic restarts which invalidate the current cycle. In this latter case this is therefore perceived as a delay of the stabilized output. The same applies when a fault affects the phase of determining the current bank, or faults occurring during the voting among temporal redundancy modules, or faults affecting the spatial redundancy modules. More details on this can be found in [4]. Tolerance of permanent faults resulting in node crashes is achieved by using the Data Stabilizing System as a dependable mechanism compliant to the recovery language approach described in Chapter 6. As explained in that chapter, this approach exploits a high level distributed application (the TIRAN backbone) and a library of error detection tools in order to detect events such as node and task crashes. User defined error recovery actions can then be attached to the error detection events so as to trigger corrective actions such as a reconfiguration of the Data Stabilizing System tasks.

### 7.4 Conclusion

In the text above a software system implementing a data stabilizing tool has been described. Such tool is to be placed in highly disturbed environments such as those typical of sub-station automation. Due to its design, based on a combination of spatial and temporal redundancy and on a cyclic restart policy, the tool proved to be capable of tolerating transient and permanent faults and to guarantee data stabilization. Initially developed on a Parsytec CC system, the tool has been then ported to several runtime systems. The most important lessons learned while developing this tool are those that brought the author of this book to the concept of *service* portability introduced in Chapter 2: A software code such as the one of the system described so far can be ported to a different environment with a moderated effort; but porting the service is indeed something else. In this case, a thorough evaluation of the new working environment is due in order to come up with proper new values for parameters such as $N_{time}$ and $N_{spat}$. A system like the Data Stabilizing tool puts this requirement in the foreground and makes it possible to perform an off-line adaptability of its service. As a consequence A is assessed as "moderate". Augmenting the approach towards acceptable degrees of A would call for the adoption of a strategy such as the one used in redundant variables (see Chapter 4). As a final remark, its single-purpose design intrinsically translates in bad SA.

## 8 An Approach to Express Recovery Blocks: The Recovery Meta-Program

The Recovery Meta-Program (RMP) (Ancona, Dodero, Giannuzzi, Clematis, & Fernandez, 1990) is a mechanism that alternates the execution of two cooperating processing contexts. The concept behind its architecture can be captured by means of the idea of a debugger, or a monitor, which:

- is scheduled when the application is stopped at some *breakpoints*,
- executes some sort of a *program*, written in a specific language,

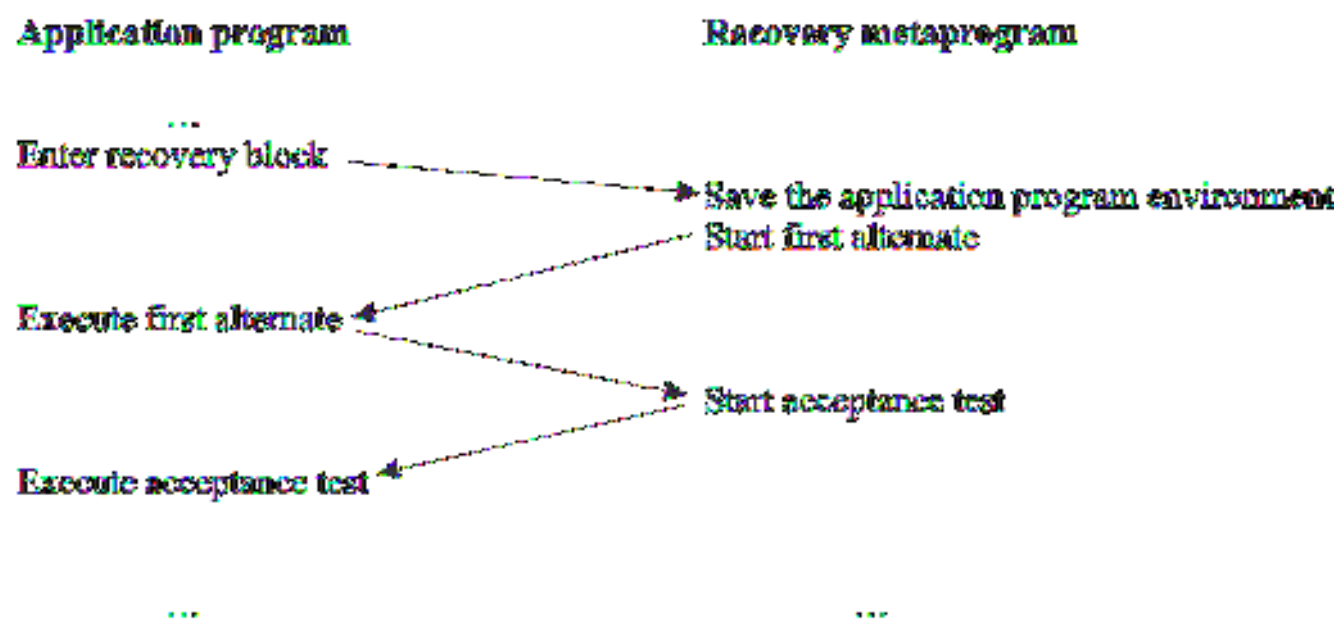

Figure 19: Control flow between the application program and RMP while executing a fault-tolerance strategy based on recovery blocks.

- and finally returns the control to the application context, until the next breakpoint is encountered.

Breakpoints outline portions of code relevant to specific fault-tolerance strategies—for instance, breakpoints can be used to specify alternate blocks or acceptance tests of recovery blocks (see Sect. 2.3)—while programs are implementations of those strategies, e.g., of recovery blocks or $N$-version programming. The main benefit of RMP is in the fact that, while breakpoints require a (minimal) intervention of the functional-concerned programmer, RMP scripts can be designed and implemented without the intervention and even the awareness of the developer. In other words, RMP guarantees a good separation of design concerns. As an example, recovery blocks are implemented, from the point of view of the functionally concerned designer, specifying alternates and acceptance tests, while the execution goes like in Fig. 19:

- When the system encounters a breakpoint corresponding to the entrance of a recovery block, control flows to the RMP, which saves the application program environment and starts the first alternate.
- The execution of the first alternate goes on until its end, marked by another breakpoint. The latter returns the control to RMP, this time in order to execute the acceptance test.
- Should the test succeed, the recovery block is exited, otherwise control goes to the second alternate, and so forth.

In RMP, the language to express the meta-programs is Hoare's Communicating Sequential Processes language (Hoare, 1978) (CSP).

**Conclusions.** In the RMP approach, all the technicalities related to the management of the fault-tolerance provisions are coded in a separate programming context. Even the language to code the provisions may be different from the one used to express the functional aspects of the application. One can conclude that RMP is characterised by optimal SC.
The design choice of using CSP to code the meta-programs influences negatively attribute SA. Choosing a pre-existent formalism clearly presents many practical advantages, though it means adopting a fixed, immutable syntactical structure to express the fault-tolerance strategies. The choice of a pre-existing general-purpose distributed programming language as CSP is therefore questionable, as it appears to be rather difficult or at least cumbersome to use it to express at least some of the fault-tolerance provisions. For instance, RMP proves to be an effective linguistic structure to express strategies such as recovery blocks and $N$-version programming (Yeung & Schneider, 2003), where the main components are coarse grain processes to be arranged into complex fault-tolerance structures. Because of the choice of a pre-existing language like CSP, RMP appears not to be the best choice for representing provisions such as, e.g., atomic actions (Jalote & Campbell, 1985). This translates in very limited SA. No A was foreseen among the design choices of RMP.
Our conjecture is that the coexistence of two separate layers for the functional and the non-functional aspects could have been better exploited to reach the best of the two approaches: Using a widespread programming language such as Java for expressing the functional aspect, while devising a custom language for dealing with non-functional requirements, e.g., a language especially designed to express error recovery strategies. This design choice has been taken in the approach described in Chapter 6, the recovery language approach.

# 9 A Hybrid Case: The RAFTNET Library for Dependable Farmer-Worker Parallel Applications

RAFTNET is a tool to compose dependable parallel applications obeying the farmer-worker data parallel paradigm. It is described here as a hybrid example of a system that appears to its users as a library, hence a single-version software fault-tolerance provision, though at run-time manages a potentially large degree of redundant components—a typical characteristics of multiple-version software fault-tolerance.

## 9.1 Introducing RAFTNET

The RAFTNET Library is another example of a library to build fault-tolerant services. The main difference between RAFTNET and a system such as the Voting Farm is that it does not provide the application programmer with a *dependable mechanism* (in this case, distributed voting), but rather it provides a *dependable structure* for a class of target applications. In more detail, RAFTNET is a library for data parallel, farmer-worker applications: Any such applications using RAFTNET makes uses of the available redundancy not only to reach higher performance but also to tolerate certain faults and disruptions that would normally jeopardize its progress. In the following the structure of RAFTNET, its models, properties, and features are described.

## 9.2 Why Dependable Parallel Applications?

Parallel computing is nowadays the only technique that can be used in order to achieve the impressive computing power needed to solve a number of challenging problems; as such, it is being employed by an ever growing community of users in spite of what are known as two main disadvantages, namely:

1. harder-to-use programming models, programming techniques and development tools—if any,—which sometimes translate into programs that don't match as efficiently as expected with the underlying parallel hardware, and

2. the inherently lower level of dependability that characterizes any such parallel hardware i.e., a higher probability for events like a node's permanent or temporary failure.

A real, effective exploitation of any given parallel computer asks for solutions which take into a deep account the above outlined problems.
Let us consider for example the synchronous farmer-worker algorithm i.e., a well-known model for structuring data-parallel applications: a master process, namely the farmer, feeds a pool of slave processes, called workers, with some units of work; then polls them until they return their partial results which are eventually recollected and saved. Though quite simple, this scheme may give good results, especially in homogeneous, dedicated environments.
But how does this model react to events like a failure of a worker, or more simply to a worker's performance degradation due e.g., to the exhaustion of any vital resource? Without substantial modifications, this scheme is not able to cope with these events—they would seriously affect the whole application or its overall performances, regardless the high degree of hardware redundancy implicitly available in any parallel system. The same inflexibility prevents a failed worker to re-enter the computing farm once it has regained the proper operational state.
As opposed to this synchronous structuring, it is possible for example to implement the farmer-worker model by de-coupling the farmer from the workers by means of an intermediate module, a dispatcher which asynchronously feeds these latter and supplies them with new units of work on an on-demand basis. This strategy guarantees some sort of a dynamic balancing of the workload even in heterogeneous, distributed environments, thus exhibiting a higher matching to the parallel hardware. The Live Data Structure computational paradigm, known from the LINDA context, makes this particularly easy to set up (see for example (Carriero & Gelernter, 1989a, 1989b; De Florio, Murgolo, & Spinelli, 1994)).
With this approach it is also possible to add a new worker at run-time without any notification to both the farmer and the intermediate module—the newcomer will simply generate additional, non-distinguishable requests for work. But again, if a worker fails or its performances degrade, the whole application may fail or its overall outcome be affected or seriously delayed. This is particularly important when one considers the inherent loss in dependability of any parallel (i.e., replicated) hardware.
Next sections introduce and discuss a modification to the above sketched asynchronous scheme, which inherits the advantages of its parent and offers new ones, namely:

- it allows a non-solitary, temporarily slowed down worker to be left out of the processing farm as long as its performance degradation exists, and
- it allows a non-solitary worker which has been permanently affected by some fault to be definitively removed from the farm,

both of them without affecting the overall outcome of the computation, and dynamically spreading the workload among the active processors in a way that results in an excellent match to various different MIMD architectures.

## 9.3 The Technique

For the purpose of describing the technique the following scenario is described: a MIMD machine consists of $n + 2$ identical "nodes" ($n > 0$), or processing entities, connected by some communication line. On each node a number of independent sequential processes are executed on a time-sharing basis. A message passing library is available for sending and receiving messages across the communication line. A synchronous communication approach is used: a sender blocks until the intended receiver gets the message. A receiver blocks waiting for a message from a specific sender, or for a message from a number of senders. When a message arrives, the receiver is awaken and is able to receive that message and to know the identity of the sender. Nodes are numbered from 0 to $n + 1$. Node 0 is connected to an input line and node $n + 1$ is connected to an output line.

- Node 0 runs:
  - a Farmer process, connected by the input line to an external producer device. From now on a camera is assumed to be the producer device. A control line wires again the Farmer to the camera, so that this latter can be commanded to produce new data and eventually send this data across the input line;
  - a Dispatcher process, yet to be described.

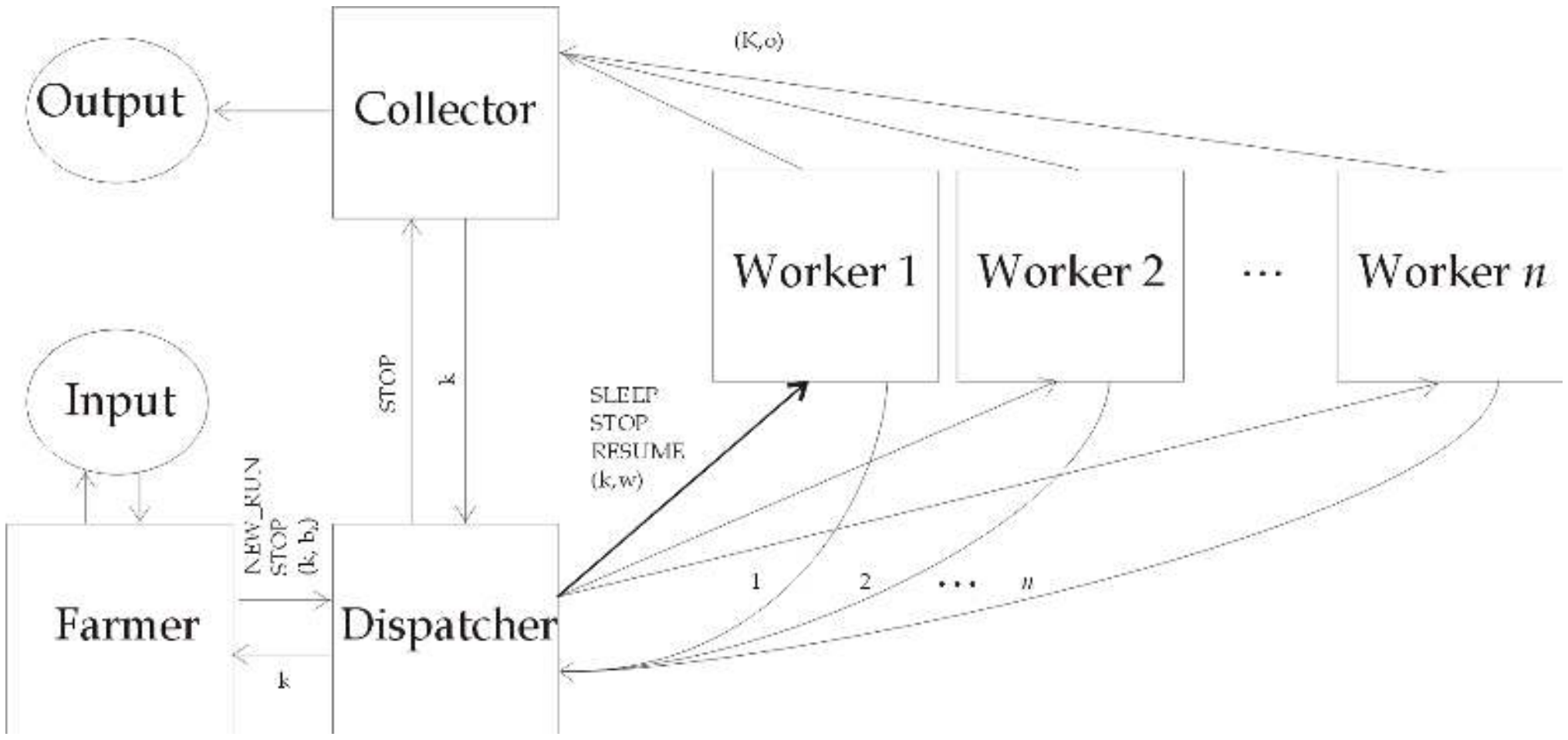

Figure 20: Summary of the interactions among the processes.

- Node $n+1$ runs a Collector process, to be described later on, connected by the output line to an external storage device e.g., a disk;
- Each of the nodes from 1 to $n$ is purely devoted to the execution of one instance of the Worker process. Each Worker is connected to the Dispatcher and to the Collector processes.

## 9.4 Interactions Between the Farmer and the Dispatcher

On demand of the Farmer process, the camera sends it an input image. Once it has received an image, the Farmer performs a predefined, static data decomposition, creating $m$ equally sized sub-images, or blocks. Blocks are numbered from 1 to $m$, and are represented by variables $b_i, 0 < i < m+1$. The Farmer process interacts exclusively with the camera and with the Dispatcher process.

- Three classes of messages can be sent from the Farmer process to the Dispatcher (see Fig. 20):

    1. a `NEW_RUN` message, which means: "a new bunch of data is available";
    2. a `STOP` message, which means that no more input is available so the whole process has to be terminated;
    3. a couple $(k, b_k), k$ in $\{1, \ldots m\}$ i.e., an integer which identifies a particular block (it will be referred from now on as a "block-id"), followed by the block itself.

- The only type of message that the Dispatcher process sends to the Farmer process is a block-id i.e., a single integer in the range $\{1, \ldots, m\}$ which expresses the information that a certain block has been fully processed by a Worker and recollected by the Collector (see Sect. 9.4.2.)

At the other end of the communication line, the Dispatcher is ready to process a number of events triggered by message arrivals. For example, when a class-3 message is received, the block is stored into a work buffer as follows:

`receive` $(k, b_k)$
$s_k =$ `DISABLED`
$w_k = b_k$

(Here, `receive` is the function for receiving an incoming message, $\vec{s}$ is a vector of $m$ integers pre-initialized to `DISABLED`, which represents some status information that will be described later on, and $\vec{w}$ is a vector of "work buffers", i.e., bunches of memory able to store any block. `DISABLED` is an integer which is not in the set $\{1, \ldots, m\}$.)
As the Farmer process sends a class-1 message, that is, a `NEW_RUN` signal, the Dispatcher processes that event as follows:

$\vec{s} = 0$
`broadcast RESUME`

that is, it zeroes each element of $\vec{s}$ and then broadcasts the `RESUME` message to the whole farm.

When the first image arrives to the Farmer process, it produces a series $(b_i)_{0<i<m+1}$, and then a sequence of messages $(i, b_i)_{0<i<m+1}$. Finally, the Farmer sends a `NEW_RUN` message.
Starting from the second image, and while there are images to process from the camera, the Farmer performs the image decomposition in advance, thus creating a complete set of $(k, b_k)$ couples. These couples are then sent to the Dispatcher on an on-demand basis: as soon as block-id $i$ is received, couple $(i, b_i)$ is sent out. This is done for anticipating the transmission of the couples belonging to the next run of the computation. When eventually the last block-id of a certain run has been received, a complete set of "brand-new" blocks is already in the hands of the Dispatcher; at that point, sending the one `NEW_RUN` message will simultaneously enable all blocks.

### 9.4.1 Interactions Between the Dispatcher and the Workers

The Dispatcher interacts with every instance of the Worker process.

- Four classes of messages can be sent from the Dispatcher to the Workers (see Fig. 20):
    1. a `SLEEP` message, which sets the receiver into a wait condition;

2. a `RESUME` message, to get the receiver out of the waiting state;
3. a `STOP` message, which makes the Worker terminate;
4. a $(k, w)$ couple, where $w$ represents the input data to be elaborated.

- Worker $j$, $0 < j < n+1$, interacts with the Dispatcher by sending it its worker-id message, i.e., the $j$ integer. This happens when Worker $j$ has finished dealing with a previously sent $w$ working buffer and is available for a new $(k, w)$ couple to work with.

In substance, Worker $j$ continuously repeats the following loop:

```
send j to Dispatcher
receive message from Dispatcher
process message
```

Clearly, `send` transmits a message. The last instruction, in dependence with the class of the incoming message, results in a number of different operations:

- if the message is a `SLEEP`, the Worker waits until the arrival of a `RESUME` message, which makes it resume the loop, or the arrival of any other message, which means that an error has occurred;
- if it is a `STOP` message, the Worker breaks the loop and exits the farm;
- if it is a $(k, w)$ couple, the Worker starts computing the value $f(w)$, where $f$ is some user-defined function e.g., an edge detector. If a `RESUME` event is raised during the computation of $f$, that computation is immediately abandoned and the Worker restarts the loop. Contrariwise, the output couple $(k, f(w))$ is sent to the Collector process.

When the Dispatcher gets a $j$ integer from Worker $j$, its expected response is a new $(k, w)$ couple, or a `SLEEP`. What rules in this context is the $\vec{s}$ vector—if all entries of $\vec{s}$ are `DISABLED`, then a `SLEEP` message is sent to Worker $j$. Otherwise, an entry is selected among those with the minimum non-negative value, say entry $l$, and a $(l, b_l)$ message is then sent as a response. Then $s_l$ is incremented by 1.
More formally, considered set $S = \{s \text{ in } \vec{s} \,|\, s \neg = \texttt{DISABLED}\}$, if $S$ is non-empty it is possible to partition $S$ according to the equivalence relation $R$ defined as follows:

$$\text{For all } (a, b) \text{ in } S \times S : aR\,b \text{ if and only if } s_a = s_b.$$

So the blocks of the partition are the equivalence classes:

$$[x] \stackrel{\text{def}}{=} \{s \text{ in } S \,|\, \text{ Exists } y \text{ in } \{1 \ldots m\} \text{ such that } (s = s_y) \text{ and } (s_y = x)\}.$$

Now, first let us consider

$$a = \min\{b \,|\, \text{ Exists } b >= 0 \text{ such that } [b] \text{ in } S/R\};$$

then $l$ is chosen in $[a]$ in some way, e.g. pseudo-randomly; finally, message $(l, b_l)$ is sent to Worker $j$, $s_l$ is incremented, and the partition is reconfigured accordingly. If $S$ is the empty set, a `SLEEP` message is generated.

In other words, entry $s_i$ when greater than or equal to 0 represents some sort of a priority identifier (the lower the value, the higher the priority for block $b_i$). The block to be sent to a requesting Worker process is always selected among those with the highest priority; after the selection, $s_i$ is updated incrementing its value by 1. In this way, the content of $s_i$ represents the degree of "freshness" of block $b_i$: it substantially counts the number of times it has been picked up by a Worker process; fresher blocks are always preferred. As long as there are "brand-new" blocks i.e., blocks with a freshness attribute of 0, these are the blocks which are selected and distributed. Note that this means that as long as the above condition is true, each Worker deals with a different unit of work; on the contrary, as soon as the last brand-new block is distributed, the model admits that a same block may be assigned to more than one Worker.

This is tolerated up to a certain threshold value; if any $s_i$ becomes greater than that value, an alarm event is raised—too many workers are dealing with the same input data, which might mean that they are all affected by the same problem e.g., a software bug resulting in an error when $b_i$ is being processed. This special case shall not be considered. Another possibility is that two or more Workers had finished their work almost at the same time thus bringing rapidly a flag to the threshold. Waiting for the processing time of one block may supply the answer.

A value of `DISABLED` for any $s_i$ means that its corresponding block is not available to be computed. It is simply not considered during the selection procedure.

### 9.4.2 Interactions Between the Workers and the Collector

Any Worker may send one class of messages to the Collector; no message is sent from this latter to any Worker (see Fig. 20).

The only allowed message is the couple $(k, o)$ in which $o$ is the fully processed output of the Worker's activity on the $k^{\text{th}}$ block.

The Collector's task is to fill a number of "slots", namely $p_i, i = 1, \ldots, m$, with the outputs coming from the Workers. As two or more Workers are allowed to process a same block thus producing two or more $(k, o)$ couples, the Collector runs a vector of status bits which records the status of each slot: if $f_i$ is `FREE` then $p_i$ is "empty" i.e., it has never been filled in by any output before; if it is `BUSY`, it already holds an output. $\vec{f}$ is firstly initialized to `FREE`.

For each incoming message from the Worker, the Collector repeats the following sequence of operations:

```
receive (k, o) from Worker
if f_k is equal to FREE
   then
```

```
            send k to Dispatcher
            p_k = o
            f_k = BUSY
            check-if-full
      else
            detect
   endif
```

where:

`check-if-full` checks if, due to the last arrival, all entries of $\vec{f}$ have become BUSY. In that case, a complete set of partial outputs has been recollected and, after some user-defined post-processing (for example, a polygonal approximation of the chains of edges produced by the Workers), a global output can be saved, and the flag vector re-initialized:

```
if f is equal to BUSY
   then
         post-process p
         save p
         f = FREE
endif
```

`detect` is a user-defined functionality—he or she may choose to compare the two $o$'s so to be able to detect any inconsistency and start some recovery action, or may simply ignore the whole message.

Note also that an acknowledgment message (the block-id) is sent from the Collector to the Dispatcher, to inform it that an output slot has been occupied i.e., a partial output has been gathered. This also means that the Farmer can anticipate the transmission of a block which belongs to the next run, if any.

### 9.4.3 Interactions Between the Collector and the Dispatcher

As just stated, upon acceptance of an output, the collector sends a block-id, say integer $k$, to the Dispatcher—it is the only message that goes from the Collector to the Dispatcher.
The Dispatcher then simply acts as follows:

```
s_k = DISABLED
send k to Farmer
```

that is, the Dispatcher "disables" the $k^{\text{th}}$ unit of work—set $S$ as defined in Sect. 9.4.1 is reduced by one element and consequently partition $S/R$ changes its shape; then the block-id is propagated to the Farmer (see Fig. 20).
On the opposite direction, there is only one message that may travel from the Dispatcher to the Collector: the STOP message that means that no more input is available and so processing is over. Upon reception of this message, the Collector stops itself, like it does any other receiver in the farm.

## 9.5 Discussion

The just proposed technique uses asynchronicity in order to efficiently match to a huge class of parallel architectures. It also uses the redundancy which is inherent to parallelism to make an application able to cope with events like e.g., a failure of a node, or a node being slowed down, temporarily or not.

- If a node fails while it is processing block $k$, then no output block will be transferred to the Collector. When no more "brand-new" blocks are available, block $k$ will be assigned to one or more Worker processes, up to a certain limit. During this phase the replicated processing modules of the parallel machine may be thought of as part of a hardware redundancy fault tolerant mechanism. This phase is over when any Worker module delivers its output to the Collector and consequently all others are possibly explicitly forced to resume their processing loop or, if too late, their output is discarded.
- If a node has been for some reason drastically slowed down, then its block will be probably assigned to other possibly non-slowed Workers. Again, the first who succeeds, its output is collected; the others are stopped or ignored.

In any case, from the point of view of the Farmer process, all these events are completely masked. The mechanism may be provided to a user in the form of some set of basic functions, making all technicalities concerning both parallel programming and fault tolerance transparent to the programmer.
Of course, nothing prevents the concurrent use of other fault tolerance mechanisms in any of the involved processes e.g., using Watchdog timers to understand that a Worker has failed and consequently reset the proper entry of vector $\vec{f}$. The ability to re-enter the farm may also be exploited committing a reboot of a failed node and restarting the Worker process on that node.

### 9.5.1 Reliability Analysis

In order to compare the original, synchronous farmer-worker model with the one described in this paper, a first step is given by observing that the synchronous model depicts a *series system* (Johnson, 1989) i.e., a system in which each element is required not to have failed for the whole system to operate. This is not the case of the model described in this paper, in which a subset of the elements, namely the Worker farm, is a *parallel system* (Johnson, 1989): if at least one Worker has not failed, so it is for the whole farm subsystem. Note how Fig. 20 may be also thought of as the reliability block diagram of this system.

Considering the sole farm subsystem, if $C_i(t), 0 < i < n+1$ is the event that Worker on node $i$ has not failed at time $t$, and $R(t)$ is the reliability of any Worker at time $t$ then, under the assumption of mutual independency between the events, one can conclude that:

$$R_s(t) \stackrel{\text{def}}{=} P(\bigcap_{i=1}^{n} C_i(t)) = \prod_{i=1}^{n} R(t) = (R(t))^n \qquad (1)$$

being $R_s(t)$ the reliability of the farm as a series system, and

$$R_p(t) \stackrel{\text{def}}{=} 1 - P(\bigcap_{i=1}^{n} \overline{C_i}(t)) = 1 - \prod_{i=1}^{n}(1 - R(t)) = 1 - (1 - R(t))^n \qquad (2)$$

where $R_p(t)$ represents the reliability of the farm as a parallel system. Of course failures must be independent, so again data-induced errors are not considered. Figure 21 shows the reliability of the farm in a series and in a parallel system as a Worker's reliability goes from 0 to 1.

**An Augmented LINDA Model.** The whole idea pictured in this paper may be implemented in a LINDA tuple space manager (see for example (Carriero & Gelernter, 1989b, 1989a)). Apart from the standard functions to access "common" tuples, a new set of functions may be supplied which deal with "book-kept tuples" i.e., tuples that are distributed to requestors by means of the algorithm sketched in Sect. 9.4.1. As an example:

`fout` (for fault tolerant `out`) may create a book-kept tuple i.e., a content-addressable object with book-kept accesses;

`frd` (fault tolerant `rd`) may get a copy of a matching book-kept tuple, chosen according to the algorithm in Sect. 9.4.1;

`fin` (fault tolerant `in`) may read-and-erase a matching book-kept tuple, chosen according to the algorithm in Sect. 9.4.1,

and so on. The ensuing augmented LINDA model results in an abstract, elegant, efficient, dependable, and transparent mechanism to exploit a parallel hardware. Chapter 9 describes another example of a dependable LINDA model.

## 9.6 Implementation Issues

The RAFTNET system has been developed for network of workstations using a subset of the MPICH routines (MPICH (MPICH, n.d.) is a portable implementation of the MPI message passing library (Gropp, Lusk, & Skjellum, 1999)). The code of this system has been also ported to a “pure” TCP/IP environment by porting the MPICH subset (Raftnet, n.d.).

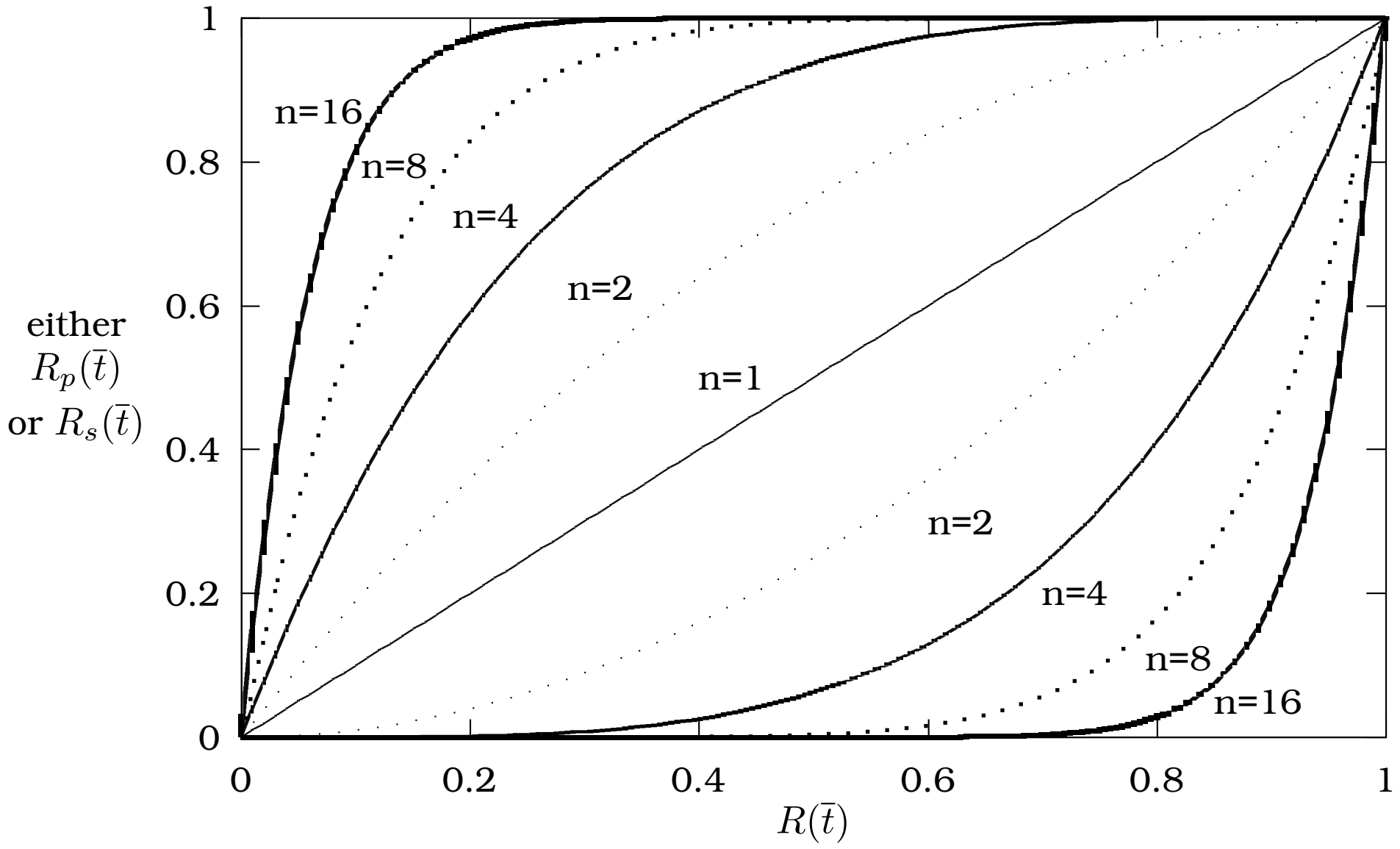

Figure 21: For a fixed value $\vec{t}$, a number of graphs of $R_p(\vec{t})$ (the reliability of the parallel system) and $R_s(\vec{t})$ (the reliability of the series system) are portrayed as functions of $R(\vec{t})$, the reliability of a Worker at time $\bar{t}$, and $n$, the number of the components. Each graph is labeled with its value of $n$; those above the diagonal portray reliabilities of parallel systems, while those below the diagonal pertain to series systems. Note that for $n = 1$ the models coincide, while for any $n > 1$ $R_p(\vec{t})$ is always above $R_s(\vec{t})$ except when $R(\vec{t}) = 0$ (no reliable Worker) and when $R(\vec{t}) = 1$ (totally reliable, failure-free Worker).

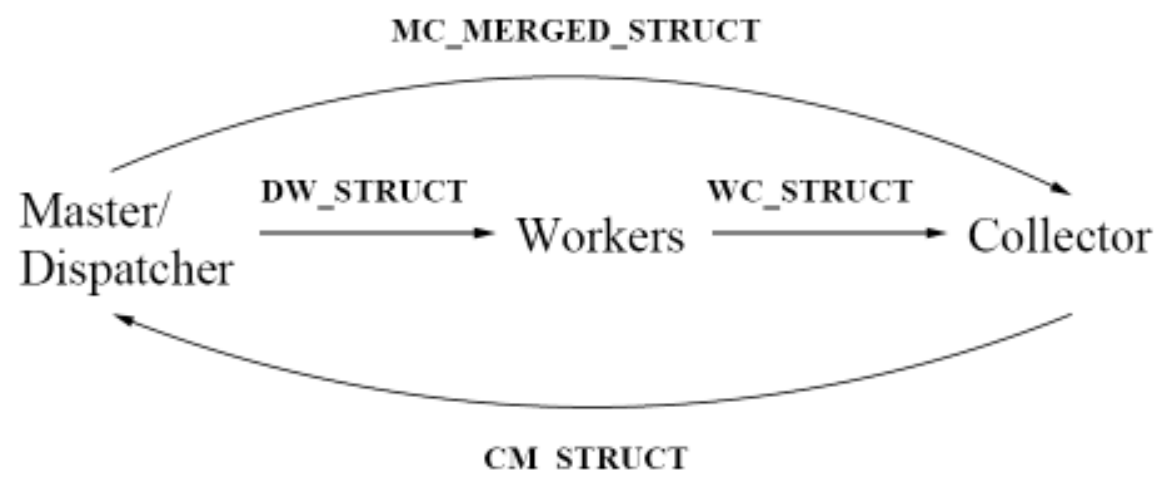

Figure 22: Data structures and data structure conversions in the masterworker library.

The implemented master-worker system acts as black box. Firstly, knowledge about the algorithm details is not required by the user application programmer. This allows a fast integration of the master-worker library in existing serial applications. Secondly, in RAFTNET data structures in the user application can take any form. They need not to be known beforehand by the master-worker algorithm. Both aspects contribute to the flexibility of the system. The resulting master-worker library can be integrated in a wide variety of programs, on the condition that a limited set of user application functions are written. Both ideas are explained in this section. The modified master-worker algorithm is implemented in the master-worker library. The application that is to be parallelized, resides in the user application library. A main process calls the master-worker library with a restricted set of arguments. Most of these arguments are pointers to functions in the user library for splitting, merging, post-processing the output data or for handling data structures. Communication between the master-worker and the user application library is solely via these functions. Writing these functions forms the main effort of the programmer for integrating the master-worker library with his/her user application library. This setup ensures that the master-worker algorithm can be used for a wide variety of programs. Furthermore, the detailed implementation of the master-worker algorithm remains a black box for the user.

The concept of the user application acting as a black box for the master-worker library, is mainly a matter of handling data structures and requires a more elaborated explanation. Often, the program flow can be regarded as a series of data structure conversions. To support a variety of applications, the master-worker library keeps track of various data structures (Fig. 22). DW STRUCT, WC STRUCT and CM STRUCT denote the input blocks, the processed blocks and the post-processed output data respectively. The prefixes are the first letters of source and destination. Conversion between these structures is performed in the user functions for processing and merging. MC MERGED STRUCT is a structure containing supporting information for merging the data. Fields in data structures are sent one at a time. Details about these structures are beforehand only known by the user library. On the other hand, the master-worker library needs to know these details for sending and receiving the data. Two functions written by the user application programmer, that are passed on via the master-worker interface, serve as a communication channel between user application library and master-worker library for these structure details (Fig. 23.)

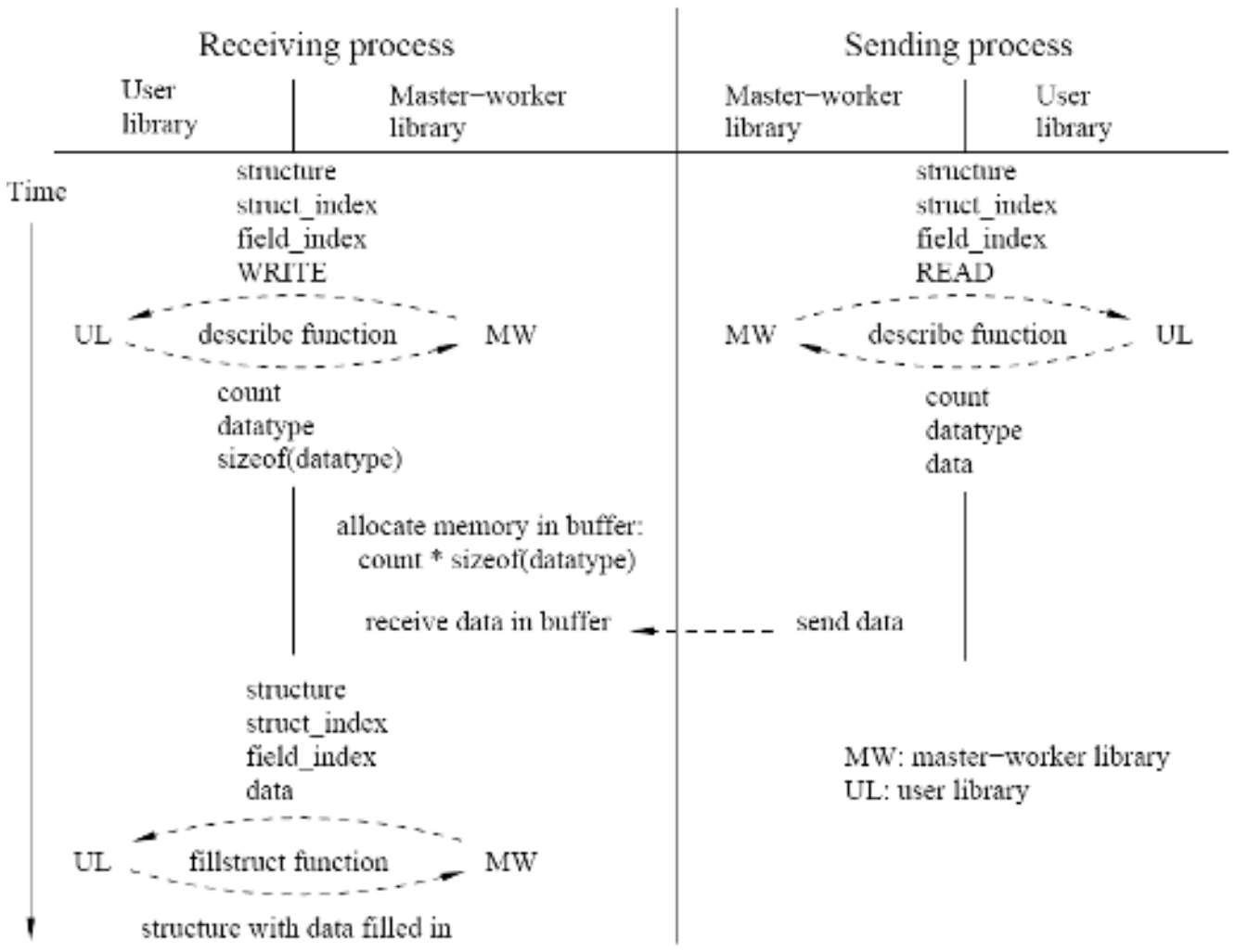

Figure 23: Process for sending and receiving one field of a data structure.

- The first function (`describe`) receives as input from the master-worker library, information such as the index of the structure (which is one of the structures in Fig. 22) and the field index. The `describe` function in the user library returns information such as the number of entries in that field (in case it is an array) and the data type. For the receiving process, the returned information allows allocating memory for the receive buffer by the master-worker library. For the sending process, also the data itself is returned to the master-worker library, which is transmitted immediately.
- After actually receiving this data, the receive process fills in the users data structure with this received field. The user function `fillstruct` receives via its arguments the structure and field index, the structure itself and the data. It returns the structure with the data field filled in.

Via this schedule, the master-worker library does not need to be aware of the data structure details—these are returned and filled in by the user application's `describe` and `fillstruct` functions. The user programmer is completely in control of the data structures that are used.

### 9.6.1 Failure semantics

Due to the characteristics of the underlying algorithm, our system compensates omission/performance and state-transition failures of its workers.

- Omission failures occur when an agreed reply to a request is missing. The request appears to be ignored. This is compensated by redistributing a request when its "freshness" allows it.
- Performance failures occur when the service is supplied, though too late with respect to some real-time interval possibly agreed upon in the specifications. This class of failures can be compensated by adopting a number of workers large enough to mask, e.g., crashing or late processing workers.
- State transition failures are also covered, as the state of the system is never lost because of a failing worker: Indeed, by construction, each state transition is atomic: A block is marked as processed only when its completion is explicitly acknowledged. Half finished blocks are simply discarded.

**Memory Allocation Delays and Failures.** A memory allocation failure can be due to either not enough available memory on that node possibly resulting from memory leaks or due to another memory-greedy program running. In the former case, memory allocation fails permanently and contribution to the system eventually ends. In the latter, the worker might be able to continue after the other program is finished or when it is in a further stage of processing.
When receiving data, the receiving process can fail to allocate memory for the receive buffer. In this case, the process sleeps a certain time slice, polling for available memory regularly. In case the failure is temporary (e.g. another memory-greedy program running on the same node), this will allow the process to continue its program cycle. For permanent failures, the process stops contribution after trying to allocate memory for a certain number times.

**Processes Sending to Failed Workers.** The algorithm handles failures of workers if error codes are returned to the master-worker library. These can fail to contribute or be delayed for quite some time. However, processes that send towards these workers are not aware of these kinds of events. Therefore, such events are handled by using appropriate send operations.
When using synchronous messages for sending a bunch of data, a sort of "handshake" is required between the sender and the receiver. When the send operation is a blocking one and the receiving process is a worker that failed meanwhile, the whole system will block. For this reason, using synchronous blocking messages for sending towards workers is avoided.
In the MPI-1 standard, no explicit asynchronous send operation is available. The standard send operation can be asynchronous, but this depends typically on the MPI implementation and—in case of MPICH—also on the length of the message that is to be transmitted. Therefore, non-blocking send operations are invoked. This is combined with standard messages, in order to exploit optimizations of buffering in the MPI implementation.

### 9.6.2 Initialization and Finalization

The master-worker library and the MPI environment are initialized and finalized with two separate calls to the master-worker library, `masterworker_init` and `masterworker_final`. The masterworker interface itself can be invoked one or more times in between. One master-worker run can e.g. use the output of a former master-worker run as input. Also, other process functions or other user functions can be used in these separate masterworker runs.

All nodes can be initialized and finalized via user application functions that are passed on to the master-worker library via the `masterworker` interface. These functions can be used for setting up or freeing global variables in these processes once. These global variables can be used during the program flow of that node, avoiding re-initialization and re-finalization after each process cycle of that node. If such initialization is not necessary, empty functions can be used.

Not all variables that are needed for processing are available at all nodes. If a set of variables is needed by the collector for merging the data, these can be transmitted in the `MC_MERGED_STRUCT` structure. Variables needed for processing a block, but changing according to the blockid ($k$), should be sent to the workers via the `DW_STRUCT`. Variables that do not change according to the blockid, should however not be transmitted with each block. These can be initialized as global variables before calling the `masterworker_init` function. All processes will possess these variables in this case.

### 9.6.3 Speeding up

Possible time-consuming operations, such as saving results, splitting input units or merging and post-processing the output, are performed in RAFTNET by separate threads. In this way, temporary delays are smoothed out over one cycle of the master-worker system for splitting, processing and postprocessing a data unit.

The same reasoning has been applied to the master and the dispatcher. These consist of separate threads on the same node. Sending a certain set of input blocks from the master to the dispatcher results in this case in only exchanging pointers. This avoids the communication cost of sending input blocks from the master to the dispatcher or output units from the dispatcher to the master.

## 9.7 Application and System Description

### 9.7.1 Parallelization Choice

Not all algorithms are good candidates to apply the master-worker on. Parallelising an application over different processors offers a real added value when code optimizations have been done on the algorithm. As such, these two approaches are complementary. The following rules of thumb were used in the decision of applying the raftnet library on the algorithms described in the rest of this section:

Algorithm locality: Splitting up the input data for parallel execution is most efficient when operations have no or little data dependency or when these operations are very local if there is a positional data dependency (e.g. pixels in an image grid).

Input and output: When data is split up, this can affect the final result. Either the result is erroneous (mostly because of the absence of locality in the algorithm), the output has noise added to it or the merged output is correct (identical to serial execution). This example is often because parallel execution only used a similar split policy that was already inherently used in the serial execution (e.g. audio MP3 encoding). In the second case, the evaluation has to be made if the noise is significant for the final outcome and if it is, can the noise be filtered out in an efficient way.

In the following, a case study is briefly described: Feature selection via corner detection and Delaunay triangulation on the reconstructed points. The testing environment consisted of a PA RISC 85000 processor with a 400 MHz clock and 1.5 MB cache, running the HP-UX 11.0 operating system. MPICH 1.2.2 has been used, which implements the MPI 1.2 standard.

### 9.7.2 Corner Detection

The corner detection algorithm (Harris & Stephens, 1988) forms the basis of, e.g., a 3D reconstruction algorithm. The features selected in this phase are matched (i.e. the corresponding corners in several images) and used for reconstruction of the 3rd dimension, which was lost in the projection to 2D while capturing. During the tests, the setup in Fig. 24 was used.
Machine A captures, via a Conexant Bt878 based card, the data onto a local disk (NVREC, n.d.). In turn, this disk is shared via NFS with the rest of the network, in particular the parallel master-worker cluster. The images are captured in PAL VideoCD format, 352 × 288, 25 fps (frames per second), to pgm frames.
The initial version in C, based on the C++ development code was not even close to be able to process the images in soft real time and only 7 fps were processed. In a following step, the algorithm was optimized by including faster and more memory aware data types and by reducing intermediate buffers and loop transformations (Leeman et al., 2005).
Due to these changes, this algorithm got a speedup of 30%, which still put it about two times slower than the minimal soft real time requirement (10–12 fps).
Since conventional code optimization techniques were exhausted, assigning a larger cycle budget to the task within the available environment was done by adding a parallelization layer. Figure 25 shows two possible parallelization schemes. One possibility is to split up the figure. In doing this, edge effects have to be accounted for, so about 7% of the pixel values need to be replicated. In the test setup, the second option was preferred: The frames are handled in a FIFO by the master and are sent as a whole to the workers. This has also the advantage that no splitting is needed at the dispatcher side, nor is there need for merging of the results (coordinate data) on the collector's side. For the integration of the master-worker algorithm, only 125 extra lines of code had to be written to use the required functionality of raftnet. Tests on the implemented system show that soft real time feature detection *in the absence of faults* is possible with 4 to 5 workers (depending on the network and worker load). Adding more workers improves further performance and, at the same time, enhances the dependability of the service.

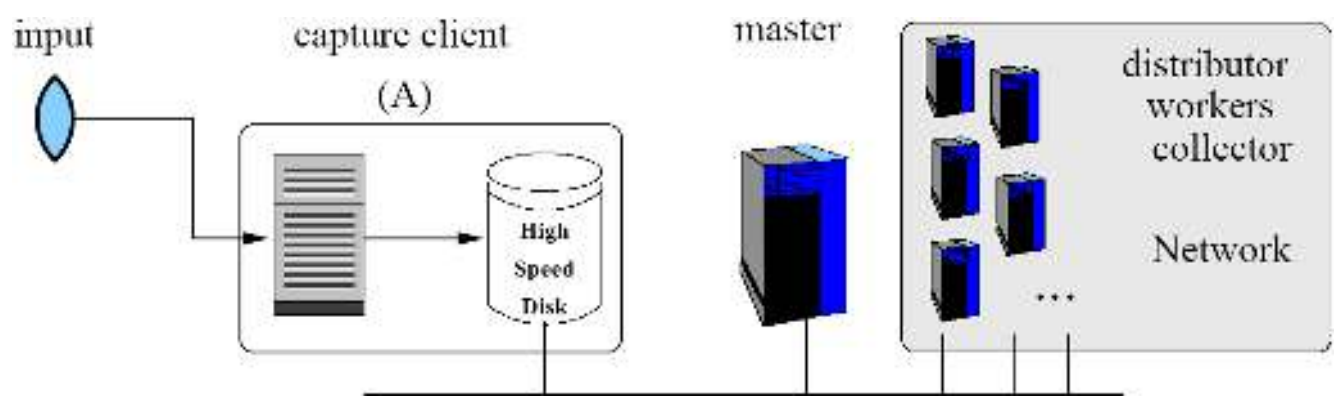

Figure 24: Capture and processing setup.

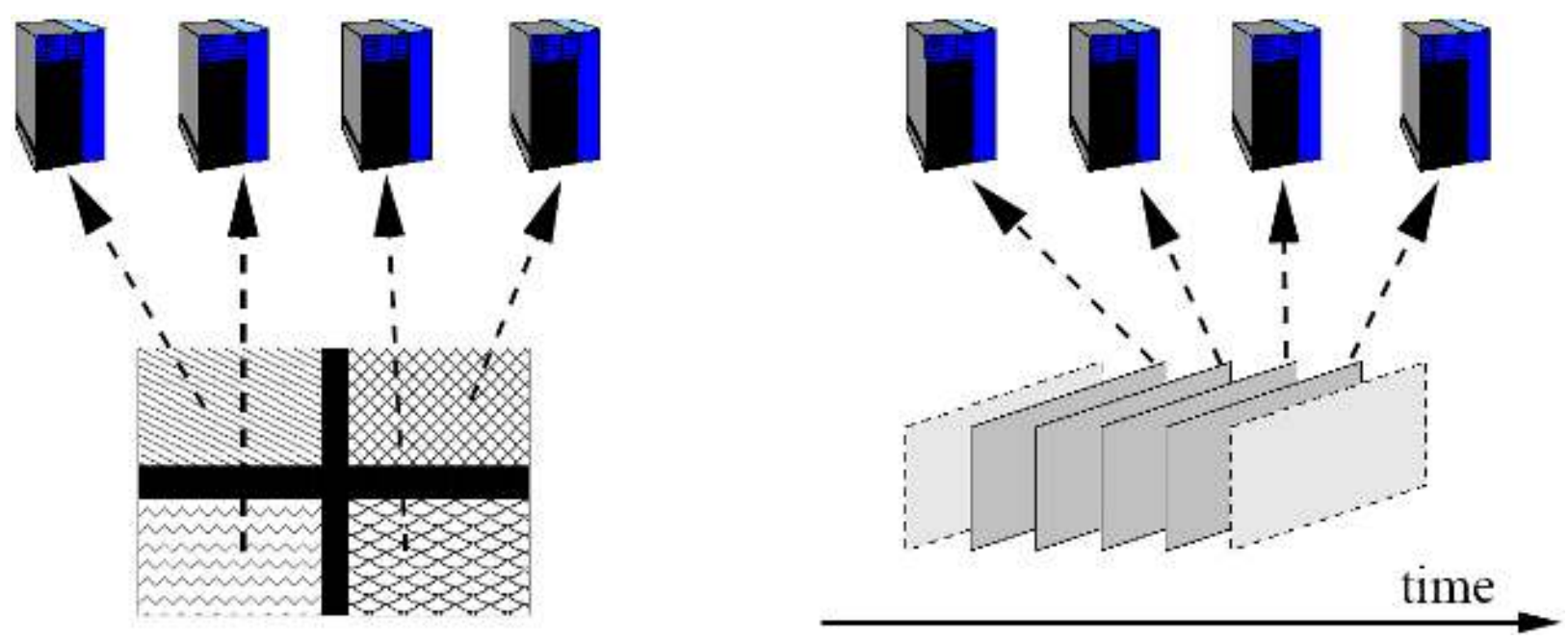

Figure 25: Image processing parallelization schemes.

| | | | |
|---|---|---|---|
| Voting Farm | Broadcast algorithm | Sect. 3.1.3 | Table 2 |
| Trap Handler | General algorithm | Sect. 5.2 | - |
| Atomic Actions | General algorithm | Sect. 6.1.1 | Table 4 |

Table 6: The algorithms of the EFTOS tools.

# 10 CONCLUSION

In this chapter single-version and multiple-version software fault-tolerance have been described. Also the main features of a few systems based on these approaches have been discussed. After this a number of examples have been treated in more detail—a flexible and easy to use mechanism for software voting; a software watchdog timer; a tool providing exception handling for the C programming language; a tool to manage transactions in C; and a system combining spatial and temporal redundancy to achieve data stabilization. Some of their algorithms have been introduced—Table 6 locates them in the chapter.

A so-called recovery meta-program—a syntactical structure for the expressions of provisions such as recovery blocks—has been also introduced. Finally a hybrid system, RAFTNET, combining aspects of both approaches, has been discussed.

Main aspect of the SV approach is the inherent mixture of the functional and the fault-tolerant code, which requires the developer to deal with the two corresponding concerns at the same time. MVSFT do not suffer from such flaw. On the other hand, as already mentioned, MVSFT have been designed expressly in order to address residual design faults. Such fault model though is intrinsically static, as the approach requires an off-line definition of the variants. This means that any service adopting the MVSFT provision assumes a static, stable environment, the typical faults of which are known as well as the extent of their consequences. Any time this is not valid anymore, the risk is higher of a coverage failure for the assumption of statistical independence becomes greater.

Despite their relatively long life, the approaches reviewed in this chapter are still being used and debated upon. This is especially true for MVSFT based on design diversity, whose effectiveness with respect to the problem of correlated failures appears to be a constant field of debate (Meulen & Revilla, 2005), while its applicability is being enlarged to domains such as security (Cox et al., 2006) and to non custom-made versions (Gashi et al., 2006).

# Notes

[1]Software rejuvenation (Huang et al., 1995) offers tools for periodical and graceful termination of an application with immediate restart, so that possible erroneous internal states, due to transient faults, be wiped out before they turn into a failure.

[2]This requirement is explained very clearly by Randell (Randell, 1975): "All fault-tolerance must be based on the provision of useful redundancy, both for error detection and error recovery. In software the redundancy required is not simple replication of programs but *redundancy of design*." Chapter 2 reported the consequences of a well known case where redundancy of design would have prevented a catastrophic failure—the Ariane 5 flight 501.

[3]Randell himself states that, given the ever increasing complexity of modern computing, there is still an urgent need for "richer forms of structuring for error recovery and for design diversity" (Randell & Xu, 1995).

[4]This weakness of NVP can be narrowed, if not solved, adopting the approach used in the so-called "voting farm" (De Florio, Deconinck, & Lauwereins, 1998c, 1998a; De Florio, 1997), a generic voting tool designed by author of this book in the framework of his participation to project "EFTOS" (see Sect. 2.1.1): such a tool works with opaque objects that are compared by means of a user-defined function. This function returns an integer value representing a "distance" between any two objects to be voted. The user may choose between a set of predefined distance functions or may develop an application-specific distance function. Doing the latter, a distance may be endowed with the ability to assess that bitwise different objects are semanticly equivalent. Of course, the user is still responsible for supplying a bug-free distance function—though is assisted in this simpler task by a number of template functions supplied with that tool.

[5]We recall that EPX, or Embedded Parix, was the operating system and message passing library of EFTOS' target platforms.

page

# Contents

# FAULT-TOLERANT PROTOCOLS USING COMPILERS AND TRANSLATORS

## 1 INTRODUCTION AND OBJECTIVES

In this chapter our survey of methods and structures for application-level fault-tolerance continues getting closer to the programming language: Indeed, tools such as compilers and translators work at the level of the language—they parse, interpret, compile or transform our programs, so they are interesting candidates for managing dependability aspects in the application layer. An important property of this family of methods is the fact that fault-tolerance complexity is extracted from the program and turned into architectural complexity in the compiler or the translator.
Apart from continuing with our survey, this chapter also aims at providing the reader with two practical examples:

- Reflective and refractive variables, i.e. a syntactical structure to express adaptive feedback loops in the application layer. This is useful to resilient computing because a feedback loop can attach error recovery strategies to error detection events.
- Redundant variables, that is a tool that allows designers to make use of adaptively redundant data structures with commodity programming languages such as C or Java. Designers using such tool can define redundant data structures in which the degree of redundancy is not fixed once and for all at design time, but rather it changes dynamically with respect to the disturbances experienced during the run time.

Both tools are new research activities that are currently being carried out by the author of this book at the PATS research group of the University of Antwerp. It is shown how through a simple translation approach it is possible to provide sophisticated features such as adaptive fault-tolerance to programs written in any language, even plain old C.

# 2 FAULT-TOLERANT PROTOCOLS USING COMPILERS AND TRANSLATORS

Our first subject is tools that work "within" the compiler: Meta-object protocols. Most of such tools are based on the concept of reflection: The ability to mirror the feature of a system by creating a causal connection between sub-systems and internal objects. In other words, events experienced by a reflected sub-system trigger events on the object representing that sub-system, and vice-versa.

## 2.1 Compiler-level Tools: Meta-object Protocols, Reflection, and Introspection

Some of the negative aspects pointed out while describing single and multiple version software approaches can be in some cases weakened, if not solved, by means of a generic structuring technique that allows to reach in some cases an adequate degree of flexibility, transparency, and separation of design concerns: the adoption of *meta-object protocols* (Kiczales, Rivières, & Bobrow, 1991). The idea is to "open" the implementation of the run-time executive of an object-oriented language like C++ or Java so that the developer can adopt and program different, custom semantics, adjusting the language to the needs of the user and to the requirements of the environment. Using meta-object protocols, the programmer can modify the behavior of a programming language's features such as methods invocation, object creation and destruction, and member access. The transparent management of spatial and temporal redundancy (Taylor, Morgan, & Black, 1980) is a context where meta-object protocols appear to be particularly adequate.
The key concept behind meta-object protocols is that of *computational reflection*, or the causal connection between a system and a meta-level description representing structural and computational aspects of that system (Maes, 1987). The protocols offer the meta-level programmer a representation of a system as a set of *meta-objects*, i.e., objects that represent and reflect properties of "real" objects, i.e., those objects that constitute the functional part of the user application. Meta-objects can for instance represent the structure of a class, or object interaction, or the code of an operation. This mapping process is called *reification* (Robben, 1999).
The causality relation of meta-object protocols could also be extended to allow for a dynamical reorganisation of the structure and the operation of a system, e.g., to perform reconfiguration and error recovery. The basic object-oriented feature of inheritance can be used to enhance the reusability of the fault-tolerance mechanisms developed with this approach.
A related concept is *introspection*, a technique that hacks the compiler to unravel and reason upon the hidden structure of our programs.

### 2.1.1 OpenC++ and Project FRIENDS

OpenC++ has been defined as a "code analysis library" (Karpov, 2008): A tool to parse and analyze C++ source code. Such tool makes use of a meta-object protocol to provide services for language extensions. In OpenC++, classes are objects as in Smalltalk, and called class metaobjects. A class metaobject translates—at compile time—expressions involving a class. In a sense, it works like a filter parsing all the expressions that mention its class. The default filter leaves the expressions unmodified, but the programmer can choose otherwise. This simple idea translates in a powerful tool for the programmer: For instance (Shigeru, 1996), shows how easy it is to perform some accounting on all methods invocations of a given class (doing this in a standard programming language such as C++ would require considerable programming effort (Leeman et al., 2005)).

As one can easily understand, such features may translate in a useful syntactical structure for the tolerance of faults. An architecture supporting this approach is the one developed in the framework of project FRIENDS (Fabre & Pérennou, 1996, 1998). Name FRIENDS is the acronym of "Flexible and Reusable Implementation Environment for your Next Dependable system". This project aims at implementing a number of fault-tolerance provisions (e.g., replication, group-based communication, synchronization, voting) at meta-level. In FRIENDS a distributed application is a set of objects interacting via the proxy model, a proxy being a local intermediary between each object and any other (possibly replicated) object.

### 2.1.2 Javassist

Javassist (Java Programming Assistant) (Javassist, n.d.) is a Java meta-library: It allows the programmer to access and manipulate the bytecode of an application, that is, the Java pseudo-code that is interpreted by the Java Virtual Machine to execute that application. This powerful feature allows to change the implementation of a class at run-time, which is known as structural reflection. Javassist can edit the bytecode either in the high-level form of the corresponding Java source level or directly as bytecodes. It is even possible to compile bytecode strings on the fly. Working with the standard Java compiler and virtual machine, Javassist offers a meta-object interface to control method invocations on base-level objects. All these features makes it a powerful tool for program transformation. One of the possible uses of Javassist is to augment Java for aspect-oriented programming[1], e.g. inserting `before`, `after`, and `around` advices. This gave raise to the GluonJ system, "a light-weight but powerful AOP framework on top of Javassist" (GluonJ, n.d.). Another interesting application would be to make use of Javassist to monitor and protect objects, for instance taking periodic "snapshots" of their state. This is but one possible way to achieve higher dependability through an approach like the one offered by Javassist, though to the best of our knowledge no research is being carried out on this topic yet.

### 2.1.3 MetaC++

MetaC++ (Strasser, 2005) is a tool that reads a C++ program and produces a tree with a representation of language constructs. The tree is exported to clients through an API or in the form of an XML file. The gathered meta-data can be used to rewrite a modified version of the original program, including all types of patches, filters, and improvements—which makes it interesting as a fault-tolerance provision. It was designed by its author primarily as a source-to-source code translator, but also to be used by source code analysis tools, stub generators, and source-to-UML converters. According to its author, future releases of MetaC++ will provide mechanisms to mark elements of the language with attributes, as in

**persistent** `class MyDatabaseObject;`

This will facilitate the set up of systems such as redundant variables, the dynamically redundant data structures described in Chapter 10.

### 2.1.4 Introspection

Data hiding and encapsulation are powerful tools to master the complexity of complex computer services, but they also represent a process that inhibits the access to complexity that is still there, and therefore can potentially jeopardize those services. The idea of introspection is to gain access into this complexity, to inspect the hidden structure of programs, and to interpret their meaning through semantic processing, the same way the Semantic Web promises to achieve with the data scattered in the Internet. Quoting its author, "introspection is a means that, when applied correctly, can help crack the code of a software and intercept the hidden and encapsulated meaning of the internals of a program". The way to achieve introspection is by instrumenting the software with data collectors producing information available in a form allowing semantic processing, such as RDF. This idea is being used in the Introspector project, which aims at instrumenting the GNU programming tool-chain so as to create a sort of semantic web of all software derived from those tools. The ultimate goal is very ambitious: "To create a super large and extremely dense web of information about the outside world extracted automatically from computer language programs" (DuPont, n.d.). Should this become possible, we would be given a tool to reason (and to let *computers* reason on our behalf!) about the dependability characteristics of the application software—imagine what could mean for instance being able to check for the presence of a hidden design goal and have the problem solved without human intervention. Some years ago this would have sounded like science fiction, now it appears that many researchers are focussing on tools that go in this direction: GASTA (Gcc Abstract Syntax Tree Analysis) (Thouvenin, 2004) for instance uses introspection to automatically annotate C code to analyze the presence of null pointer design faults). GCC-XML (King, 2004) has similar goals. A more advanced example appears to be XOGASTAN (XML-Oriented Gcc Abstract Syntax Tree ANalyzer) (Antoniol, Di Penta, Masone, & Villano, 2004), which uses the abstract syntax tree produced by the GNU compiler while processing a C file and translates it into XML. Another of their tools then can read the XML file and analyze it.
So let us hope this may become part of our future.

**Conclusions.** Meta-object protocols and reflection are indeed a promising system structure for embedding different non-functional concerns in the application-level of a computer program. They work at *language* level, providing means to modify the semantics of basic object-oriented language building blocks like object creation and deletion, calling and termination of class methods, and so forth. This appears to match perfectly to a proper subset of the possible fault-tolerance provisions, especially those such as transparent object redundancy, which can be straightforwardly managed with the meta-object approach. When dealing with these fault-tolerance provisions, meta-object protocols and reflection provide a perfect separation of the design concerns, i.e., optimal SC. Some other techniques, specifically those who might be described as "the most coarse-grained ones", such as distributed recovery blocks (Kim & Welch, 1989), appear to be less suited for being efficiently implemented via meta-object protocols. These techniques work at distributed, i.e., macroscopic, level.
The above situation reminds the author of this book of another one, regarding the "quest" for a novel computational paradigm for parallel processing able of dealing effectively with the widest class of problems, the same way the Von Neumann paradigm does for sequential processing, though with the highest degree of efficiency and the least amount of changes in the original (sequential) user code. In that context, the concept of computational *grain* came up: Some techniques were inherently looking at the problem "with coarse-grained glasses," i.e., at macroscopic level, while others were considering the problem exclusively at microscopic level. Meta-object protocols are an example of the latter approaches: They provide a small grain access to a program's atomic structures and provide means to modify their behavior.
Meta-object protocols look at the problem from a microscopic point of view. Some of the hitherto developed fault-tolerance techniques fit perfectly with this approach—for instance, object redundancy can be easily managed with the meta-object approach, and, for instance, Robben (Robben, 1999) describes meta-programs for checkpointing, active replication, and transaction processing.
It is still unclear whether this set is general enough to host, *efficaciously*, *many* forms of fault-tolerance, as it is remarked for instance in (Randell & Xu, 1995; Lippert & Videira Lopes, 2000) ("what reflective capabilities are needed for what form of fault-tolerance, and to what extent these capabilities can be provided in more-or-less conventional programming languages, and allied to other structuring techniques [like RB and NVP] remain to be determined"). It is therefore difficult to establish a qualitative assessment of attribute SA for meta-object protocols and reflection. Judging from the evidence acquired to date one can cautiously assess their SA as "$\geq$ average" (average or more.)
The run-time management of libraries of meta-object protocols may be used to reach satisfactory values for attribute A. To the best of our knowledge this feature has not been exploited for reaching higher dependability in any system supporting meta-object protocols and reflection .
As evident, the target application domain is mostly the one of object-oriented applications written with languages extended with a meta-object protocol, such as Open C++.
One can conclude that meta-object protocols and reflection offer an elegant system structure to embed a certain set of non-functional services (including some fault-tolerance provisions) in an object-oriented program.
Computational reflection is a relatively old concept, though still actively investigated in various forms, especially because it is one of the possible ways to realize aspect-orientation, which is being given an even larger attention by researchers these days. OpenC++ is still referenced and investigated (Karpov, 2008). MetaC++ does not appear to have received wide attention. On the contrary, introspection appears to be spreading with projects such as GASTA, XOGASTAN and GCC-XML.
Both computational reflection and introspection are conceptual ingredients that can be used to craft fault-tolerance systems, though cannot provide an answer of their own but must be coupled with other approaches. No fault-tolerance provisions based on introspection surfaced to date, so it is difficult to assess the value of the structural attributes for introspection for the time being.

## 2.2 Enhancing Programs through Translations

### 2.2.1 LINDA Systems

The Linda (Carriero & Gelernter, 1989b, 1989a) approach adopts a special model of communication, known as *generative communication* (Gelernter, 1985). According to this model, communication is still carried out through messages, though messages are not sent to one or more addressees for them to be retrieved from local mailboxes—on the contrary, messages are included in a distributed (virtual) shared memory, called tuple space, where every Linda process has equal read/write access rights. In generative communication, the sender of a message cannot know who will read any sent message—if any eventually will. This means that traditional terms such as client-server or peer-to-peer, which are normally used to describe relationships between communicating entities, do not apply to generative communication. A tuple space is some sort of a shared relational database for storing and withdrawing special data objects called tuples, sent by the Linda processes. Tuples are basically lists of objects identified by their contents, cardinality and type. Two tuples match if (1) they have the same number of objects, (2) if the objects are pairwise compatible for what concerns their types, and (3) if the memory cells associated to the objects are bitwise equal. A Linda process inserts, reads, and withdraws tuples via blocking or non-blocking primitives. Read operations can be performed by supplying a template tuple—that is, a prototype tuple consisting of constant fields and of fields that can assume any value. A process trying to access a missing tuple via a blocking primitive enters a wait state that continues until any tuple matching its template tuple is added to the tuple space. This allows processes to synchronize. When more than one tuple matches a template, the choice of which actual tuple to address is done in a non-deterministic way. Concurrent execution of processes is supported through the concept of "live data structures": Tuples requiring the execution of one or more functions can be evaluated on different processors—in a sense, they become active, or "alive". Once the evaluation has finished, a (no more active, or passive) output tuple is entered in the tuple space.

Parallelism is implicit in Linda—there is no explicit notion of network, number and location of the system processors, though Linda has been successfully employed in many different hardware architectures and many applicative domains, resulting in a powerful programming tool that sometimes achieves excellent speedups without affecting portability issues. Unfortunately the model does not cover the possibility of failures—for instance, the semantics of its primitives are not well defined in the case of a processor crash, and no fault-tolerance means are part of the model. Moreover, in its original form, Linda only offers single-op atomicity (Bakken & Schlichting, 1995), i.e., atomic execution for only a single tuple space operation. With single-op atomicity it is not possible to solve problems arising in two common Linda programming paradigms when faults occur: Both the distributed variable and the replicated-worker paradigms can fail (Bakken & Schlichting, 1995). As a consequence, a number of possible improvements have been investigated to support fault-tolerant parallel programming in Linda. Apart from design choices and development issues, many of them implement stability of the tuple space (via replicated state machines (Schneider, 1990) kept consistent via ordered atomic multicast (Birman, Schiper, & Stephenson, 1991)) (Bakken & Schlichting, 1995; Xu & Liskov, 1989; Patterson, Turner, Hyatt, & Reilly, 1993), while others aim at combining multiple tuple-space operations into atomic transactions (Bakken & Schlichting, 1995; Anderson & Shasha, 1991; Cannon & Dunn, 1992). Other techniques have also been used, e.g., tuple space checkpoint-and-rollback (Kambhatla, 1991). The author of this book also proposed an augmented Linda model for solving inconsistencies related to failures occurring in a replicated-worker environment (see Chapter 9) and an algorithm for implementing a resilient replicated worker scheme for message-passing farmer-worker applications. As seen in Chapter 3, this algorithm can mask failures affecting a proper subset of the set of workers (De Florio, Deconinck, & Lauwereins, 1999).

Linda can be described as a parallel processing extension that can be added to any existing programming language. The greater part of these extensions requires a preprocessor translating the extension in the host language. This is the case, e.g., for FT-Linda (Bakken & Schlichting, 1995), PvmLinda (De Florio, Murgolo, & Spinelli, 1994), C-Linda (Berndt, 1989), and MOM (Anderson & Shasha, 1991). A counterexample is, e.g., the POSYBL system (Schoinas, 1991), which implements Linda primitives with remote procedure calls, and requires the user to supply the ancillary information for distinguishing tuples.
Linda systems allow dealing with redundant data and redundant processes in a transparent way, so they are in general characterized by good SC. Though available in many flavours, typically Linda systems focus on a few fault-tolerance solutions, hence SA is assessed as "average/potentially good". Adaptivity is not supported in the available Linda systems, so their A is deemed as insufficient. The Linda *approach*, on the other hand, may make use of its high transparency to achieve fault-tolerance applications whose redundancy changes with the experienced or forecast faults—which would bring to good A.
Very popular some years ago, Linda appears not to be an emergent fault-tolerance the solution these days.

#### 2.2.2 The Porch Translator

Porch (Strumpen, 1998) is a source-to-source translator that converts a sequential C program into an equivalent one that supports the saving and reloading of "portable checkpoints". This means that Porch allows to dump a checkpoint on machine $m$ and restart it on machine $n$, even though $m$ and $n$ have different computer architecture. Clearly this can be used to achieve task migration and reconfiguration. Porch allows to execute checkpoints periodically, or by sending signals e.g. through the UNIX "kill" command.
Porch suffers from several limitations: Rolling back a checkpoint does not preserve open file descriptors and sockets; only initialized pointers are stored in the checkpoints, hence uninitialized pointers may change their value.
Porch is functionally not dissimilar from other checkpoint and rollback systems such as psncLibCkpt (Meyer, 2003), ckpt (Zandy, n.d.) or Dynamite (Iskra et al., 2000). Its added value is the use of a translator, which improves its SC. Clearly SA is limited to a single class of fault-tolerance provisions. No adaptivity is foreseen in Porch, so A is set to insufficient.

Next section describes an application-level approach to fault-tolerance employing translators. The approach is based on so-called Reflective and refractive variables. Available as a rough prototype, a system supporting this approach is currently being developed by the author of this book.

```
#include <stdio.h>
int main(void)
{
        Ref_t int cpu;
        Ref_t int tcpTxRate;

        cpu=0;
        while (1) {
                printf("&cpu == %x, cpu == %d\n",
                        &cpu, cpu);
                if (cpu > 90) break;
                sleep(1);
        }

        tcpTxRate = 70;
}
```

Figure 1: A simple example of the use of reflective and refractive variables.

# 3 AN EXAMPLE: REFLECTIVE AND REFRACTIVE VARIABLES

The idea behind reflective and refractive variables is to use memory access as an abstraction to perform concealed tasks. Reflective and refractive variables are volatile variables whose identifier links them with an external device, such as a sensor, or an RFID, or an actuator. In reflective variables, memory cells get asynchronously updated by service threads that interface those external devices. The well-known concept of reflection is used, because those variables "reflect" the values measured by those devices. In refractive variables, on the contrary, write requests trigger a request to update an external parameter, such as the data rate of the local TCP protocol entity or the amount of redundancy to be used in transmissions. We use to say that write accesses "refract" (that is, get redirected (Institute for Telecommunication Sciences, n.d.)) onto corresponding external devices.

The reflective and refractive variables model does not require any special language: Figure1 is an example in the C language. The portrayed program declares two variables: "cpu", a reflective integer, which reports the current percentage of usage of the local CPU as an integer number between 0 and 100, and "tcpTxRate", a reflective *and refractive* integer, which reports *and sets* the send rate parameter of the TCP layer. The code periodically queries the CPU usage and, when that reaches a value greater than 90%, it requests to change the TCP send rate. Note that the only non standard C construct used in the example is attribute "Ref_t", which specifies that a corresponding declaration is reflective or refractive or both. Through a translation process, discussed in detail in Sect. 3, this code is instrumented so as to include the logics required to interface the cpu and the TCP external devices. Figure3 shows this simple code in action on our development platform—a Pentium-M laptop running Windows XP and the Cygwin tools.

One can observe that through the reflective and refractive variable model the design complexity is partitioned into two well defined and separated

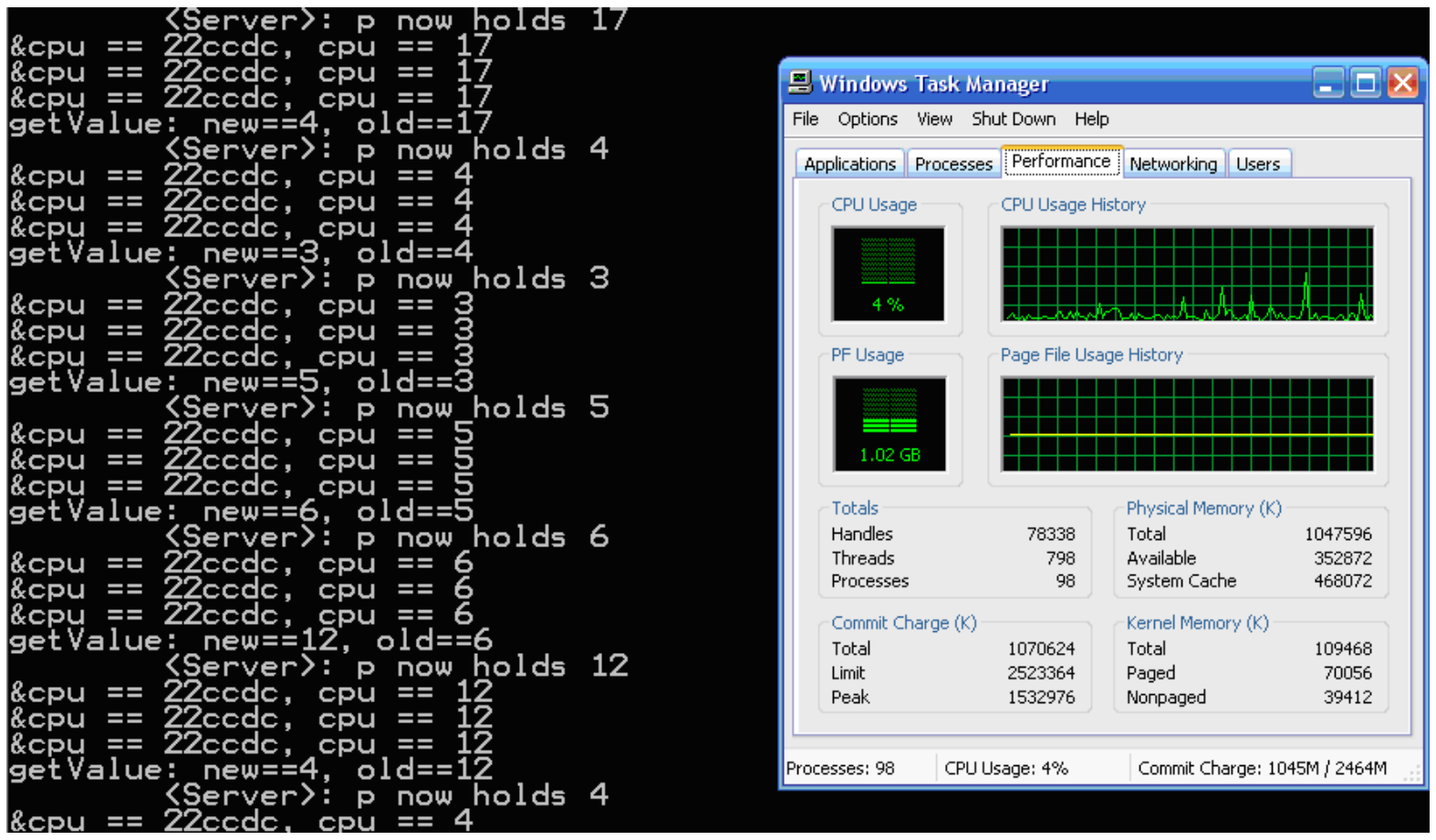

Figure 2: An excerpt from the execution of the code in Fig. 2.

components: The code to interface external devices is specified "elsewhere" (Sect. describes where and how) while the functional code is specified in a familiar way, in this case as a C code reading and writing integer variables. The result is a structured model to express tasks such as cross-layered optimization, adaptive or fault-tolerant computing in an elegant, non intrusive, and cost-effective way. Such model is characterized by strong separation of design concerns (that is, excellent SC), for the functional strategies are not to be specified aside with the layer functions. Only instrumentation is required, and this can be done once and for all. This prevents spaghetti-like coding for both the functional and the non-functional aspects, and translates in improved maintainability and enhanced efficiency. The reflective and refractive variable model provides the designer also with another attribute: a variable, be it a reflective / refractive variables or a "common" one, can be tagged as being "*redundant*". Redundant variables are variables whose contents get replicated several times so as to protect them from memory faults. Writing to a redundant variable means writing to a number of replicas, either located strategically[2] on the same processing node, or on remote nodes—when available and the extra overhead be allowed. Reading from a redundant variable actually translates in reading from each of its cells and performing majority voting (through an approach such as the one described in Chapter 3). The result of this process is monitored by a special device, called Redundance. Redundance measures the amount of votes that differ from the majority vote, and uses this as a measure of the disturbance in the surrounding environment. Under normal situation, Redundance triplicates the memory cells of redundant variables. This corresponds to tolerating up to one memory fault in the cells associated to a redundant variable. Under more critical situations, the amount of redundancy should change. This is indeed what actually happens: The component that manages redundant variables

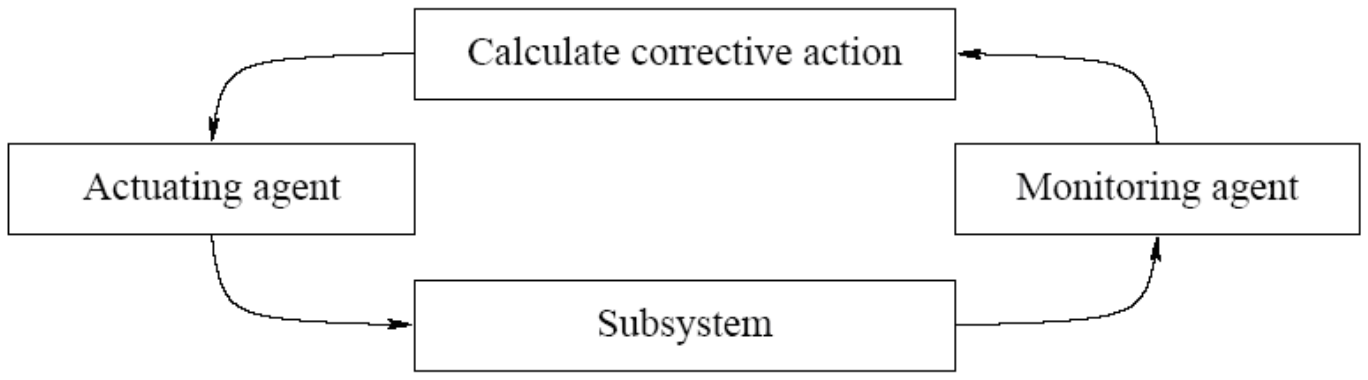

Figure 3: General structure of feedback loops.

```
#include <stdio.h>
int main(void)
{
        int a;
        Ref_t int Redundance;
        redundant int myProtectedInteger;

        while (1) {
                sleep(5);
                printf("&Redundance == %x, Redundance == %d\n",
                        &Redundance, Redundance);
                if (cpu > 90) break;
        }

        tcpTxRate = 70;
        myProtectedInteger = 1; // writes are replicated Redundance times
        a = myProtectedInteger; // reads are managed through voting
}
```

Figure 4: Redundant variables.

declares the integer reflective variable "ref_t int redundance". The latter is set asynchronously by the Redundance device, which adjusts the corresponding memory cells[3] with a number representing the ideal degree of redundancy with respect to the current degree of disturbances.

The reflective and refractive variables model can also be used in contexts such as cross-layer optimization. In general, it provides an application-level construct to manage feedback loops.

Feedback loops (see Fig. 2)—a well known concept from system theory—are ideal forms to shape our systems so as to be adaptive-and-dependable (Van Roy, 2006). As we have mentioned already in Chapter 2, such property is an important pre-requisite for the welfare of our computer-dominated societies and economies: In the cited paper Van Roy explains their relevance to future software design. Reflective and refractive variables provide a straightforward syntactical structure and software architecture for the expression of feedback loops. This structure is used, e.g., to implement redundant variables. The main advantage in this case is that, instead of taking a design decision once and for all, one lets a system parameter change as needed, zeroing in on the optimum. The use of reflective and refractive variables simplifies the design of our solution, which also enhances maintainability. But probably the most important consequence of that design choice is that our solution does not assume a fixed, immutable fault model, but lets it change with the actual faults being experienced—which is precisely the property pointed out in Chapter 2 as a key requirement for modern dependable software.

Figure4 shows how simple it is to use a redundant variable: No syntactic differences can be noticed. The required logic is "hidden" in the translation process. More detail about redundant variables is provided in Sect. 4.

### 3.1 Implementation

The core of the reflective and refractive variables architecture is a LEX analyzer[4] that translates the input source code into two source files, one with an augmented version of the original code and one server-side to monitor and drive the external devices. This process is considered in Fig. 5, an excerpt from the translation of the code in Fig. 4. Let us review the resulting code in more detail (please note that item x in the following list refer to lines tagged as "// *x*" in the code):

1. First the translator removes the occurrences of attributes "ref_t" and "redundant".
2. Then it performs a few calls to function "aopen". This is to open the associative arrays "reflex" and "rtype". As well known, an associative array generalizes the concept of array so as to allow addressing items by non-integer indexes. The arguments to "aopen" are functions similar to "strcmp", from the C standard library, which are used to compare index objects. The idea is that these data structures create links between the name of variables and some useful information (see below).
3. There follow a number of "awrites", i.e., associations are created between variables identifiers and two numbers: The corresponding variables' address and an internal code representing its type and attributes.
4. Then "Server", the thread responsible to interface the external devices, is spawned.
5. Besides a write access into refractive variable tcpTxRate, the translator places a call to function "CalltcpTxRate". In general, after a call to refractive variable $v$, the call "Call$v$(&$v$)" is produced.
6. Similarly, a write access to redundant variable $w$, of type $t$, is followed by a call to "RedundantAssign_$t$(&$w$)".
7. Finally, reading from redundant variable $w$, of type $t$, is translated into a call to function "RedundantRead_$t$(&$w$)".

It is the responsibility of the designer to make sure that proper code for functions "Call$v$(&$v$)" is produced. Functions "RedundantAssign_$t$(&$w$)" and "RedundantRead_$t$(&w)" are automatically generated through a template-like approach—the former performs a redundant write, the latter a redundant read plus majority voting. For voting, an approach similar to that in Chapter 3 has been followed (that is, voting farms).
As already mentioned, the "Server" thread is the code responsible to monitor and interface the external devices. Its algorithm is quite simple (see Fig. 6):

```
int main(void)
{
        int a;
        /* Ref_t */  int tcpTxRate;                          // 1
        /* Ref_t */  int Redundance;
        /* redundant */  int myProtectedInteger;

        reflex = aopen(acmp), rtypes = aopen(acmp);          // 2

        awrite(reflex, "tcpTxRate", (void*)&tcpTxRate);      // 3
        awrite(rtypes, "tcpTxRate", (void*)129);

        awrite(reflex, "Redundance", (void*)&Redundance);
        awrite(rtypes, "Redundance", (void*)129);

        awrite(reflex, "myProtectedInteger", (void*)&myProtectedInteger);
        awrite(rtypes, "myProtectedInteger", (void*)65);

        pthread_create(&t, NULL, Server, (void*) reflex);    // 4


        while (1) {
                sleep(5);
                printf("&cpu == %x, cpu == %d\n", &cpu, cpu);
                if (cpu > 90) break;
        }

        tcpTxRate = 70;                                      // 5
        CalltcpTxRate(&tcpTxRate);

        myProtectedInteger = 1;                              // 6
        RedundantAssign_int(&myProtectedInteger);

        a = RedundantRead_int(&myProtectedInteger);          // 7
}
```

Figure 5: Abridged version of the main function of the translated code.

the code continuously waits for a sensor update (lines tagged with "// 1"), then retrieves the address and type of the corresponding reflective variable (in "// 2") and finally updates that variable ("// 3").
The complexity to interface external devices is charged to function "getValue", an excerpt of which is shown in Fig. 7. The core of "getValue" is function "cpu", which returns the amount of CPU currently being used.

## 3.2 Other Uses of Reflective and Refractive Variables

As mentioned at the beginning, our research on this topic has begun only recently. We are in the process of making use of our approach in several real-life applications—most likely the subject of future publications. In what follows a few contexts where our variables could provide effective and low-cost solutions are discussed.

### 3.2.1 Management of Concurrency

> As cleverly explained e.g. by Gates in (Gates, 2007), a well known challenge in robotics is *concurrency*, defined in the cited paper as "how to simultaneously handle all the data coming in from multiple sensors and send the appropriate commands to the robot's motors". The conventional approach, i.e., making use of a long loop that first reads all the data from the sensors, then processes the input and finally controls the robot is not adequate enough. Because of this, the robot control could be using stale values, which could bring to disastrous consequences. As Gates mentions in the cited paper, this is a scenario that applies not only to

```
int Server(void)
{
        char sensor[80];    /* the name of the sensor */
        void *p;            /* the address of a reflective var */
        int type;           /* its type */
        mytypes_t object;   /* the new value of that varable */

        while (1) {
            /* wait for sensor update */
            if ( getValue(sensor, &object) == -1 ) continue;      // 1

            /* message received */
            p = (void*)aread(reflex, sensor);                     // 2
            type = (int) aread(rtypes, sensor);

            if (p != NULL)                                        // 3
            {
                if (type & IsChar)   *( (char*)p )   = object.c;
                if (type & IsShort)  *( (short*)p )  = object.s;
                if (type & IsInt)    *( (int*)p )    = object.i;
                if (type & IsLong)   *( (long*)p )   = object.l;
                if (type & IsFloat)  *( (float*)p )  = object.f;
                if (type & IsDouble) *( (double*)p ) = object.d;
            }
        }
}
```

Figure 6: The Server code.

```
int getValue(char *sens, mytypes_t *obj) {
  static int oldcpuvalue, newcpuvalue;

  // first sensor: cpu
  newcpuvalue = cpu();
  printf("getValue: new==%d, old==%d\n", newcpuvalue, oldcpuvalue);
  if (newcpuvalue != oldcpuvalue) {
        obj->i = oldcpuvalue = newcpuvalue;
        strcpy(sens, "cpu");
        return 0;
  }

  // second sensor ...
  // etc

  // no value available: return -1
  return -1;
}
```

Figure 7: Function getValue interfaces all the external devices that are connected to reflective and refractive variables.

```
#include <stdio.h>
int main(void)
{
       int beep(void);
       Ref_t char* rfid =
          "ISBN 90-5682-266-7"
       with alarm = beep;

       while (1) ;
}
```

Figure 8: RR var to localize objects with RFID tags on them.

robotics but also to all those fields such as distributed, parallel and resilient computing where data and control often need to be effectively orchestrated under strict real-time constraints and despite the occurrence of faults. “To fully exploit the power of processors working in parallel, the new software must deal with the problem of concurrency”, Gate says. It is our belief that an approach such as the one offered by reflective and refractive variables could result in an effective syntactic structure for that: A control loop using reflective variables, for instance, would not need to specify a reading order for the input variables, which are updated asynchronously, as new values need to replace old ones. Within that loop one could attach e.g. the asynchronous management of failures.

### 3.2.2 Localizing Hidden Assets

Extending our translator it can be possible to allow writing programs such as the one in Fig. 8.

At first sight the program may sound meaningless, as it only declares a function and a reflective variable, “rfid”, and does not seem to perform any useful action. “Behind the lines”—a nice feature offered by translators—what happens is that surrounding RFID tags reflect their content onto variable “rfid”. Data stored into that variable is compared with the initialization value (in this case, an ISBN number). In case of a match, function “beep” is called. Now imagine running this code onto your PDA while walking through the lanes of a large library such as the Vatican Library in search for a “lost” or misplaced book. When in reach of the searched item, the PDA starts beeping[5]. Or imagine that, thanks to international regulations, all “companies” building antipersonnel mines be obliged by law to embed RFID tags into their “products”. When activated, these tags and a program as simple as the one in Fig. 8 could easily prevent dreadful events that continuously devastate the lives of too many a human being. Changing context, the same approach could be used to speed up the localizing of people trapped in the ruins of a fallen building after an earthquake, and so forth. This could be sensible in a highly seismic region.

**Conclusions.** This section introduced a translation system that allows making use of reflection in a standard programming language such as C. The same translator supports "refraction", that is the control of external devices through simple memory write accesses. These two features are used to realize redundant data structures. As well known, redundancy is a key property in fault-tolerance. The Shannon theorem teaches us that through any unreliable channel it is possible to send data reliably by using a proper degree of redundancy. This famous result can be read out in a different way: For each degree of unreliability, there is a minimum level of redundancy that can be used to tolerate any fault. Our approach uses reflective and refractive variables to attune the degree of redundancy required to ensure data integrity to the actual faults being experienced by the system. This provides an example of adaptive fault-tolerant software.
Reflective and refractive variables can be used to express problems in cross-layer optimization, but also in contexts where concurrency calls for expressive software structures, e.g. robotics. Localization problems could also be solved through a very simple scheme. Other fields where our tool is being exercised include personalized healthcare (De Florio, Vaerenbergh, & Blondia, 2007) and global adaptation frameworks to enhance the quality of experience of mobile services (Sun, De Florio, Ning, & Blondia, 2007).
By construction, reflective and refractive variables exhibit a good amount of transparency, which translates in good SC. SA is limited to those fault-tolerance techniques that can be expressed in terms of the variables; some variables could be used e.g. to express processing and communication entities, and allow them to migrate transparently; other variables could change their state to represent the detection of faults, and so forth. As authors we are convinced of their merits, but we must be objective and state that no such provisions exist so far—with the exception of the ones described in next section. So a fair assessment for SA could be "limited/potentially good". The hidden implementation of reflective and refractive variables allows to achieve a good A with an approach such as the one described in the next section.

# 4 ADAPTIVE DATA INTEGRITY THROUGH DYNAMICALLY REDUNDANT DATA STRUCTURES

Changes, they use to say, are the only constant in life. Everything changes rapidly around us, and more and more key to survival is the ability to rapidly adapt to changes. This consideration applies to many aspects of our lives. Strangely enough, this nearly self-evident truth is not always considered in computer science with the seriousness that it calls for: The assumptions that are drawn for our systems often do not take into due account that e.g., the run-time environments, the operational conditions, or the available resources *will* vary. Software is especially vulnerable to this threat, and with today's software-dominated systems controlling crucial services in nuclear plants, airborne equipments, health care systems and so forth, it becomes clear how this situation may potentially lead to catastrophes.
Let us consider in particular the viewpoint of the software fault-tolerance engineer and the design goal of enhanced data integrity: It is clear that protecting data against memory faults requires the definition of a fault model stating the possible fault scenarios that our software is going to experience. The current practice is to take this as a static choice, which means that our data integrity provisions (DIP) have a fixed range of admissible events to address and tolerate. This translates into two risks:

1. overshooting, i.e., over-dimensioning the DIP with respect to the actual threat being experienced, and

2. undershooting, namely underestimating the threat in view of an economy of resources.

Note how those two risks turn into a crucial dilemma to the designer: Wrong choices here can lead to either unpractical, too costly designs or cheap but vulnerable provisions.
A sensible example of the problem just stated is given by redundant data structures (RDS). Here redundancy and voting are used to protect memory from possible transient or permanent corruptions. The common choice would be that of adopting a static fault model assumption, such as e.g. "during any mission, up to 3 faults shall affect the replicas", which translates into using a 7-redundant cell to address the worst possible scenario. This is precisely a case of the above stated dilemma.
Our starting point in the research reported herein was the question "Is this the right approach?," or "Does it make sense in the first place to fix, once and for all, a set of possible conditions affecting memory modules?" We don't think so. First of all, memory technology (like all technologies) changes, which means that while yesterday our software was running atop CMOS chips, today the common choice e.g. for airborne applications is SDRAMs because of speed, cost, weight, power and simplicity of design (Ladbury, 2002). But CMOS memories for airborne applications mostly experience single bit errors (Oey & Teitelbaum, 1981), while SDRAMS are known to be subject to several classes of faults, including so-called "single-event effects", i.e. single faults affecting whole chips[6] (Ladbury, 2002). Furthermore the cited paper remarks how even from lot to lot error/failure rates vary more than one order of magnitude. A static choice of the fault model cannot take all this into proper account.
A second argument is that run-time environments change, often because the application is mobile but sometimes also because of external events affecting e.g. temperature, radiation, electro-magnetic interference, or cosmic rays (think of a wireless sensor network to assist a fire brigade, or of a spaceborne application circulating around the sun). Also in these cases a static choice would be unpractical, as the nature of faults may change during a same mission.

```
#include <stdio.h>
int main(void)
{
        int a;
        redundant int myProtectedInteger;

        myProtectedInteger = 1;
        a = myProtectedInteger;
}
```

Figure 9: A simple example of use of redundant variables.

Our conjecture is that the solution to this problem should come by considering the nature of faults as a dynamic system, i.e., a system evolving in time, and by modeling faults as a function $F(t)$. Consequently, any fault-tolerance provision that be able to solve effectively the problem of over-dimensioning and under-dimensioning should function as an adaptation feedback loop (Van Roy, 2006) in which the resources are allocated after the estimated values of $F(t)$, derived by monitoring sensible environmental variables. This section reports on the design of a tool compliant to that model.
Our tool allows designers to make use of adaptively redundant data structures with commodity programming languages such as C or Java. Designers using such tool can define redundant data structures where the degree of redundancy changes dynamically with respect to the disturbances experienced at run-time. Our approach attunes the degree of redundancy required to ensure data integrity to the actual faults being experienced by the system and provides an example of adaptive fault-tolerance software provision.
In what follows a description of our tool and design issues are given.

### 4.1 Dynamically Redundant Data Structures

Our tool is a translator that loosely parses a source code performing some transformations as reported in the rest of this section. A translator was developed in the C programming language together with the Lex lexical analyzer generator and the YACC syntactic analyzers generator (Levine, Mason, & Brown, 1992). The reported version supports the C syntax though the same principles can be easily applied to any other language. Our translator performs a simple task—it allows the programmer to tag variables with a keyword, "redundant," and then instruments the memory accesses to tagged variables. Figure 9 shows how this is done in practice with a very simple example whose translation is provided in Fig. 10. Let us review the resulting code in more detail (please note that item x in the following list refer to lines tagged as "// *x*" in the code):

1. First the translator removes the occurrences of attribute "redundant".
2. Then it performs a few calls to function "aopen". This is to open the associative arrays "redun" and "rtype". As well known, an associative

```
#include <stdio.h>
#include <pthread.h>
#include "assoc.h"
#include "sensors.h"
#include "redundance.h"
int acmp(const void*a, const void*b) { return strcmp(a, b); }
int icmp(const void*a, const void*b) { return a-b; }

int Server(void);

ASSOC *rtypes, *redun;
pthread_t t;
static int _Redundance = REDUNDANCE;
int main(void)
{
        int a;
        /* redundant */  int myProtectedInteger;      // 1

        redun = aopen(icmp), rtypes = aopen(acmp);    // 2

        awrite(redun, "myProtectedInteger",           // 3
              (void*)&myProtectedInteger);
        awrite(rtypes, "myProtectedInteger",
               (void*)64);

        pthread_create(&t, NULL, Server, NULL);       // 4

        myProtectedInteger = 1;                       // 5
        RedundantAssign_int(&myProtectedInteger);

        a = RedundantRead_int(&myProtectedInteger);   // 6
}
```

Figure 10: An excerpt from the translation of the code in Fig. 9. Variable "_Redundance" represents the current amount of redundancy, initially set to "REDUNDANCE" (that is, 3).

array generalizes the concept of array so as to allow addressing items by non-integer indexes. The arguments to "aopen" are functions similar to "strcmp", from the C standard library, which are used to compare index objects. The idea is that these data structures create links between the name of variables and some useful information (see below).

3. There follow a number of "awrites", i.e., associations are created between variables' identifiers and two numbers: the corresponding variables' address and an internal code representing its type and attributes (code 64 means "redundant int").

4. Then the "Server" thread, responsible to allocate replicas and to monitor and adapt to external changes, is spawned.

5. A write access to redundant variable $w$, of type $t$, is followed by a call to "RedundantAssign_$t$(&$w$)".

6. Finally, reading from redundant variable $w$, of type $t$, is translated into a call to function "RedundantRead_$t$(&$w$)".

The strategy to allocate replicas is a research topic on its own. Solutions range from naïve simple strategies such as allocating replicas into contiguous cells—which makes them vulnerable to burst faults—to more sophisticated strategies where replicas get allocated e.g. in different memory banks, or different memory chips, or even on different processors. Clearly each choice represents a trade-off between robustness and performance penalty. In our current version replicas are separated by strides of variable length.
The core of our tool is given by functions "RedundantAssign_$t(\&w)$" and "RedundantRead_$t$(&w)", which are automatically generated for each type $t$ through a template-like approach. The former function performs a redundant write, the latter a redundant read plus majority voting. For voting, an approach similar to that in (De Florio, Deconinck, & Lauwereins, 1998) is followed.
What differs our tool from classical libraries for redundant data structures such as the one in (Taylor et al., 1980) is the fact that in our system the amount of replicas of our data structures changes dynamically with respect to the observed disturbances. A monitoring tool is assumed to be available to assess the probability of memory corruptions of the current environment. An example of such monitoring tool, which estimates that probability by measuring for each call to "RedundantRead_$t$" the risk of failure $r$, is provided. Quantity $r$ may be defined for instance as follows: If our current redundancy is $2n+1$, and if the maximum set of agreeing replicas after a "RedundantRead_$t$" is $m$ $(0 < m < 2n+2)$, then

$$r = \begin{cases} (2n+1-m)/n & \text{if } m > n \\ 1 & \text{otherwise.} \end{cases} \tag{1}$$

For instance if redundancy is 7 and $m = 6$, that is if only one replica differs, then $r = 1/3$. Clearly the above choice of $r$ lets risk increase linearly with the number of replicas not in agreement with the majority. Other formulations for $r$ and for the monitoring tool[7] are possible and likely to be more effective than the ones taken here—also a matter for future research.
Our strategy to adjust redundancy is also quite simple: If $r > 0.5$, redundancy is increased by 2; if $r = 0$ for 1000 consecutive calls to "RedundantRead_$t$", redundancy is decreased by 2. Lower bound and upper bound for redundancy have been set to 3 and 11 respectively.
Each request for changing redundancy is reflected by the "Server" thread into variable "_Redundance" through the scheme introduced in (De Florio & Blondia, 2007).
In Chapter 10 it is described how our tool behaves when memory faults are injected. It us shown how, despite the above naïve design choices, our tool already depicts valuable results.

## 4.2 Conclusions

As already pointed out, redundancy is a key asset to achieve fault-tolerance. The Shannon theorem teaches us that through any unreliable channel it is possible to send data reliably by using a proper degree of redundancy. This famous result can be read out in a different way: For each degree of unreliability characterizing a run-time environment there is a minimum level of redundancy that can be used to tolerate any fault. Unfortunately environments change, e.g. because of external events, or because assets dissipate, or because the service is mobile. Hence, there is no static allocation of redundancy that can accommodate for any possible scenario: A highly redundant system will withstand no more faults than those considered at design time, and will allocate a large amount of resources even when no faults are threatening the service.
Denying this truth is not a wise choice: As Einstein said, the rule should be "Make everything as simple as possible, but not simpler". Likewise, hiding complexity is good, but hiding too much can lead to disasters—history of computing is paved with dreadful examples unfortunately. The main lesson learned by carrying out our research is that this problem *must* be addressed, and that this complexity must not be neglected, but isolated into architectural components. This section introduced a translation-based approach that has precisely these design goals—addressing the problem of complex fault behaviors and providing an architectural solution that make it as simple as possible for the designer to concentrate on their functional design goals.
Our tool provides programmers of commodity languages such as C or Java with adaptively redundant data structures. Designers using our tool can define redundant data structures in which the degree of replication is not fixed once and for all at design time, but changes dynamically with respect to the disturbances experienced during the run time. In Chapter 10 the performance of our tool is reported. Such analysis is carried out through a fault injector called scrambler, which is also briefly introduced.
Redundant variables are a recent research project of our group at the University of Antwerp. Future work will be devoted to further developing and enhancing both those tools, e.g. by experimenting with other adaptation strategies, other replica allocation strategies, and introducing more complex fault injection scenarios such as single event effects. The global effect of adaptive dependable provisions on power consumption shall also be investigated in the near future.

# 5 CONCLUSIONS

Focus of this chapter has been approaches that reach fault-tolerance by means of compilers and translators. Such tools are not only used to bring "syntactic sugar" and possibly enhance SA and SC: It was shown how compilers can become another important "place" where one may place application-level fault-tolerant protocols that operate at the level of the programming language through so-called meta-object protocols. The reader is also provided of what we consider to be a glimpse into the future of our discipline: As cleverly stated by Torres-Pomales,

> "we use computer languages to try to capture the essence of software, but the concepts are so intricate that they generally defy

> attempts to completely visualize them in a practical manner and require the use of techniques to simplify relationships and enable communication among designers.” (Torres-Pomales, 2000)

This is exactly the starting point of the approaches based on introspection. It is our belief that such approaches, combined with knowledge management, semantic processing, and service orientation, may provide us with an important tool to deal effectively with software complexity.

page

# FAULT-TOLERANT PROTOCOLS USING FAULT-TOLERANCE PROGRAMMING LANGUAGES

## 1 INTRODUCTION AND OBJECTIVES

The programming language itself is the focus of this chapter: Fault-tolerance is not embedded in the program (as it is the case e.g. for single-version fault-tolerance), nor around the language (through compilers or translators); on the contrary, fault-tolerance is provided through the syntactical structures and the run-time executives of fault-tolerance programming languages. Also in this case a significant part of the complexity of dependability enforcement is moved from each single code to the architecture, in this case the programming language.
Many cases exist of fault-tolerance programming languages; this chapter proposes a few of them, considering three cases: Object-oriented languages, functional languages, and hybrid languages. In particular it is discussed the case of Oz, a multi-paradigm programming language that achieves both transparent distribution and translucent failure handling.

## 2 FAULT-TOLERANT PROTOCOLS USING CUSTOM PROGRAMMING LANGUAGES

Another approach is given by working at language level enhancing a pre-existing programming language or developing an ad hoc distributed programming language so that it hosts specific fault-tolerance provisions. The following two sections cover these topics.

### 2.1 Object-oriented approaches

#### 2.1.1 The Arjuna Distributed Programming System

Arjuna (Arjuna Technologies, ltd.) is an object-oriented system for portable distributed programming in C++ (Shrivastava, 1995). It can be considered as a clever blending of useful and widespread tools, techniques, and ideas—as such, it is a good example of the evolutionary approach towards application-level software fault-tolerance. It exploits remote procedure calls (Birrell & Nelson, 1984) and UNIX daemons (Haviland & Salama, 1987). On each node of the system an *object server* connects client objects to objects supplying services. The object server also takes care of spawning objects when they are not yet running (in this case they are referred to as "passive objects"). Arjuna also exploits a "naming service", by means of which client objects request a service "by name". This transparency effectively supports object migration and replication. Arjuna offers the programmer means for dealing with atomic actions (via the two-phase commit protocol) and persistent objects. Unfortunately, it requires the programmers to explicitly deal with tools to save and restore the state, to manage locks, and to declare in their applications instances of the class for managing atomic actions. As its authors state, in many respects Arjuna asks the programmer to be aware of several complexities—as such, it is prejudicial to transparency and separation of design concerns (insufficient SC). On the other hand, its good design choices result in an effective, portable environment.
Arjuna provides the programmer with "a computation model in which application programs manipulate persistent objects under the control of atomic actions" (Shrivastava, 1995). This concise definition reflects the opinions of many fault-tolerance language designers, that is, that coupling the object and atomic action model with object persistency provides a good solution to designing fault-tolerant systems. In Arjuna a failure appears to the programmer as a persistent object becoming unavailable. Actual reasons for this event may be e.g. a crash of the object repository, or a network partitioning. Accordingly, avoiding failures in Arjuna is achieved by any method directed to increasing the availability of its objects. The main approach used for this in Arjuna is object replication. Like in any replication scheme, clearly this calls for consistency protocols among the replicas.
As mentioned above, Arjuna uses an object server to turn passive objects into active ones, when a reference to any such object is detected in the system. Binding is the name of the operation that creates a local instance of any valid Arjuna object and creates one such reference. The object server then attaches the latest (committed) state of the named object to the local instance.

An example (from the cited reference) follows:

```
{       /* scope of o1,o2 and act is between the C++ block ("automatic variables") */
MyArjunaObject       o1(Name-A); /* o1 is bound to object whose global name is "A" */
MySecondArjunaObject o2(Name-B); /* o2 to "B" */
AtomicAction act;                /* o1 and o2 will be used in atomic action act */
act.Begin();                     /* act starts */
o1.method1(...);
o2.method1(...);                 /* invocations, possibly changing state of o1 and o2 *
o2.method2(...);
.......
act.End();                       /* act commits: if successful, "A" and "B" get updated
```

```
}       /* o1 and o2 are deallocated, which breaks bindings to A and B */
```

As it is often the case for ALFT based on programming languages, Arjuna builds on top of a few design axioms that determine its effectiveness but also limit its general usability. Hence SA is assessed as “average”. No support for adaptivity exists, though transparency replication could have been exploited to reach a better A (limited/potentially good).

### 2.1.2 Programming languages based on Compositional Filters

Composition Filters (CF) are a concept developed and explored at the University of Twente, The Netherlands. Originally CF were embedded into a custom programming language called Sina. Later the concept was enucleated from Sina and made available as extensions to existing languages such as C++ (Glandrup, 1995) or Java (Wichman, 1999). The latter reference provides a very nice and clear metaphor explaining what CF are. In what follows the main reference is (Wichman, 1999) and the Java implementation of composition filters, ComposeJ. Composition filters allow extending the object oriented model with the possibility to model concerns such as coordinated behavior, dynamic inheritance, delegation and multiple interfaces/views in a reusable, extensible and adaptable way. The main idea is that of introducing input and output filters, i.e. components that intercept and manipulate message transmission and reception. An input filter set filters incoming messages and an output filter set filters outgoing messages. Filters in a filter set orderly evaluate and possibly manipulate the messages before passing them to the target object. All messages sent to the object have to pass its input filter set. All messages sent by the object have to pass its output filter set. A special object (the filter object) provides the filters as its methods, which are not publicly available to the filtered object. Filter set are like a block of selection statements: An ordered cascade of conditions, each of which checks whether the incoming message belongs or not to some category. If so, the message is intercepted (”accepted”, in the CF lingo) and processed. As a consequence of this processing, the message may change. Then the message is passed to the next filter in the set and this goes on until the last filter in the set is encountered. A message that is not accepted by the last filter raises an exception. Clearly this procedure can be added to any hosting language, as it is a procedural addition that has no impact on the language grammar.
The input and output filter set are declared within a class through the inputfilters and outputfilters keywords, for instance,
inputfilters filter1 : Error = ..(will be explained shortly) filter2 : Dispatch =(will be explained shortly)
The general structure of a filter element is

FilterName : FilterType = Conditions ConclusionOperator MatchingPattern Parameters

"Conditions" is a Boolean that rejects the message when false. If true, the second and the third sections come into play: if "ConclusionOperator" is an inclusion, represented by '`=>`', then one accepts all messages that verify the matching pattern; if it is an exclusion, represented by '`~>`', then one accepts all message that *did not* verify a previous inclusion operator. "Parameters" is an optional section that depends on the filter being used. If a message is accepted, the filter's accept handler is executed, otherwise the next filter element is evaluated. If there are no more elements left in the current filter, the filter's reject handler is executed.
Filters belong to different types:

- Dispatch: any accepted message is delegated to one or more objects; rejected messages go to the next filter in the set.
- Error: any accepted message does to next filter; rejected ones raise an exception.
- Substitution: the accepted message is substituted with another one as specified in section "Parameters"; rejected messages go to the next filter.
- Send: accepted messages are passed to any object; rejected messages go to the next filter.
- Meta: any accepted message is reified and delegated to a "meta-object" (that is, an object describing and monitoring the behavior of another object); rejected messages go to the next filter.
- Wait: accepted messages go to the next filter, while rejected ones are blocked until a certain condition is met—which is clearly useful for concurrency control and synchronization.

For the sake of brevity these filters shall not be discussed in detail, but it is clear that they have the potential to compose powerful fault-tolerance mechanisms based, e.g., on replication and reflection. As an example, a Dispatch input filter could be used to redirect an input message to several replicas, while a Substitution or a Meta filter could be used to do voting among the output objects produced by the replicas; a Wait filter could make sure that all replicas synchronize, and so forth.
Composition Filters and Aspect Oriented Programming (described in Chapter 7) have many points in common, as they represent two paths towards reaching similar goals, e.g., separation of concerns. The main difference between the two approaches is in the fact that Composition Filters work on a per object basis, while AOP languages provide a system-wide (application-wide) specification which is integrated with the class hierarchies by means of a pre-processor (the aspect weaver, see Chapter 7).
Composition Filters provide "by construction", so to say, excellent sc. As it is often the case in this category of fault-tolerance provisions, sa is limited to those fault-tolerance techniques that can be expressed in terms of composition filters (average SA). All CF implementations we are aware of work at compile-time, which impacts negatively on A.

#### 2.1.3 FT-SR

FT-SR (Schlichting & Thomas, 1995) is basically an attempt to augment the SR (Andrews & Olsson, 1993) distributed programming language with mechanisms to facilitate fault-tolerance. FT-SR is based on the concept of fail-stop modules (FSM). A FSM is defined as an abstract unit of encapsulation. It consists of a number of threads that export a number of operations to other FSMs. The execution of operations is atomic. FSM can be composed so to give rise to complex FSMs. For instance it is possible to replicate a module $n > 1$ times and set up a complex FSM that can survive to $n - 1$ failures. Whenever a failure exhausts the redundancy of a FSM, be that a simple or complex FSM, a failure notification is automatically sent to a number of other FSMs so to trigger proper recovery actions. This feature explains the name of FSM: as in fail-stop processors, either the system is correct or a notification is sent and the system stops its functions. This means that the computing model of FT-SR guarantees, to some extent, that in the absence of explicit failure notification, commands can be assumed to have been processed correctly. This greatly simplifies program development because it masks the occurrence of faults, offers guarantees that no erroneous results are produced, and encourages the design of complex, possibly *dynamic* failure semantics (see Chapter 1) based on failure notifications. Of course this strategy is fully effective only under the hypothesis of perfect failure detection coverage—an assumption that sometimes may be found to be false.
FT-SR exloits much of the expressive power of SR to offer fault-tolerance to the programmer; the only additions added by FT-SR are:

- Automatic generation of failure notifications, when a resource is destroyed due to failure or explicit termination, and
- so-called higher-level fail-stop atomic objects.

**Automatic generation of failure notifications.** In FT-SR a failure is detected by the language runtime system. The focus herein is on the ways the application programmer instructs a notification.
FT-SR offers so-called synchronous and asynchronous failure notifications:

- Synchronous notification is one that is attached to a method's invocation and specifies a backup method to be executed as soon as the primary has been detected as failed. This is done very simply from the point of view of the programmer: He or she has just to add the identifier of the backup method, as in

```
call { task1.primary, task2.backup }.
```

- Asynchronous failure notifications are used when one wants to make sure that, whenever a certain condition takes place during a certain time interval, a given "alarm" or reactive measure will be instructed. This is specified by issuing the `monitor` statement, as in

  ```
  monitor task1 send task2(arguments),
  ```

  after which the FT-SR run-time executive starts monitoring `task1` with `task2(arguments)` as reactive measure. This mean that, in case `task1` fails during monitoring, method `task2(arguments)` is implicitly invoked with the current value of its arguments.

**Higher-level fail-stop atomic objects.**

FT-SR makes use of replication and error recovery techniques to build more complex fault-tolerance mechanisms (Schlichting & Thomas, 1992). Replication is used to create a group of replicated resources that appear to the user as a single, more resilient or more performant resource. A similar concept has been adopted in Ariel for tasks (see Chapter 6). Replication is available to the FT-SR programmer through the `create` statement—an augmented version of the SR statement with the same name. As an example,

```
task1_cap := create (i := 1 to N) task2() on remote_node_caps[i]
```

creates a replicated task consisting of N replicas of task2, with replica i to be executed on the processing node specified in `remote_node_caps[i]`.

Once the replicated task is created, all operations to that task are managed accordingly: Sending messages becomes a multicast and the same applies for invoking methods. The system guarantees consistent total order, but the programmer can choose otherwise. Sending is managed through atomic broadcast. When performing a call to a method in a replicated task,it is the run-time system that makes sure that only a single result is returned to the caller. No voting is done by the run-time system in this case: the failure semantic assumptions of fail-silent processes allows the system to just return the first result becoming available.

Another important feature offered by FT-SR is restartability—the ability to instruct the automatic restart of a failed entity in a healthier location of the system's. The syntax for doing so is very simple:

```
restart entity on somewhere_else.
```

The entity may be replicated, in which case the programmer can make use of a syntax similar to that of `create`:

```
restart (i := 1 to N) task2() on remote_node_caps[i].
```

Restarted FT-SR entities are not…restarted from scratch: they retain their state—a useful property which is not available, e.g., with the RESTART recovery code of Ariel (again, see Chapter 6 for more details on Ariel).
Finally, FT-SR offers persistency (also called "implicit restart" by its authors): any entity tagged with the `persistent` attribute when declared is automatically restarted on any of a certain number of backup nodes ("backup virtual machines" in FT-SR lingo). This allows to compose easily a stable storage resource, which in turn is an important requirement to build even more complex and advanced fault-tolerance protocols.
FT-SR places fault-tolerance in the foreground of system design, which translates in bad SC. It offers several constructs, with sufficient SA. No support for A is part of the language.

#### 2.1.4 ARGUS

Argus (Liskov, 1988) is a distributed object-oriented programming language and operating system. Argus was designed to support application programs like banking systems. To capture the object-oriented nature of such programs, it provides a special kind of objects, called guardians, which perform user-definable actions in response to remote requests. To solve the problems of concurrency and failures, Argus allows computations to run as atomic transactions. Argus' target application domain is clearly the one of transaction processing.
Like in Arjuna, Argus builds on top of a few fault-tolerance design axioms, which limits Argus' SA. Explicit, non-transparent support translates in insufficient SC. No support for adaptivity has been foreseen in Argus.

#### 2.1.5 The Correlate Language

The Correlate object-oriented language (Robben, 1999) adopts the concept of *active object*, defined as an object that has control over the synchronization of incoming requests from other objects. Objects are active in the sense that they do not process immediately their requests—they may decide to delay a request until it is accepted, i.e., until a given precondition (a guard) is met—for instance, a mailbox object may refuse a new message in its buffer until an entry becomes available in it. The precondition is a function of the state of the object and the invocation parameters—it does not imply interaction with other objects and has no side effects. If a request cannot be served according to an object's precondition, it is saved into a buffer until it becomes servable, or until the object is destroyed. Conditions like an overflow in the request buffer are not dealt with in (Robben, 1999). If more than a single request becomes servable by an object, the choice is made non-deterministically.

Correlate uses a communication model called "pattern-based group communication"—communication goes from an "advertising object" to those objects that declare their "interest" in the advertised subject. This is similar to Linda's model of generative communication, introduced in Chapter 4. Objects in Correlate are autonomous, in the sense that they may not only react to external stimuli but also give rise to autonomous operations motivated by an internal "goal". When invoking a method, the programmer can choose to block until the method is fully executed (this is called synchronous interaction), or to execute it "in the background" (asynchronous interaction). Correlate supports meta-object protocols. It has been effectively used to offer transparent support for transaction, replication, and checkpoint-and-rollback. The first implementation of Correlate consists of a translator to plain Java plus an execution environment, also written in Java.
Correlate bears several similarities with Composition Filters and reaches the same values of the structural attributes.

### 2.1.6 Fragmented Objects

Fragmented Objects (FO) are an extension of objects for distributed environments (Makpangou, Gourhant, Narzul, & Shapiro, 1994). FO do not reside integrally in one processing node, but are decomposed into chunks called "fragments", consisting of data and methods, which may reside on different nodes. The logics for the distribution of fragments is part of the objects themselves. The client of a FO must have access to at least one fragment.
FO offer an abstract view and a concrete view: In the abstract view they appear as a single, shared object. In the concrete view, the designer can decompose the objects into fragments and can deploy them on different machines. He or she may also control the communications among fragments. All these aspects are specified through a custom programming language, FOG (an extension of C++), a toolbox, and a compiler also responsible for object serialization.
The key aspect of FO with respect to dependability is the full transparency that they provide their users with: In particular there is no way to distinguish between a local object and a local fragment. This paves the way to the transparent adoption of dependability methods based on replication and reconfiguration (Reiser, Kapitza, Domaschka, & Hauck, 2006). In particular the amount of redundancy used could be made adaptively tracking the current disturbances—with an approach similar to the redundant variables described in Chapter 4. This translates in potentially good A. Also SC is good in FO, due to the ingenious separation between abstract and concrete views. SA appears to be somewhat limited due to specifics of the approach.
The interest around FO has never abated—examples include the adaptive fragmented objects of FORMI (Kapitza, Domaschka, Hauck, Reiser, & Schmidt, 2006).

## 2.2 Functional Languages

### 2.2.1 Fault-Tolerance Attribute Grammars

The system models for application-level software fault-tolerance encountered so far all have their basis in an imperative language. A different research trend exists, which is based on the use of functional languages. This choice translates in a program structure that allows a straightforward inclusion of fault-tolerance means, with high degrees of transparency and flexibility. Functional models that appear particularly interesting as system structures for software fault-tolerance are those based on the concept of *attribute grammars* (Paakki, 1995). This paragraph briefly introduces the model known as FTAG (fault-tolerant attribute grammars) (Suzuki, Katayama, & Schlichting, 1996), which offers the designer a large set of fault-tolerance mechanisms. A noteworthy aspect of FTAG is that its authors explicitly address the problem of providing a syntactical model for the widest possible set of fault-tolerance provisions and paradigms, developing coherent abstractions of those mechanisms while maintaining the linguistic integrity of the adopted notation. This means that optimizing the value of attribute SA is one of the design goals of FTAG.

FTAG regards a computation as a collection of pure mathematical functions known as *modules*. Each module has a set of input values, called inherited attributes, and of output variables, called synthesized attributes. Modules may refer to other modules. When modules do not refer any other module, they can be performed immediately. Such modules are called primitive modules. On the other hand, non-primitive modules require other modules to be performed first—as a consequence, an FTAG program is executed by decomposing a "root" module into its basic submodules and then applying recursively this decomposition process to each of the submodules. This process goes on until all primitive modules are encountered and executed. The execution graph is clearly a tree called *computation tree*. This scheme presents many benefits, e.g., as the order in which modules are decomposed is exclusively determined by attribute dependencies among submodules, a computation tree can be mapped onto a parallel processing means straightforwardly.

The linguistic structure of FTAG allows the integration of a number of useful fault-tolerance features that address the whole range of faults—design, physical, and interaction faults. One of these features is called *redoing*. Redoing replaces a portion of the computation tree with a new computation. This is useful for instance to eliminate the effects of a portion of the computation tree that has generated an incorrect result, or whose executor has crashed. It can be used to implement easily "retry blocks" and recovery blocks by adding ancillary modules that test whether the original module behaved consistently with its specification and, if not, give rise to a "redoing", a recursive call to the original module.

Another relevant feature of FTAG is its support for *replication*, a concept that in FTAG translates into a decomposition of a module into $N$ identical submodules implementing the function to replicate. The scheme is known as *replicated decomposition*, while involved submodules are called *replicas*. Replicas are executed according to the usual rules of decomposition, though only one of the generated results is used as the output of the original module. Depending on the chosen fault-tolerance strategy, this output can be, e.g., the first valid output or the output of a demultiplexing function, e.g., a voter. It is worth remarking that no syntactical changes are needed, only a subtle extension of the interpretation so to allow the involved submodules to have the same set of inherited attributes and to generate a collated set of synthesized attributes.

FTAG stores its attributes in a stable object base or in primary memory depending on their criticality—critical attributes can then be transparently retrieved from the stable object base after a failure. Object versioning is also used, a concept that facilitates the development of checkpoint-and-rollback strategies.

FTAG provides a unified linguistic structure that effectively supports the development of fault-tolerant software. Conscious of the importance of supporting the widest possible set of fault-tolerance means, its authors report in the cited paper how they are investigating the inclusion of other fault-tolerance features and trying to synthesize new expressive syntactical structures for FTAG—thus further improving attribute SA.

FTAG also exhibits "by construction" a good separation of concerns (SC). No support for A is known to exist for FTAG.

Unfortunately, the widespread adoption of this valuable tool is conditioned by the limited acceptance and spread of the functional programming paradigm outside the academia.

## 2.3 A Hybrid Case: Oz

Oz is defined by its authors as "a multiparadigm programming language". The main reasons for that is that Oz offers, by construction, several features common to programming paradigms such as logic, functional, imperative, and object-oriented programming. Another important feature of Oz is that it provides the programmer with a network-transparent distributed programming model that facilitates considerably the development of distributed fault-tolerant applications. The Oz programming system, called Mozart, was designed by the so-called Mozart Consortium.

Thanks to its rich model, Oz allows to solve, to some extent, the transparency conundrum of distributed computing: Indeed distributed computing approaches either choose to mask all complexity providing an illusion of a fully synchronous system where all failures and disruptions are masked, or go for a fully translucent system where everything is made known and reflected onto the system controller. Oz solves this and shows that "network transparency is not incompatible with entity failure reflection" (Collet & Mejías, 2007). The idea is that the language gives the illusion of a single memory space shared by distributed processing nodes called sites. Full transparency is achieved for this:

It is simply not possible to tell whether a method or an entity is local or distributed. But this is not true for all aspects of distribution—in particular, site crashes and partial failures are made translucent and reflected in the language. The mechanism offered by Oz to handle partial failures is asynchronous failure detectors, managed through failure listeners: All entities produce streams of events that reflect the sequential occurrence of their fault states. Any Oz task can become a failure listener, that is, it can hook to such streams and be informed of all the faults experienced by any other tasks. This means that fault detection is intrinsically managed by Oz and Mozart. Error recovery can then be managed by guarded actions, a little like in ARIEL, the error recovery language described in next chapter. This facilitates considerably the development of asynchronous failure detectors with one of the algorithms described in Chapter 8.

The trade off between transparency and translucency in Oz leads to a satisfactory SC. Its multiparadigmic nature should translate in a good score for SA, though no concrete evidence to this appears to exist. Oz has been used to express adaptive control loops (Van Roy, 2006) for self-management, so it proved to exhibit good A.

# 3 CONCLUSION

Several examples of fault-tolerance protocols embedded in custom programming languages have been shown. This class of methods can achieve satisfactory degrees of SC. The language designer has the widest syntactic freedom which is necessary to achieve good values of SA, but often, as a result of the design choices, the programmer is confined to a limited amount of possibilities. Attribute A depends on specific characteristics of the language, e.g. being able to select dynamically the error recovery strategy when the environment and its faults change.

Most of the programming languages discussed so far is not being supported anymore, a noteworthy exception being Oz, whose most widespread platform, Mozart, is a strategic research path at the University of Louvain-la-neuve, Belgium.

Next chapter is devoted to a special case of a fault-tolerance programming language: The ARIEL error recovery language.

page

# THE RECOVERY LANGUAGE APPROACH

## 1 INTRODUCTION AND OBJECTIVES

After having discussed the general approach of fault-tolerance languages and their main features, the focus is now set on one particular case: The ARIEL[1] recovery language. It is also described an approach towards resilient computing based on ARIEL and therefore dubbed the "recovery language approach" ($\mathcal{R}\varepsilon\mathcal{L}$). In this chapter first the main elements of $\mathcal{R}\varepsilon\mathcal{L}$ are introduced in general terms, coupling each concept to the technical foundations behind it. After this a quite extensive description of ARIEL and of a compliant architecture are provided. Target applications for such architecture are distributed codes, characterized by non-strict real-time requirements, written in a procedural language such as C, to be executed on distributed or parallel computers consisting of a predefined (fixed) set of processing nodes.
Reason for giving special emphasis to ARIEL and its approach is not in their special qualities but more on the fact that, due to the first-hand experience of the author, who conceived, designed and implemented ARIEL in the course of his studies, it was possible for him to provide the reader with what may be considered as a sort of practical exercise in system and fault modeling and in application-level fault-tolerance design, recalling and applying several of the concepts introduced in previous chapters.

## 2 THE ARIEL RECOVERY LANGUAGE

This section casts the basis of a general approach in abstract terms, while a particular instance of the herein presented concepts is described in Section 3 as a prototypic distributed architecture supporting a fault-tolerance linguistic structure for application-level fault-tolerance. System and fault models are drawn. The approach is also reviewed with respect to the structural attributes (SC, SA and A) and to the approaches presented in Chapter 3, 5, and 6.
The structure of this section is as follows:

- Models are introduced in Sect. 2.1.
- Key ideas, concepts, and technical foundations are described in Sect. 2.2.
- Section 2.4 shows the workflow corresponding to using $\mathcal{R}\varepsilon\mathcal{L}$.

- Sect. 2.5 summarizes the positive values of the structural attributes SA, SC, and A for $\mathcal{RεL}$.

## 2.1 System and Fault Models

This section introduces the system and fault models that will be assumed in the rest of this chapter.

### 2.1.1 System Assumptions

In the following, the target system is assumed to be a distributed or parallel system. Basic components are nodes, tasks, and the network.

- A node can be, e.g., a workstation in a networked cluster or a processor in a MIMD parallel computer.
- Tasks are independent threads of execution running on the nodes.
- The network system allows tasks on different nodes to communicate with each other.

Nodes can be commercial-off-the-shelf (COTS) hardware components with no special provisions for hardware fault-tolerance. It is not mandatory to have memory management units and secondary storage devices.
A general-purpose operating system is required on each node. No special purpose, distributed, or fault-tolerant operating system is required.
The number $N$ of nodes is assumed to be known at compile time. Nodes are addressable by the integers in $\{0, \ldots, N-1\}$. For any integer $m > 0$ let us call the set of integers $\{0, \ldots, m-1\}$ as $I_m$. Let us furthermore refer to the node addressed by integer $i$ as $n_i, i$ in $I_N$.
Tasks are pre-defined at compile-time: in particular for each $i$ in $I_N$, it is known that node $n_i$ is to run $t_i$ tasks, up to a given node-specific limit. No special constraints are posed on the task scheduling policy.
On each node, say node $i$, tasks are identified by user-defined unique local labels—integers greater than or equal to zero. Let us call $I_{n_i}$ the set of labels for tasks to be run on node $n_i$, $i$ in $I_N$. The task with local label $j$ on node $i$ will be also referred to as $n_i[j]$.

The system obeys the *timed asynchronous distributed system model* (Cristian & Fetzer, 1999) already introduced in Chapter 2. As already mentioned, such model allows a straightforward modeling of system partitioning—as a consequence of sufficiently many omission or performance communication failures, correct nodes may be temporarily disconnected from the rest of the system during so-called "periods of instability" (Cristian & Fetzer, 1999).
Moreover it is assumed that, at reset, tasks or nodes restart from a well-defined, initial state—partial-amnesia crashes (defined in Chapter 1) are not considered.
A message passing library is assumed to be available, built on the datagram service. Such library offers asynchronous, non-blocking multicast primitives.

The adoption of a library like ISIS (see Chapter 3) is suggested, in order to inherit the benefits of its reliable multicast primitives.
As clearly explained in (Cristian & Fetzer, 1999), the above hypotheses match well to nowadays distributed systems based on networked workstations—as such, they represent a general model with no practical restriction.
The following assumptions characterize the user application:

1. (When $N > 1$ nodes are available): the target application is distributed on the system nodes.

2. It is written or is to be written in a procedural language such as, e.g., C or C++.

3. The service specification includes non-safety-critical dependability goals—safety-critical systems may also be addressed, but in this case the crash failure semantics assumption *must be supported with a very high coverage* (Mortensen, 2000) This would require:

   - Extensive self-checking (by means of, e.g., signature checking, arithmetic coding, control flow monitoring, or dual processors).
   - Statistical estimation of the achieved coverage, by means of proper fault injection.

4. Inter-process communication takes place by means of the functions in the above mentioned message passing library. Higher-level communication services, if available, must be built atop that message passing library too.

The reason behind the third assumption is that, forcing communication through a single virtual provision, namely the functions for sending and for receiving messages, allows a straightforward implementation of mechanisms for task isolation. This concept is explained in more detail in Sect. 3.2.9.

### 2.1.2 Fault Model

As suggested in Chapter 2, any effective design including dependability goals requires provisions, *located at all levels*, to avoid, remove, or tolerate faults. Hence, as an *application-level* structure, $\mathcal{R}\varepsilon\mathcal{L}$ is complementary to other approaches addressing fault-tolerance at *system level*, i.e., hardware-level and OS-level fault-tolerance. In particular, a system-level architecture such as GUARDS (Powell et al., 1999), that is based on redundancy and hardware and operating system provisions for systematic management of consensus, appears to be particularly appropriate for being coupled with $\mathcal{R}\varepsilon\mathcal{L}$, which offers application-level provisions for NVP and replication (see later on).
The main classes of faults addressed by $\mathcal{R}\varepsilon\mathcal{L}$ are those of accidental, permanent or temporary design faults, and temporary, external, physical faults. Both value and timing failures are considered. The architecture addresses one fault at a time: The system is ready to deal with new faults only after having recovered from the present one.

## 2.2 Key Ideas and Technical Foundations

The design of $\mathcal{R}\varepsilon\mathcal{L}$ tries to capture, *by construction*, some of the positive aspects of most of the approaches so far surveyed.
Some of the key design choices of $\mathcal{R}\varepsilon\mathcal{L}$ are:

- The adoption of a fault-tolerance toolset.
- The separation of the configuration of the toolset from the specification of the functional service.
- The separation of the system structure for the specification of the functional service from that for error recovery and reconfiguration.

These concepts and their technical foundations are illustrated in the rest of this section.

## 2.3 Adoption of a Fault-Tolerance Toolset

A requirement of $\mathcal{R}\varepsilon\mathcal{L}$ is the availability of a fault-tolerance **toolset**, to be interpreted herein as the conjoint adoption of:

- A set of fault-tolerance tools addressing error detection, localisation, containment and recovery, such as the ones in SwIFT (Huang, Kintala, Bernstein, & Wang, 1996) or EFTOS (Deconinck, De Florio, Lauwereins, & Varvarigou, 1997; Deconinck, Varvarigou, et al., 1997). As seen in detail in Chapter 3, fault-tolerance services provided by the toolset include, e.g., watchdog timers and voting. Such tools are called **basic tools** (BT) in what follows.
- A "**basic services library**" (BSL) is assumed to be present, providing functions for:
  - intra-node and remote communication;
  - task management;
  - access to the local clock;
  - application-level assertions;
  - functions to reboot or shut down a node.

  This library is required to be available in source code so that it can be instrumented, e.g., with code to forward information transparently to some collector (described in what follows). Information may include, for instance, the notification of a successful task creation or any failure of this kind. This allows to create fault streams as in Oz (Chapter 5). If supported, meta-object protocols (see Chapter 4) may also be used to implement the library and its instrumentation. It is also suggested that the functions for sending messages work with opaque objects that reference either single tasks or groups of tasks. In the first case, the function would perform a plain “send”, while in the second case it would perform a multicast. This would increase the degree of transparency.

- A distributed component serving as a sort of backbone controlling and monitoring the toolset and the user application. Let us call this application “the **Backbone**”. It is assumed that the Backbone has a component *on each node of the system* and that, through some software (and, possibly, hardware) fault-tolerance provisions, *it can tolerate crash failures of up to all but one node or component.* An application such as the EFTOS DIR net discussed in Chapter 3 may be used for this.

  Notifications from the BSL and from the BT are assumed to be collected and maintained by the Backbone into a data structure called “the **database**” (DB). The DB therefore holds data related to the current *structure* and *state* of *the system, the user application, and the Backbone.* A special section of the DB is devoted to keeping track of *error notifications*, such as, for instance, “divide-by-zero exception caught while executing task 11” sent by a trap handling tool like the one discussed in Chapter 3. If possible, error detection support at hardware or kernel level may be also instrumented in order to provide the Backbone with similar notifications. The DB is assumed to be stored in a reliable storage device, e.g., a stable storage device, or replicated and protected against corruption or other unwanted modifications.

- Following the hypothesis of the timed asynchronous distributed system model (Cristian & Fetzer, 1999), a time-out management system is also assumed to be available. This allows an application to define *time-outs*, namely, to schedule an event to be generated a given amount of “clock ticks” in the future (Cristian & Schmuck, 1995). Let us call this component the “**time-out manager**” (tom).

A prototype of a $\mathcal{R}\varepsilon\mathcal{L}$-compliant toolset has been developed within the European ESPRIT project “TIRAN”. Section 3.2 describes its main components.

### 2.3.1 Configuration Support Tool

The second key component of $\mathcal{R}\varepsilon\mathcal{L}$ is a tool to support *fault-tolerance configuration*, defined herein as the deployment of customized instances of fault-tolerance tools and strategies. $\mathcal{R}\varepsilon\mathcal{L}$ envisages a **translator** to help the user configure the toolset and his / her application. The translator has to support a custom **configuration language** especially conceived to facilitate configuration and therefore to reduce the probability of fault-tolerance design faults—the main cause of failure for fault-tolerant software systems (Laprie, 1998; Lyu, 1998).
As an output, the translator could issue, e.g., C or C++ header files defining configured objects and symbolic constants to refer easily to the configured objects. Recompilation of the target application is therefore required after each execution of the translator.
Configuration can group a number of activities, including:

- Configuration of system and application entities,
- configuration of the basic tools,
- configuration of replicated tasks,
- configuration for retry blocks,
- configuration for multiple-version software fault-tolerance.

The above configuration activities are now briefly described.

### 2.3.2 Configuration of System and Application Entities

One of the tasks of a configuration language is to declare the key entities of the system and to define a global naming scheme in order to refer to them. Key entities are nodes, tasks, and groups of tasks.
For each node $n_i, 0 <= i < N$ and for each task $n_i[j], 0 <= j < I_{n_i}$, a unique, global-scope identifier must be defined by the user. Let us call this identifier the task's *unique-id*. This can be done by editing a configuration script with rules of the form

$$\text{task}_t = n_i[j], \tag{1}$$

which assigns unique-id $t$ to $n_i[j]$, that is, task number $j$ on node $i$. Similarly one could define groups of task with rules of the form

$$\text{group}_g = \{\overline{u}\}, \tag{2}$$

where $\overline{u}$ is a list of comma-separated integers representing unique-ids. Such rule defines then a group named $g$ and made of the tasks corresponding to the mentioned unique-ids. The translator would then turn a configuration script containing rules of this kind into a header file to be compiled with the target application and, if necessary, in configuration files expected by the toolset.

### 2.3.3 Configuration of the Fault-Tolerance Tools in the Toolset

Specific instances of the tools in the toolset can be statically configured by means of the translator. For instance, in the case of a watchdog timer, configuration can specify:

- The unique-id of the watching task as well as that of the watched task.
- The initial expected frequency of "heartbeats" to be sent from the watched task to the watchdog.
- The actions to be taken when an expected heartbeat is not received in time,

and so forth. As an example, a watchdog timer may be configured in high level terms as follows:

```
WATCHDOG  TASK 10  WATCHES TASK 14
  HEARTBEATS EVERY 100 MS
  ON ERROR WARN TASK 18
END WATCHDOG
```

Ideally, the output of the translator should be a configuration file for the BSL to associate transparently the creation of the configured instance of the watchdog to the creation of the watched task. Doing like this, the only non-functional code to be intruded in the watched task can be the function call corresponding to sending the heartbeat to the watchdog. A generic "`HEARTBEAT`" method, with no arguments, can be used. This can be a symbolic name properly defined in a header file written by the translator and automatically included when compiling the user application. Even when instrumenting is not possible, the watched task can start the watchdog by means of a symbolic name properly defined in the above mentioned header file. This kind of translations can be applied to most of the tools of a library such as SwIFT or EFTOS that have been described in Chapter 3. Note how the minimal or absent code intrusion provides an optimal SC. The adoption of a custom, ad hoc language for the expression of the configuration concerns can be used to reach high values for SA and compile-time A.

**Configuration of Replicated Tasks.** The translator and the BSL may be used to implement replicated tasks, i.e., multiple instances of the same task that perform like a more dependable entity. The goal of the translator is to mask this choice and any other fault-tolerance technicalities, including, in this specific case, replication. This can be obtained, e.g., by solving separately the following sub-problems, both at syntactical and at semantic level:

1. Replication and forwarding of the input value.
2. Execution support of a fault-tolerance strategy.
3. Output management.

Problem 1 can be solved by defining a group-of-tasks object that, once passed to the BSL function for sending messages, triggers a multicast of the same message to the whole group, as suggested in Sect. 2.3.2.
Problem 2 can be solved in various ways—for instance, through a temporal redundancy scheme, executing the involved tasks one after the other, on the same node, or via spatial redundancy, executing tasks in parallel. Increased dependability may be obtained via a number of software techniques, implementing schemes such as active replication or primary-backup replication (Guerraoui & Schiper, 1997). As noticed in the cited paper, each scheme has both positive and negative aspects and requires solving specific problems. All these problems are low-level design issues that can be made transparent to the user of an $\mathcal{RεL}$ system. Other choices and options, e.g., which type and degree of replication and which redundancy scheme to adopt, that would result in non-functional code intrusion, can be also made transparent to the user by means of the translator (see Sect. 2.3.3), hence increasing configurability.
Problem 3, depending on the adopted scheme, can be as simple as sending a message (when in primary-backups mode) or could require special processing. For instance, in the case of active replication, two sub-problems would call for specific treatment:

3.1. Routing the outputs produced by the base tasks.

3.2. De-multiplexing, i.e., production of a unique output value from the values routed in sub-problem 3.1.

Sub-problem 3.1 can be solved, e.g., by a proper combination of pipelining and redirection. Other approaches may be used when the OS does not support the above tools. De-multiplexing, i.e., solution of sub-problem 3.2, can be for instance the result of a *voting procedure* performed among the outputs produced by the base tasks. Also in this case, the availability of a translator and an appropriate syntax rule could guarantee an almost complete separation of design concerns. The configuration of a replicated task requires the specification of:

- The unique-id of a task globally representing a set of replicas.
- The unique-ids of the replicas.
- The replication method and its parameters.
- The actions to be taken when an output is produced by the replicated task.
- The actions to be taken when an error occurs.

In the case of replicated tasks, the translator is also responsible for the set up of a proper run-time executive. The latter would then be responsible, at run-time, for the orchestration of the services required by the replicated tasks—task management, distributed voting, and so forth. Proper calls to the BSL and to instances of the basic tools may be used for this. A voting tool such as the EFTOS voting farm, described in Chapter 3, appears to be a natural choice, as it addresses many of the required issues.
It is worth noting how, also in this case, full transparency is reached: A client of a replicated task would have no way to tell whether its server be simple or replicated—possibly apart from some performance penalty and a higher quality of service. This translates into optimal SC. The same applies to SA and A for the reasons mentioned in Sect. 2.3.3.

**Configuration for Retry Blocks.** Transparent support for redoing (see e.g. language FTAG in Chapter 5)—another important fault-tolerance provisions—can be provided via "retry blocks". Again, a proper run-time executive is to be produced by the translator. The problems to be solved at this level include

1. reversibility of a failed task and
2. input replication.

The first problem can be solved by implementing some "recovery cache" (as in the recovery block technique described in Chapter 3): That is a mechanism to checkpoint the state of the calling task before entering the retry block and to roll it back to its original value in case the acceptance test fails. This may be done transparently or with the intervention of the user. In the latter case, one restricts the size of the recovery cache and reduces the corresponding overheads, at the same time increasing the code intrusion.
The second problem could be solved via an "input cache", i.e., a mechanism that:

- Intercepts the original input message.
- Stores the original message into some stable means.
- And forwards the saved original message to each new retry instance.

Transparent adoption of an input cache can also be realized, e.g., by means of pipelining and redirection (when the OS supports these).
The configuration of a retry block requires the specification of:

- The unique-id of a task to be retried in case of errors.
- An acceptance test, in the form of the name of a function returning a Boolean value or of a task that, upon termination, returns a Boolean value[2].
- A threshold $r$ representing the maximum number of retries.
- The actions to be taken when the base task fails for $r$ times in a row.

**Configuration for Multiple-Version Software Fault-Tolerance.** Compile-time support towards multiple-version software fault-tolerance can be provided by the translator through syntax rules and techniques similar to those described for task replication.
The configuration of a provision for MV requires the specification of:

- The unique-id of a task representing the provision.
- The unique-id's of the tasks running the versions.

- A set of thresholds representing time-outs on the execution of the version tasks[3].
- The name of the user-specified function to be executed by each version task.
- A method to de-multiplex the multiple outputs produced by the versions into a single output value.
- Possible arguments to the de-multiplexing method.
- The unique-id of a task to be notified each time an execution cycle is successfully completed.
- The unique-id of a task to be notified each time an execution cycle fails.

Support towards consensus recovery blocks (Scott, Gault, & McAllister, 1985) may be provided in a similar way. Acceptance tests should be specified as described in Sect. 2.3.3.

**An Example Scenario and Some Conclusions.** A possible compile-time and run-time scenario is now described for the case of the configuration of multiple-version software fault-tolerance. This is done in order to provide the reader with a more concrete view of the kind of support supplied by a configuration language.
It is assumed that the OS supports pipelining and stream redirection. It is also assumed that, by agreement, the user tasks that are going to be used as NVP versions forward a single output value onto the standard output stream. Finally, user tasks are assumed to be side effect-free.
Once fed with a configuration script, the translator writes a number of source files for the tasks corresponding to the employed versions. Each of these source files, set up from some template file, specify how to:

- Set up a configured instance of a distributed voting tool (such as the EFTOS voting farm).
- Redirect standard output streams.
- Execute one of the version tasks.

During the execution, when a client needs to access a service supplied by the provisions, it simply sends a request message to the corresponding task. The client does not need to know that the latter is actually an “NVP task”, that is, a group. Through the BSL, this sending turns into a multicast to the version tasks. These tasks, which in the meanwhile have transparently set up a distributing voting tool,

- get their input,
- compute some user-specified function,

- produce an output,
- and (by the above agreement) write that output into their standard output stream.

This output, which was already redirected through a piped stream to the template task, is fed into the voting system. The latter eventually produces an output that, in case of success, is sent to some output task with the notification of successful completion of a processing cycle.
Note that the client task of such an "NVP task" is completely unaware of the context in which it is running, with full transparency and separation of design concern. One can conclude that the adoption of a configuration strategy like the one just sketched can lead to an optimal SC. The variety of fault-tolerance provisions that can be supported by configuration and the adoption of a linguistic environment separated from the functional application layer can be exploited by programming language designers in order to attain optimal values for SA and A as well.

### 2.3.4 Recovery Languages

Two of the three key concepts of $\mathcal{R}\varepsilon\mathcal{L}$ have been described, namely the adoption of a fault-tolerance toolset and that of a configuration language. This section now introduces the third component of $\mathcal{R}\varepsilon\mathcal{L}$. This component supports *two* application layers, namely:

- The functional application layer, i.e., the one devoted to the fulfillment of the functional requirements and to the specification of an intended user service, supported by the run-time modules of conventional programming languages such as, e.g., C or C++.
- An ancillary application layer, specifically devoted to error processing and fault identification, to be switched in either *asynchronously*, when errors are detected in the system, or when the user *synchronously* signals particular run-time conditions, such as a failed assertion, or when the control flow runs into user-defined breakpoints.

In the following the **service language** is the language constituting the front layer and **recovery language** the one related to error processing.
Note how a recovery language can be specified in a separate script. The latter can then be translated into pseudo-code to be interpreted at run-time, or, e.g., into plain C to be compiled with the user application. In the first case, the functional code and the recovery code are fully separated, also at run-time, which can be exploited to reach optimal A and to extend or reduce the set of faults to be tolerated. This feature could be used to realize the vision of a fault model structured as a dynamic system, $F(t)$, as described in Chapter 2. The second case eliminates the overhead of interpreting the pseudo-code. The same **translator** used for fault-tolerance configuration may be used to generate the pseudo-code or the C source code.
The general strategy is as follows:

- During the lifetime of the application, *in the absence of errors*, the front layer controls the progress of the service supply while the Backbone collects and maintains in the DB the data concerning the state of each component of the application, the state of each node of the system, the state and progress of public and private resources, and so forth.

- *As soon as an error is detected* by a basic tool, such as a watchdog timer, the latter transparently forwards a notification of this event to the Backbone, which awakes the ancillary layer by enabling a module to interpret or execute the recovery code. Let us call this module **RINT**, for recovery interpreter.

A possible syntactical structure for the recovery language is that of a list of **guarded actions**, i.e., statements in the form

$$g : a,$$

where $g$ is a **guard**, i.e., a Boolean expression on the contents of the DB, and $a$ is one or more **actions** (to be specified later on). Hence, adequate paradigms for the recovery language could be that of *procedural* or *logic programming languages*. The following two paragraphs describe guards and actions in more detail.

**Guards** represent conditions that require recovery. As just said, they are Boolean expressions made of basic queries called **atoms**. Possible atoms may express conditions such as:

- Task $t$ has been detected as faulty.

- Task $t$ has been detected as faulty by error detection tool $d$.

- For $m$ times in a row, task $t$ has been detected as faulty.

- A time-out concerning task $t$ has expired.

- An $N$-version task has signaled that no full consensus has been reached, or in more detail:

  - For $m$ times in a row, the same version, say task $t'$, has been found in minority with respect to the other versions;
  - for $m$ times in a row, version $t'$ did not produce any output within its deadline;
  - and so forth.

- Node $n$ is down, or, in more detail:

  - Node $n$ has crashed;
  - Node $n$ is unreachable;

- No sign of life from node $n$ in the last $s$ seconds, as measured by the local clock;
  - and so forth.

- Task $t$ could not be restarted;
- Some of the tasks in group $g$ are faulty.

and so forth. Most of the atoms in a Boolean clause will require a DB query or proper actions on DB fields—for instance, conditions such as "task $t$ is affected by a transient fault" require the adoption, within the DB, of a thresholding statistical technique such as $\alpha$-count (Powell et al., 1999; Bondavalli, Chiaradonna, Di Giandomenico, & Grandoni, 1997), which is capable of distinguishing between transient and permanent/intermittent faults. In particular, $\alpha$-count (described in Sect. 3.2.4) can be straightforwardly "wired" into the DB management system within the Backbone, because that mechanism is based on adjusting some counters according to the contents in a stream of error messages forwarded by a set of error detectors.

**Actions** are local or remote commands to be executed when their guard is evaluated as true. Actions can specify, for instance, recovery or reconfiguration services. A special case of action can be also another guarded action—this allows to have nested guarded actions that can be represented as a tree. The execution of an inner guard is, again, its evaluation. When a parent guard is evaluated as false, all its actions are skipped, including its child guards. Actions can include, for instance:

- Switching tasks in and out of a fault-tolerance structure such as, e.g., an "NVP task" (see Sect. 2.3.3).
- Synchronizing groups of tasks.
- Isolating[4] groups of tasks.
- Instructions to roll the execution of a task or a group of tasks back to a previously saved checkpoint,

and so forth. Actions have a system-wide scope and are executed on any processing node of the system where a non-faulty component of the Backbone is running. They may include commands to send control signals to specific components of the Backbone or to user tasks. Basic tools such as a distributed voting tool like the EFTOS voting farm (discussed in Chapter 3) can be instructed so that they respond to that signal supplying transparent support towards graceful degradation, task switching, and task migration, as described for instance in (De Florio, Deconinck, & Lauwereins, 1998). Furthermore, the ability to distinguish between transient and permanent/intermittent faults can be exploited, in order to avoid unnecessary reconfigurations or other costly or redundancy-consuming actions, attaching different strategies to these two cases.

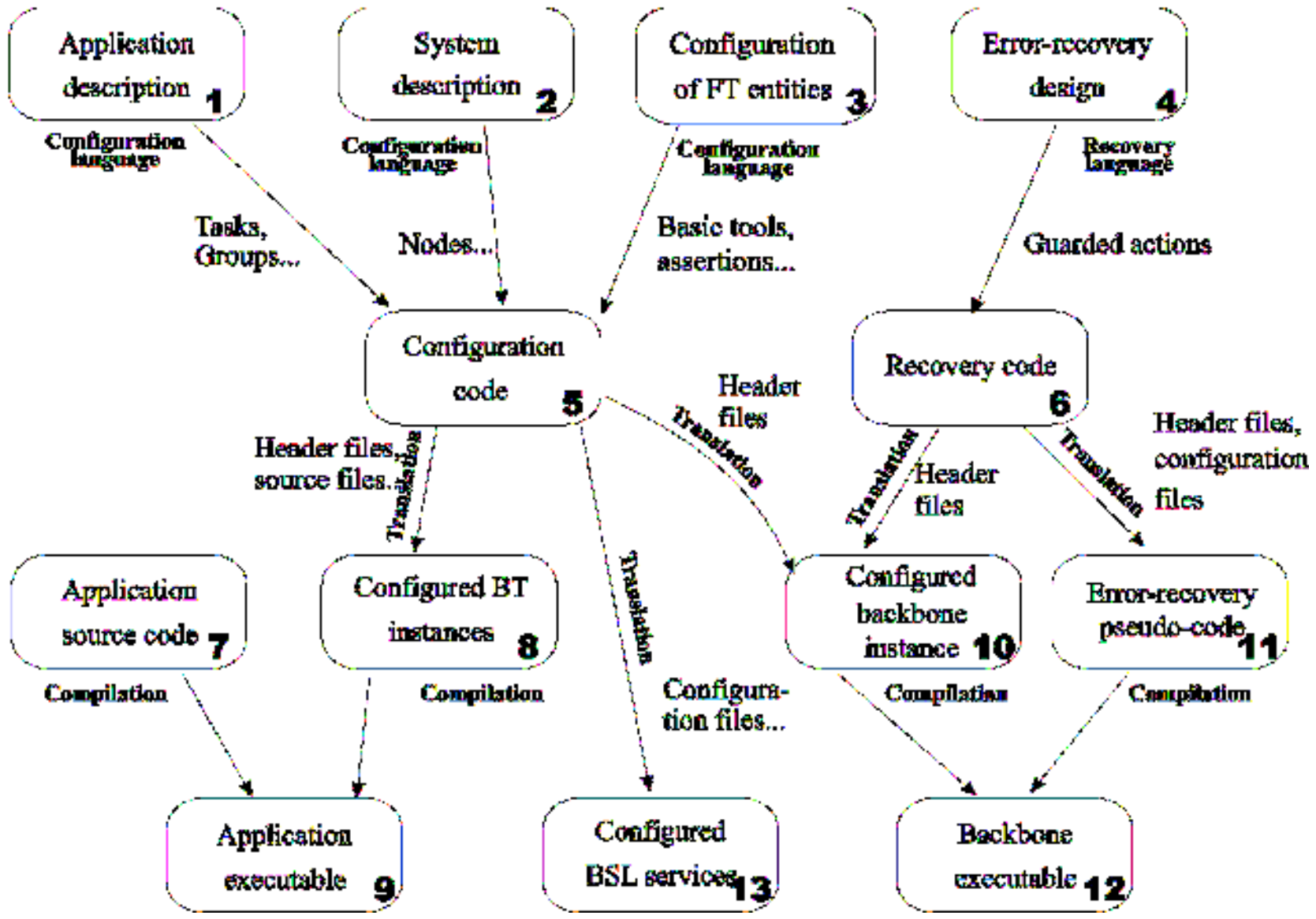

Figure 1: A workflow diagram for $\mathcal{R}\varepsilon\mathcal{L}$. Labels refer to usage steps and are described in Sect. 2.4.

## 2.4 Workflow

This section describes the workflow corresponding to the adoption of the $\mathcal{R}\varepsilon\mathcal{L}$ approach.
Figure 1 summarizes the workflow. The following basic steps have been foreseen:

- In the first steps (labels 1 and 2 in the cited figure), the designer describes the key application and system entities, such as tasks, groups of tasks, and nodes. The main tool for this phase is the configuration language.
- Next (step 3), the designer configures the basic tools and the fault-tolerance provisions he / she has decided to use. The configuration language is used for this. The output of steps 1–3 is the configuration code.
- Next (step 4), the designer defines what conditions need to be caught, and what actions should follow each caught condition. The resulting list is coded as a number of guarded actions via a recovery language.
- The configuration code and the recovery code are then converted via the translator into a set of C header files, C fragments, and system-specific configuration files (steps 5 and 6). These files represent: configured instances of the basic tools, of the system and of the application;

initialization files for the communication management functions; user preferences for the backbone; and the recovery pseudo-code.

- On steps 7–9, the application source code and the configured basic tool instances are compiled in order to produce the executable code of the application.
- Next, the backbone is compiled on steps 10–12.
- Finally, on step 13, the communication management services of the BSL are configured in order to allow the proper management of multicasting and other communication services.

## 2.5 Conclusions

In this section an approach to application-level software fault-tolerance has been introduced. Such an approach provides its users with a linguistic structure to express application-level error recovery concerns and to configure a number of fault-tolerance provisions *outside the functional application layer.* The latter can be provided by any procedural language.
The $\mathcal{R}\varepsilon\mathcal{L}$ approach addresses dependability goals for distributed or parallel applications written in any procedural language. Its target hardware platforms include distributed and parallel computers consisting of a set of processing nodes known at compile time. Within this application and system domain, the novel approach can be qualitatively assessed as reaching optimal values of the structural attributes. In particular, with respect to the application-level software fault-tolerance approaches reviewed in Chapter 3, 4, and 5:

- Error recovery is expressed in a separate programming context—thus good SC can be obtained.
- The executable code is separable—that is, at run-time, the portion of the executable code devoted to the fault-tolerance aspects is clearly distinct from the functional code.
- Run-time adaptability to diverse environmental conditions may be obtained without requiring the entire executable code to be recompiled. This and previous items bring to a good A.
- A large number of well-known fault-tolerance strategies are supported in a straightforward way (i.e. good SA).
- The $\mathcal{R}\varepsilon\mathcal{L}$ approach addresses a wide class of applications: That of distributed applications, with non-strict real-time requirements, written in a procedural language, to be executed on distributed or parallel computers consisting of a set of processing nodes known at compile time.
- Our approach reaches both the benefits of the evolutionary approaches, which base themselves on standards, and those of "revolutionary" approaches, exploiting ad-hoc, non-standard solutions.

- It can host other provisions and approaches.
- As shown in Chapter 9, in same cases $\mathcal{R}\varepsilon\mathcal{L}$ can be adopted with minimal programming effort and minimal adaptation of a pre-existing, non fault-tolerant application.

Finally, one can note that the language-based approaches such as those in Chapter 5 differ considerably from $\mathcal{R}\varepsilon\mathcal{L}$. Despite their many positive characteristics, their very axioms—using a custom syntax for both the functional and the fault-tolerance concerns—may restrict significantly their usability. The lack of standards is further detrimental to their diffusion. On the contrary, the choice to support a standard language for the functional aspects and an ad-hoc syntactical structure for the fault-tolerance aspects allows reach optimal SA with no impact on the usability.
Next chapter describes a prototype architecture based on the $\mathcal{R}\varepsilon\mathcal{L}$ approach.

# 3 A DISTRIBUTED ARCHITECTURE BASED ON THE RECOVERY LANGUAGE APPROACH

This section describes a prototypic architecture based on $\mathcal{R}\varepsilon\mathcal{L}$ that has been designed in the context of the European ESPRIT project TIRAN.
The structure of the section is as follows: first, project TIRAN is described in Sect. 3.1. The main components of the TIRAN architecture are covered in Sect. 3.2. Section 3.2.6 describes the TIRAN configuration and recovery language. Section 5 closes this part and draws a few conclusions.

## 3.1 The ESPRIT Project TIRAN

TIRAN is the name of the European ESPRIT project 28620. Its name loosely derives from "Tailorable fault-tolerance frameworks for embedded applications" (Botti et al., 1999). The main objective of TIRAN was to develop a software *framework*[5] providing fault-tolerant capabilities to automation systems. Application-level support to fault-tolerance was included by means of a $\mathcal{R}\varepsilon\mathcal{L}$ architecture, which is described in the rest of this chapter. The framework provides a library of software fault-tolerance provisions that are parametric and support an easy configuration process. TIRAN builds on top of EFTOS, inheriting several of the concepts and tools developed in that project and described in Chapter 3.
Using the TIRAN framework, application developers are allowed to select, configure and integrate provisions for fault masking, error detection, isolation and recovery among those offered by the library. Goal of the TIRAN project was to provide a tool that significantly reduced the development times and costs for dependable systems. TIRAN's target market segment was that of non-safety-critical[6] distributed real-time embedded systems (Botti et al., 2000).
TIRAN explicitly adopted formal techniques to support requirement specification and predictive evaluation. This, together with the intensive testing on pilot applications, was exploited in order to:

- Guarantee the correctness of the framework.
- Quantify the fulfillment of real-time, dependability and cost requirements.
- Provide guidelines to the configuration process of the users.

Most of the TIRAN framework was designed for being platform independent. A single version of the framework has been written in the C programming language making use of a BSL designed by the TIRAN consortium and initially developed by one of the partners for the development platform (Microsoft Windows NT). Within TIRAN, a number of target platforms and systems compliant to the model sketched in Sect. 2.1 have been selected. Porting the TIRAN code to these platforms was mainly obtained through the porting of the BSL. As mentioned already, this does not automatically translate in *porting the service* expected by the BSL; the latter was obtained by proper verification and validation of its properties on the target platforms.

Such platforms included:

- A set of Mosaic020[7] boards running the Virtuoso microkernel (Systems, 1998; Mosaic020, 1998).
- A set of VME boards based on PowerPC processors and running the VxWorks operating system (VxWorks, 1999). An MMU is available on these boards, which translates in special hardware support for memory protection.
- A set of DEC Alpha boards running the TEX kernel. These boards have no MMU and address spaces of tasks are not protected. Proper protection is guaranteed by TEX, which does not allow any dynamic memory allocation. All addresses are therefore known at compile time, so a preprocessing tool could be used in principle to prevent memory contamination.
- Clusters of Windows-CE personal computers (PCs).

The final version of the TIRAN framework, running on all the above systems, was demonstrated in November 2000 at the final TIRAN review meeting and at the TIRAN workshop (Thielemans, 2000).

Some elements of the TIRAN framework (namely those derived from EFTOS) are also available on Parsytec CC and Parsytec Xplorer systems based on the PowerPC processor and the EPX kernels (Parsytec, 1996a, 1996b) and on a proprietary hardware board based on the DEC Alpha chip and running the TEX kernel.
The main components of the TIRAN framework are:

1. The TIRAN toolset, the components of which are discussed in Sect. 3.2.

2. The TIRAN configuration and recovery language ARIEL, described in Sect. 3.2.6. In particular:

   - Its translator is dealt with in Sect. 3.2.9.
   - The run-time executive of ARIEL, which is sketched in Sect. 3.2.9.

As mentioned already TIRAN was built on top of ESPRIT project EFTOS—described in full detail in Chapter 3. In particular TIRAN adopted and improved the EFTOS voting farm (De Florio et al., 1998), also described in the mentioned chapter, so as to function as the distributed voting system of the TIRAN framework.
Figure 2 draws the TIRAN architecture and locates its main components into it. Two types of edges are used in the picture:

- The thinner, directed edge represents relation "<Sends>": If such an edge goes from entity *a* to entity *b* then it means that *a* sends a *data* message to *b* or requests a service provided by *b* through a *control* message or a method invocation.

- The thicker edge is only used between the RINT and ARIEL entities and means that *the former implements the latter* through the process of interpreting a pseudo-code called "r-codes" (detailed information about this is available in Sect. 3.2.9).

The central, whiter layers represent the TIRAN framework. In particular:

- The level 0 hosts the TIRAN BSL, which gives system-independent access to the services provided by the underlying run-time system (see Sect. 2.3). The user application, the BT, the Backbone, the Time-out Manager, and the dependable mechanisms (DMs, see Sect. 3.2.3), all make use of the BSL, e.g., to create tasks or to send messages through mailboxes.

- Level 1 services are provided by the BT for error detection and fault masking (level 1.1) and by the BT addressing error isolation, recovery and reconfiguration (level 1.2). Both services are supplied by node-local (simplex) provisions. Section 3.2.1 describes in more detail the BT. The edge from BT to Backbone represents the sending of error detection or diagnostic messages. The edge from Backbone to BT represents control messages, such as, for instance, a request to modify a parameter of a watchdog, or a request to reboot the local node (see Fig. 4).

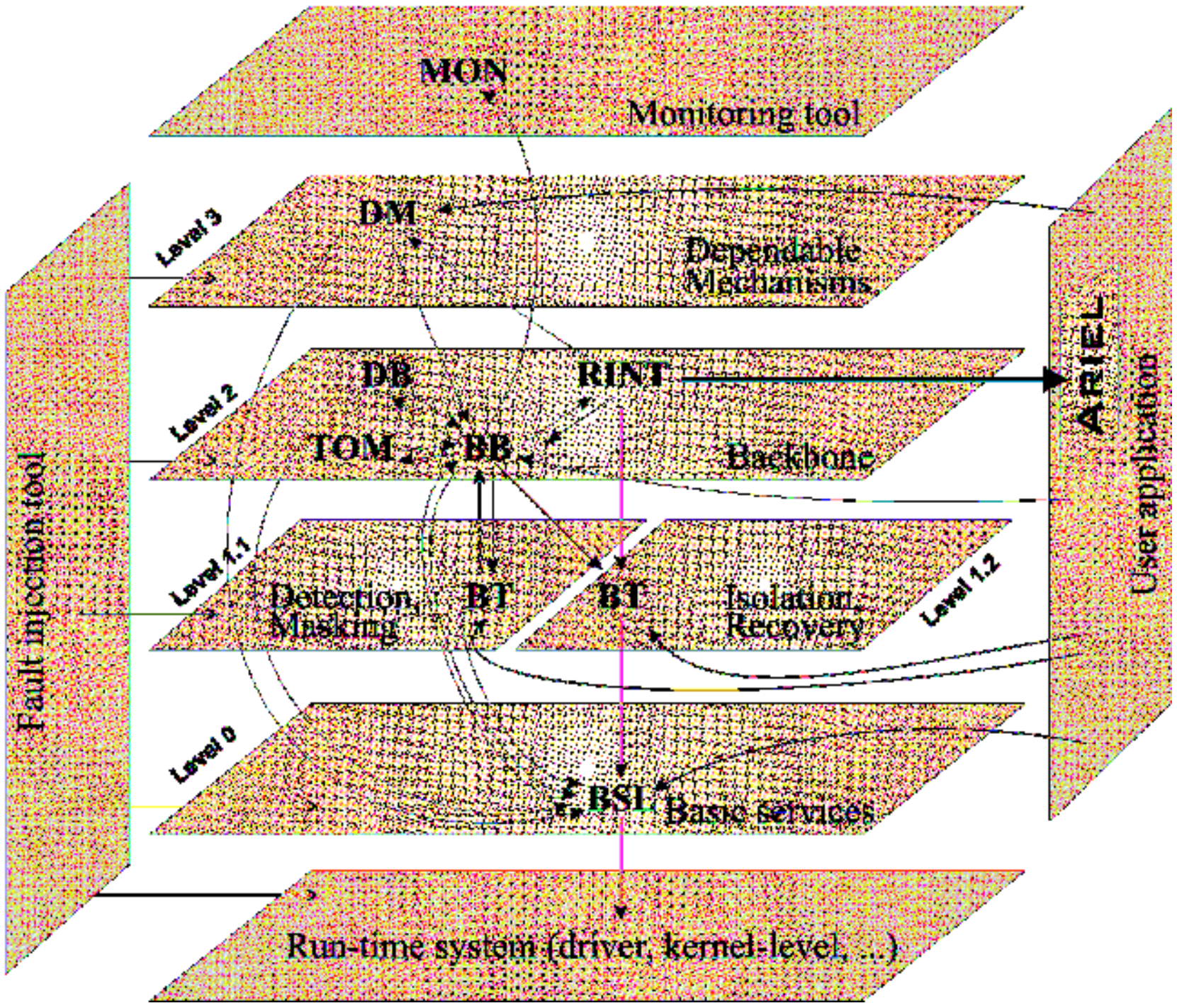

Figure 2: A representation of the TIRAN elements. The central, whiter layers constitute the TIRAN framework. This same structure is replicated on each processing node of the system.

- Level 2 hosts the TIRAN Backbone, including the Time-out Manager component, the DB management functions, and the recovery interpreter, RINT. In Fig. 2, the thicker edge connecting RINT to ARIEL means that RINT actually implements (executes) the ARIEL programs. Note the control and data messages that flow from Backbone to Time-out Manager, DB, and RINT. RINT also sends control messages to the isolation and recovery BT. Data messages flow also from Backbone to the monitoring tool.
- Dependable mechanisms, i.e., high-level, distributed fault-tolerance tools exploiting the services of the Backbone and of the BT, are located at level 3. These tools include the distributed voting tool, the distributed synchronization tool, the stable memory tool, and the distributed memory tool, described in Sect. 3.2.3. Dependable Mechanisms receive notifications from RINT in order to execute reconfigurations such as introducing a spare task to take over the role of a failed task (see Table 5).

The layers around the TIRAN framework in Fig. 2 represent (from the layer at the bottom and proceeding clockwise):

- The run-time system.
- A provision to inject software faults at all levels of the framework and in the run-time system.
- The monitoring tool, for hypermedia rendering of the current state of the system within the windows of an Internet browser.
- The functional application layer and the recovery language application layer (the latter is represented as the box labeled ARIEL).

Figure 3 pictures the key elements of the TIRAN architecture within the workflow diagram in Fig. 1.
The following sections describe some of the TIRAN components in more detail.

## 3.2 Some Components of the TIRAN Framework

This section describes some key components of the TIRAN framework, namely: The TIRAN toolset, i.e., the BT, the BSL, the Backbone, and the Time-out Manager service. It also briefly enumerates the TIRAN level-3 mechanisms.

### 3.2.1 The TIRAN Basic Tools

The TIRAN Basic Tools represent a layer of node-local services for error detection, fault masking, error recovery, and isolation. The TIRAN BT include:

1. A watchdog timer.

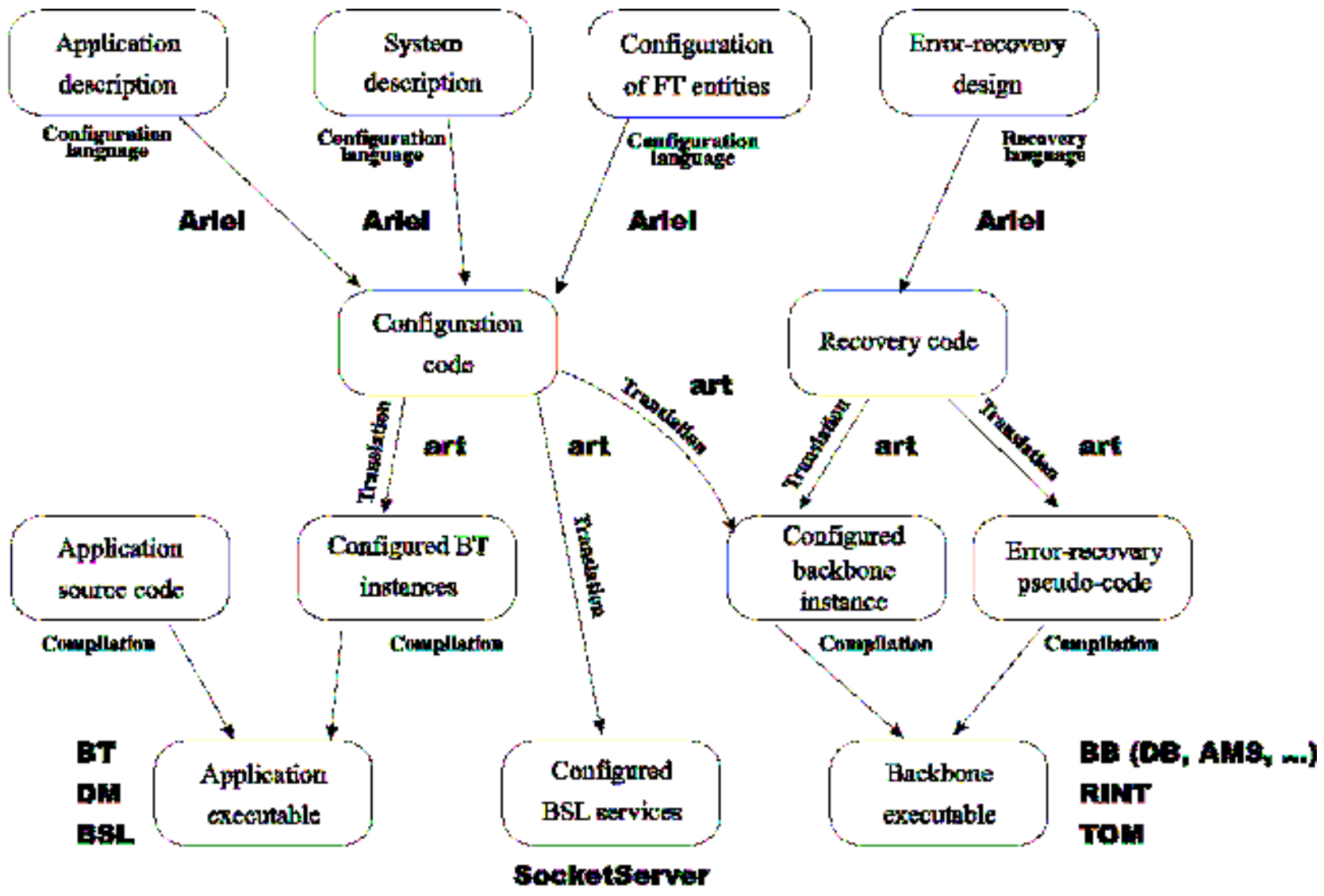

Figure 3: The main components of the $\mathcal{RεL}$ architecture developed in TIRAN are located within the workflow diagram of Fig. 1. A so-called "SocketServer" has been also depicted, i.e., a BSL component managing message transmission and dispatching.

2. A node-local voting system.
3. An input replicator, to be used with the voting system.
4. An output collector, to be used with the voting system.
5. A tool responsible for node shutdowns.
6. A tool responsible for node reboots.
7. A tool responsible for isolating a task.

Some of the BT (such as the level 1.1 BT with numbers 1–4) are application-level software tools, others (for instance, the level-1.2 BT number 5–7) provide hooks to lower-level services. Most of these tools have been designed and developed by the TIRAN partners and have been especially conceived in order to meet hard real-time requirements. A deeper description of some of these tools can be found in (Deconinck, Botti, Cassinari, De Florio, & Lauwereins, 1998; Deconinck et al., 1999). As a side effect of using these tools, the TIRAN Backbone (see Sect. 3.2.4) is transparently notified of a number of events related to fault-tolerance.

> Notifications generally describe the state of a given component and have the form of a 4-tuple of integers
>
> $$(c, t, u, p),$$

in which $c$ identifies a specific condition, $t$ is a label that identifies a class of BT, $u$ is the unique-id of either a task, or a group of tasks, or a node, and $p$ is a possibly empty list of optional arguments.

Unlike in Mozart (Chapter 5), where an entity can register to a fault stream and be notified of all faults affecting any entity, within TIRAN an entity must explicitly set up such fault stream by calling function `RaiseEvent`. Such function forwards notifications to the Backbone. It has the following prototype:

```
int RaiseEvent(int condition, int actor, TIRAN_TASK_ID uniqueId,
                           int nargs, ...).
```

### 3.2.2 The TIRAN Basic Services Library

The TIRAN Basic Services Library offers specific services such as communication, task creation and management, access to the local hardware clock, management of semaphores, and so forth. It supports multicasting: Messages are sent to so-called "*logicals*," i.e., groups of tasks. It was designed by the TIRAN Consortium and developed by various partners for the different target platforms and operating systems. Specific adaptation layers may be designed for mapping existing communication libraries, such as MPI2, to the TIRAN BSL.
As a side effect of using some of the BSL functions, the TIRAN Backbone (see Sect. 3.2.4) can be transparently notified of events such as a successful spawning of a task, or the failure state of a communication primitive. BSL notifications are similar to BT notifications—4-tuples of integers

$$(c, t, u, p),$$

where $c$ and $p$ are as in Sect. 3.2.1, while in this context $t$ is a label that specifies the class of BSL services and $u$ is the unique-id of the task that experienced condition $c$.
A full description of the TIRAN BSL can be found in (Calella et al., 1999).

### 3.2.3 The TIRAN Dependable Mechanisms

As already mentioned, the TIRAN Dependable Mechanisms are distributed, high-level fault-tolerance provisions that can be used to facilitate the development of dependable services. The TIRAN Dependable Mechanisms include:

- A tool to enhance data integrity by means of a stabilizing procedure (data is only allowed to pass through the system boundaries when it is confirmed a user-defined number of times). This is known as Stable Memory Tool and is described in detail in Chapter 3.

- A Distributed Synchronization Tool, i.e., a software tool for synchronizing a set of tasks. A description of this tool can be found in (Calella et al., 1999).
- A so-called Distributed Memory Tool that creates and manages distributed replicas of memory cells. Several features of this tool have been described in Chapter 3.
- A “Redundant Watchdog”, i.e., a distributed system exploiting multiple instances of the watchdog basic tool in order to enhance various dependability properties (see Chapter 9).
- The TIRAN distributed voting tool (DV), an adaptation of the EFTOS voting farm described in Chapter 3.

### 3.2.4 The TIRAN Backbone

The TIRAN Backbone is the core component of the fault-tolerance distributed architecture described in this chapter. The main objectives of the TIRAN Backbone include:

1. Gathering and maintaining error detection information produced by TIRAN BT and BSL.
2. Using this information at error recovery time.

In particular, as described in Sect. 3.2.1, the TIRAN BT focusing on error detection and fault masking forward their deductions to the Backbone. The Backbone maintains these data in the TIRAN DB, replicated on multiple nodes of the system. Incoming data are also fed into an $\alpha$-count (Powell et al., 1999) filter (see Sect. 3.2.4). This mechanism allows identifying statistically the nature of faults—whether a given fault is transient or whether it is permanent or intermittent. A set of private functions to query the current state of the DB is available within the Backbone. Other private functions request remote services such as, for instance, rebooting a node or spawning a certain task on a given node.
When the underlying architecture is built around a host computer controlling a number of target boards by means of a custom OS, the latter may execute remote commands on the boards. On the contrary, when the OS is general-purpose and node-local, remote services can be executed on any node of the system by sending command messages from one component to the other of the Backbone—for instance, a “reboot” message may be sent from the component of the Backbone on node 0 to the component on node 2. On receipt, the latter may execute the reboot command by forwarding a local request to a BT addressing error isolation or recovery, e.g., the tool managing the reboot service. Figure 4 represents this scheme. The above remote services can be the basis for more complex, system-wide recovery and reconfiguration actions.

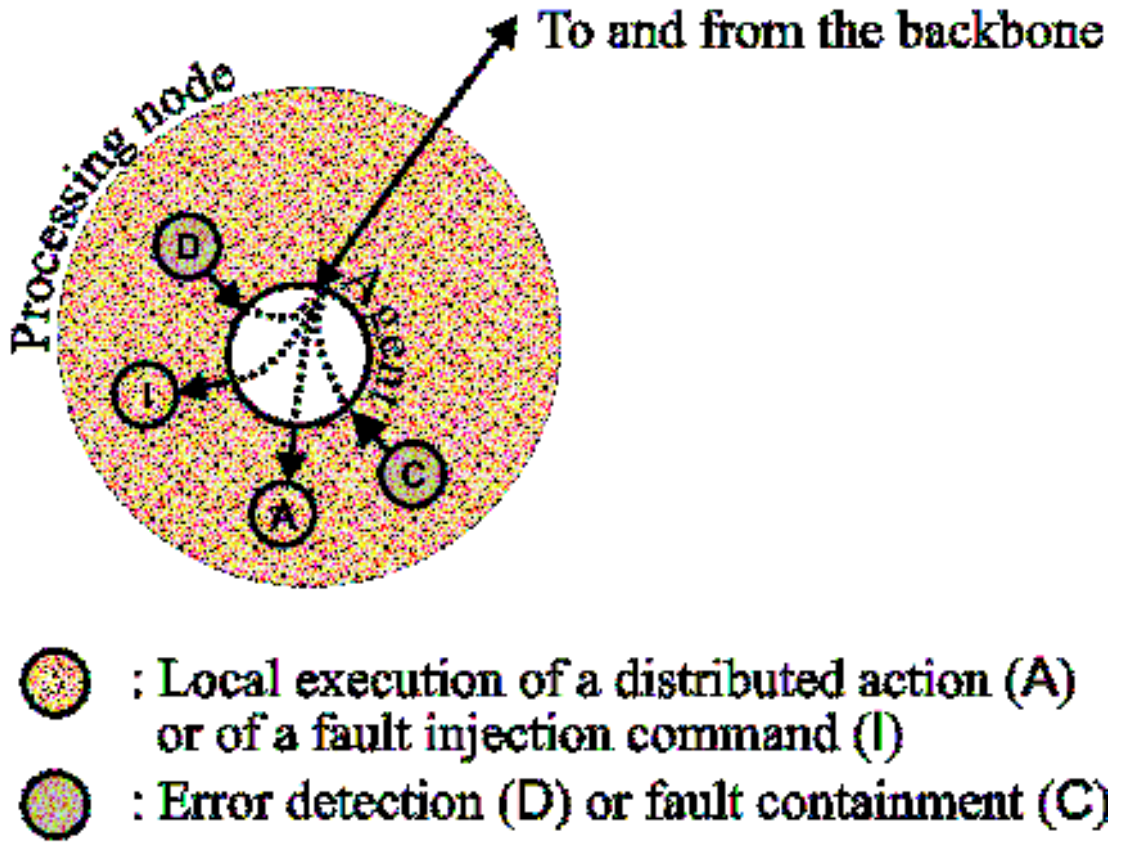

Figure 4: On each processing node exactly one component of the Backbone is located. This component is the intermediary of the Backbone on that node. In particular, it gathers information from error detection and fault containment BT (grey circles) and forwards requests to error containment and recovery BT (light grey circles). These latter execute recovery actions and possibly, at test time, fault injection requests.

The TIRAN Backbone consists of two core components. In the absence of system partitioning, within the system there is exactly one **manager**, holding the main copy of the DB and with recovery management responsibilities. On other nodes, Backbone components called **assistants** deal with DB replicas, forward local deductions to the manager and can take over the role of the manager should the latter or its node fail.
A key point in the effectiveness of this approach is *guaranteeing that the Backbone itself tolerates internal and external faults.* A custom distributed algorithm has been designed to detect the failures of up to all but one of the components of the Backbone or the nodes of the system. The same algorithm also tolerates system partitioning during the "periods of instability" (see the timed-asynchronous distributed system model (Cristian & Fetzer, 1999), briefly described in Chapter 2). This procedure has been called the "algorithm of mutual suspicion" (AMS) since each component of the Backbone continuously questions the correctness of all the other valid components. It is described in detail in Chapter 7.

**The $\alpha$-Count Fault Identification Mechanism.** It is a well known fact that reconfiguration is a costly activity, in the sense that it always results in redundancy consumption. This translates in a possibly graceful, though actual degradation of the quality of service of a system. This may have drastic consequences, especially in systems where service degradations are static and irreversible. For instance, in satellites and space probes, rapid exhaustion of redundancy may severely affect the duration of the useful mission hours and hence reliability (Inquiry, 1996). Two important issues towards solving the problem just stated are:

1. Understanding the nature of faults, and in particular identifying faults as permanent (and thus actually requiring reconfiguration) with respect to transient ones.

2. Tolerating transient faults with less redundancy consuming techniques, possibly not based on reconfiguration.

Issue 1 means that the adopted fault-tolerance mechanism is required not only to locate a component subject to an error, but also to assess the nature of the originating fault. This implies processing additional information and unfortunately translates also in a larger delay in fault diagnosis. Despite this larger delay, in same cases the benefits of techniques based on the above issues may be greater than its penalties in performance and latency (Bondavalli et al., 1997).
A number of techniques have been devised to assess the character of faults—some of them are based on tracking the occurrences and the frequency of faults and adopting thresholds of the kind "a device is diagnosed as affected by a permanent fault when four failures occurs within 24 hours". This and other similar heuristics are described in (Lin & Siewiorek, 1990). A fault identification mechanism, called $\alpha$-count, has been described in (Powell et al., 1999) and generalized in (Bondavalli, Chiaradonna, Di Giandomenico, & Grandoni, 2000). $\alpha$-count is also based on thresholds. The basic idea is that each system component to be assessed is to be "guarded" by an error detection device. This device should forward its deductions, in the form of a binary digit, to a device called $\alpha$-count filter. For each incoming judgement, the filter would then perform the $\alpha$-count technique and issue an assessment about the nature of the fault leading to the detected error.

> More formally, given $n$ "guarded" components $u_i, 0 <= i < n$, the authors of the strategy call $J(i, L)$ the $L$-th judgement on $u_i$. Judgement 0 means success ($u_i$ is "healthy"), 1 means failure ($u_i$ is faulty). A score vector, $\alpha(i), 0 <= i < n$, is initially set to zero and updated, for each judgement $L > 1$, as follows:
>
> $$\alpha(i, L) = \alpha(i, (L-1)) K if J(i, L) = 0$$
>
> $$\alpha(i, L) = \alpha(i, (L-1)) + 1 if J(i, L) = 1, \qquad (3)$$
>
> being $0 <= K <= 1$. When $\alpha(i, L)$ becomes greater than a certain threshold $\alpha_T$, $u_i$ is diagnosed as affected by a permanent or intermittent fault. This event may be signaled to a reconfigurator or another fault passivation mechanism. The authors of $\alpha$-count show that their mechanism is asymptotically able to identify all components affected by permanent or intermittent faults—if the threshold $\alpha_T$ is set to any finite positive integer $A$, and if the

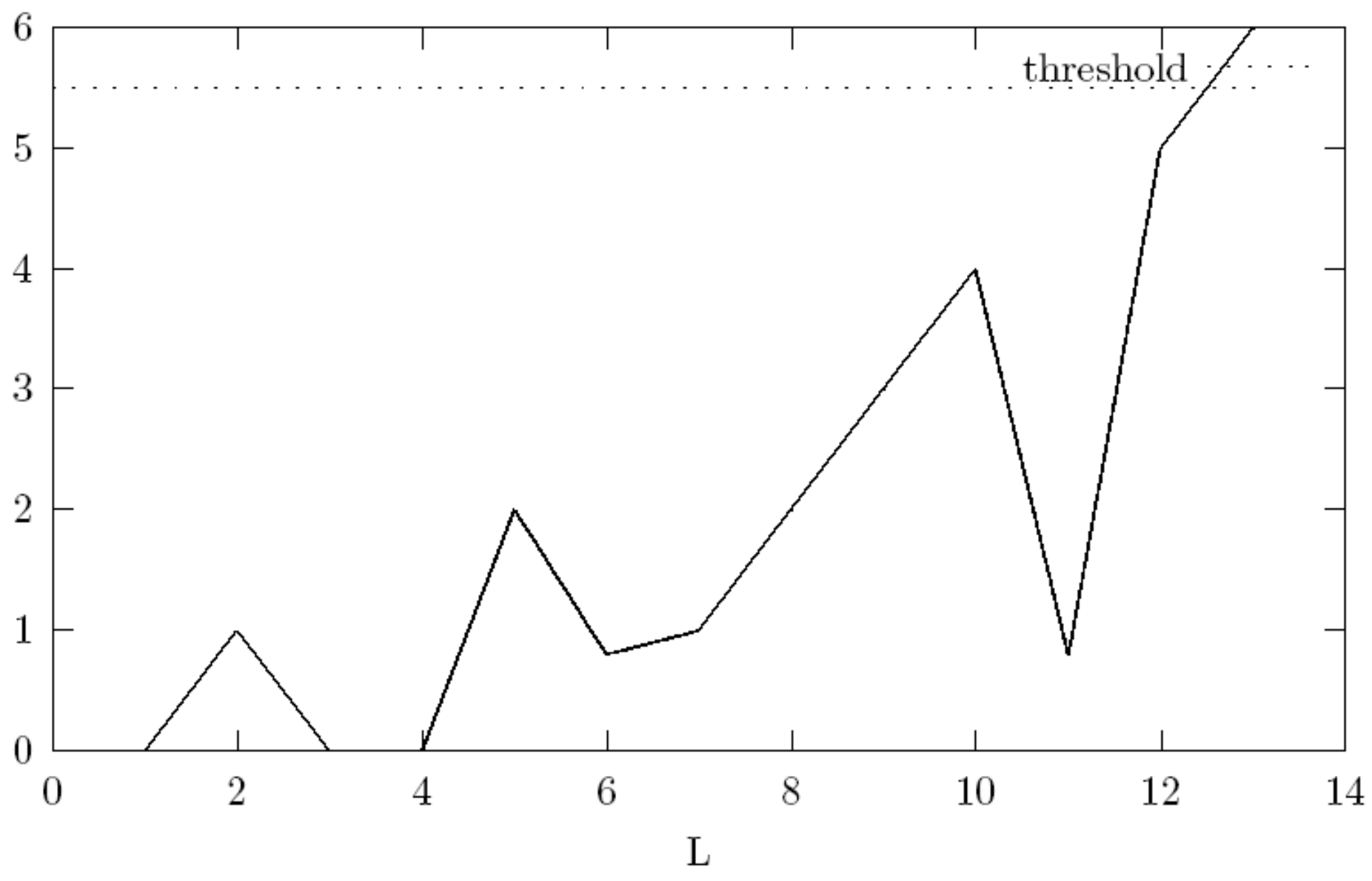

Figure 5: The picture shows how an $\alpha$-count filter, $\alpha(1, L)$, is updated when it receives 14 error detection notifications ($0 <= L <= 13$) from the Backbone. Notifications are in this case chosen randomly. Note that, on the last notification ($L = 13$), a threshold (in this case, 5.5) is reached and the corresponding fault is assessed as permanent or intermittent.

> component is indeed affected by a permanent or intermittent fault, they prove that $\alpha(i, L)$ will eventually become greater than or equal to $A$ (Bondavalli et al., 1997). The authors also prove that similar results can be reached with some variants of formula (3). Figure 5 shows how that counter evolves when random-valued notifications (either positive or negative assessments) are sent to the Backbone.

It is worth noting how the mechanism described above requires an approach slightly similar to the one of the $\mathcal{RεL}$ toolset: in both cases, a fault stream has to flow from a periphery of detection tools to a collector. This collector is the $\alpha$-count filter in the one case and the TIRAN Backbone and its DB in the other one. This makes it straightforward adopting $\alpha$-count in TIRAN: The filter can simply be fed with data coming from the detection BT just before this data in inserted into the DB. The added value of this approach is that it is possible to set up a function that returns the current estimation of the nature of a fault (see Sect. 3.2.9). Clearly, a requirement of this technique is that each detection tool not only forward notifications of erroneous activity, but also confirmations of normal behavior. This may have negative consequences on the performance of the approach, as it may result in many communication requests. In the prototype version developed by the author, each notification from any detection tool to the backbone is tagged with an integer label starting at 0 and incremented by 1 at each new deduction. This way, only negative assessments can be sent—in fact, if, for instance, the backbone receives from the same detection tool two consecutive negative assessments labeled respectively, e.g., 24 and 28, then this implicitly means that there have been three positive assessments in between, and the $\alpha$-counter can be updated accordingly—though possibly with some delay. This strategy has been conceived by the author of this book.
Figure 6 shows the $\alpha$-counter in action—a task has reached its threshold and has been declared as affected by a permanent or intermittent fault.
Clearly a system like Mozart (Chapter 5) may also easily adopt a tool like $\alpha$-count.

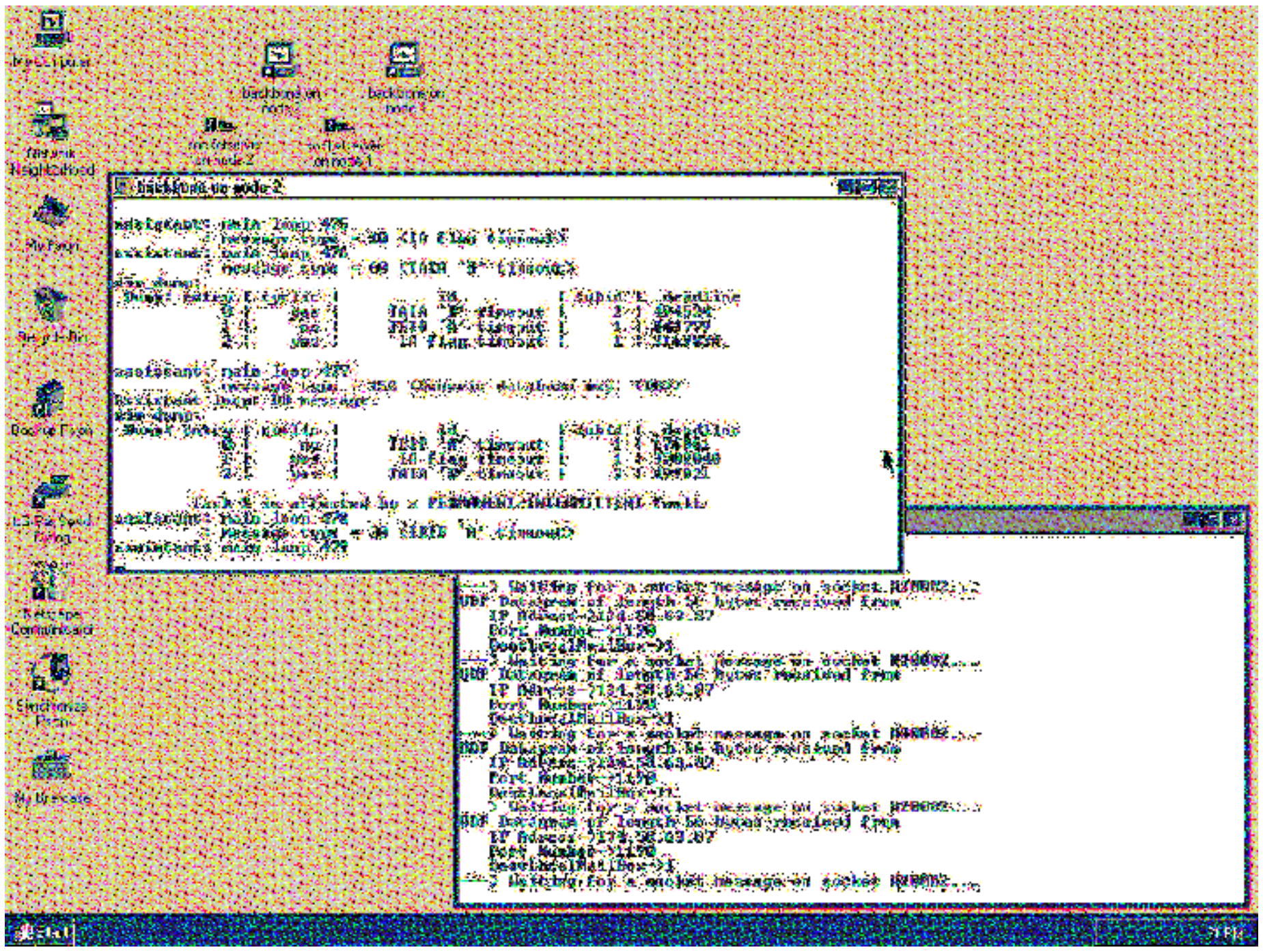

Figure 6: An assistant running on a node of a cluster of PCs. The upper window hosts the assistant, the lower one its SocketServer (see Fig. 3). In the former one, note how the $\alpha$-counter detects that task 2 is affected by a permanent or intermittent fault.

### 3.2.5 The TIRAN Time-Out Manager

As already mentioned, the main assumption of $\mathcal{REL}$ is the adoption of the timed asynchronous distributed system model (Cristian & Fetzer, 1999), the promising model for solving problems such as dynamic membership (Cristian

& Schmuck, 1995) in distributed systems that has been introduced in Chapter 2. The availability of a class of functions for managing time-outs is an important requirement for that model. The TIRAN Time-out Manager (De Florio, Deconinck, & Lauwereins, 2000) fulfils this need—it is basically a mechanism for managing lists of *time-outs*, that is objects that postpone a function call for a certain number of local clock units. Time-outs are ordered by clock-units-to-expiration, and time-out lists are managed in such a way that only the head of the list needs to be checked for expiration. When the specified amount of time elapses for the head of the list, its function—let us call it "alarm"—is executed and the object is either thrown out of the list or renewed (in this second case, a time-out is said to be "cyclic"). A special thread monitors and manages one or more of such lists, checking for the expiration of the entries in the time-out list.
Within the strategy of the TIRAN Backbone, a Time-out Manager task is available on each node of the system, spawned at initialization time by the local Backbone component. Time-out Manager is used in this context to translate time-related clauses, such as, "$p_a$ seconds have elapsed", into message arrivals. In other words, each Time-out Manager instance of the Backbone may be represented as a function $a$ from $C$ to $M$ such that, for any time-related clause $c$ in $C$:

$$a(c) = \text{message "clause } c \text{ has elapsed" is in } M.$$

This homomorphism is useful because it allows deal with the entire set of possible events—both messages and time-related events—as *the arrival of a homogeneous set of messages*. Hence, a single multiple-selection statement such as the C language `switch` can be used to manage all cases, which translates into a simpler and more straightforward implementation for error detection protocols such as the AMS.
The Time-out Manager uses a well-known algorithm for managing its time-outs (Tanenbaum, 1996). Once the first time-out is entered, the manager creates a linked-list of time-outs and polls the top of the list. For each new time-out to be inserted, a proper position in the list is found and the list is modified accordingly, as described in Fig. 7. If the top entry expires, a user-defined alarm function is invoked. This is a general mechanism that allows associate any event with the expiring of a time-out. In the case of the backbone, the Time-out Manager component on node $k$ sends a message to Backbone component on the same node—the same result may also be achieved by sending that component a UNIX signal (Haviland & Salama, 1987). Special time-outs are defined as "cyclic", i.e., they are automatically renewed at each new expiration, after invoking their alarm function. A special function renews a time-out, i.e., it deletes and re-enters that entry. It is also possible to temporarily suspend[8] a time-out and re-enable it afterwards.
Time-out Manager exploits multiple alarm execution threads in order to reduce the congestion that is due to concurrent execution of alarms and the consequent run-time violations. A description of this strategy can be found in (De Florio et al., 2000).

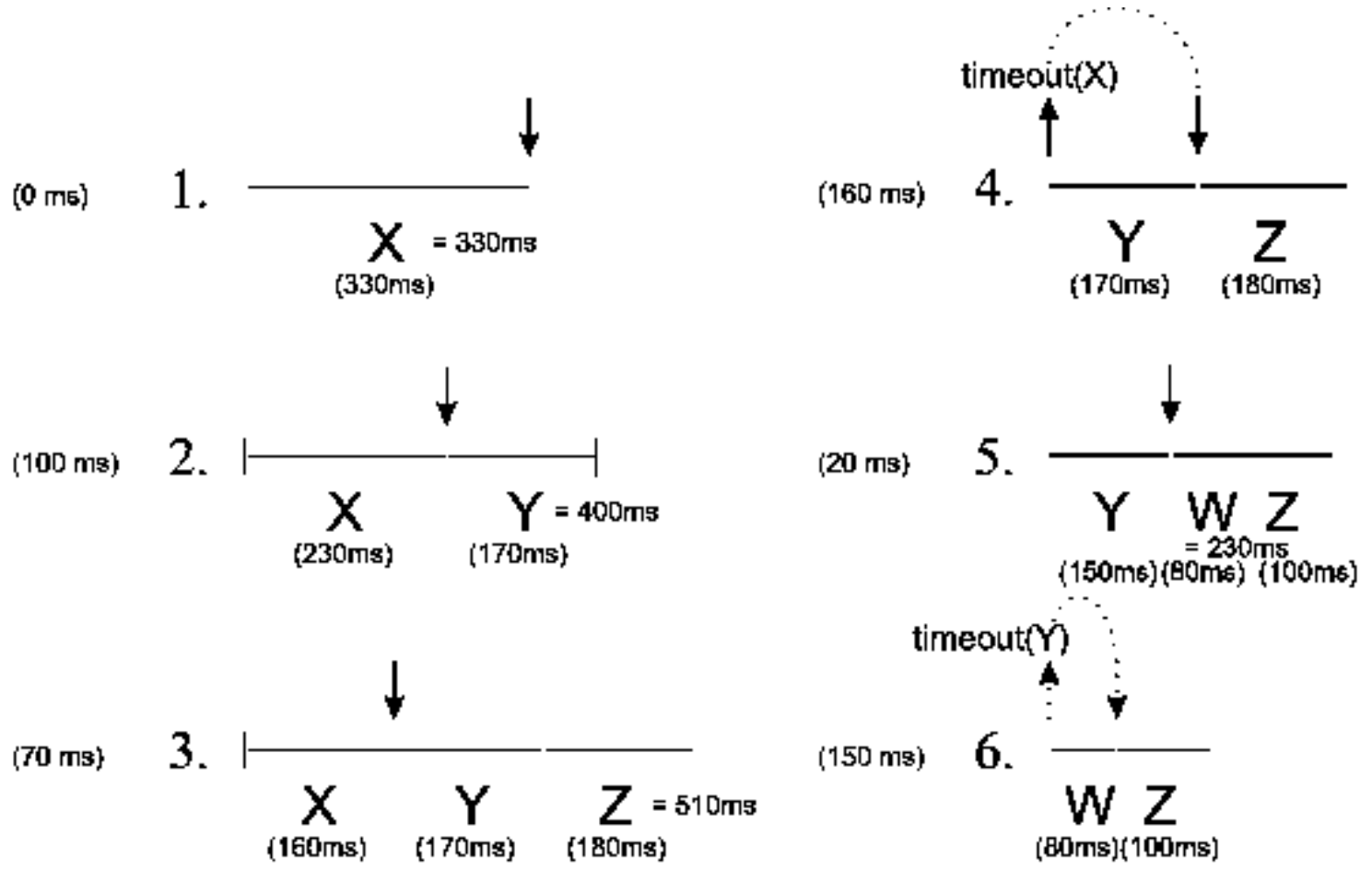

Figure 7: How the alarm manager works: in 1., a 330ms alarm called X is inserted in the list. In 2., after 100ms, X has been reduced to 230ms and a 400ms alarm, called Y, is inserted (its value is 170, i.e., 400-230). Another 70ms have passed in 3., so X has been reduced to 160ms. At that point, a 510ms alarm, Z is inserted–it goes at the third position. In 4., after 160ms, alarm X occurs–Y becomes then the top of the list; its decrement starts. In 5. another 20ms have passed and Y is at 150ms–at that point a 230ms alarm, called W is inserted. Its position is in between Y and Z, therefore the latter is adjusted. In 6., after 150ms, Y expires and W goes on top.

The TIRAN Time-out Manager is a fundamental building block for the TIRAN dependability strategies, but it is a useful tool also for other approaches: Chapter 7 describes one of them, that is failure detection protocols.

### 3.2.6 The Ariel Configuration and Recovery Language

> ARIEL My master through his art foresees the danger
> That you, his friend, are in; and sends me forth—
> For else his project dies—to keep them living.
> (Shakespeare, *The Tempest*, Act II, Scene I)

Within TIRAN, a single syntactical structure—provided by the ARIEL language (De Florio, Deconinck, & Lauwereins, 2001; Deconinck, De Florio, & Botti, 2001)—has been devised by the author of this book as both a configuration and a recovery language. This chapter describes this language and its compile-time and run-time tools. The structure of the section is as follows:

- The general characteristics of ARIEL are described in Sect. 3.2.7.
- ARIEL as a configuration language is introduced in Sect. 3.2.8.
- Section 3.2.9 is on ARIEL as a recovery language.

ARIEL *bears its name after the character with the same name of Shakespeare's last Comedy,* The Tempest. *In fact, in the Tempest, spirit Ariel is a magic creature, the invisible assistant of Prospero, the Duke of Milano. While Prospero plays his role in the foreground of the Comedy—orchestrating a strategy to regain possession of his dukedom, which had been usurped by Prospero's brother Antonio—Ariel faithfully serves Prospero performing his magic in the background, exploiting his powers to execute Prospero's commands when his "Master through his art foresees the danger" (see above quote). In a sense, the role of the* ARIEL *language is somewhat similar to the one of Prospero's ally. Its run-time support is given by the agent that is sent forth when the Backbone, through its nervous terminals—the error detection basic tools—senses a potentially dangerous condition. The author of this book thinks that the name Ariel captures well this similarity. This is also the reason that led to the choice of "$\mathcal{R}\varepsilon\mathcal{L}$" as an abbreviation for "*Re*covery* L*anguage": indeed, the spelling of that word is "*[a:*]-[i:]-[el]*".*

### 3.2.7 General Characteristics of ARIEL

ARIEL is a declarative language with a simple syntax somewhat similar to that of a UNIX shell. One instruction per line is allowed. Comments are like in the C shell (“#” starts a comment which ends at next new line or at end-of-file). Names are not case-sensitive. ARIEL deals with five basic types: “*nodes*”, “*tasks*”, “*logicals*”, integers, and real numbers. A node is a uniquely identifiable processing node of the system, e.g., a processor of a MIMD supercomputer. A task is a uniquely identifiable process or thread in the system. A logical is a uniquely identifiable collection of tasks, possibly running on different nodes. Nodes, tasks, and logicals are generically called *entities*. Entities are uniquely identified via non-negative integers; for instance, `NODE3` or `N3` refer to processing node currently configured as node number 3.
Integer symbolic constants can be “imported” from C language header files through the statement `INCLUDE`. For instance, if the C language header file `"vf.h"` contains a define statement such as:

```
#define PROC_NUM 4,
```

then it is possible to use that symbolic constant wherever an integer is expected in the language. To de-reference a symbolic constant imported via `INCLUDE`s a “brace-operator” has been defined—for instance, under the above assumptions the following valid ARIEL statement:

```
NPROCS = {PROC_NUM}
```

(described later on) is equivalent to

```
NPROCS = 4.
```

An ARIEL script basically consists of two parts:

- A part dealing with configuration. This is described in Sect. 3.2.8.
- A part containing the guarded actions which constitute the user-defined error recovery strategy. They are described in Sect. 3.2.9.

### 3.2.8 ARIEL as a Configuration Language

Special linguistic support has been designed by the author of this book while taking part in the TIRAN project. Aim of this linguistic support is to facilitate the configuration of the instances of the framework elements, of the system and application parameters, and of the fault-tolerance provisions. Let us call these elements a *framework instance*.
Once the user has configured a framework instance with ARIEL, the TIRAN ARIEL translator “`art`” must be used to translate these high level specifications into the actual C language calls that set up configured tasks such as, for instance, user-configured watchdogs. The output of the translator is a set of C files that need to be compiled with the user application.
A subset of the fault-tolerance provisions described in Chapter 6 is supported by the version of ARIEL described in this book. The rest of this section describes the process of configuring a framework instance with ARIEL.

**System and Application Parameters.** The ARIEL configuration language can be used to define and configure the target system and application entities, e.g., nodes, tasks, and group of tasks. The rules defined in Sect. 2.3.2 are coded as follows:

- Rule (1), as for instance in

  $$\text{task}_3 = n_0[8],$$

  is coded as

  ```
  TASK 3 = "TMR.EXE" IS NODE 0, TASKID 8.
  ```

  In other words, the above statement declares that task 8, local to node 0, is to be globally referred to as "task 3". String "`TMR.EXE`" may also be used to refer symbolically[9] to $\text{task}_3$. More complex rules are possible—for instance,

  ```
  TASK [1,3] = "Triple" IS NODE 0, TASKID [6,8]
  ```

  is equivalent to

  ```
  TASK 1 = "Triple1" IS NODE 0, TASKID 6,
  TASK 2 = "Triple2" IS NODE 0, TASKID 7,
  TASK 3 = "Triple3" IS NODE 0, TASKID 8.
  ```

- Rule (2), as in

  $$\text{group}_{10} = \{5, 6, 7\},$$

  is coded as

  ```
  LOGICAL 10 = "TMR" IS TASK 5, TASK 6, TASK 7 END LOGICAL.
  ```

  (String "`TMR`" may also be used to refer symbolically to $\text{group}_{10}$). The example defines a logical to be globally referred to as "logical 10", symbolically known as `TMR`, and corresponding to the three tasks whose unique-id are 5, 6, and 7.

Let us assume the following lines have been written in a file called "`test.ariel`":

```
TASK 5 = "TMR_LEADER" IS NODE 0, TASKID 8
TASK 6 = "TMR2" IS NODE 1, TASKID 8
TASK 7 = "TMR3" IS NODE 2, TASKID 8
LOGICAL 10 = "TMR" IS   TASK 5, TASK 6, TASK 7   END LOGICAL
```

File `test.ariel` can be translated by executing program `art` as follows:

```
bash-2.02$ art -i test.ariel
Ariel translator, v2.0g 03-Aug-2000, (c) K.U.Leuven 1999,
2000.
Parsing file test.ariel...
...done (4 lines in 0.01 CPU secs, or 400 lines per CPU sec.)
Output written in file .rcode.
Logicals written in file LogicalTable.csv.
Tasks written in file TaskTable.csv.
Alpha-count parameters written in file alphacount.h.
```

The bold typefaced string is the command line. What follows is the output printed to the user screen. The italics-highlighted strings are the names of two configuration files that are written by art on successful execution. These two files declare tasks and logicals in the format expected by the BSL and by other ancillary components (e.g., on the version for Windows NT, these tables are also used by a "SocketServer[10]").

The user can also define a number of other parameters of the TIRAN world like, for instance:

- $N$, i.e., the total number of processing nodes in the system.
- $t_i$, i.e., the number of tasks running on node $n_i$.
- Task-specific parameters of the $\alpha$-count fault identification mechanism supported by the TIRAN Backbone.

As an example, the following lines declare a system consisting of two nodes, each of which has to host 10 user tasks, and define the $\alpha$-count parameters of task 5:

```
NPROCS = 2
NUMTASKS 1 = 10
NUMTASKS 2 = 10
ALPHA-COUNT 2 IS  threshold = 3.0, factor = 0.4 END ALPHA-COUNT.
```

The output produced by the art translator is given in this case by a number of header files:

```
bash-2.02$ art -i test.ariel -s
Ariel translator, v2.0g 03-Aug-2000, (c) K.U.Leuven 1999,
2000.
Parsing file test.ariel...
...done (8 lines in 0.01 CPU secs, or 800 lines per CPU sec.)
Output written in file .rcode.
Logicals written in file LogicalTable.csv.
Tasks written in file TaskTable.csv.
static version
Preloaded r-codes written in file trl.h.
Time-outs written in file timeouts.h.
Identifiers written in file identifiers.h.
Alpha-count parameters written in file alphacount.h.
```

Again, the command has been given in bold typeface and relevant lines have been highlighted using the italics typeface. The "-s" option, for "static", requests the writing of a number of header files. The produced header files contain definitions like the following one, from file "timeouts.h":

```
/* Number of available nodes
 */
#define MAX_PROCS                 2
```

These are the first few lines of the output file "alphacount.h":

```
/****************************************************************
 *                                                              *
 * Header file alphacount.h                                     *
 *                                                              *
 * This file contains the parameters of the alphacount filter   *
 * (factor and threshold)                                       *
 * Written by art (v2.0g 03-Aug-2000) on Wed Aug 16 15:46:40 2000 *
 * (c) Katholieke Universiteit Leuven / ESAT / ACCA - 2000. *
 *                                                              *
 ****************************************************************/

#ifndef    __ALPHA_COUNT__
#define    __ALPHA_COUNT__

#include "DB.h"

alphacount_t alphas[] = {
               { 0.000000, 0.000000, 0 }, /* entry 0 */
               { 0.000000, 0.000000, 0 }, /* entry 1 */
               { 0.400000, 3.000000, 1 }, /* entry 2 */
               { 0.000000, 0.000000, 0 }, /* entry 3 */
```

(Note how, at entry 2, the threshold and factor of the $\alpha$-count related to task 2 have been entered in an array. The latter will then be used by the TIRAN backbone when updating the $\alpha$-count filters—see Sect. 3.2.4).

**Backbone Parameters.** ARIEL can be used to configure the Backbone. In particular, the initial role of each Backbone component must be specified through the `DEFINE` statement. For instance, the following two lines configure a system consisting of four nodes and place the Backbone manager on node 0 and three assistants on the other nodes:

```
DEFINE 0 = MANAGER
DEFINE 1-3 = ASSISTANTS
```

A number of backbone-specific parameters can also be specified via the ARIEL configuration language. These parameters include, for instance, the frequency of setting and checking the <I'm Alive> flag of the backbone. Values are specified in microseconds. A complete example can be seen in Table 1. Again, the `art` translator changes the above specifications into the appropriate C language settings as expected by the TIRAN Backbone.

```
   # Specification of a strategy in the recovery language Ariel
   # Include files
1  INCLUDE "phases.h"
2  INCLUDE "vf.h"

   # Definitions
   #     After keyword 'DEFINE', the user can specify
   #     an integer, an interval, or a list, followed by
   #     the equal sign and a backbone role, that may be
   #     ASSISTANT(s) or MANAGER
3  NPROCS = 4
4  Define 0 = MANAGER
5  Define 1-3 = ASSISTANTS

   # Time-out values for the Backbone and the <I'm Alive> mechanism
6  MIA_SEND_TIMEOUT = 800000 # Manager Is Alive -- manager side
7  TAIA_RECV_TIMEOUT = 1500000 # This Assistant Is Alive -- manager side

8  MIA_RECV_TIMEOUT = 1500000 # Manager Is Alive -- backup side
9  TAIA_SEND_TIMEOUT = 1000000 # This Assistant Is Alive -- backup side

10 TEIF_TIMEOUT = 1800000 # After this time a suspected node is assumed
   # to have crashed.

11 I'M_ALIVE_CLEAR_TIMEOUT = 900000 # <I'm Alive> timeout -- clear IA flag
12 I'M_ALIVE_SET_TIMEOUT = 1400000 # <I'm Alive> timeout -- set IA flag

   # Number of tasks
13 NUMTASKS 0 = 11 # node 0 is to be loaded with 11 tasks
14 NUMTASKS 1 = 10
15 NUMTASKS 2 = 10
16 NUMTASKS 3 = 10

17 TASK [0,10] IS NODE 0, TASKID [0,10]
18 TASK [11,20] IS NODE 1, TASKID [1,10]
19 TASK [21,30] IS NODE 2, TASKID [1,10]
20 TASK [31,40] IS NODE 3, TASKID [1,10]

21 LOGICAL 1 IS TASK 1, TASK 2, TASK 3 END LOGICAL
```

Table 1: An excerpt from an ARIEL script: configuration part. Line numbers have been added for the sake of clarity.

**Basic Tools.** ARIEL can be used to configure statically the TIRAN tools. The current prototypic version can configure only one tool, the TIRAN watchdog. The following syntax is recognized by `art` to configure it:

```
WATCHDOG 10 WATCHES TASK 14
  HEARTBEATS EVERY 100 MS
  ON ERROR WARN TASK 18
END WATCHDOG.
```

The output in this case is a C file that corresponds to a configured instance of a watchdog. The application developer needs only to send heartbeats to that instance, which can be done as follows:

```
HEARTBEAT 10.
```

**Configuring Multiple-Version Software Fault-Tolerance.** As described in Sect. 2.3.3, it is possible to design syntax rules to support the configuration of the software fault-tolerance provisions described in Chapter 3. This section describes the solution provided by ARIEL in order to support $N$-version programming (Avižienis, 1985), and a possible syntax to support consensus recovery blocks (Scott et al., 1985) and retry blocks (Huang & Kintala, 1993). The following is an example that shows how it is possible to define an "$N$-version task" with ARIEL:

```
#include "my_nvp.h"
N-VERSION LOGICAL {NVP_LOGICAL}
  VERSION 1 IS TASK{VERSION1}  TIMEOUT {VERSION_TIMEOUT}
  VERSION 2 IS TASK{VERSION2}  TIMEOUT {VERSION_TIMEOUT}
  VERSION 3 IS TASK{VERSION3}  TIMEOUT {VERSION_TIMEOUT}
METRIC "nvp_comp"
ON SUCCESS TASK{NVP_OUTPUT}
ON ERROR   TASK{NVP_ERROR} {NVP_LOGICAL}
VOTING ALGORITHM IS MAJORITY
END N-VERSION
```

The `art` translator, once fed with the above lines, produces three source files for tasks the unique-id of which is `{VERSION1}`, `{VERSION2}` and `{VERSION3}`. These source files consist of code that

- sets up the TIRAN distributed voting tool (described in Sect. 3.2.3) using metric function

  `int nvp_comp(const void*, const void*)`, setting the voting algorithm to majority voting, and so forth;

- redirects standard output streams;

- executes a user task, e.g., task `{VERSION3}`.

By agreement, each user task (i.e., each version) has to write its output onto the standard output stream.
During run-time, when the user needs to access a service supplied by an NVP logical, it simply sends a message to entity `{NVP_LOGICAL}`. This translates into a multicast to tasks `{VERSION1}`, `{VERSION2}` and `{VERSION3}`. These tasks, which in the meanwhile have transparently set up a distributing voting tool,

- get their input,
- compute a generic function,
- produce an output
- and (by the above stated agreement) they write the output onto their standard output stream.

The latter, which had been already redirected through a piped stream to the template task, is fed into the voting system. Such system eventually produces an output that goes to task `{NVP_OUTPUT}`.
A time-out can also be set up so to produce an error notification when no output is sent by a version within a certain deadline.
Table 2 shows one of the three files produced by the ARIEL translator when it parses the script of Sect. 3.2.8. Note how this file basically configures an instance of the TIRAN DV tool described in Sect. 3.2.3. Note also how all technicalities concerning:

- The API of the tool,
- input replication,
- the adopted voting strategy,
- output communication,

and so forth are fully made transparent to the designer, who needs only be concerned with the functional service. This allows the fault-tolerance designer to modify all the above mentioned technicalities with no negative relapses on the tasks of the application designer, and even to deploy different strategies depending on the particular target platform. This can be exploited in order to pursue performance design goals.
Support towards consensus recovery block may be provided in a similar way, e.g., as follows:

```
#include "my_crb.h"
CONSENSUS RECOVERY BLOCK LOGICAL {CRB_LOGICAL}
  VARIANT 1 IS TASK{VARIANT1}  TIMEOUT {VARIANT_TIMEOUT}
    ACCEPTANCE TEST TASK{ACCEPT1}
  VARIANT 2 IS TASK{VARIANT2}  TIMEOUT {VARIANT_TIMEOUT}
    ACCEPTANCE TEST TASK{ACCEPT2}
```

```
#include "TIRAN.h"
/* Task 101 of NVersion Task 20
   Version 1 / 3
 */
int TIRAN_task_101(void)
{
    TIRAN_Voting *dv; size_t size;
    double task20_cmp(const void*, const void*);

    dv = TIRAN_VotingOpen(task20_cmp);
    if (dv == NULL)
    {
        RaiseEvent(TIRAN_ERROR_VOTING_CANTOPEN,TIRAN_DV,101,0);
        TIRAN_exit(TIRAN_ERROR_VOTING_CANTOPEN);
    }

/* voting task description: which tasks and which versions */
/* constitute the n-version task */
    TIRAN_VotingDescribe(dv, 101, 1, 1);
    TIRAN_VotingDescribe(dv, 102, 2, 0);
    TIRAN_VotingDescribe(dv, 103, 3, 0);

    TIRAN_VotingRun(dv);

/* output should be sent to task 40 */
    TIRAN_VotingOutput(dv, 40);
    TIRAN_VotingOption(dv, TIRAN_VOTING_IS_MAJORITY);

/* redirect stdout into a pipe input stream */
    TIRAN_pipework();

/* execute the version */
    task_101();

    size = read(0, buff, MAX_BUFF);
    if (size > 0)
    {
/* forward the input buffer to the local voter of this version */
        TIRAN_VotingInput(dv, buff, size);
    }
    else
    {
/* signal there's no input */
        TIRAN_VotingInput(dv, NULL, 0);
        RaiseEvent(TIRAN_ERROR_VOTING_NOINPUT,TIRAN_DV,101,0);
        TIRAN_NotifyTask(60, TIRAN_ERROR_VOTING_NOINPUT);
    }
}


/* EOF file TIRAN_task_101.c */
```

Table 2: Translation of the N-Version Task defined in Sect. 3.2.8.

```
1  REPLICATED   TASK 10   IS   TASK 101, 102, 103
2    METHOD IS MODULAR REDUNDANCY
3      VOTING ALGORITHM IS MAJORITY
4      METRIC "int_cmp"
5    END METHOD
6    ON SUCCESS TASK 20
7    ON ERROR   TASK 30
8  END REPLICATED
```

Table 3: A possible syntax rule for compile-time management of replicated tasks.

```
     VARIANT 3 IS TASK{VARIANT3}  TIMEOUT {VARIANT_TIMEOUT}
       ACCEPTANCE TEST TASK{ACCEPT3}
   METRIC "crb_comp"
   ON SUCCESS TASK{CRB_OUTPUT}
   ON ERROR   TASK{CRB_ERROR} {CRB_LOGICAL}
   VOTING ALGORITHM IS MAJORITY
   END CRB
```

This is syntactically similar to the previous example, but the user is asked to supply tasks for the execution of acceptance tests. Other possibilities might also be considered, e.g., supplying a function name corresponding to the acceptance test, in order to avoid the overhead of spawning a task for that purpose.

As described in Sect. 2.3.3, it is possible to design syntax rules to support the configuration of replicated tasks. Table 3 shows the syntax recognized by ARIEL for this. Lines have been numbered to ease the discussion. Line 1 specifies which tasks constitute the replicated task. Line 2 defines the method of replication (NMR in this case). Sub-options of the chosen method, at lines 3 and 4, specify the type of voting to perform when de-multiplexing the multiple outputs of the base modules of the replicated task (here, majority voting[11]). Line 6 specifies the task the output of the replicated task is to be sent to. Line 7 specifies which task to notify in case of failure, e.g., when no majority can be found.

ARIEL supports redoing as described in Table 4. Note how, in that table, task 10 has to complete within a user-defined deadline (line 2) fulfilling an acceptance test (executed by task 20 at line 3). This allows regard run-time violations as actual faults. Note also that both the latter task and task 10 could be also transparently replicated by means of the technique depicted in Sect. 2.3.3. An upper bound on the number of retries is set at line 4: should task 10 fail three times in a row, task 30 would be notified of this event and could execute some error recovery scheme.

```
RETRY TASK 10
  TIMEOUT  100ms
  ACCEPTANCE TEST TASK 20
  RETRIES  3
  ON ERROR   TASK 30
END RETRY
```

Table 4: A possible syntax rule for compile-time management of retry blocks.

#### 3.2.9 ARIEL as a Recovery Language

The same linguistic structure that realizes the TIRAN configuration language is used also to host the structure in which the user defines his or her error recovery strategies. Recovery strategies are collections of *sections* with the following syntax[12]:

```
section :  if elif else fi ;

if      :  IF '[' guard ']' THEN actions ;

elif    :
        |  ELIF '[' guard ']'
           THEN actions elif ;

else    :
        |  ELSE actions ;

fi      :  FI ;
```

where non-terminals `guard` and `actions` are the syntactical terms defined respectively in Sect. 2.3.4 and Sect. 2.3.4.

**Ariel's guards.** An excerpt of the context-free grammar rules for guards follows:

```
status  :FAULTY | RUNNING | REBOOTED | STARTED | ISOLATED
         | RESTARTED | TRANSIENT ;

entity  :GROUP | NODE | TASK ;

expr    :status entity
        |'(' expr ')'
        |expr AND expr
        |expr OR expr
        |NOT expr
        |ERRN '(' entity ')' comp NUMBER
        |PHASE '(' entity ')' comp NUMBER ;

comp    :EQ | NEQ | GT | GE | LT | LE ;
```

The following conditions and values have been foreseen:

**Faulty.** This is true when an error notification related to a processing node, a group of tasks, or a single task, can be found in the TIRAN DB.

**Running.** True when the corresponding entity is active and no error has been detected that regards it.

**Rebooted** (only applicable to nodes). This means that a node has been rebooted at least once during the run-time of the application.

**Started** (not applicable to nodes). This checks whether a waiting task or group of task has been started.

**Isolated.** This clause is true when its argument has been isolated from the rest of the application through a deliberate action.

**Phase** (only applicable to tasks). It returns the current value of an attribute set by any task via the public function `RaiseEvent`. This value is forwarded to the Backbone to represent its current "phase" or state (e.g., an identifier referring to its current algorithmic step, or the outcome of a test or of an assertion). For instance, a voting task could inform the Backbone that it has completed a given algorithmic step by setting a given integer value after each step (this approach is transparently adopted in the EFTOS voting tool and is described in more detail in (De Florio et al., 1998)). Recovery block tests can take advantage of this facility to switch back and try an alternate task when a primary one sets a "failure" phase or when a guarding watchdog expires because a watched task sent it no signs of life. This condition returns an integer symbol that can be compared via C-like arithmetic operators.

**Restarted** (not applicable to nodes). This returns the number of times a given task or group has been restarted. It implies **started**.

**Transient** is true when an entity has been detected as faulty and the current assessment of the $\alpha$-count fault identification mechanism (see Sect. 3.2.4) is "transient". It implies **faulty**.

Furthermore, it is possible to query the number of errors that have been detected and pertain to a given entity. Complex guards can be built via the standard logic connectives and parentheses. As an example, the following guard:

```
FAULTY TASK{MASTER} AND ERRN(TASK{MASTER}) > 10 AND
RESTARTED TASK{MASTER}
```

checks whether the three conditions:

- the task, the unique-id of which is the value of the symbolic constant `MASTER`, has been detected as faulty;

- more than 10 errors have been associated to that task;
- that task has been restarted,

are all true.

**Ariel's actions.** "Actions" can be attached to the THEN or ELSE parts of a section. In the current implementation of the language, these actions allow to start, isolate, restart, terminate a task or a group of tasks, to isolate or reboot a node, to invoke a local function. Moreover, it is possible to multicast messages to groups of tasks and to purge events from the DB.
A section of the context-free grammar for Ariel's actions follows:

```
actions   :
          |  actions action ;

action    :
          |  section
          |  recovery_action ;

recovery_action
          :  STOP entity
          |  ISOLATE entity
          |  START entity
          |  REBOOT entity
          |  RESTART entity
          |  ENABLE entity
          |  SEND NUMBER TASK
          |  SEND NUMBER GROUP
          |  WARN entity ( condition )
          |  REMOVE PHASE entity FROM ERRORLIST
          |  REMOVE ANY entity FROM ERRORLIST ;
          |  CALL NUMBER
          |  CALL NUMBER '(' list ')'

condition :  ERR NUMBER entity ;
```

As suggested in Sect. 2.3.4, a special case of action is a section, i.e., another guarded action. This allows specify hierarchies (trees) of sections. During the run-time evaluation of the recovery strategies, a branch shall only be visited when its parent clause has been evaluated as true.
The following actions are supported:

**Stop** terminates a task or a group of tasks, or initiates the shutdown procedure of a node[13].

**Isolate** prevents an entity to communicate with the rest of the system[14].

**Reboot** reboots a node (via the TIRAN_Reboot_Node basic tool).

```
IF [ FAULTY (GROUP{My_Group}) AND NOT TRANSIENT (GROUP{My_Group}) ]
THEN
      STOP TASK@
      SEND {DEGRADE} TASK~
FI
```

Table 5: This section queries the state of group {My_Group} and, if any of its tasks have been detected as affected by a permanent or intermittent fault, it stops those tasks and sends a control message to those considered as being correct so that they reconfigure themselves gracefully degrading their overall service.

**Start** spawns (or, in static environments, awakes) a task or a group.

**Restart** is reverting a task or group of tasks to their initial state or, if no other means are available, stopping that entity and spawning a clone of it.

**Enable** awakes a task or group, or boots a node.

**Send** multicasts (or sends) signals to groups of tasks (or single tasks).

**Warn** informs a task or group of tasks that an error regarding an entity has been detected. Action "`WARN` $x$" is equivalent to action "`SEND {WARN}` $x$"

**Remove** purges records from the section of the DB collecting the errors or the phases of a given entity.

Custom actions and conditions may be easily added to the grammar of ARIEL[15].

When actions are specified, it is possible to use meta-characters to refer implicitly to a subset of the entities involved in the query. For instance, when the first atom specifies a group of tasks, `STOP TASK@1` means "terminate those tasks, belonging to the group mentioned in the first atom of the guard, which fulfill that condition", while `WARN TASK~2` means "warn those tasks, belonging to the group mentioned in the second atom of the guard, that *do not fulfil* the second condition". If one does not specify any number, as in `STOP TASK@`, then the involved group of tasks is the one that fulfils the whole clause. Table 5 shows an example of usage of this feature. Metacharacter "star" (`*`) can be used to refer to any task, group, or processing node in the system. Actions like `STOP TASK*` or `STOP GROUP*` are trapped by the translator and are not allowed. Metacharacter "dollar" (`$`) can be used to refer in a section to an entity mentioned in an atom. For instance, `STOP GROUP$2` means "stop the group mentioned in the second atom of the clause".

A larger excerpt of ARIEL's grammar has been given in Chapter 4.

**Compile-time Support for Error-Recovery.** Once fed with a recovery script, the `art` translator produces a binary pseudo-code, called the **r-code**. In the current version, this r-code is written in a binary file and in a C header file as a statically defined C array, as in Table 6. As can be seen in that table, the r-code is made of a set of "triplets" of integers, given by an opcode and two operands. These are called "r-codes".

This header file needs to be compiled with the application. Run-time error recovery is carried out by the RINT module, which basically is an r-code interpreter. This module and its recovery algorithm are described in the following section. The rest of this section describes how to translate an ARIEL script into the r-code. Within this section and the following one the simple script of Table 7 will be used as an example.

The following scenario is assumed: a triple modular redundancy (TMR) system consisting of three voting tasks, identified by integers `{VOTER1}`, `{VOTER2}`, and `{VOTER3}` is operating. A fourth task, identified as `T{SPARE}`, is available and waiting. It is ready to take over one of the voting tasks should the latter fail. The failed voter signals its state to the backbone by entering phase `HAS_FAILED` through some self-diagnostic module (e.g., assertions or control-flow monitoring). The spare is enabled when it receives a `{WAKEUP}` message and it requires the identity of the voter it has to take over. Finally, it is assumed that once a voter receives a control message with the identity of the spare, it has to initiate a reconfiguration of the TMR such that the failed voter is switched out of and the spare is switched in the system.

Table 7 shows a recovery section that specifies what to do when task `{VOTER1}` fails. The user needs to supply a section like the one in lines 8–12 for each voting task.

Once the specification has been completed, the user can translate it, by means of the `art` program, into a pseudo-code whose basic blocks are the r-codes (see Table 8). A textual representation of the r-codes is also produced (see Table 9).

Other than syntax errors, `art` catches a number of semantic inconsistencies which are reported to the user—as an example, a non-sense request, such as asking the phase of a node, gives rise to the following response:

```
bash-2.02$ art -i .ariel -s
Ariel translator, v2.0f 25-Jul-2000, (c) K.U.Leuven 1999, 2000.
Parsing file .ariel...
        Line 76: semantic error: Can only use PHASE with tasks
        if-then-else: ok
...done (85 lines.)
1 error detected --- output rejected.
```

**The ARIEL Recovery Interpreter.** This section briefly describes the ARIEL recovery interpreter, RINT. Basically RINT is a virtual machine executing r-codes. Its algorithm is straightforward: Each time a new error or burst of errors is detected,

- it executes the r-codes one triplet at a time;

```
/*********************************************************
 *                                                       *
 *  Header file trl.h                                    *
 *                                                       *
 *  Hardwired set of r-codes for Ariel file        ariel *
 *  Written by art (v2.0g 03-Aug-2000) on Fri Aug 18 2000 *
 *  (c) Katholieke Universiteit Leuven 1999, 2000.       *
 *                                                       *
 *********************************************************/
#ifndef _T_R_L__H_
#define _T_R_L__H_

#include "rcode.h"
#define RCODE_CARD 15 /* total number of r-codes */

rcode_t rcodes[] = {
/*line#*/    /* opcode */    /* operand 1 */ /* operand 2 */
/*0*/        { R_INC_NEST,      -1,          -1 },
/*1*/        { R_STRPHASE,       0,          -1 },
/*2*/        { R_COMPARE,        1,        9999 },
/*3*/        { R_FALSE,         10,          -1 },
/*4*/        { R_STOP,          18,           0 },
/*5*/        { R_PUSH,          18,          -1 },
/*6*/        { R_SEND,          18,           3 },
/*7*/        { R_PUSH,           0,          -1 },
/*8*/        { R_SEND,          18,           3 },
/*9*/        { R_PUSH,           3,          -1 },
/*10*/       { R_SEND,          18,           1 },
/*11*/       { R_PUSH,           3,          -1 },
/*12*/       { R_SEND,          18,           2 },
/*13*/       { R_DEC_NEST,      -1,          -1 },
/*14*/       { R_OANEW,          1,          -1 },
/*15*/       { R_STOP,          -1,          -1 },
};
```

Table 6: The beginning of header file `trl.h`, produced by `art` specifying option "`-s`". Array `rcodes` is statically initialized with the r-code translation of the recovery strategy in the ARIEL script.

```
INCLUDE "my_definitions.h"

TASK {VOTER1} IS NODE {NODE1}, TASKID {VOTER1}
TASK {VOTER2} IS NODE {NODE2}, TASKID {VOTER2}
TASK {VOTER3} IS NODE {NODE3}, TASKID {VOTER3}
TASK {SPARE} IS NODE {NODE4}, TASKID {SPARE}

IF [ PHASE (T{VOTER1}) == {HAS_FAILED} ]
THEN
      STOP T{VOTER1}

      SEND {WAKEUP} T{SPARE}
      SEND {VOTER1} T{SPARE}

      SEND {SPARE} T{VOTER2}
      SEND {SPARE} T{VOTER3}
FI
```

Table 7: Another excerpt from a recovery script: after the declarative part, a number of "sections" like the one portrayed in here can be supplied by the user.

- if the current r-code requires accessing the DB, a query is executed and the state of the entities mentioned in the arguments of the r-code is checked;
- if the current r-codes requires executing actions, a request for execution is sent to the Backbone.

  RINT plays an important role within the $\mathcal{R}\varepsilon\mathcal{L}$ architecture—its main objective is establishing and managing a *causal connection* between the entities of the Ariel language (identifiers of nodes, tasks, and groups of tasks) and the corresponding components of the system and of the target application. This causal connection is supported by the Backbone and its DB. In particular, each atom regarding one or more entities is translated at run-time into one or more DB queries. Under the hypothesis that the DB reflects—with a small delay—the actual state of the system, the truth value of the clauses on the entities of the language will have a large probability to tally with the truth value of the assertions on the corresponding components of the system. Furthermore, by means of RINT, symbolic actions on the entities are translated into actual commands on the components. These commands are then managed by the Backbone as described in Fig. 4.

The RINT task is available and disabled on each Backbone assistant while it is enabled on the Backbone manager. Only one execution process is allowed. RINT has the architecture of a stack-based machine—a run-time stack is used

```
bash-2.02$ art -i ariel -v -s
Ariel translator, v2.0g 03-Aug-2000, (c) K.U.Leuven 1999, 2000.
Parsing file ariel...
[ Including file ‘my_definitions.h’ ...9 associations stored. ]
substituting {VOTER1} with 0
substituting {NODE1} with 1
substituting {VOTER2} with 1
substituting {NODE2} with 2
substituting {VOTER3} with 2
substituting {NODE3} with 3
substituting {SPARE} with 3
substituting {NODE4} with 4
substituting T{VOTER1} with T0
substituting {HAS_FAILED} with 9999
substituting {WAKEUP} with 18
substituting T{SPARE} with T3
substituting T{VOTER2} with T1
substituting T{VOTER3} with T2
if-then-else: ok
...done (17 lines in 0.02 CPU secs, or 850.000 lines per CPU sec.)
Output written in file .rcode.
Tasks written in file TaskTable.csv.
Preloaded r-codes written in file trl.h.
Time-outs written in file timeouts.h.
Identifiers written in file identifiers.h.
```

Table 8: The `art` program translates the section mentioned in Table 7 into r-codes. The "`-i`" option is used to specify the input filename, "`-v`" sets the verbose mode, while "`-s`" allows create three header files containing, among other things, an array of pre-loaded r-codes (see Table 6).

```
Art translated Ariel strategy file: .... ariel
into rcode object file : ............... .rcode

 line              rcode     opn1   opn2
----------------------------------------------
00000           SET_ROLE      0    Manager
00001           SET_ROLE      1    Assistant
00002           SET_ROLE      2    Assistant
00003           SET_ROLE      3    Assistant
00004                 IF
00005     STORE_PHASE...   Thread      0
00006         ...COMPARE       ==   9999
00007              FALSE     10
00008               STOP   Thread      0
00009            PUSH...     18
00010            ...SEND   Thread      3
00011            PUSH...      0
00012            ...SEND   Thread      3
00013            PUSH...      3
00014            ...SEND   Thread      1
00015            PUSH...      3
00016            ...SEND   Thread      2
00017                 FI
00018    ANEW_OA_OBJECTS      1
00019               STOP
```

Table 9: A textual representation of the r-code produced when translating the recovery section in Table 7.

during the evaluation of clauses. In a future release of RINT, the run-time stack shall also be used as a means for communicating information between actions. For any r-code being executed, a message will be broadcast to the non-faulty assistants. Next r-code will only be executed when all the non-faulty assistants acknowledge the receipt of this message and update their stack accordingly. This allows maintain a consistent copy of the current status of the run-time stack on each assistant. Should the manager fail while executing recovery, the new manager would then be able to continue recovery seamlessly, starting from the last r-code executed by the previous manager.

# 4 INTEGRATING RECOVERY STRATEGIES INTO A PRIMARY SUBSTATION AUTOMATION SYSTEM

The DepAuDE architecture provides an approach to integrate fault tolerance support into distributed embedded automation applications. It allows error recovery to be expressed in terms of recovery strategies, i.e., the lightweight code fragments described in Chapter 6. At run time, the middleware orchestrates their execution. This section reports on the integration of diferent recovery scripts into a distributed run-time environment applied to the embedded automation system of a primary substation. An instrumented automata-based design environment allows the application to be deployed on a heterogeneous platform with several real-time operating systems. While the middleware detects the errors and selects the correct recovery scripts to be executed, the application functionality is maintained through system reconfiguration or graceful degradation. The added value comes from the flexibility to modify recovery strategies without requiring major modifications to the application, while tolerating the same physical faults as in the dedicated hardware solutions.

## 4.1 Introduction

Industrial distributed embedded systems such as those used in the control and automation of electrical energy infrastructures rely on off-the-shelf components and protocols to ensure cost-efficient exploitation (Caird, 1997; Dy-Liacco, 2002). As a particular application can be deployed on a variety of hardware targets (with different sets of sensors and actuators attached) and within different environments (e.g. with different levels of electro-magnetic interference), flexibility is needed both to instantiate the application functions appropriately and to react adequately to disturbances to the information and communication infrastructure on which the application is running. For instance, system reconfiguration and recovery may be different, depending on which I/O devices are connected to the different parts of the distributed controllers. More generally, adaptability is required to modify fault tolerance strategies depending on the environment. The DepAuDE architecture deploys a set of middleware modules to provide fault tolerance by exploiting the embedded systems distributed hardware and by separating functional behavior from the recovery strategy, i.e., the set of actions to be executed when an error is detected.

This architecture has been integrated in an innovative demonstrator of a Primary Substation Automation System, i.e. the embedded hardware and software in a substation for electricity distribution, connecting high voltage lines (HV) to medium voltage (MV) lines over transformers. The Primary Substation Automation System requires protection, control, monitoring and supervision capabilities. It is representative of many applications with dependability requirements in the energy field (Gargiuli, Mirandola, & et al., 1981). As mentioned already in Chapter 4, the major source of faults in the system is electro-magnetic interference caused by the process itself (opening and closing of HV/MV switchgear) in spite of the attention paid to designing for electromagnetic compatibility. Software and hardware faults in the automation system have to be considered as well. These cause errors in communication, execution and memory subsystems.

In the ongoing renewal of electric infrastructures in Europe, utility companies are replacing their dedicated hardware-based fault tolerance solutions by commercial, interconnected platforms. This trend is motivated by the growing need for more functionality: Development of new, dedicated (hardware-based) solutions is considered too expensive and not flexible enough to keep up with the evolving requirements of the liberalized electricity market. The required dependability is reached by exploiting hardware redundancy in the distributed platform, combined with software-implemented fault tolerance solutions at the application and at the middleware level. Although software-based fault tolerance may have less coverage than hardware-based solutions, this was not considered inhibitive, because the physical (electrical, non-programmable) protection in the plant continues to act, as a last resort, as a safeguard for non-covered faults[16]. Besides, high-quality software engineering and extensive on-site testing remain important to avoid the introduction of design faults that could hamper mission-critical services. This section presents the experience of integrating this DepAuDE software architecture into an application developed with the ASFA distributed run-time environment, a prototypic embedded control system for a primary substation. According to the approach proposed here, support for allocation of tasks to components, for reactions to detected errors and for maintainability of the fault tolerance strategy is accomplished by using the configuration-and-recovery language ARIEL introduced in Chapter 6, which was developed in the framework of project TIRAN and later became is a key component of the DepAuDE architecture. In what follows we describe in Sect. 4.2 the DepAuDE assumptions and models, in Sect. 4.3 the Primary Substation Automation System application and in Sect. 4.3 the instantiation of the DepAuDE architecture therein. Section 4.4 concludes with a qualitative evaluation of the experience.

## 4.2 Assumptions and Models

DepAuDE relies on the following assumptions:

- **Fault model and failure semantics**: A single physical fault affects execution or communication entities (tasks, nodes, links). Experiments confirm that EMI affects only the entity to which the responsible I/O element is connected (Gargiuli et al., 1981). Depending on the underlying hardware and RTOS (for instance, whether a memory management unit is available or not), a fault containment region is a task or node. Crash failure semantics (fail-silent behavior) is assumed for the fault containment region.

- **System model**: A synchronous system model is assumed (i.e. known and bounded processing delays, communication delays, clock differences and clock drifts, as explained in more detail in Chapter 2). As mentioned already, this is realistic for the set of targeted real-time automation applications, because of their implementation on dedicated systems.

- Communication, provided by a Basic Services Library similar to the one designed in the project TIRAN, is assumed to be perfect (no lost messages, no duplicates, keeping message order). In order to increase the coverage for this assumption, a set of mechanisms such as the EFTOS Fault Tolerant Communication tasks described in Chapter 3 can be employed or developed in a similar way.

- The communication mechanism targets groups of tasks, that is, a multicast service is prescribed to be available, whose behavior is assumed to be atomic: When a message is sent to a group of tasks, the either all the non-crashed processes receive it or none of them. If this assumption coverage is too low, dedicated atomic multicast support and group membership functions can be added.

The DepAuDE middleware supports the reintegration of the basic services library and of the DepAuDE fault tolerance mechanisms (a concept also derived from the TIRAN project). Furthermore it can reload application tasks. The application in itself is responsible for reintegrating these restarted tasks into the ongoing execution, as no checkpoint/restore mechanisms are included.

## 4.3 Case Study

The Local Control Level module is a Primary Substation Automation System component providing control and protection functions for the primary substation, as well as an interface to the operator and—over the inter-site network—to remote control systems and remote operators. The Local Control Level controls the switches to the two HV/MV transformers, the switch connecting the Red MV bar (on the left) to the Green MV bar (on the right), as well as switches local to the MV lines (Figure 8). The pilot application implements two functions from the Local Control Level module: automatic power resumption (function1) and parallel transformers (function2). Function1 allows automatic power resumption when a HV/MV transformer goes down, e.g. triggered by internal protection (temperature too high, oil alarm, ...). It disconnects the MV lines connected to the busbar of the transformer, computes the load carried by the transformer just before the event happened, and if possible, causes the remaining transformer to take the entire load, as e.g. in the following scenario:

- (Initially) Red transformer carries 32 MVA (8 lines of 4 MVA) and Green transformer 24 MVA (8 x 3 MVA); the switches connecting the Red and Green bars to the transformers are closed; the switch connecting the Green MV bar to the Red MV bar is open.
- (Anomaly) An internal protection mechanism shuts down the Green transformer, and its power drops from 24 MVA to zero. The switch connecting the Green bar to the Green transformer opens. (The switch connecting the Red bar to the Red transformer remains closed and the switch connecting the two bars remains open.)
- (Reaction) The switch connecting the Green bar to the Red bar receives the command to close. It closes 1 execution cycle (100 ms) later and the load carried by the Red transformer rises to 56 MVA.

Function2 (parallel transformers) consists of a series of automatic actions, assisting remote operators. As an example, an operator can request to switch on a transformer and function2 translates this request into a specific sequence of commands. Such a re-insertion scenario may be applied some time after transformer exclusion.

**System Setup** The Primary Substation Automation System application has been developed using a proprietary, automata-based, design environment based on the specification technique ASFA. Application development consists of several steps:

- Function1 and function2 are extracted from the Primary Substation Automation System application and specified through the ASFA Graphical Editor, obtaining a tabular description of the pilot application.
- These ASFA tables are processed by the ASFA-C Translator, producing a target-independent C-code version of the application, and by the ASFA Partitioner, allowing an application to be mapped to a single task or decomposed into a set of tasks (Ciapessoni & Maestri, 2001). The single task version has been used for the functional test of the application on a single host node, while a four-task version was selected for testing on a distributed system.

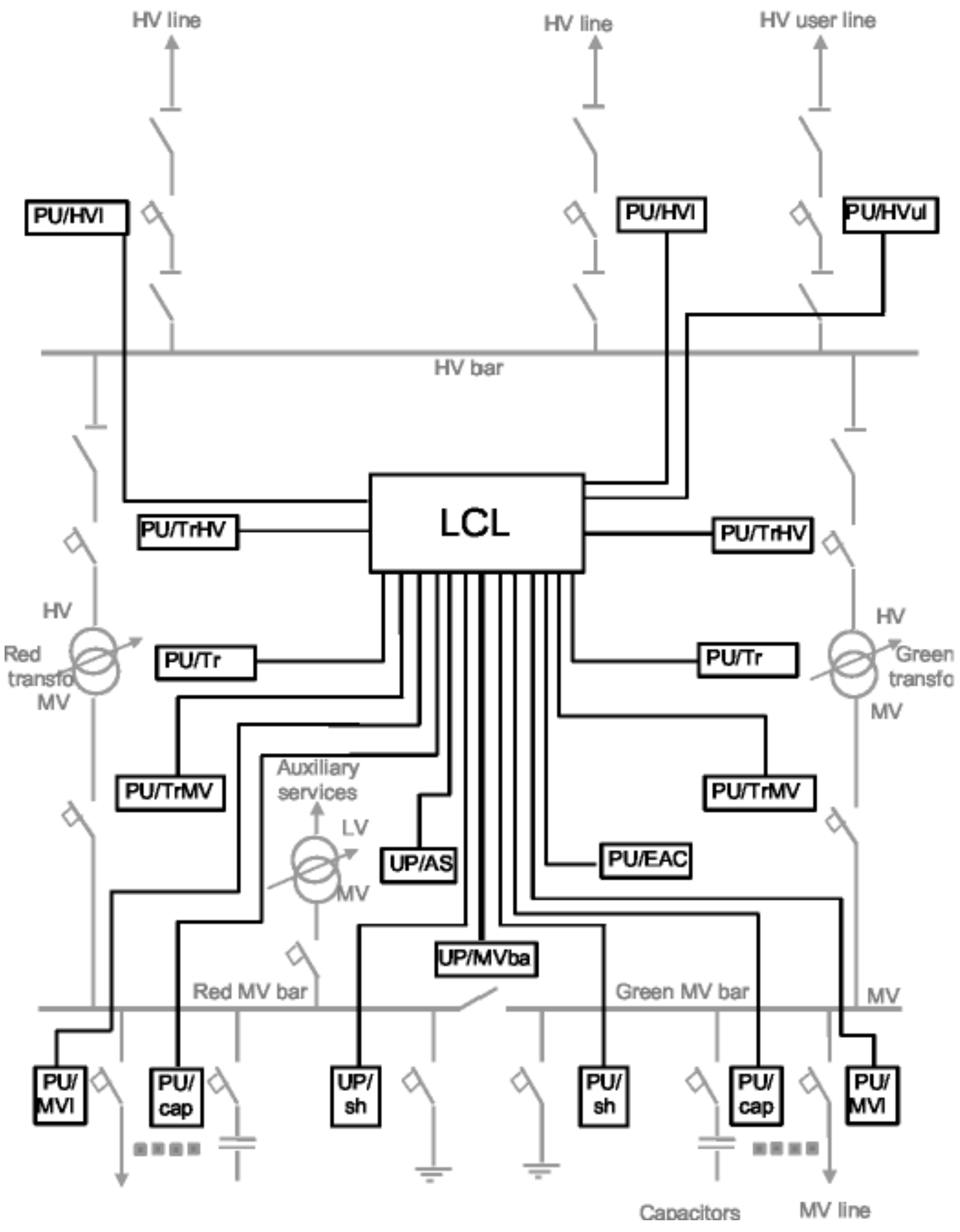

Figure 8: Electric circuit (grey lines) and control architecture (black lines) of the Primary Substation.

| | N1<br>VxWorks<br>*VMIC IPC* | N2<br>VxWorks<br>*INOVA IPC* | N3<br>RMOS32<br>*Siemens PLC* | N4<br>Linux<br>*BB PC* | N5<br>WinNT<br>*OC PC* |
|---|---|---|---|---|---|
| BSL tasks | * | * | * | * | * |
| Backbone | | | | * | |
| LAN Monitor | * | * | * | * | * |
| Operator Console | | | | | * |
| BSW | * | * | * | | |
| Executive | * | * | * | | |

Table 10: Allocation of middleware tasks to nodes. PU stands for "Peripheral Unit".

- At run time, the Distributed Execution Support Module, composed of Basic Software (BSW) and Executive, enforces cyclic execution, typical for PLC-based automation systems (PLC=Programmable Logic Controller). Robust execution is ensured by cyclically refreshing the I/O image and the non-protected memory areas, while the applications state is safeguarded by hardware or software mechanisms [5]. The Basic Software takes care of synchronization and exception handling, while the Executive supplies the RTOS interface and a set of ASFA-specific library functions.

A peculiarity of the ASFA environment is that the application code is automatically obtained by translating the automata-based specification. Besides reducing the probability of introducing coding errors, this approach provides portability to all platforms supported by the Distributed Execution Support Module.
As shown in Figure 9, this pilot application was deployed on a distributed system consisting of three dedicated heterogeneous ("target") processors for the automation functions and two standard PCs for support functions, interconnected by an Ethernet switch:

- N1 and N2: two industrial PCs (VMIC and INOVA), with VxWorks as RTOS;
- N3: A Siemens SIMATIC M7, that is, an extended PLC with I/O modules, with RMOS32 as real-time operating system;
- N4: A number of Linux-based standard PCs, hosting the DepAuDE Backbone (an enhanced version of the EFTOS DIR net and of the TIRAN Backbone described in previous chapters);
- N5: A Windows-NT PC with Operator Console functions.

For inter-site connections (not considered here), an additional node hosts the gateway software.

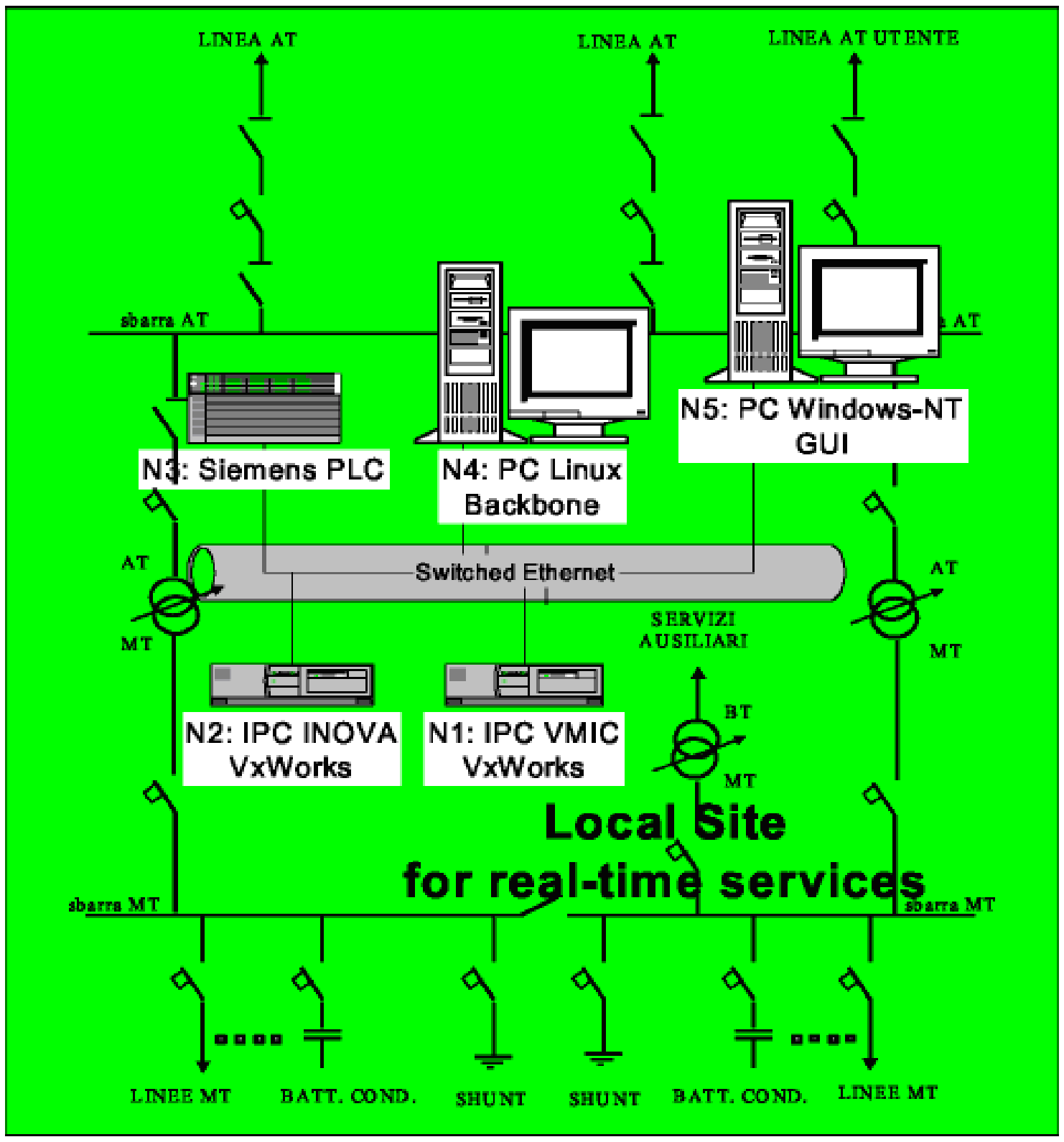

Figure 9: Primary Substation Automation System hardware architecture.

The pilot application runs on this heterogeneous hardware equipment; input and output from/to the field is simulated. Synchronization signals, for cyclic application execution, are generated by the internal clock of one of the nodes (in a real set-up, they are obtained from an independent, external device). The following assumptions are made for the *target* nodes:

- All three target nodes (N1, N2 and N3 in Table 10) are attached to I/O components on the field (the PU boxes in Fig. 8, that is, peripheral units).
- The target node N3 handles the synchronization signal. In order to provide a backup solution in case of faults on N3, synchronization interrupts are also available at N1 and N3.

**Instantiating the DepAuDE Mechanisms on the Primary Substation Automation System.**
The run-time components of the DepAuDE framework are integrated into the Primary Substation Automation System pilot application (see Table 10). The fault containment region is a node.

- An RMOS32 and a VxWorks implementation of the Basic Services Library tasks run on the target nodes (N1, N2, N3); a Linux and WinNT version runs on the host nodes (N4, N5).
- A LAN Monitor—that is, a fault-tolerant mechanism used for detecting crashed or isolated nodes—is present on all nodes.
- The DepAuDE Backbone task, responsible for the execution of the recovery strategies as described in Chapter 6, is allocated to N4.
- Each of the three target nodes hosts an instance of the ASFA Distributed Execution Support Module, composed of Basic Software and Executive. Each instance of the Basic Software is able to act as master (BSW_M, on the master node) or slave (BSW_S, on the slave nodes). The role is chosen depending on the specific system configuration. All BSW_S entities make up the BSW_SLAVE_GROUP. The configuration with highest performance (see below) requires BSW_M to be allocated to N3 and the BSW_S processes to run on N1 and N2. Executive process instances are identical on each processing node and they compose the EXECUTIVE_GROUP.

The allocation of the application tasks depends on the partitioning of the two Local Control Level functions (function1 and function2), among which there is no communication. Function2 consists of a single task, PARALLEL_TRS, function1 (automatic power resumption) consists of three tasks: two tasks (BUSBAR1 and BUSBAR2) handle low-level, I/O dependent, computations relative to the MV lines attached to each busbar; one task, STRAT, coordinates the whole function and performs no field I/O. There is no communication between the two BUSBAR tasks, while both communicate with STRAT. The basic constraint for allocating tasks to nodes is that a task that controls a specific plant component should be allocated to a processor attached to that plant component (due to I/O paths). As both functions of the pilot application control the same set of field components (same transformers and switches), all target nodes are assumed to be connected to that portion of the field. We assume that target node N2 provides better computing performance than N1.
The start-up configuration is the optimal distribution of application tasks onto the heterogeneous hardware. The most performant configuration, Config_0 in Table 2, does not require off-node communication among the application tasks:

- no application task is allocated to N3, whose Basic Software acts as master and handles communication with the remote control center;
- PARALLEL_TRS runs on N1;
- BUSBAR1, BUSBAR2, and STRAT are allocated to N2.

| | N1 | N2 | N3 |
|---|---|---|---|
| Config_0 | PARALLEL_TRS | STRAT, BUSBAR1, BUSBAR2 | - |
| Config_1 | *CRASHED* | PARALLEL_TRS STRAT, BUSBAR1, BUSBAR2 | - |
| Config_2 | PARALLEL_TRS | *CRASHED* | STRAT BUSBAR1, BUSBAR2 |
| Config_3 | STRAT, BUSBAR1, BUSBAR2 | - | *CRASHED* |

Table 11: Different configurations to allocate active Primary Substation Automation System application tasks to target nodes.

- Each application task has at least one standby replica task_Ri on a different target node Ni (i=1 . . . 3).

#### 4.3.1 Recovery Strategy of the Primary Substation Automation System

In order to cope with temporary and permanent physical faults affecting the information and communication infrastructure of the Primary Substation Automation System, an appropriate recovery strategy has been designed and coded as a set of Ariel recovery scripts. Such strategy combines different kinds of error detection mechanisms, error recovery and system reconfiguration. Reconfiguration is statically associated to the crash of a single node. If two nodes crash simultaneously no reconfiguration is possible. The following scripts are examples of recovery actions.

**Example 1.** If a slave node (e.g., N1) crashes, the LAN Monitor detects this event and notifies the Backbone executing the following Ariel code:

```
IF
  [FAULTY NODE{N1} AND RUNNING NODE{N2} AND RUNNING NODE{N3} AND
   PHASE(TASK{BSW_M}) == {NEW_CYCLE_PH}]
THEN
  ISOLATE NODE{N1}
  SEND {CONFIG_1} TASK{BSW_MSG_M}
  SEND {CONFIG_1} GROUP{BSW_SLAVE_GROUP}
  RESTART GROUP{EXECUTIVE_GROUP}
  RESTART TASK{PARALLEL_TRS_R2}
FI
```

If the guard of the above script is fulfilled, application tasks are reconfigured as of CONFIG_1 from Table 11. CONFIG_1 maintains the full Primary

Substation Automation System functionality by transferring Parallel_TRS to N2, actually activating its spare replica. This node is able to cope with the whole computational load, as it does not need to perform communication requested by the BSW_M's functions. To avoid undesired interference by the Backbone during critical phases of BSW_M activity, a condition on the current execution phase (`PHASE (TASK{BSW_M} ) == {NEW_CYCLE_PH}`) must be satisfied in conjunction with the crash test. The ISOLATE NODE action corresponds to informing other nodes that they may not accept any message from the isolated peer—even if it comes back to life—until the isolation is undone.

**Example 2.** If a target node (e.g. N2) crashes during a different execution phase of the master Basic Software, then this error is notified by the BSW_M to the Backbone (through function `RaiseEvent (RE_BSW_error)`) , causing the execution of the following ARIEL code:

```
IF [EVENT {RE_BSW_error}]
  THEN
  IF [FAULTY NODE{N2} AND RUNNING NODE{N3}] THEN
    ISOLATE NODE{N2}
    SEND {CONFIG_2} TASK{BSW_MSG_M}
    SEND {CONFIG_2} TASK{BSW_MSG_S1}
    RESTART GROUP{EXECUTIVE_GROUP}
    RESTART TASK{BUSBAR1_R3}, TASK{BUSBAR2_R3}, TASK{STRAT_R3}
    RESTART TASK{PARALLEL_TRS_R1}
  FI
FI
```

Hence the system is reconfigured as in Config_2: The spare replicas of BUSBAR1, BUSBAR2 and STRAT are activated on N3.

**Example 3.** In case of a fault on target node N3 (where BSW_M is running), the following ARIEL code is executed, triggered by error detection by the LAN Monitor and subsequent notification to the Backbone:

```
IF
  [FAULTY NODE{N3} AND RUNNING NODE{N1} AND RUNNING NODE{N2}]
THEN
  ISOLATE NODE{N3}
  SEND {CONFIG_3} GROUP{BSW_SLAVE_GROUP}
  SEND {BACKUP_MASTER} TASK{BSW_MSG_S2}
  RESTART GROUP{EXECUTIVE_GROUP}
  STOP TASK{PARALLEL_TRS}
  RESTART TASK{STRAT_R1}, TASK{BUSBAR1_R1}, TASK{BUSBAR2_R1}
FI
```

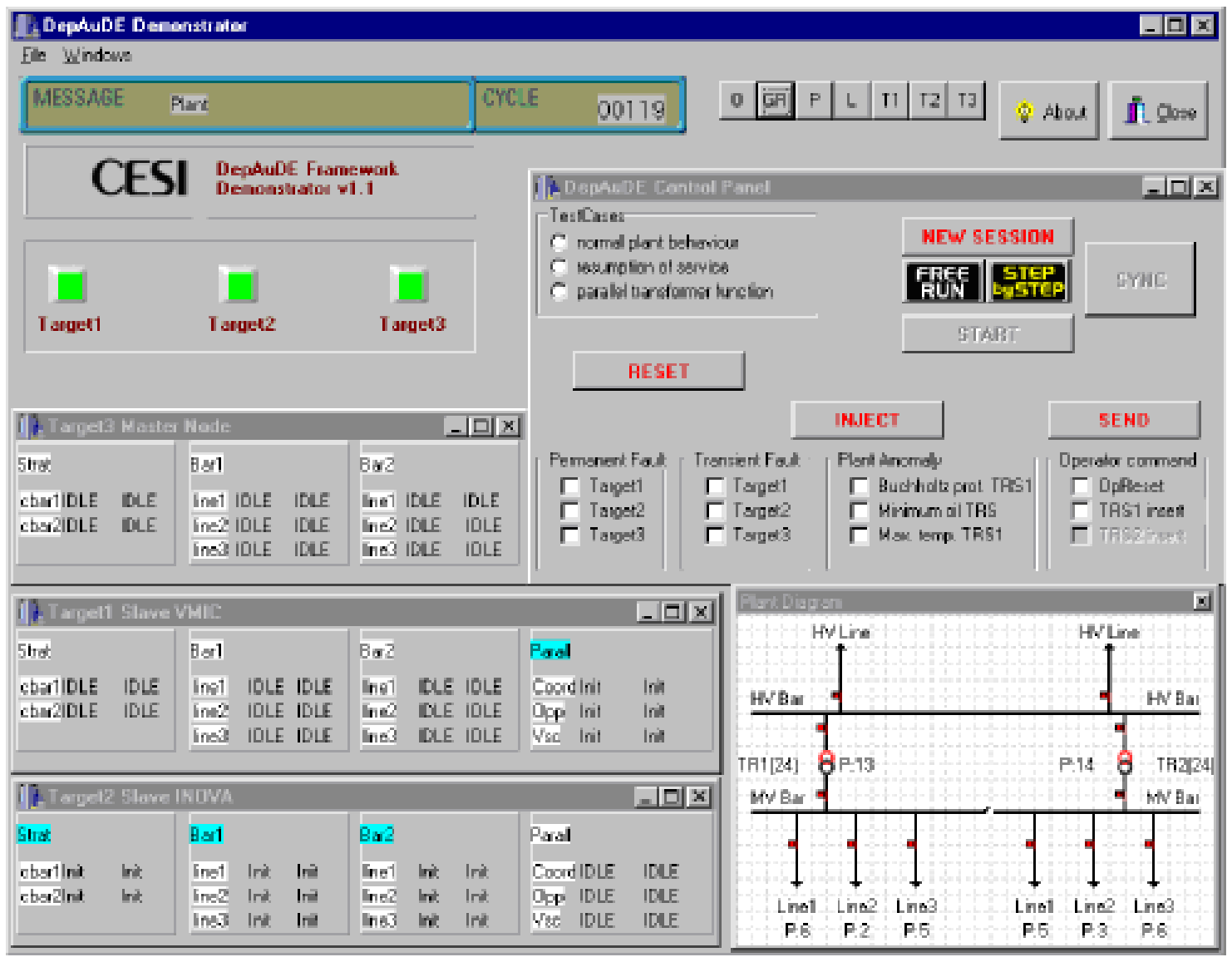

Figure 10: User interface for application supervision and active task allocation.

As a consequence, the function of master node is transferred to N2 and the application tasks of N2 are moved to N1. As N1 cannot support both application functions simultaneously, PARALLEL_TRS is disabled, thus proceeding to a graceful degradation of the automation system (config_3).

**Evaluation.** Other recovery strategies, such as restarting all tasks on a node after a transient fault, or shutting down the system when reconfiguration is not possible, have also been coded in ARIEL and implemented. We did not provide recovery strategies associated with a crash of N4 or N5, because as they are not target nodes, they are not concerned with the automation control function itself; so even if they crash, the application is not endangered. In a real deployment they could be replicated or could backup each other.
Figure 10 shows the user interface of the pilot application demonstrator.

## 4.4 Summary and Lessons Learned

The lack of flexibility that is inherent to dedicated hardware-based fault tolerance solutions makes their adoption not cost-effective in cases where similar functionality has to be deployed in several sites, each characterized by a slightly different environment. This section presented the integration of a fault-tolerance architecture based on the recovery language approach into the distributed automation system of a primary substation. The deployment of this fault tolerance middleware allows different recovery strategies to be integrated on a heterogeneous platform. Given the generality of the methods and techniques used the designed solution is applicable to a wide class of process automation systems.
The following points summarize the lessons learned:

- The ASFA design environment with automatic code generation provides several advantages: less development time, absence of coding errors, portable application code and possibilities for application partitioning. It is straightforward to interface it to IEC 61850-compliant Intelligent Electronic Devices (IED).
- The implementation effort required to integrate the DepAuDE Basic Services Library into an ASFA application was limited (about 2400 lines of code for the RMOS and VxWorks targets). The communication mechanism supplied by the DepAuDE Basic Services Library provided transparent inter-process communication among ASFA application tasks. The grouping of tasks revealed itself as a useful tool when implementing the standby replicas. Inter-processor communication among application tasks strongly influences application performance and reconfiguration time in case of faults. Therefore inter-processor data flow should be avoided if possible, or at least minimized.
- The deployment of the DepAuDE architecture allowed integrating several recovery strategies on a heterogeneous platform. The separation between functional and error recovery programs provides flexibility to modify recovery strategies without requiring major modifications to the application, while tolerating the same physical faults as in the dedicated hardware solutions.

# 5 CONCLUSION

The recovery language approach and ARIEL, a configuration-and-recovery language, have been discussed. First the concepts have been treated in general terms; then, a particular context has been chosen and the concepts in that context have been deployed in the form of a prototypic architecture developed in the framework of a European project, TIRAN. Finally, a real-life example from the domain of electrical automation system has been discussed. ARIEL and its run-time system provide the user with a fault-tolerance linguistic structure that appears to the user as a sort of second application-level especially conceived and devoted to address the error recovery concerns. Designed by the author of this book while taking part to several European research projects, ARIEL is currently being used as a linguistic structure to express adaptive feedback loops.

# Notes

[1]To be pronounced as "[a:*]-[i:]-[el]."

[2]For instance, via the `exit` function call from the C standard libraries.

[3]Note how, according to the hypothesis of adherence to the timed-asynchronous distributed system model, such thresholds are known because all services are timed.

[4]"Virtual" isolation of a task can be obtained, when the task obeys the fourth application assumption at page 164, "disactivating" the corresponding BSL communication descriptors.

[5]To be intended herein as a set of software libraries, distributed components, and formal techniques.

[6]Safety-critical systems, i.e., computer, electronic or electromechanical systems whose failure may cause injury or death to human beings, such as an aircraft or nuclear power station control system (FOLDOC, 2000), were not covered within TIRAN. This allowed the crash failure semantics assumption (see Sect. 2.1.1) to be satisfied with less strict coverage, which translates into lower development costs for the framework.

[7] Mosaic020 boards have been developed in the framework of the ESPRIT project Dipsap II (Dipsap-II, 1997) by DASA/DSS and Eonic Systems. They are based on the Analog Devices ADSP-21020 DSP and the SMCS communication chip. The SMCS chip complies with the IEEE 1355 standard (IEEE, 1995) and has hardware support for detecting transmission and connection errors and for higher level system protocols such as reset-at-runtime.

[8]A time-out is said to be suspended when, on expiration, no alarm is executed. The corresponding entry stays in the time-out list and obeys its rules—in particular, if the time-out was cyclic, on expiration the entry is renewed.

[9]An associative array (see, for instance, (De Florio, 1996)) may then be used to de-reference an entity through its symbolic name.

[10]In TIRAN lingo, a SocketServer is a task run on each node of the system, which is used by the TIRAN BSL for managing off node communication (via UDP sockets) and for local dispatching of remote messages. This is a well-known technical solution which is used, e.g., in PVM, where a single component, `pvmd` (PVM daemon), is launched on each node of a PVM cluster to manage global tasks (Geist et al., 1994).

[11]A number of voting techniques have been generalized in (Lorczak, Caglayan, & Eckhardt, 1989) to "arbitrary $N$-version systems with arbitrary output types using a metric space framework". To use these algorithms, a metric function can be supplied by the user so to compare any two votes. Such a function is called a "metric" and has a fixed prototype, the one of function `strcmp` of the C standard library function—in the example, a function called `int_cmp` is selected. The object code of this function must be available and addressable when compiling the target application. See Chapter 3 for more details on this.

[12]Here and in the following, context-free grammars are used in order to describe syntax rules. The syntax used for describing those rules is that of the YACC (Johnson, 1975) parser generator. Appendix A to Chapter 11 describes YACC in detail. Terminal symbols such as `GT` are in capital letters. They are considered as intuitive and their definition (in this case, string "`>`") in general will not be supplied.

[13]These services are obtained via specific function calls to the level-1.2 BT (see the edge from RINT to those BT in Fig. 2). Such BT, in turn, can either execute, through the BSL, a kernel-level function for stopping processes—if available—or send a termination signal to the involved processes. The actual choice is taken transparently, and RINT only calls one or more times either a `TIRAN_Stop_Task` or a `TIRAN_Stop_Node` function.

[14]This service is obtained as described in the previous footnote. Depending on the characteristics of the adopted platform, isolation can be reached either through support at the communication driver or kernel level, or as follows: when a task opens a connection, a reference to a given object describing the connection is returned to both the user code and the

local component of the Backbone. Action `ISOLATE` simply substitutes this object with another one, the methods of which prevent that task to communicate. This explains the third application-specific assumption of Sect. 2.1.1.

[15]For instance, condition "`DEADLOCKED`" and action "`CALL`" (see the Appendix of Chapter 11) were added to test the inclusion in ARIEL, respectively, of a provision for checking whether two tasks are in a deadlock (see (Efthivoulidis et al., 1998) for a description of this provision) and of a hook to the function call invocation method. These two provisions were easily introduced in the grammar of ARIEL.

[16]As we have seen in the case of the Therac-25 linear accelerator in Chapter 2, it is often a good design choice not to remove hardware safeguards. . .

page

# FAULT-TOLERANT PROTOCOLS USING ASPECT ORIENTATION

## 1 INTRODUCTION AND OBJECTIVES

This chapter resumes our survey of application-level fault-tolerance protocols considering approaches based on aspect-oriented programming.
Aspect-compliant programming languages allow to treat a source code as a pliable web that the designer can weave so as to specialize or optimize towards a certain goal without having to recode it. This useful property allows separate concerns, bound complexity and enhance maintainability. Aspect programs may be used for different objectives, including non-functional properties such as dependability. To date it is not known whether aspect-orientation will actually provide satisfactory solutions for fault-tolerance in the application layer. Some researchers believe this is not the case (Kienzle & Guerraou, 2002)—at least for some fault-tolerance paradigm. Some preliminary studies have been carried out (for instance in (Lippert & Videira Lopes, 2000)), but no definitive word has been said on the matter. It is our belief that, at least for some paradigms, aspects may reveal themselves as invaluable tools to engineer the application-level of fault-tolerance services. For this reason their approach is described in this chapter.

## 2 FAULT-TOLERANT PROTOCOLS THROUGH ASPECT ORIENTATION

### 2.1 General Ideas

Aspect-oriented programming (AOP) (Kiczales et al., 1997) is a programming methodology and a structuring technique that explicitly addresses, at application level, the problem of the best code structure to express different, possibly conflicting design goals such as high performance, optimal memory usage, and dependability.
Indeed, when coding a non-functional service within an application—for instance an application-level error handling protocol—using either a procedural or an object-oriented programming language, one is required to decompose the original goal, in this case a certain degree of dependability, into a multiplicity of fragments scattered among a number of procedures or objects.

This happens because those programming languages only provide abstraction and composition mechanisms to cleanly support the *functional* concerns. In other words, specific non-functional goals, such as high performance, cannot be easily captured into a single unit of functionality among those offered by a procedural or object-oriented language, and must be *fragmented* and *intruded* into the available units of functionality. As already observed, this code intrusion is detrimental to maintainability and portability of both functional and non-functional services (the latter called “aspects” in aspect-oriented terminology). Such aspects tend to crosscut the system’s class and module structure rather than staying, well localized, within one of these unit of functionality, e.g., a class. This increases the complexity of the resulting systems.
The main idea of aspect-oriented programming is to use:

1. A “conventional” programming language (that is, a procedural, object-oriented, or functional programming language) to code the basic functionality. The resulting program is called a *component program*. The basic functional units of the component program are called *components*.

2. A so-called *aspect-oriented language* to implement given aspects by defining specific interconnections (“aspect programs” in aspect-oriented lingo) among the components in order to address various systemic concerns.

3. An *aspect weaver*, that takes as input both the aspect and the component programs and produces with those (“weaves”) an output program (“tangled code”) that addresses specific aspects.

The weaver first generates a data flow graph from the component program. In this graph, nodes represent components, and edges represent data flowing from one component to another. Next, it executes the aspect programs. These programs edit the graph according to specific goals, collapsing nodes together and adjusting the corresponding code accordingly. Finally, a code generator takes the graph resulting from the previous step as its input and translates it into an actual software package written, e.g., for a procedural language such as Java. This package is only meant to be compiled and produce the ultimate executable code fulfilling a specific aspect like, e.g., higher dependability.
In a sense, aspect-oriented programming systematically automatizes and supports the process to adapt an existing code so that it fulfils specific aspects. Aspect-oriented programming may be defined as a software engineering methodology supporting those adaptations in such a way as to guarantee that they do not destroy the original design and do not increase complexity. The original idea of aspect-oriented programming is a clever blending and generalization of the ideas that are at the basis, for instance, of optimizing compilers, program transformation systems, meta-object protocols, and even literate programming (Knuth, 1984).

## 2.2 AspectJ and Aspectwerkz

### 2.2.1 AspectJ

AspectJ is the first and probably most wide-spread aspect-oriented language (Kiczales, 2000; Lippert & Videira Lopes, 2000). Developed as a Xerox PARC project, AspectJ can be defined as an aspect-oriented extension to the Java programming language. AspectJ provides the programmer with the following constructs:

- Join points, i.e. points in a program where additional behavior (logging for example) can be attached. They represent relevant points in a program's dynamic call graph. Join points mark the code regions that can be manipulated later one by an aspect weaver (see above). In AspectJ, these points can be
  - method executions,
  - constructor calls,
  - constructor executions,
  - field accesses, and
  - exception handlers.

  Join points can be expressed through pointcuts (see below).

- Pointcuts are a way to express join points in a program. Let us suppose for example one would need to achieve transparent logging of all the methods whose names start with string "do" . In AspectJ this can be expressed as pointcut

  doMethod(): execution(* do*(*) );

- An advice is the additional code (the actual logging, for instance) that has to be executed at the join point. In the logging example it can be implemented as follows:

  after() returning:doMethod() //log something .

- Inter-type declarations provide a mechanism to change the structure of existing classes. It is possible to add methods, fields and even interfaces to existing classes without changing the class itself. For instance if one needs to log how many times a method is executed, it is possible to add a counter field to the class.

AspectJ is in a sense an extension to Java and in this regard it has two important properties. The first property is that all valid Java programs are valid AspectJ programs and the second is that after transformation of the AspectJ program, it becomes a valid Java program that can be run in the Java Virtual Machine (jvm). In order for AOP to work, AspectJ has an AspectJ-compiler (ajc) that weaves the aspects into the code. This weaving can happen at three different times:

- Compile-time: ajc will compile the classes from source and produce the woven classes as output. This is the simplest approach and only necessary if the aspects are required for the code to compile (which would not be a good approach).
- Post-compile-time: ajc will weave existing class- and jar-files with the aspects. Post-compiletime weaving allows great flexibility as it enables us to add aspects after the original code has been compiled.
- Load-time: The weaving only happens when the class loader loads the class in the environment. This requires the support from a weaving class loader and weaving agent by adding aspectjweaver.jar to the classpath. Load-time weaving allows the most flexibility as the aspects to be woven need only be known at runtime.

Another extension to Java is AspectJ's support of the Design by Contract methodology (Meyer, 1997), where *contracts* (Hoare, 1969) define a set of pre-conditions, post-conditions, and invariants, that *determine how to use* and *what to expect* from a computational entity.
A study has been carried out on the capability of AspectJ as an aspect-oriented programming language supporting exception detection and handling (Lippert & Videira Lopes, 2000). It has been shown how AspectJ can be used to develop so-called "plug-and-play" exception handlers: libraries of exception handlers that can be plugged into many different applications. This translates into better support for managing different configurations *at compile-time*. This addresses one of the aspects of attribute A defined in Chapter 2.

### 2.2.2 Dynamic weaving of aspects

As mentioned in (Hiltunen, Taïani, & Schlichting, 2006), dynamic behavior through run-time weaving is being recognized more and more as a key property in modern aspect-oriented architectures. While early versions of AspectJ could only define joinpoints as static source code locations it is now possible to define joinpoints on the occurrence of run-time conditions. This allows activate or deactivate aspects dynamically. Many researchers believe that through this powerful syntactical framework it will be possible to tackle effectively dynamic adaptation and self-configuration.
An interesting example of this trend is AspectWerkz (Bonér & Vasseur, 2004; Vasseur, 2004), defined by its authors as "a dynamic, lightweight and high-performant AOP framework for Java" (Bonér, 2004). AspectWerkz utilizes bytecode modification to weave classes at project build-time, class load time or runtime. This capability means that the actual semantics of an AspectWerkz code may vary dynamically over time, e.g., as a response to environmental changes. This translates in a useful structure to create dependable service whose fault model changes over time, as discussed in Chapter 2, hence characterized by an excellent A.
Recently the AspectJ and AspectWerkz projects have agreed to work together as one team to produce a single aspect-oriented programming platform building on their complementary strengths and expertise.

### 2.3 Variations on the Main Theme: AspectC++ and GluonJ

A recent project, AspectC++ (AspectCpp, n.d.), proposes an aspect-oriented implementation of C++ which appears to achieve most of the positive properties of the other Java-based approaches and adds to this efficiency and good performance.
AspectJ and Aspectwerkz are not the only aspect-oriented languages focusing on Java: Another example is GluonJ (GluonJ, n.d.; Chiba & Ishikawa, 2005), whose primary design goal is simplicity—a fundamental property for truly dependable systems. Quoting his author, GluonJ "provides simple but powerful AOP constructs by using annotations in regular Java. Developers can use GluonJ as a compile-time AOP system or a load-time AOP system". GluonJ in particular provides its programmer with a mechanism for extending an existing class, called refinement. As mentioned in Chapter 4, GluonJ has been written on top of Javassist, the Java Programming Assistant meta-library briefly introduced in that chapter.

## 3 CONCLUSION

Figure 1 synthesizes the main characteristics of AOP: it allows decompose, select, and assemble components according to different design goals. This has been represented by drawing the components as pieces of a jigsaw puzzle created by the aspect program and assembled by the weaver into the actual source code. AOP addresses explicitly code re-engineering, which in principle should allow to reduce considerably **maintenance costs**.
AOP is a relatively recent approach to software development. AOP can in principle address any application domain and can use a procedural, functional or object-oriented programming language as component language. The isolation and coding of aspects requires extra work and expertise that may be well paid back by the capability of addressing new aspects while keeping a single unmodified and general design.
For the time being it is not yet possible to tell whether AOP will spread out as a programming paradigm among academia and industry the way object-oriented programming has done since the Eighties. The many qualities of AOP are currently being quantitatively assessed, both with theoretical studies and with practical experience, and results seem encouraging. Furthermore, evidence of an increasing interest in AOP is given by the large number of research papers and conferences devoted to this interesting subject. From the point of view of the dependability aspect, one can observe that AOP exhibits optimal SC ("by construction", in a sense (Kiczales & Mezini, 2005)), and that recent results show that attribute A can in principle reach good values when making use of run-time weaving (Vasseur, 2004), often realized by dynamic bytecode manipulation. This work by Ostermann (Ostermann, 1999) is an interesting survey on this subject.
The adequacy at fulfilling attribute SA is indeed debatable also because, to date, no fault-tolerance aspect languages have been devised[1]—which may possibly be an interesting research domain. Positive values for SA have been questioned, at least in the case of fault-tolerance paradigms such as transactions, which appear to be difficult to be "aspectized" (Kienzle & Guerraou, 2002) especially because of the "very syntactic-only nature" of the manipulations supported by the approach. Possibly an evolution of aspects allowing semantic processing, a little like in Introspector (Chapter 4), could provide us with the ultimate tool to master the application-level complexity of fault-tolerance expression.

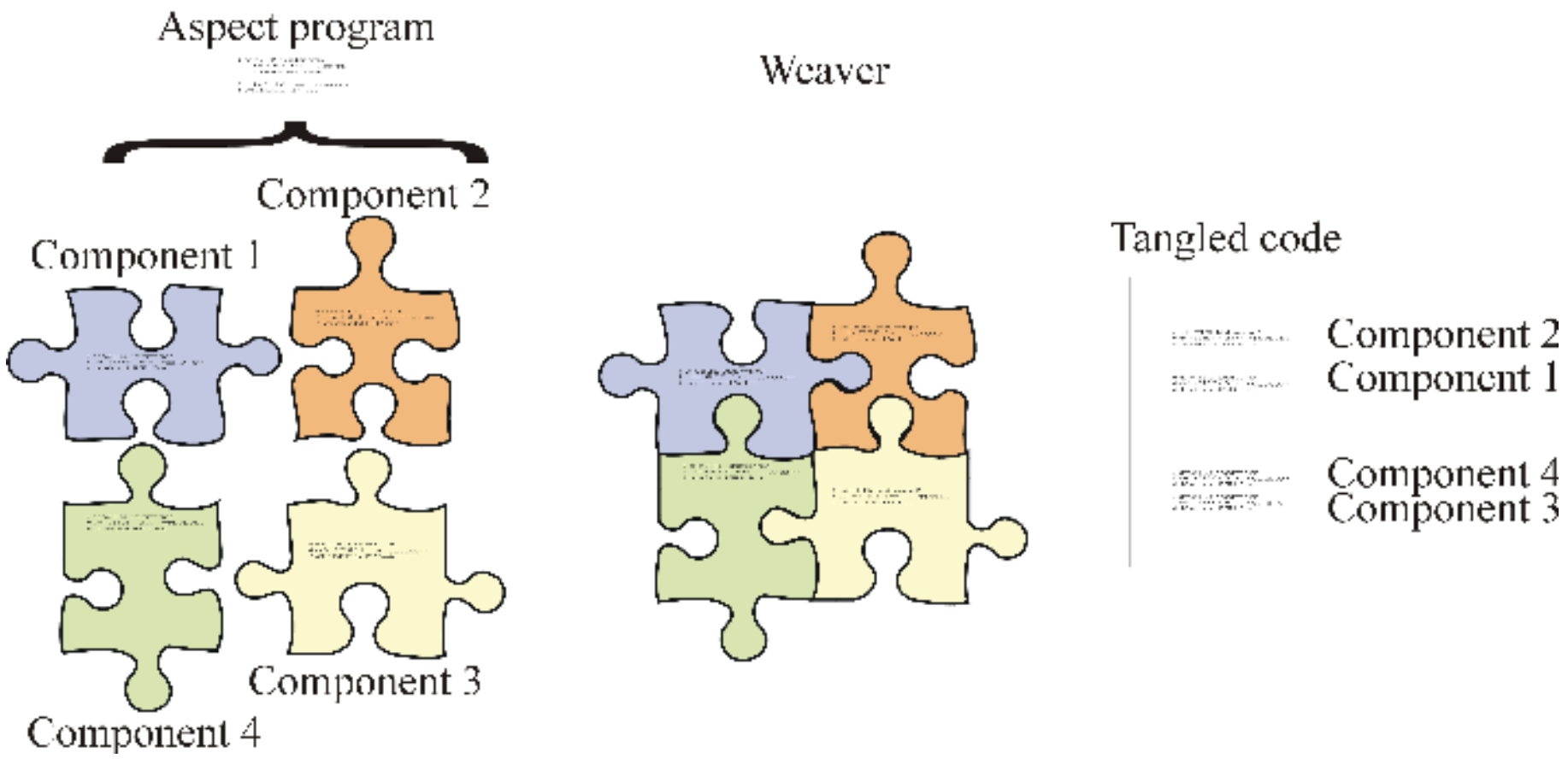

Figure 1: A fault-tolerant program according to an AOP system.

# Notes

[1]For instance, AspectJ only addresses exception error detection and handling. Remarkably enough, the authors of a study on AspectJ and its support to this field conclude (Lippert & Videira Lopes, 2000) that "whether the properties of AspectJ [documented in this paper] lead

to programs with fewer implementation errors and that can be changed easier, is still an open research topic that will require serious usability studies as AOP matures".

page

# FAILURE DETECTION PROTOCOLS IN THE APPLICATION LAYER

## 1 INTRODUCTION AND OBJECTIVES

Failure detection is a fundamental building block to develop fault-tolerant distributed systems. Accurate failure detection in asynchronous systems (Chapter 2) is notoriously difficult, as it is impossible to tell whether a process has actually failed or it is just slow. Because of this, several impossibility results have been derived—see for instance the well known paper (Fischer, Lynch, & Paterson, 1985). As a consequence of these pessimistic results, many researchers have devoted their time and abilities to understanding how to reformulate the concept of system model in a fine-grained alternative way. Their goal was being able to tackle problems such as distributed consensus with the minimal requirements on the system environment. This brought to the theory of unreliable failure detectors for reliable systems, pioneered by the works of Chandra and Toueg (Chandra & Toueg, 1996). This chapter introduces these concepts and the formulation of failure detection protocols in the application layer. In particular a linguistic framework is proposed for the expression of those protocols. As a case study it is described the algorithm for failure detection used in the EFTOS DIR net and in the TIRAN Backbone—that is, the fault-tolerance managers introduced respectively in Chapter 3 and Chapter 6.

## 2 FAILURE DETECTION PROTOCOLS IN THE APPLICATION LAYER

In Chapter 2 the concept of system model was briefly introduced together with the main features of the classical asynchronous and synchronous system models. The former model, also known as "time-free" system model, is the one that is implicitly used by most non real-time services: For such systems there is no bound for the time required to execute any computation or communication step, which means that there is no way to tell whether a certain part of the system is slow or if it has failed. How to distinguish between these two cases? The answer found by researchers is failure detectors. As cleverly expressed by Michel Raynal, failure detectors may be considered as a sort of distributed oracle for failure detection. This oracle observes the system and draws its conclusions about failures, informing those who query it. It could be regarded as a sort of middleware service for failure detection. Failure detectors are characterized by two properties:

- Completeness, which is the actual detection of failures, and
- Accuracy, which tells how reliable a failure detector can be in its assessments.

In a sense, completeness and accuracy represent two coordinates by means of which the spectrum of all possible failure detectors can be drawn. This spectrum represents also a two-dimensional set of possible system models, a set which is much more detailed and fine-grained than the linear one hitherto available to researchers:

> In other words, before failure detectors, the researchers had a sort of interval defined by its two extremes, the asynchronous model ("I ask nothing, so I get nothing") and the synchronous model ("I ask too much, so I can't get it"). Partial synchronous systems are points vaguely identified within that interval. With failure detectors everything changes and one can talk of system model $(c, a)$, where $(c, a)$ are the completeness and accuracy of the minimal failure detector $FD_{(c,a)}$ that can be implemented in a system obeying that model. This view has revolutionized the research on dependable distributed systems.

Several and very important have been the consequences of the introduction of failure detectors. Among them the following ones are highlighted herein:

- Famous impossibility results such as the impossibility to solve consensus in a fully asynchronous system (Fischer et al., 1985), which had puzzled the research community for years, have been now understood and tamed. The correct way to solve it is to ask which is the weakest of failure detectors the underlying asynchronous system has to be equipped with in order for the consensus problem to be solved. This in turns means to be able to identify which minimal set of additional mechanisms must be available in an asynchronous system in order implement that failure detector, thus making it able to support consensus. The same reasoning can be applied to any other distributed computing problem. Clearly this is a powerful conceptual and practical tool to reason about reliable systems.
- Another important consequence of introducing a theory of failure detectors is that a failure detector also reveals what are the minimal requirements necessary for a service to migrate from a system to another while keeping the desired quality of service. One of the key messages of this book is that it is important to distinguish between porting a code and porting the service that that code is meant to offer, and that the history of the relations between human beings and computers is paved of cases where erroneous software reuse has led to dreadful disasters. Failure detectors provide the designers with powerful "lens" through which the differences among systems and the consequences of migrations are put in the foreground. It is a pity that the awareness of the role of failure detectors in dependable software development has not been fully recognized and exploited yet in the ICT community at large.

| Construct | NFD-E | $\varphi$ | FD | GMFD | $\mathcal{D} \in \Diamond\mathcal{P}$ | $\mathcal{HB}$ | $\mathcal{HB}$-pt |
|---|---|---|---|---|---|---|---|
| Repeat periodically | no | no | yes | no | yes | yes | yes |
| Upon $t =$ current time | yes | no | yes | yes | no | no | no |
| Upon receive message | yes | yes | yes | yes | yes | yes | yes |
| Concurrency management | yes | yes | no | no | yes | yes | yes |

Table 1: Syntactical constructs used in several failure detector protocols. $\varphi$ is the accrual failure detector discussed in (Hayashibara, 2004; Hayashibara et al., 2004). $\mathcal{D}$ is the eventually perfect failure detector of (Chandra & Toueg, 1996). $\mathcal{HB}$ is the Heartbeat detector of (Aguilera et al., 1999). $\mathcal{HB}$-pt is the partition-tolerant version of the Heartbeat detector. By "Concurrency management" it is meant co-routines, threading or forking.

- Finally, from a theoretical point of view, failure detectors create a partial ordering among problems, which is also very important to better understand and compare the threats of dependability.

Failure detection protocols are often described by their authors making use of informal pseudo-codes of their conception. Often these pseudo-codes use syntactical constructs such as repeat periodically (Chandra & Toueg, 1996)(Aguilera, Chen, & Toueg, 1999)(Bertier, Marin, & Sens, 2002), at time $t$ send heartbeat (Chen, Toueg, & Aguilera, 2002; Bertier et al., 2002), at time $t$ check whether message has arrived (Chen et al., 2002), or upon receive (Aguilera et al., 1999), together with several variants (see Table 1). Such syntactical constructs are not often found in COTS programming languages such as C or C++, which leads us to the problem of translating the protocol specifications into running software prototypes using one such standard language. Furthermore the lack of a formal, well-defined, and standard form to express failure detection protocols often leads their authors to insufficiently detailed descriptions. Those informal descriptions in turn complicate the reading process and exacerbate the work of the implementors, which becomes time-consuming, error-prone and at times frustrating.

Several researchers and practitioners are currently arguing that failure detection should be made available as a network service (Hayashibara et al., 2004; Renesse, Minsky, & Hayden, 1998). No such service exists to date. Lacking such tool, it is important to devise methods to express in the application layer of our software even the most complex failure detection protocols in a simple and natural way.

In the following one such method is introduced, which is based on the class of time-outs (i.e., objects that postpone a certain function call by a given amount of time) that has been introduced in Chapter 6. As mentioned already, this feature allows convert time-based events into non time-based events such as message arrivals. It also allows easily express the constructs in Table 1 in standard C[1]. In some cases, our class allows get rid of concurrency management requirements such as co-routines or thread management libraries. The formal character of our method allows rapid-prototype the algorithms with minimal effort. This is proved through a Literate Programming (Knuth, 1984) framework that produces from a same source file both the description meant for publication and a software skeleton to be compiled in standard C or C++.

The rest of this chapter is structured as follows: Section 2.1 introduces our tool. In Sect. 2.1.2 our tool is put to work and used to express three classical failure detectors. Section 2.2 is a case study describing a software system built with our tool. Our conclusions are drawn in Sect. 2.3.

## 2.1 TOM: A Time-outs Management System

This section briefly describes the architecture of our time-out management system (TOM). The TOM class appears to the user as a couple of new types and a library of functions. Table 2 provides an idea of the client-side protocol of our tool.

To declare a time-out manager, the user needs to define a pointer to a `TOM` object and then call function `tom_init`. Argument to this function is an alarm, i.e., the function to be called when a time-out expires:

```
int alarm(TOM *);  tom = tom_init( alarm );
```

The first time function `tom_init` is called a custom thread is spawned. That thread is the actual time-out manager.

At this point the user is allowed to define his or her time-outs. This is done via type `timeout_t` and function `tom_declare`; an example follows:

```
timeout_t t;  tom_declare(&t,TOM_CYCLIC, TOM_SET_ENABLE, TID,
                    TSUBID, DEADLINE).
```

In what above, time-out `t` is declared as:

- A cyclic time-out (renewed on expiration; as opposed to `TOM_NON_CYCLIC`, which means "removed on expiration"),

```
1.  /* declarations */
    TOM *tom;
    timeout_t t1, t2, t3;
    int my_alarm(TOM*), another_alarm(TOM*);
2.  /* definitions */
    tom ← tom_init(my_alarm);
    tom_declare(&t1, TOM_CYCLIC, TOM_SET_ENABLE, TIMEOUT1, SUBID1, DEADLINE1);
    tom_declare(&t2, TOM_NON_CYCLIC, TOM_SET_ENABLE, TIMEOUT2, SUBID2, DEADLINE2);
    tom_declare(&t3, TOM_CYCLIC, TOM_SET_DISABLE, TIMEOUT3, SUBID3, DEADLINE3);
    tom_set_action(&t3, another_alarm);
3.  /* insertion */
    tom_insert(tom, &t1), tom_insert(tom, &t2), tom_insert(tom, &t3);
4.  /* control */
    tom_enable(tom, &t3);
    tom_set_deadline(&t2, NEW_DEADLINE2);
    tom_renew(tom, &t2);
    tom_delete(tom, &t1);
5.  /* deactivation */
    tom_close(tom);
```

Table 2: An example of usage of the TOM class. In **1.** a time-out list pointer and three time-out objects are declared, together with two alarm functions. In **2.** the time-out list and the time-outs are initialized, and an alarm differing from the default one is attached to time-out `t3`. Insertion is carried out at point **3.** At **4.**, some control operations are performed on the list, namely, time-out `t3` is enabled, a new deadline value is specified for time-out `t2` which is then renewed to activate the changing, and time-out `t1` is deleted. The whole list is finally deactivated in **5.**

- enabled (only enabled time-outs "fire", i.e., call their alarm on expiration; an alarm is disabled with `TOM_SET_DISABLE`),
- with a deadline of `DEADLINE` local clock ticks before expiration.

A time-out `t` is identified as a couple of integers, in the above example `TID` and `TSUBID`. This is done because in our experience it is often useful to distinguish instances of classes of time-outs. `TID` is used for the class identifier and `TSUBID` for the particular instance. A practical example of this is given in Sect. 2.2. Once defined, a time-out can be submitted to the time-out manager for insertion in its running list of time-outs—see (De Florio & Blondia, 2006) for further details on this. From the user point of view, this is managed by calling function

`tom_insert( TOM *, timeout_t * ).`

Note that a time-out might be submitted to more than one time-out manager.

After successful insertion an enabled time-out will trigger the call of the default alarm function after the specified deadline. If that time-out is declared as `TOM_CYCLIC` the time-out would then be re-inserted.
Other control functions are available: a time-out can be temporarily suspended while in the time-out list via function

```
tom_disable( TOM *, timeout_t * )
```

and (re-)enabled via function

```
tom_enable( TOM *, timeout_t * ).
```

Furthermore, the user is allowed to specify a new alarm function via `tom_set_action`) and a new deadline via `tom_set_deadline`; he or she can delete a time-out from the list via

```
tom_delete( TOM *, timeout_t * ),
```

and renew[2] it via

```
tom_renew( TOM *, timeout_t * ).
```

Finally, when the time-out management service is no longer needed, the user should call function

```
tom_close( TOM * ),
```

which possibly halts the time-out manager thread should no other client be still active.

### 2.1.1 Requirements

A fundamental requirement of our model is that processes must have access to some local physical clock giving them the ability to measure time. The availability of means to control the priorities of processes is also an important factor to reducing the chances of late alarm execution. Another assumption is that the alarm functions are small grained both in CPU and I/O usage so as to not interfere "too much" with the tasks of the TOM. Finally, asynchronous, non-blocking primitives to send and receive messages are assumed to be available.

### 2.1.2 Discussion

In this section it is shown that the syntactical constructs in Table 1 can be expressed in terms of our class of time-outs. This is done by considering three classical failure detectors and providing their time-out based specifications in the cweb Literate Programming framework (Knuth, 1984).
Let us consider the classical formulation of eventually perfect failure detector $\mathcal{D}$ (Chandra & Toueg, 1996). The main idea of the protocol is to require each task to send a "heartbeat" to its fellows and maintain a list of tasks suspected to have failed. A task identifier $q$ enters the list of task $p$ if no heartbeat is received by $p$ during a certain amount of time, $\Delta_p(q)$, initially set to a default value. This value is increased when late heartbeats are received.
The basic structure of $\mathcal{D}$ is that of a co-routine with three concurrent processes, two of which execute a task periodically while the third one is triggered by the arrival of a message:

*Every process p executes the following*:

$output_p \leftarrow 0$
**for** all $q \in \Pi$
    $\Delta_p(q) \leftarrow$ default time interval
**cobegin**
    —— *Task 1:* **repeat periodically**
        send "$p$-is-alive" to all

    —— *Task 2:* **repeat periodically**
        **for** all $q \in \Pi$
            **if** $q \notin output_p$ and $p$ did not receive "$q$-is-alive" during
                the last $\Delta_p(q)$ ticks of $p$'s clock **then**
                    $output_p \leftarrow output_p \cup \{q\}$

    —— *Task 3:* **when** receive "q-is-alive" for some $q$
        **if** $q \in output_p$
            $output_p \leftarrow output_p - \{q\}$
            $\Delta_p(q) \leftarrow \Delta_p(q) + 1$
**coend**.

The **repeat periodically** in *Task 1* is called a "multiplicity 1" repeat, because indeed a single action (sending a "$p$-is-alive" message) has to be tracked, while the one in *Task 2* is called a "multiplicity $q$" repeat, which requires to check $q$ events.
Our reformulation of the above code is as follows:

*Every process p executes the following:*

```
timeout_t t_task1, t_task2[NPROCS];
task_t p, q;
for (q=0; q<NPROCS; q++) {
       Δ_p[q] = DEFAULT_TIMEOUT;
       output_p[q] = TRUST;
}

/* ``⇝'' is our symbol for the ``address-of'' operator */
tom_declare(⇝t_task1, TOM_CYCLIC,
             TOM_SET_ENABLE, p, 0, Δ_p[q]);
tom_set_action(⇝t_task1, action_Repeat_Task1);
tom_insert(⇝t_task1);

for (q=0; q<NPROCS; q++) {
       if (p ≠ q) {
       tom_declare(t_task2+q, TOM_CYCLIC,
                    TOM_SET_ENABLE, q, 0, Δ_p[q]);
       tom_set_action(t_task2+q, action_Repeat_Task2);
       tom_insert(⇝t_task1);
}

getMessage(⇝m);
switch (m.type)  {
       TASK1;
       TASK2;
       TASK3;
}
```

where tasks and actions are defined as follows:

```
TASK1 ≡ case REPEAT_TASK1:
                sendAll(I_AM_ALIVE);
        break;
TASK2 ≡ case REPEAT_TASK2:
                q = m.id;
                if (output_p[q] ≡ TRUST)
                        output_p[q] = SUSPECT;
        break;
TASK3 ≡ case I_AM_ALIVE:
                q = m.sender;
                if (output_p[q] ≡ SUSPECT)   {
                        output_p[q] = TRUST;
                        Δ_p(q) = Δ_p(q) + 1;
                }
        break;

action_Repeat_Task1() {
        message_t m;
        m.type = REPEAT_TASK1;
        Send(m, p);
}

action_Repeat_Task2(timeout_t *t) {
        message_t m;
        m.type = REPEAT_TASK2;
        m.id = t->id;
        Send(m, p);
}
```

The following observations can be drawn:

- Our syntax is less abstract than the one adopted in the classical formulation. Indeed it was deliberately chosen a syntax very similar to that of programming languages like C and its derivatives. Behind the lines, a similar semantics is also assumed.
- Our syntax is more strongly typed: We have deliberately chosen to define (most of) the objects our code deals with.
- Set-wise operations such as union, complement or membership have been systematically avoided by transforming sets into arrays as, e.g., in

$$output_p \leftarrow output_p \mathbf{U} \{q\},$$

which was changed into

$$output_p[q] = \texttt{PRESENT}.$$

- The abstract constructs `repeat periodically` have been systematically rewritten as one or more time-outs (depending on their multiplicity). Each of these time-out has an associated action that sends one message to the client process, $p$. This means that

  1. time-related event "it's time to send $p$-is-alive to all" becomes event "message `REPEAT_TASK1` has arrived."
  2. time-related events "it's time to check whether $q$-is-alive has arrived" becomes event "message (`REPEAT_TASK2`, id=$q$) has arrived."

- Due to the now homogeneous nature of the possible events (message arrivals) a single process may manage those events through a multiple selection statement (a switch). The requirement for a co-routine has been removed.

Through the Literate Programming approach and a compliant tool such as cweb (Knuth & Levy, 1993; Knuth, 1984) it is possible to further improve our reformulation. As well known, the cweb tool allows have a single source code to produce a pretty printable TEX documentation and a C file ready for compilation and testing. In our experience this link between these two contexts can be very beneficial: testing or even simply using the code provides feedback on the specification of the algorithm, while the improved specification may reduce the probability of design faults and in general increase the quality of the code.
Figure 1 and Figure 2 respectively show a reformulation for the $\mathcal{HB}$ failure detector for partitionable networks (Aguilera et al., 1999) and for the group membership failure detector (Raynal & Tronel, 1999) produced with cweb. A full description of these protocols is out of the scope of this book—for that the reader is referred to the above cited articles. The focus here is mainly on the syntactical constructs used in them and our reformulations, which include simple translations for the syntactical constructs in Table 1 in terms of our time-out API. A case worth noting is that of the group membership failure detector: Here the authors mimic the availability of a cyclic time-out service but intrude its management in their formulation—which could be avoided using our approach.

## 2.2 A Development Experience: The DIR net

As mentioned in Chapter 3 and Chapter 6, at the core of the software fault tolerance strategies of several European projects (Deconinck, De Florio, Varvarigou, Verentziotis, & Botti, 2002; Botti et al., 2000; Deconinck, De Florio, Dondossola, & Szanto, 2003) there is a distributed application

**1. Code of the $\mathcal{HB}$ failure detector for partitionable networks.**
Aguilera, Chen and Toueg, Theoretical Computer Science n.1, 1999.

```
#define HEARTBEAT 1
#define ITTB 2
#define SOME_PERIOD 100000
#define FOREVER 1
#define PRESENT 1
#define ABSENT 0
  ⟨Initialisation 2⟩
```

**2.** Every process $p$ executes the following:

```
⟨Initialisation 2⟩ ≡
  main()
  {
    timeout_t τ_t2b;
    message_t m;
    for (q = 0; q < NPROCS; q++) {
      D_p[q] = 0;
      path[q] = ABSENT;
    }
    tom_declare(&τ_t2b, TOM_CYCLIC, TOM_SET_ENABLE, 1, 1, 1);
    tom_set_action(&τ_t2b, actionItsTimeToBroadcast);     /* sends ITTB */
    tom_set_deadline(&τ_t2b, SOME_PERIOD);     /* every 100000 ticks */
    tom_insert(&nextb);
    do {
      getMessage(&m);     /* sets m.date */
      switch (m.type) {
        ⟨Task1 3⟩
        ⟨Task2 4⟩
      }
    } while (FOREVER);
  }
```

This code is used in section 1.

**3.** Task 1

```
⟨Task1 3⟩ ≡
case ITTB: D_p[p] = D_p[p] + 1;
  m.type = HEARTBEAT, m.path = p;
  for (q = 0; q < NPROCS; q++)
    if (isneighbor(q, p)) sendMessage(m, q);
  break;
```

This code is used in section 2.

**4.** Code of Task 2

```
⟨Task2 4⟩ ≡
case HEARTBEAT:
  for (q = 0; q < NPROCS; q++)
    if (m.path[q] ≠ ABSENT) D_p[p] = D_p[p] + 1;
  m.path[p] = m.path[p] + 1;
  for (q = 0; q < NPROCS; q++)
    if (isneighbor(q, p) ∧ m.path[q] ≤ PRESENT) sendMessage(m, q);
  break;
```

This code is used in section 2.

**5.** Extra functions

```
  int actionItsTimeToBroadcast()     /* sends ITTB to caller */
  {
    sendMessage(ITTB, p);
  }
```

Figure 1: Reformulation of the $\mathcal{HB}$ failure detector for partitionable networks.Special symbols such as $\tau$ and $\mathcal{D}_p$ are caught by cweb and translated into legal C tokens via its "@f" construct.The expression $m.path[q]$ $\leq$`PRESENT` means "$q$ appears at most once in *path*".

**1. Code of a group membership failure detector.**
Raynal and Tronel, Distributed Systems Engineering **6** (1999) 95–102.

```
#define ITTS 1
#define ITTB 2
#define I_AM_ALIVE 3
  ⟨Initialisation 2⟩
```

**2.** Every process p executes the following:

```
⟨Initialisation 2⟩ ≡
  main()
  {
    timeout_t τ_nextt, τ_bcast;
    date_t nextTimeout_i, timeout_i[NPROCS], nextBroadcast_i;
    boolean_t groupFailure_i;
    message_t m;
    task_t j;
    groupFailure_i = False;
    nextBroadcast_i = getCurrentDate();
    for (q = 0; q < NPROCS; q++) {
      timeout_i[q] = MAX_DATE;
      r_i[q] = 0;
    }
    nextTimeout_i = min(timeout_i, NPROCS);
    tom_set_action(&τ_nextt, actionItsTimeToStop);        /* sends message ITTS */
    tom_set_deadline(&τ_bcast, T_r);
    tom_insert(&τ_nextt);
    tom_set_action(&τ_bcast, actionItsTimeToBroadcast);   /* sends message ITTB */
    tom_set_deadline(&τ_bcast, T_e);
    tom_insert(&τ_bcast);
    while (¬groupFailure_i) {
      getMessage(&m);      /* sets m.date */
      j = m.sender;
      switch (m.type) {
        ⟨Task1 3⟩
        ⟨Task2 4⟩
        ⟨Task3 5⟩
      }
    }
  }
This code is used in section 1.
```

**3.** Task 1

```
⟨Task1 3⟩ ≡
case ITTB: sendMessageAll(I_AM_ALIVE, b_i);    /* send "i is alive" to all */
  tom_insert(&τ_bcast);     /* a cyclic timeout could have been used here */
  b_i = b_i + 1;
  break;
This code is used in section 2.
```

**4.** Code of Task 2

```
⟨Task2 4⟩ ≡
case ITTS: groupFailure_i = True;
  break;
This code is used in section 2.
```

**5.** Code of Task 3

```
⟨Task3 5⟩ ≡
case I_AM_ALIVE: timeout_i[j] = m.date + T_r;
  nextTimeout_i = min(timeout_i, NPROCS);
  tom_set_deadline(&τ_nextt, nextTimeout_i);
  tom_insert(&τ_nextt);
  r_i[j] = r_i[j] + 1;
  break;
This code is used in section 2.
```

**6.** Ancillary functions.

```
  int actionItsTimeToStop()      /* sends message ITTS */
  {
    sendMessage(ITTS, i);
  }
  int actionItsTimeToBroadcast()      /* sends message ITTB */
  {
    sendMessage(ITTB, i);
  }
```

Figure 2: Reformulation of the group membership failure detector.

called "DIR net" (De Florio, 1998) or "Backbone". In this section that application is described (let us call it just as the DIR net) and it is reported on how it was designed and developed by means of the TOM system.

The DIR net has been described as a fault-tolerant network of failure detectors connected to other peripheral error detectors (called Dtools in what follows). Objective of the DIR net is to ensure consistent fault tolerance strategies throughout the system and play the role of a backbone handling information to and from the Dtools (De Florio, Deconinck, & Lauwereins, 2000).

The DIR net consists of four classes of components. Each processing node in the system runs an instance of a so-called I-am-alive Task (IAT) plus an instance of either a DIR Manager (DIR-$\mathcal{M}$), or a DIR Agent (DIR-$\mathcal{A}$), or a DIR Backup Agent (DIR-$\mathcal{B}$). A DIR-$\mathcal{A}$ gathers all error detection messages produced by the Dtools on the current processing node and forwards them to the DIR-$\mathcal{M}$ and the DIR-$\mathcal{B}$'s. A DIR-$\mathcal{B}$ is a DIR-$\mathcal{A}$ which also maintains its messages into a database located in central memory. It is connected to DIR-$\mathcal{M}$ and to the other DIR-$\mathcal{B}$'s and is eligible for election as a DIR-$\mathcal{M}$. A DIR-$\mathcal{M}$ is a special case of DIR-$\mathcal{B}$. Unique within the system, the DIR-$\mathcal{M}$ is the one component responsible for running error recovery strategies—see (De Florio, 2000; De Florio & Deconinck, 2002) for a description of the latter. Let us use DIR-$x$ to mean "the DIR-$\mathcal{M}$ or a DIR-$\mathcal{B}$ or a DIR-$\mathcal{A}$."

An important design goal of the DIR net is that of being fault-tolerant. This is accomplished also through a failure detection protocol that will be described

| Time-out | Caller | Action | Cyclic? |
|---|---|---|---|
| $t_{\text{IA_SET}}$ | DIR-$x$ | On TimeNow + $d_{\text{IA_SET}}$ do send $m_{\text{IA_SET_ALARM}}$ to Caller | Yes |
| $t_{\text{IA_CLR}}$ | IAT | On TimeNow + $d_{\text{IA_CLR}}$ do send $m_{\text{IA_CLR_ALARM}}$ to IAT | Yes |

Table 3: Description of time-outs $t_{\text{IA_SET}}$ and $t_{\text{IA_CLR}}$.

| Message | Receiver | Explanation | Action |
|---|---|---|---|
| $m_{\text{IA_SET_ALARM}}$ | DIR-$x$ | Time to set IAF | IAF ← `TRUE` |
| $m_{\text{IA_CLR_ALARM}}$ | IAT $k$ | Time to check IAF | **if** (IAF ≡ `FALSE`) SendAll($m_{\text{TEIF}}$, $k$) **else** IAF ← `FALSE`, |

Table 4: Description of messages $m_{\text{IA_SET_ALARM}}$ and $m_{\text{IA_CLR_ALARM}}$.

in the rest of this section.

#### 2.2.1 The DIR net failure detection protocol

Our protocol consists of a local part and a distributed part. Each of them is realized through our TOM class.

**DIR net protocol: local component.** As already mentioned, each processing node hosts a DIR-$x$ and an IAT. These two components run a simple algorithm: they share a local Boolean variable, the I'm Alive Flag (IAF). The DIR-$x$ has to set periodically the IAF to `TRUE` while the IAT periodically has to check that this has indeed occurred and reverts IAF to `FALSE`. If the IAT finds the IAF set to `FALSE` it broadcasts message $m_{\text{TEIF}}$ ("this entity is faulty").
The just mentioned cyclic tasks can be easily modeled via two time-outs, $t_{\text{IA_SET}}$ and $t_{\text{IA_CLR}}$, described in Table 3 and Table 4 (TimeNow being the system function returning the current value of the clock register).
Note that the time-outs' alarm functions do not clear/set the flag—doing so a hung DIR-$x$ would go undetected. On the contrary, these functions trigger the transmission of messages that once received by healthy components trigger the execution of the meant actions.
The following is a pseudo-code for the IAT algorithm:

```
The IAT k executes as follows:

timeout_t t_IA_CLR;
msg_t activationMessage, m;

tom_declare(~>t_IA_CLR, TOM_CYCLIC,
            TOM_SET_ENABLE, IAT_CLEAR_TIMEOUT, 0, d_IA_CLR);
tom_set_action(~>t_IA_CLR, actionSend m_IA_CLR_ALARM);
tom_insert(~>t_IA_CLR);
```

```
Receive(activationMessage);

forever {
      Receive(m);
      if (m.type ≡ m_IA_CLR_ALARM)
            if (IAF ≡ TRUE) IAF ← FALSE;
            else SendAll(m_TEIF, k); delete_timeout(⇝t_IA_CLR);
}

actionSend m_IA_CLR_ALARM() { Send(m_IA_CLR_ALARM, IAT k); }
```

The time-out formulation of the DIR-$x$ is given in next section.

### 2.2.2 DIR net protocol: distributed component

The resilience of the DIR net to crash faults comes from the DIR-$\mathcal{M}$ and the DIR-$\mathcal{B}$'s running a distributed algorithm of failure detection.

**Algorithm DIR-$\mathcal{M}$.** Let us call `mid` the node hosting the DIR-$\mathcal{M}$ and $b$ the number of processing nodes that host a DIR-$\mathcal{B}$. The DIR-$\mathcal{M}$ has to send cyclically a $m_{\text{MIA}}$ ("Manager-Is-Alive") message to all the DIR-$\mathcal{B}$'s each time time-out $t_{\text{MIA_A}}$ expires—this is shown in the right side of Fig. 3. Obviously this is a multiplicity $b$ "repeat" construct, which can be easily managed through a cyclic time-out with an action that signals that a new cycle has begun. In this case the action is "send a message of type $m_{\text{MIA_A_ALARM}}$ to the DIR-$\mathcal{M}$."
The manager also expects periodically a $(m_{\text{TAIA}}, i)$ message ("This-Agent-Is-Alive") from each node where a DIR-$\mathcal{B}$ is expected to be running. This is easily accomplished through a vector of $(t_{\text{TAIA_A}}, \vec{\imath})$ time-outs. The left part of Fig. 3 shows this for node $i$. When time-out $(t_{\text{TAIA_A}}, p)$ expires it means that no $(m_{\text{TAIA}}, p)$ message has been received within the current period. In this case the DIR-$\mathcal{M}$ enters a so called "suspicion period". During such period the manager has to try to tell a late DIR-$\mathcal{B}$ from a crashed one. This is done by inserting a non-cyclic time-out, namely $(t_{\text{TEIF_A}}, p)$.
During the suspicion period only one out of the following three events may take place:

1. A late $(m_{\text{TAIA}}, p)$ is received.
2. A $(m_{\text{TEIF}}, p)$ from IAT at node $p$ is received.
3. Nothing comes in and the time-out expires.

In case 1, one gets out of the suspicion period, concludes that DIR-$\mathcal{B}$ at node $p$ was simply late and goes back waiting for the next $(m_{\text{TAIA}}, p)$. A wrong deduction at this point is possible and will be detected in one of next cycles. Adjustments of the deadlines are possible but not dealt with here for the sake of simplicity of description.

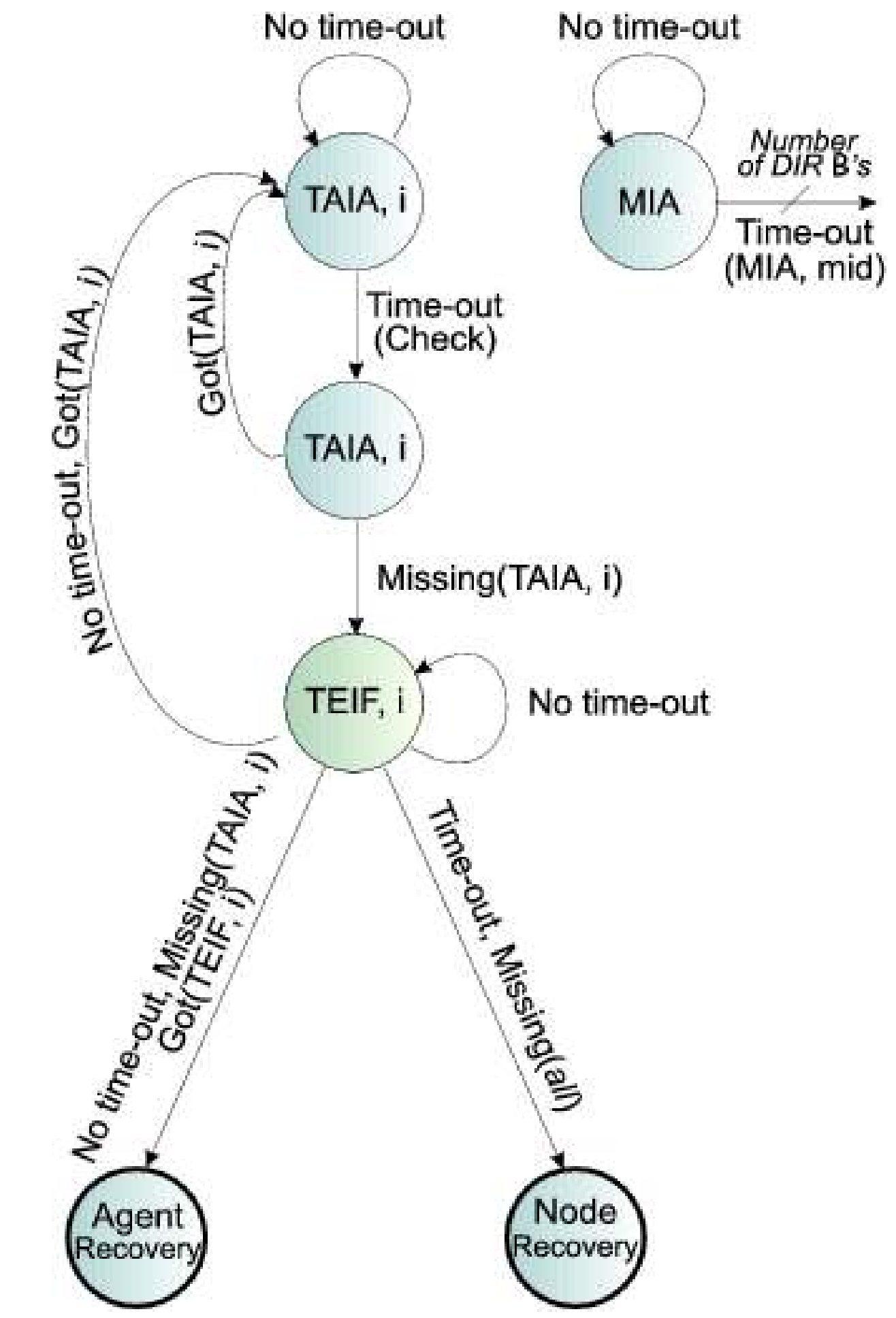

Figure 3: The algorithm of the DIR-$\mathcal{M}$.

If 2 is the case the remote component is assumed to have crashed though its node is still working properly, as the IAT on that node still gives signs of life. Therefore an error recovery step is initiated. This includes sending a "`WAKEUP`" message to the remote IAT so that it spawns another DIR-$\mathcal{B}$ on that node.
In case 3 the entire node is assumed to have crashed and node recovery is initiated.
Underlying assumption of our algorithm is that the IAT is so simple that if it fails then the whole node can be safely assumed to have failed.

**Algorithm DIR-$\mathcal{B}$.** This algorithm is also divided into two concurrent tasks. In the first one DIR-$\mathcal{B}$ on node $i$ has to cyclically send $(m_{\text{TAIA}}, i)$ messages to the manager, either in piggybacking or on expiring of time-out $t_{\mathrm{TAIA_B}}$. This is represented in the right side of Fig. 4.
The DIR-$\mathcal{B}$'s in turn periodically expect a $m_{\mathrm{MIA}}$ message from the DIR-$\mathcal{M}$. As evident when comparing Fig. 3 with Fig. 4, the DIR-$\mathcal{B}$ algorithm is very similar to the one of the manager: also DIR-$\mathcal{B}$ enters a suspicion period when its manager does not appear to respond quickly enough—this period is managed via time-out $t_{\mathrm{TEIF_B}}$, the same way as in DIR-$\mathcal{M}$. Also in this case one can get out of this state in one out of three possible ways: either

1. a late $(m_{\mathrm{MIA_B_ALARM}}, \mathrm{mid})$ is received, or
2. a $(m_{\mathrm{TEIF}}, \mathrm{mid})$ sent by the IAT at node mid is received, or
3. nothing comes in and the time-out expires.

In case 1 one gets out of the suspicion period, concludes that manager was simply late, goes back to normal state and starts waiting for the next $(m_{\mathrm{MIA}}, \mathrm{mid})$ message. Also in this case, a wrong deduction shall be detected in next cycles. If 2, one concludes that the manager has crashed though its node is still working properly, as its IAT acted as expected. Therefore a manager recovery phase is initiated similarly to the DIR-$\mathcal{B}$ recovery step described in Sect. 2.2.2. In case 3 the node of the manager is assumed to have crashed, elect a new manager among the DIR-$\mathcal{B}$'s, and perform a node recovery phase.
Table 5 summarizes the DIR-$\mathcal{M}$ and DIR-$\mathcal{B}$ algorithms.

| **Time-out** | **Caller** | **Action** | **Cyclic?** |
|---|---|---|---|
| $t_{\mathrm{MIA_A}}$ | DIR-$\mathcal{M}$ | Every $d_{\mathrm{MIA_A}}$ do send $m_{\mathrm{MIA_A_ALARM}}$ to DIR-$\mathcal{M}$ | Yes |
| $t_{\mathrm{TAIA_A}}[i]$ | DIR-$\mathcal{M}$ | Every $d_{\mathrm{TAIA_A}}$ do send $(m_{\mathrm{TAIA_A_ALARM}}, i)$ to DIR-$\mathcal{M}$ | Yes |
| $t_{\mathrm{TEIF_A}}[i]$ | DIR-$\mathcal{M}$ | On TimeNow + $d_{\mathrm{TEIF_A}}$ do send $(m_{\mathrm{TEIF_A_ALARM}}, i)$ to DIR-$\mathcal{M}$ | No |
| $t_{\mathrm{TAIA_B}}$ | DIR-$\mathcal{B}$ $j$ | Every $d_{\mathrm{TAIA_B}}$ do send $m_{\mathrm{TAIA_B_ALARM}}$ to DIR-$\mathcal{B}$ $j$ | Yes |
| $t_{\mathrm{MIA_B}}$ | DIR-$\mathcal{B}$ $j$ | Every $d_{\mathrm{MIA_B}}$ do send $m_{\mathrm{MIA_B_ALARM}}$ to DIR-$\mathcal{B}$ $j$ | Yes |
| $t_{\mathrm{TEIF_B}}$ | DIR-$\mathcal{B}$ $j$ | On TimeNow + $d_{\mathrm{TEIF_B}}$ do send $m_{\mathrm{TEIF_B_ALARM}}$ to DIR-$\mathcal{B}$ $j$ | No |

| **Message** | **Receiver** | **Explanation** | **Action** |
|---|---|---|---|
| $(m_{\mathrm{TAIA}}, i)$ | DIR-$\mathcal{M}$ | DIR-$\mathcal{B}$ $i$ is OK | (Re-)Insert or renew $t_{\mathrm{TAIA_A}}[i]$ |
| $m_{\mathrm{MIA_A_ALARM}}$ | DIR-$\mathcal{M}$ | A new heartbeat is required | Send $m_{\mathrm{MIA}}$ to all DIR-$\mathcal{B}$'s |
| $m_{\mathrm{TAIA_A_ALARM}}$ | DIR-$\mathcal{M}$ | Possibly DIR-$\mathcal{B}$ $i$ is not OK | Delete $t_{\mathrm{TAIA_A}}[i]$, insert $t_{\mathrm{TEIF_A}}[i]$ |
| $(m_{\mathrm{TEIF}}, i)$ | DIR-$\mathcal{M}$ | DIR-$\mathcal{B}$ $i$ crashed | Declare DIR-$\mathcal{B}$ $i$ crashed |
| $(m_{\mathrm{TEIF_A_ALARM}}, i)$ | DIR-$\mathcal{M}$ | Node $i$ crashed | Declare node $i$ crashed |
| $m_{\mathrm{MIA}}$ | DIR-$\mathcal{B}$ $j$ | DIR-$\mathcal{M}$ is OK | Renew $t_{\mathrm{MIA_B}}$ |
| $m_{\mathrm{TAIA_B_ALARM}}$ | DIR-$\mathcal{B}$ $j$ | A new heartbeat is required | Send $(m_{\mathrm{TAIA}}, j)$ to DIR-$\mathcal{M}$ |
| $m_{\mathrm{MIA_B_ALARM}}$ | DIR-$\mathcal{B}$ $j$ | Possibly DIR-$\mathcal{M}$ is not OK | Delete $t_{\mathrm{MIA_B}}$, insert $t_{\mathrm{TEIF_B}}$ |
| $m_{\mathrm{TEIF}}$ | DIR-$\mathcal{B}$ $j$ | DIR-$\mathcal{M}$ crashed | Declare DIR-$\mathcal{M}$ crashed |
| $m_{\mathrm{TEIF_B_ALARM}}$ | DIR-$\mathcal{B}$ $j$ | DIR-$\mathcal{M}$'s node crashed | Declare DIR-$\mathcal{M}$'s node crashed |

Table 5: Time-outs and messages of DIR-$\mathcal{M}$ and DIR-$\mathcal{B}$.

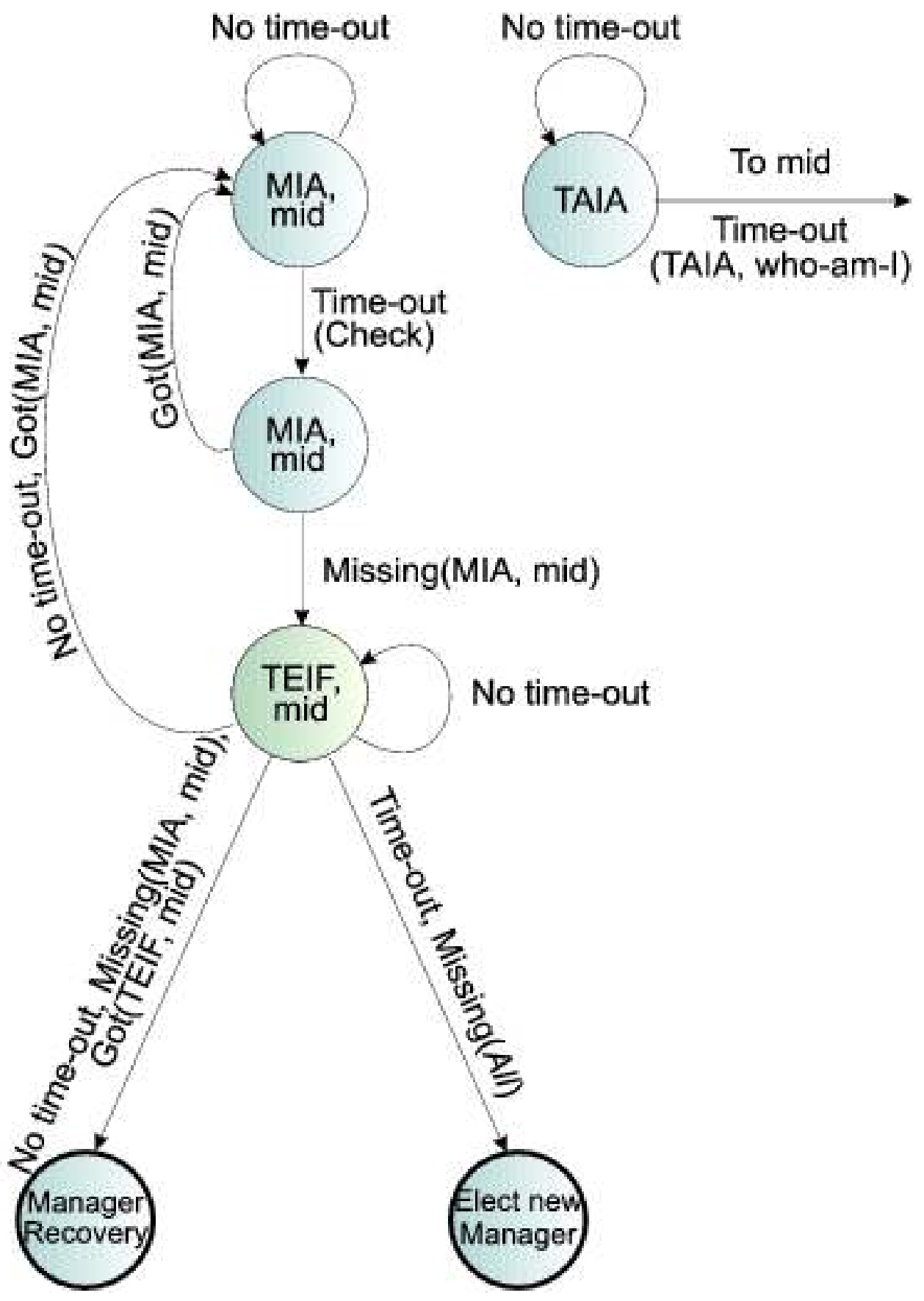

Figure 4: Algorithm DIR-$\mathcal{B}$.

The DIR net was developed using the Windows TIRAN libraries (Botti et al., 2000) and the cweb system of structured documentation. Figures 5–8 show the DIR net at work in a system consisting of just a DIR-$\mathcal{M}$ and a DIR-$\mathcal{B}$. Two ancillary processes called SocketServers are used to manage UDP communication.

#### 2.2.3 Special services

**Configuration.** The management of a large number of time-outs may be an error prone task. To simplify it, the configuration language of Chapter 6 was used. Figure 9 shows an example of configuration script to specify the structure of the DIR net (in this case, a four node system with three DIR-$\mathcal{B}$'s deployed on nodes 1–3 and the DIR-$\mathcal{M}$ on node 0) and of its time-outs. A translator produces the C header files to properly initialize an instance of the DIR net.

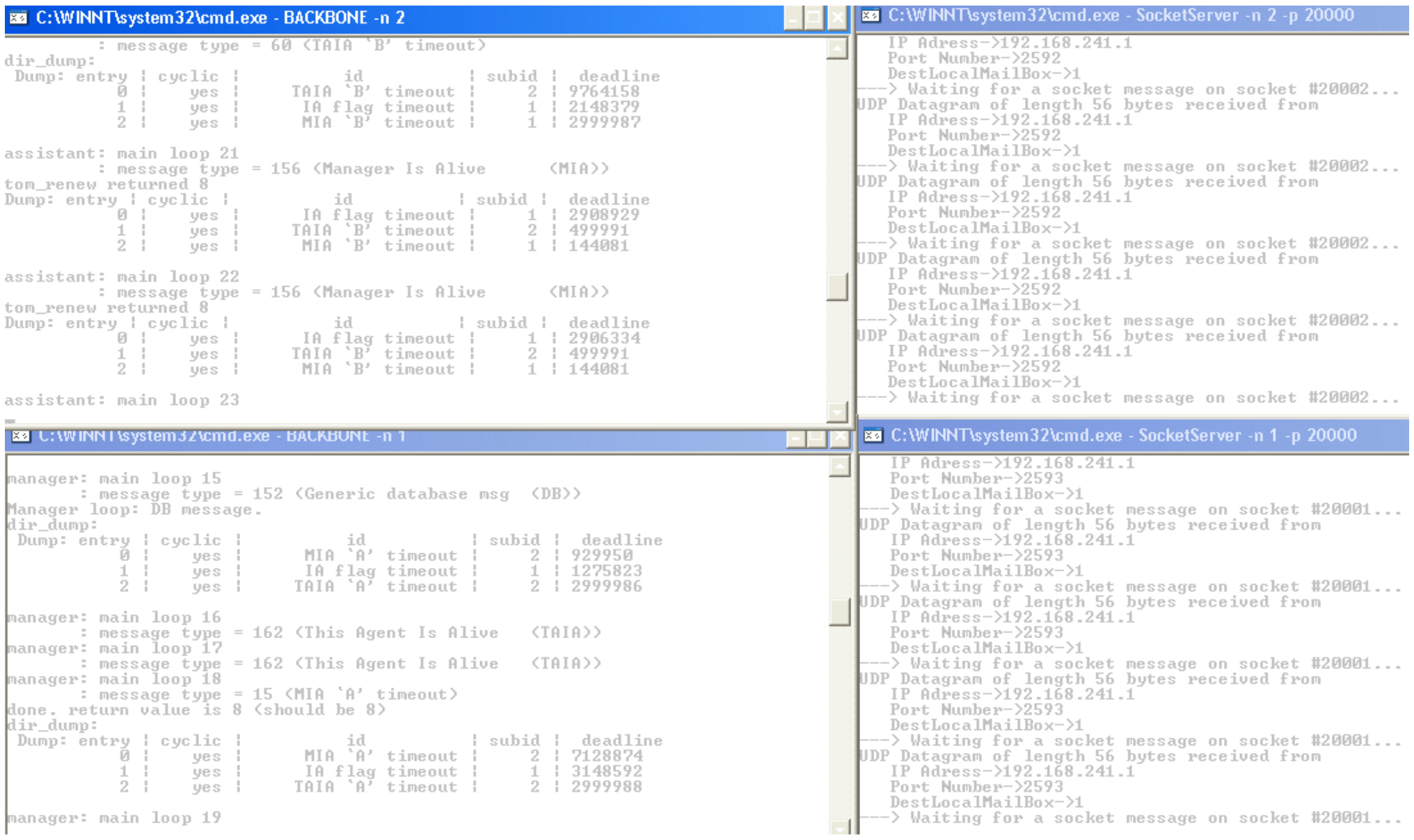

Figure 5: A DIR net consisting of a DIR-$\mathcal{B}$ (top left picture) and a DIR-$\mathcal{M}$ (bottom left). Pictures on the right portray the "SocketServers", i.e. ancillary processes managing communication.

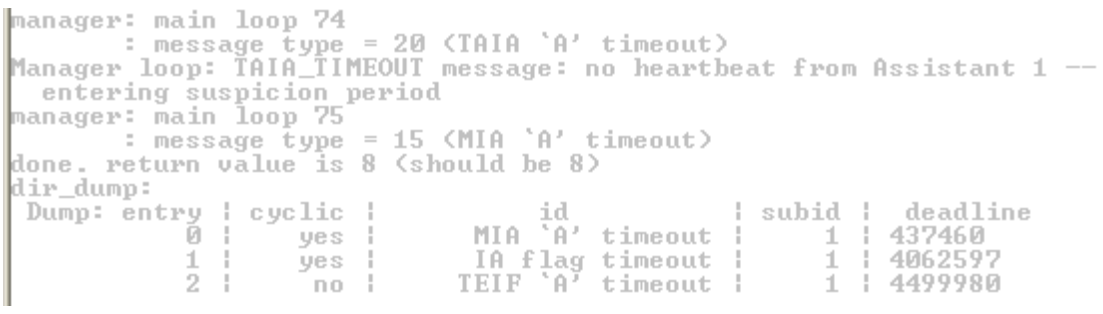

Figure 6: Time-out $t_{\text{TAIA_A}}$ expires, meaning that expected message $m_{\text{TAIA}}$ has not been received within the corresponding deadline $d_{\text{TAIA_A}}$. As a result, $t_{\text{TEIF_A}}$ is inserted. This corresponds to action "At time TimeNow + $d_{\text{TEIF_A}}$ check whether $m_{\text{TEIF}}$ has arrived".

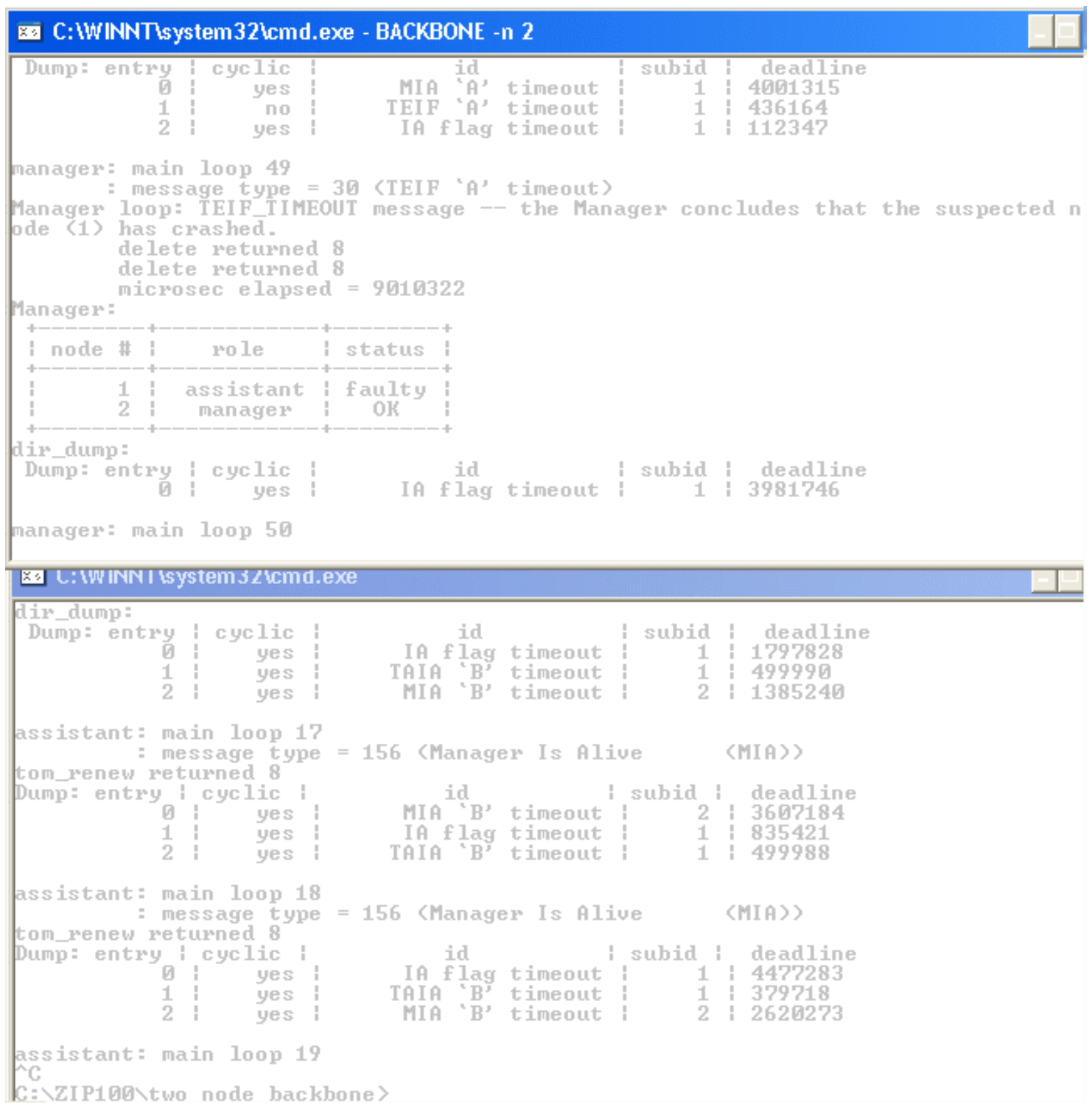

Figure 7: A DIR-$\mathcal{B}$ is stopped. Time-out $t_{\text{TEIF_A}}$ expires and as a result, the assistant is tagged as "faulty".

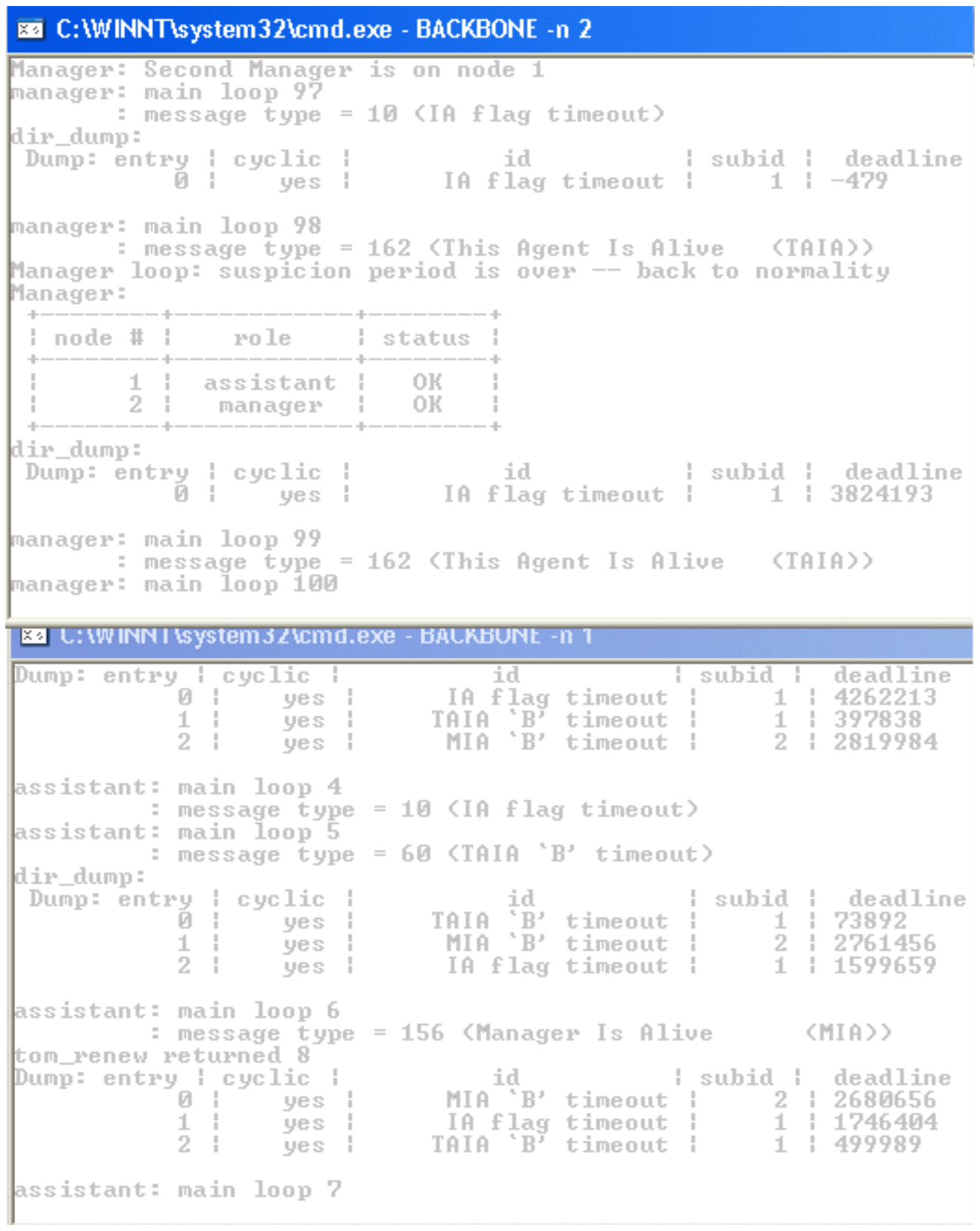

Figure 8: The DIR-$\mathcal{B}$ is restarted. As a result, it begins sending $m_{\text{TAIA}}$ messages. Upon receiving, the manager tags the assistant as "OK."

```
# include files
#       defines are importable from include files via #include statements
INCLUDE "my_definitions.h"
INCLUDE "../BACKBONE.H"
# definitions
#       definitions start with the 'DEFINE' keyword, followed
#       by an integer, an interval, or a list, followed
#       by the equal sign and a role, that may be
#       ASSISTANTs or MANAGER
NPROCS = 4
DEFINE 2-4 = ASSISTANTS
DEFINE 1 = MANAGER

# NPROCS = 2
# DEFINE 2 = ASSISTANT

MIA_SEND_TIMEOUT = 800000   # Manager Is Alive -- manager side
TAIA_RECV_TIMEOUT = 1800000  # This Agent Is Alive timeout -- manager side

MIA_RECV_TIMEOUT = 1500000   # Manager Is Alive -- backup side
TAIA_SEND_TIMEOUT = 1000000  # This Agent Is Alive timeout -- backup side

TEIF_TIMEOUT     = 1800000   # after this time a suspected node is assumed
                             # to have crashed.

I'M ALIVE_CLEAR_TIMEOUT = 900000 # I'm Alive timeout -- clear IA flag
I'M ALIVE_SET_TIMEOUT = 1400000  # I'm Alive timeout -- set and checks IA flag

REQUEST_DB_TIMEOUT = 2000000
REPLY_DB_TIMEOUT = 4000000

MID_TIMEOUT = 1000000 # if a TEIF is receaved, up to MID_TIMEOUT ticks
                      # are allowed for reintegrating a new manager,
                      # otherwise, the node of the manager is considered
                      # to be dead.
```

Figure 9: An excerpt from the configuration script of the DIR net.

#### 2.2.4 Fault injection

Time-outs may also be used to specify fault injection actions with fixed or pseudo-random deadlines. In the DIR net this is done as follows. First the time-out is defined:

```
#ifdef INJECT
        tom_declare(&inject, TOM_NON_CYCLIC, TOM_SET_ENABLE,
                    INJECT_FAULT_TIMEOUT, i, INJECT_FAULT_DEADLINE);
        tom_insert(tom, &inject);
#endif
```

The alarm of this time-out sends the local DIR-$x$ a message of type "INJECT_FAULT_TIMEOUT". Figure 10 shows an excerpt from the actual main loop of the DIR-$\mathcal{M}$ in which this message is dealt with.

**Fault tolerance.** A service such as TOM is indeed a single-point-of-failure in that a failed TOM brings the local DIR net components to the impossibility to perform their failure detection protocols. Such a case would be indistinguishable from that of a crashed node by the other DIR net components. As well known from, e.g., (Inquiry, 1996), a single design fault in TOM's implementation could bring the system to a global failure. Nevertheless, the isolation of a *service* for time-out management may pave the **18.** This loop is the real core of the manager. It has to deal with a number of messages coming from the timeout manager, its fellow backups, the recovery thread, the remote I'm Alive Tasks. The core of the fault-tolerant strategy of the DIR net is in here.

```
⟨manager loop ⟨waiting for incoming messages⟩ 18⟩ ≡
  while (1) {
    ⟨wait for an incoming message 53⟩
    tom_dump(tom);
    switch (message.type) {
    case INJECT_FAULT_TIMEOUT:
      LogError(EC_ERROR, "Manager␣loop", "Fault␣injection");
      tom_close(tom);      /* the time-out manager is detached */
      break;
    case IA_FLAG_TIMEOUT:
      LogError(EC_ERROR, "Manager␣loop", "IA_FLAG_TIMEOUT␣message␣->␣clear␣IA-flag.");
        /* time to clear the IA-flag! */
      ⟨clear IA-flag 16⟩
      break;
    case MIA_TIMEOUT:
      LogError(EC_ERROR, "Manager␣loop",
           "MIA_TIMEOUT␣message␣(time␣to␣send␣a␣MIA␣to␣Backup␣%d).", message.subid);
        /* time to send a MIA to a backup */
      ⟨send MIA to backup subid 19⟩
      tom_dump(tom + message.subid);
      tom_renew(tom, mia + message.subid);
      break;
```

Figure 10: An excerpt from the cweb source of the DIR net: the beginning of Code Section 18 (loop of the DIR-$\mathcal{M}$).

way for a cost-effective adoption of multiple-version software fault tolerance techniques (Lyu, 1998) such as the well known recovery block (Randell & Xu, 1995) or $N$-version programming (Avižienis, 1995). No such technique has been adopted in the current implementation of TOM.

## 2.3 Conclusions

A tentative *lingua franca* for the expression of failure detection protocols was introduced. TOM encloses the advantages of being simple, elegant and not ambiguous. Obvious are the many positive relapses that would come from the adoption of a standard, semi-formal representation with respect to the current Babel of informal descriptions—easier acquisition of insight, faster verification, and greater ability to rapid-prototype software systems.

Given the current lack of a network service for failure detection, the availability of standard methods to express failure detectors in the application layer is an important asset: a tool like the one described in this chapter isolates and crystallizes a part of the complexity required to express failure detection protocols. This complexity becomes transparent of the designer—which saves development times and costs—and eligible for cost-effective optimizations.

Separation of design concerns is not among the design goals of our tool, which provides the programmer with a library. Hence, our tool is characterized by bad . is not applicable, as TOM only targets one fault-tolerance provisions, i.e. failure detectors. No is considered, but in principle the server side of our system could be designed with adaptivity in mind, e.g., masking the actual number of failure detectors being used. Similar considerations apply to the DIR net.

As a final remark it is noteworthy to remark how, at the core of our design choices, is the selection of C and literate programming, which proved to be invaluable tools to reach our design goals. Nevertheless these choices may turn into intrinsic limitations for the expressiveness of the resulting language. In particular, they enforce a syntactical and semantic structure, that of the C programming language, which may be regarded as a limitation by those researchers who are not accustomed to that language. At the same time we would like to remark also that those very choices allow us a straightforward translation of our constructs into a language like Promela (Holzmann, 1991), which resembles very much a C language augmented with Hoare's CSP (Hoare, 1978). Accordingly, our future work in this framework shall include the adoption of the Promela extension of Prof. Bošnački, which allows for the verification of concurrent systems that depend on timing parameters (Bošnački & Dams, 1998). Interestingly enough, this version of Promela includes new objects, called discrete time countdown timers, which are basically equivalent to our non-cyclic time-outs. Our goal is to come up with a tool that generates from the same literate programming source (1) a pretty printout in TEX, (2) C code ready to be compiled and run, and (3) Promela code to verify some properties of the protocol.

# 3 CONCLUSION

The role of failure detection in dependable computing has been explained. Several examples of failure detection protocols have been described in a uniform way based on the time-out management system used in TIRAN to realize the algorithm used by the Backbone to detect and tolerate crash failures of nodes and Backbone components. That algorithm has been also described with time-outs as a case study.

Failure detectors are being studied intensively by many research groups throughout the world. Future work in our group at the University of Antwerp will focus on the expression of failure detection services for service-oriented architectures.

# Notes

[1] In Table 1, NFD-E refers to (Chen et al., 2002),$\varphi$ to (Hayashibara, 2004), FD to (Bertier et al., 2002), GMFD to (Raynal & Tronel, 1999), $\mathcal{D} \in \Diamond\mathcal{P}$ to (Chandra & Toueg, 1996), $\mathcal{HB}$ to (Aguilera et al., 1999), and $\mathcal{HB}$-pt to (Aguilera et al., 1999).

[2] Renewing a time-out means removing and re-inserting it.

page

# HYBRID APPROACHES

## 1 INTRODUCTION AND OBJECTIVES

This chapter describes some hybrid approaches for application-level software fault-tolerance. All the approaches reported in the rest of this chapter exploit the recovery language approach introduced in Chapter 6 and couple it with other tools and paradigms described in other parts of this book. Objective of this chapter is to demonstrate how $\mathcal{R}\varepsilon\mathcal{L}$ can serve as a tool to further enhance some of the application-level fault-tolerance paradigms introduced in previous chapters.
But why hybrid approaches in the first place? The main reason is that joining two or more concepts and their "system structures" (Randell, 1975), that is, the conceptual and syntactical axioms used in disparate application-level software fault-tolerance provisions, one comes up with a tool with better Syntactical Adequacy (the SA attribute introduced in Chapter 2). As already mentioned, a wider syntactical structure can facilitate the expression of our codes, while on the contrary awkward structures often lead to clumsy, buggy applications. Hybrid approaches are often more versatile and may also inspire brand new designs. A drawback of hybrid approaches is that they are modifications of existing designs. The extra design complexity must be carefully added to prevent the introduction if design faults in the architecture.
Which approaches to make use of is an important design choice. In some cases this may be under the control of the programmer: An example could be for instance, using a fault-tolerance library such as EFTOS to create a library of fault-tolerance meta-objects with OpenC++. Section 3 and Sect. 3 are other examples of this case. More ambitious goals would definitely require the design of new frameworks or architectures, such as it is the case for the model described in next section and the system in Sect. 5.
The first case is that of $\mathcal{R}\varepsilon\mathcal{L}$inda, a model coupling $\mathcal{R}\varepsilon\mathcal{L}$ with generative communication, introduced in Sect. 2. Section 3 then shows how to use $\mathcal{R}\varepsilon\mathcal{L}$ to enhance the resiliency of N-modular redundant systems by using spare components. A similar method is used in Sect. 4 to come up with flexible and dependable watchdog timers. Finally, Cactus is introduced—a system coupling a sort of recovery language with single-version software fault-tolerance mechanisms.

# 2 A DEPENDABLE PARALLEL PROCESSING MODEL BASED ON GENERATIVE COMMUNICATION AND RECOVERY LANGUAGES

As already remarked, redundancy is an important ingredient at the basis of both parallel computing and dependable computing. In other words, the availability of redundancy and a proper management of the available redundancy can be exploited to reach both functional objectives, such as high performance, and non-functional goals such as the ones at the core of this book, for instance high availability and data integrity.

Indeed, the truly extended use of parallel computing technologies asks for effective software engineering techniques to design, develop and maintain *dependable* parallel software—that is, it requires expressive programming models capable to make effective use of the available redundancy for capturing *both* the above classes of goals. This happens because any computing service based on parallel computing technologies *must* be designed in such a way as (1) to face and counterbalance the inherently lower level of dependability that characterizes any parallel hardware, due to the considerably higher probability for events such as a node's permanent or temporary failure in any such highly redundant system; and (2) to match the involved algorithms to the target hardware platform, so as to exploit optimally the available redundancy and reach, e.g., the expected performance.

One can conclude that, in order to be able to set up parallel computing services actually capable to provide their users with the expected high quality of service, it is important that the designers have at their disposal effective structuring techniques and software engineering methodologies for developing and maintaining *dependable parallel services*: lacking these techniques, the complexity pertaining to the management of the purely functional aspects, i.e., the service, the aspects related to the performance, i.e., the parallelism, and those related to fault tolerance, would jeopardize the design of the service increasing both its development times and costs and the probability of introducing design faults. Considering all the above and the main lessons learned exposed in this book, one can conclude that a simple and coherent structuring technique for dependable and parallel services, providing straightforward and effective means to control their complexity, would be a key factor for the wide-spreading of parallel computing technologies.

This chapter describes a novel software engineering methodology and structuring technique, $\mathcal{RE}\mathcal{L}$inda, whose main goals include separation of design concerns, dynamic adaptability of the service to varying environmental conditions, and expressiveness. Such technique provides a parallel programming model that couples the **recovery language approach** described in Chapter 5 and **generative communication** (Gelernter, 1985) i.e., the well-known model of communication of systems such as, for instance, Linda (Carriero & Gelernter, 1989) or JavaSpaces (Oaks & Wong, 2000) briefly introduced in Chapter 4. $\mathcal{RE}\mathcal{L}$inda provides its users with a single model for both parallel and dependable programming, and it is our conjecture that this single software framework provides both the elegance of generative communication and the effectiveness of recovery languages and, as such, is a valuable design tool for the composition of parallel, and dependable, services.
A prototypic implementation of some components of a $\mathcal{RE}\mathcal{L}$inda architecture took place in the framework of the ESPRIT project TIRAN, also introduced in Chapter 5. The key characteristic of the reported technique is that it allows structure the target application into three distinct components, respectively responsible for (1) the functional service, including its performance aspects, (2) the management of the fault-tolerance provisions, and (3) the adaptation to the current environmental conditions. As such it provides another solution to the problem of an optimal expression of fault-tolerance protocol whose fault model is not fixed once and for all but changes with the actual faults being experienced.
In what follows the $\mathcal{RE}\mathcal{L}$inda approach, its models and elements are described.

## 2.1 The $\mathcal{RE}\mathcal{L}$inda Approach

This section describes the $\mathcal{RE}\mathcal{L}$inda approach. It is structured into two parts. First, the system and application models are provided in Sect. 2.1.1. After that, in Sect. 2.1.2, the basic ideas behind $\mathcal{RE}\mathcal{L}$inda are exposed. A key component is then described in §2.1.3, which also points out the key relation between the recovery language approach and $\mathcal{RE}\mathcal{L}$inda.

### 2.1.1 System and Application Models

The target system for $\mathcal{RE}\mathcal{L}$inda is assumed to be a distributed or parallel system. Basic components of this system are the nodes, the tasks, and the communication network. In particular,

- a node can be, e.g., a workstation in a networked cluster or a processor in a MIMD parallel computer. The number of available nodes is $n$, $n > 0$. Nodes can be commercial-off-the-shelf hardware components with no special provisions for hardware fault-tolerance.
- Tasks are independent threads of execution running on the nodes.
- The network system allows tasks on different nodes to communicate with each other with the same properties described in Chapter 5 for the TIRAN architecture.

A general-purpose operating system (OS) is required on each node. No special purpose, distributed, or fault-tolerant OS is assumed to be required. The system is assumed to obey the *timed asynchronous distributed system model* (Cristian & Fetzer, 1999) (see Chapter 2).
The following assumptions characterize the user application:

- the service is supplied by a parallel application;
- it is written or is to be written in a procedural or object-oriented language such as, e.g., C or Java;

- the service specification includes non-safety-critical dependability goals;
- the target application is characterized by soft real-time requirements. In particular, performance failures (Cristian, 1991) may occasionally show up during error recovery;
- communication is based on the GC model. User tasks communicate exclusively through functions with the semantics of the Linda primitives—in(), out(), rd(), and their non-blocking versions, inp() and rdp().
- task management and, in particular, task creation and termination, is done respectively via a function with the semantics of the Linda primitive eval() and via function lave(), which terminates the matching active tuples.

### 2.1.2 $\mathcal{RE}\mathcal{L}$inda

Herein the elements of the $\mathcal{RE}\mathcal{L}$inda approach and some key components of its architecture are described.
To the performance-oriented designer, $\mathcal{RE}\mathcal{L}$inda appears as a standard Linda-like system. In addition to standard tuples, created by the user via function out(), $\mathcal{RE}\mathcal{L}$inda supports a special class of tuples, called *error notifications* (ENs). Error notifications are special, read-only tuples[1] that represent error detections such as, for instance "divide-by-zero exception caught while executing task 11," sent by a trap handling tool, and have the following structure:

$$(\text{“ERROR”}, m, t, e, v[1 \ldots e]),$$

where $m$ is an integer that identifies an error detector (see later on), $t$ is an integer that identifies a task found in error by $m$, $e$ is an integer, $e \geq 0$, and $v$ is a vector of $e$ integers describing the current error detection. By agreement, neither function out() nor function eval() are allowed specify tuples with the same structure as the error notifications. This means that the programmer has no *direct* way to emit an error notification[2]. Indeed, error notifications can only be produced by:

- A set of tools addressing error detection and fault masking, collectively called "error detectors" (EDs). EDs are tools similar to those of SwIFT (see Chapter 3) or the TIRAN Dtools (see Chapter 5), and provide services such as, e.g., watchdog timers and software voting (a system like the EFTOS Voting Farm could be used for that, see Chapter 3 (De Florio, Deconinck, & Lauwereins, 1998)[3];
- a failing eval();
- or a failing assert(), the latter being a function with the semantics of the C standard library function with the same name (Kernighan & Ritchie, 1988).

On the other hand, each user task has free read-only access to the error notifications, and hence can query the current status of any task $t$—according to its error detectors. This can be used to unblock potentially dangerous deadlock conditions, such as two

Table 1: Scheme of execution of TS when it processes request $r$.

```
Begin request (r) :
    if tuple type (r) ≠ EN then
        insert plain tuple (r)
        return SUCCESS
    endif
    if sender = ED then
        update α-count (r)
        insert EN tuple (r)
        awake recovery interpreter (r)
        return SUCCESS
    endif
    return FAILURE
```

tasks where each of them is waiting for the other to become available for processing—by simply specifying, as an unblocking condition, the availability of the error notification describing that the other task is being found in error.
As it is for "plain" Linda systems, also the $\mathcal{REL}$inda architecture foresees a tuple space manager (TS). The latter is a distributed application, constituted of $c$ components, where $1 \leq c \leq n$, each of which is located on a different node of the system. As a whole, the TS executes the algorithm of mutual suspicion, that is, the fault tolerance distributed algorithm described in Chapter 5 which allows tolerate up to $c - 1$ crash failures affecting either its components or the nodes hosting them.
Tuples are stored by the tuple space manager into a database that is replicated on each of its $c$ components. Each time a new request for insertion of either an error notification or a "plain" tuple reaches the tuple space manager, it is processed according to the following scheme, also summarized in Table 1:

- If the request pertains to a "plain" tuple, the latter is inserted in the tuple space. This implies updating the remote copies of the database.
- Otherwise, if the ES was sent by an error detector,
  - first, a field of the database, called $\alpha$-counter is updated (see below for more details on this);
  - then the error notification is inserted as a plain tuple; and
  - (3) a RINT recovery interpreter is started (see Chapter 5 for a description of RINT).

Field $\alpha$-counter is a task-specific variable that is processed according to the threshold-based statistical technique known as $\alpha$-count and described in Chapter 5. As explained there, this technique is capable of assessing whether a task is affected by a transient vs. a permanent or intermittent fault; as such, it allows express fault-aware error recovery strategies.
Apart from managing tuples and error notifications, the tuple space manager also keeps track of the current *structure* and *state* of *the system, the user application, and itself.* These data are used at error recovery time as described in the following text.

### 2.1.3 $\mathcal{REL}$inda Error Recovery Component

The execution of the recovery actions is done via a fixed (i.e., special-purpose) scheme, portrayed in the sequence diagram of Fig. 1: as soon as an error is detected, a notification describing that event is sent to a distributed entity responsible for the collection and the management of these notifications. Clearly, in $\mathcal{R}\varepsilon\mathcal{L}$inda this entity is the TS and the above mentioned notifications are the ENs.
Immediately after storing each notification, the guards of the recovery actions are evaluated. Guards evaluation is done by querying the database of the TS. When a guard is found to be true, its corresponding actions are executed, otherwise they are skipped.
As a conclusive remark $\mathcal{R}\varepsilon\mathcal{L}$inda may be described as a *specialization* (in object-oriented terminology) of the general abstract class $\mathcal{R}\varepsilon\mathcal{L}$ addressing parallel programming.

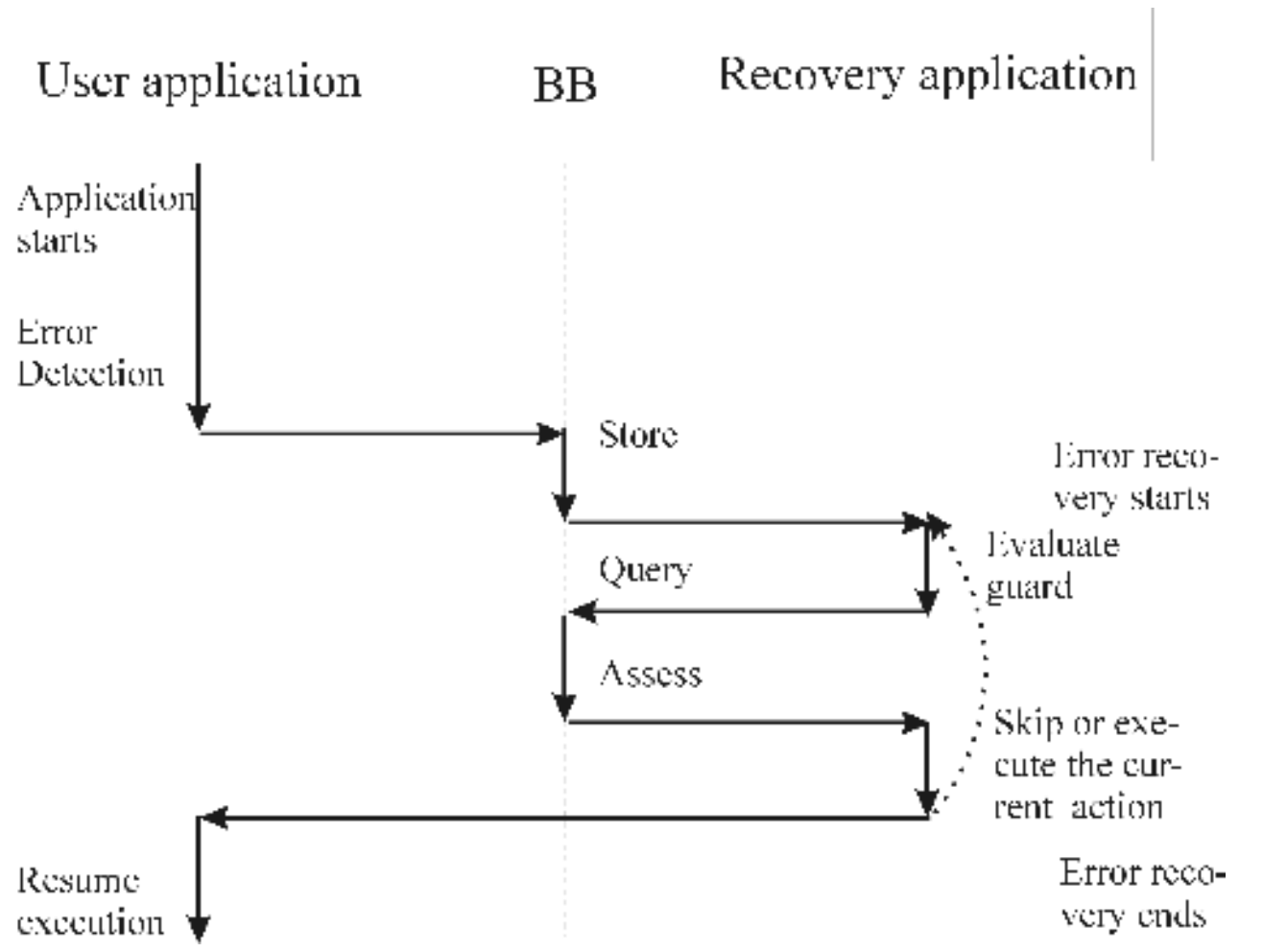

Figure 1: Scheme of execution of a $\mathcal{R}\varepsilon\mathcal{L}$inda-compliant distributed application: together with the application, two special-purpose tasks are running—a distributed tuple space manager, which stores both the error notifications sent by the error detectors and the "plain" tuples requested via out() or eval(), and a "recovery application", i.e., a task responsible for the execution of a set of recovery actions. The shown sequence diagram describes the execution of the user-specified recovery actions.

## 2.2 Conclusions

This section has exposed the key ideas behind a novel approach called $\mathcal{R}\varepsilon\mathcal{L}$inda, coupling Generative Communication with the the $\mathcal{R}\varepsilon\mathcal{L}$ approach introduced in Chapter 6. As such, $\mathcal{R}\varepsilon\mathcal{L}$inda is an attempt to explicitly address both the dependability and the performance aspects making use of a single design framework. $\mathcal{R}\varepsilon\mathcal{L}$inda offers a high degree of separation of design concerns, such that the functional and the FT aspects do not conflict with each other. Its dynamic adaptability to varying environmental conditions, which may obtained as described in (De Florio, 2000) through dynamic linking of the r-code, is another important feature of $\mathcal{R}\varepsilon\mathcal{L}$inda. SA in $\mathcal{R}\varepsilon\mathcal{L}$inda is limited to those fault-tolerance provisions that better match its model; in particular, the strong link with parallel computing makes it especially in line with methods such as MVSFT (see Chapter 3).
Several other important topics related to $\mathcal{R}\mathcal{E}\mathcal{L}$inda's components could be investigated in the future. Such topics include, for instance, the inclusion in the TS of the concept of so-called "book-kept tuples," i.e., tuples that are distributed to requesters by means of the algorithm of RAFTNET for dependable farmer worker applications described in Chapter 3. This strategy translates into effective and elegant support to highly performing and highly dependable data-parallel applications. Another research path is the extension of these schemes towards **persistent object stores** (Oaks & Wong, 2000) and CORBA-compliant or service-oriented middleware. Other strategies to further increase the dependability of $\mathcal{R}\mathcal{E}\mathcal{L}$inda applications are, e.g., the possibility to combine multiple tuple space operations into atomic transactions (Bakken & Schlichting, 1995; Anderson & Shasha, 1991; Cannon & Dunn, 1992). Other techniques that could be investigated to further enhance the dependability of the tuple space could be e.g., tuple space checkpoint-and-rollback (Kambhatla, 1991).

# 3 ENHANCING A TIRAN DEPENDABLE MECHANISM

As observed by the designers of GUARDS (Powell et al., 1999), many embedded systems may benefit from an hardware architecture based on *redundancy* and *consensus* in order to fulfill their dependability requirements. As pointed out in Chapter 2, such a hardware architecture needs to be coupled with some application-level strategy or mechanism in order to guarantee end-to-end tolerance of faults[4]. The software architecture described in this book may support in many ways the class of applications eligible for being embedded in a GUARDS system instance. The main tool for this is the TIRAN Distributed Voting mechanism (derived from the EFTOS Voting Farm described in Chapter 3). This section describes how it is possible to further increase the dependability of the TIRAN Distributed Voting mechanism by using ARIEL as both a configuration and a recovery language and making use of the TIRAN framework. An assessment of these enhancements is reported in Chapter 10.
The system realizes a sophisticated $N$-version programming executive that implements the software equivalent of an NMR system. Because of this assumption, the user is required to supply four versions of the same software, designed and developed according to the $N$-version approach (Avižienis, 1985). This requirement can be avoided if it is possible to assume safely that no design faults resulting in correlate failures affect the software.
To simplify the case study it is assumed that the service provided by any of the versions or instances is *stateless*. When this is not the case, the management of the spare would also call for a forward recovery mechanism—e.g., the spare would need to acquire the current state of one of the non-faulty tasks. Furthermore, when using pipelines of N-version tasks under strict real-time requirements, further techniques (currently not part of the prototype presented in this work) would be required in order to restore the state of the spare with no repercussion on the real-time goals (see (Bondavalli, Di Giandomenico, Grandoni, Powell, & Rabéjac, 1998) for an example of these techniques).
The enhanced TIRAN Distributed Voting mechanism described in what follows is representative of the class of applications that are best eligible for being addressed via an application-level fault-tolerance structure such as NVP. This section describes how $\mathcal{REL}$ provides to those applications two additional features, namely, support of spares and support of fault-specific reconfiguration, which are not part of plain NVP systems.
A system of four nodes is assumed. Nodes are identified by the symbolic constants `NODE1`, ..., `NODE4`, defined in the header file `nodes.h`. Header file `my_definitions.h` contains a number of user definitions, including the unique-id of each version task (`VERSION1`, ..., `VERSION4`) and the local identifier of each version task (in this case, on each node the same identifier is used, namely `VERSION`). In the same file also the time-outs of each version are defined (`TIMEOUT_VERSION1`, ..., `TIMEOUT_VERSION4`).
Let us consider the following ARIEL script:

```
INCLUDE "nodes.h"
INCLUDE "my_definitions.h"

TASK {VERSION1} IS NODE {NODE1}, TASKID {VERSION}
TASK {VERSION2} IS NODE {NODE2}, TASKID {VERSION}
TASK {VERSION3} IS NODE {NODE3}, TASKID {VERSION}
TASK {VERSION4} IS NODE {NODE4}, TASKID {VERSION}

N-VERSION TASK {TMR_PLUS_ONE_SPARE}
VERSION 1 IS TASK {VERSION1} TIMEOUT {TIMEOUT_VERSION1}ms
VERSION 2 IS TASK {VERSION2} TIMEOUT {TIMEOUT_VERSION2}ms
VERSION 3 IS TASK {VERSION3} TIMEOUT {TIMEOUT_VERSION3}ms
VERSION 4 IS SPARE TASK {VERSION4} TIMEOUT {TIMEOUT_VERSION4}ms
VOTING ALGORITHM IS MAJORITY
METRIC "tmr_cmp"
ON SUCCESS TASK 20
ON ERROR TASK 30
END N-VERSION

IF [ PHASE (T{VERSION1}) == {HAS_FAILED} || FAULTY T{VERSION1} ]
THEN
      STOP T{VERSION1}

      SEND {WAKEUP} T{VERSION4}
      SEND {VERSION1} T{VERSION4}

      SEND {VERSION4} T{VERSION2}
      SEND {VERSION4} T{VERSION3}
FI
```

```
IF [ PHASE (T{VERSION2}) == {HAS_FAILED} || FAULTY T{VERSION2} ]
THEN
      STOP T{VERSION2}

      SEND {WAKEUP} T{VERSION4}
      SEND {VERSION2} T{VERSION4}

      SEND {VERSION4} T{VERSION1}
      SEND {VERSION4} T{VERSION3}
FI

IF [ PHASE (T{VERSION3}) == {HAS_FAILED} || FAULTY T{VERSION3} ]
THEN
      STOP T{VERSION3}

      SEND {WAKEUP} T{VERSION4}
      SEND {VERSION3} T{VERSION4}

      SEND {VERSION4} T{VERSION2}
      SEND {VERSION4} T{VERSION1}
FI
```

The above script consists of two parts—one for managing the configuration of tasks and tools, the other for describing a recovery strategy. The following two subsections describe the two parts.

## 3.1 Configuration

The configuration section of the script just shown defines four tasks, each running on a different node of the system. This decision has been taken to reduce the probability of a common source for multiple failures, in case of a crash of a node. It is worth noting that this design decision—the physical location of the tasks—*is made outside the application code* and can be changed with no repercussion on it, thus allowing location transparency and high SC.
Three of the four tasks are then configured as versions of a TIRAN Distributed Voting mechanism (derived from the EFTOS Voting Farm described in Chapter 3). The fourth task (`VERSION4`) is configured as a spare. This means that, when task `TMR_PLUS_ ONE_SPARE` is launched, the first three tasks arrange themselves as components of a TIRAN Distributed Voting mechanism, with each task assisted by a local voter as described in the aforementioned section. On the contrary, the fourth task is blocked waiting for the arrival of a special "wakeup" signal.
Feeding the ARIEL translator with the configuration script as described in Chapter 6, a number of source codes are produced, including:

- The basic user tasks of a TIRAN Distributed Voting mechanism consisting of three modules, an example of which can be seen in Chapter 6 when configuring multiple-version software fault-tolerance.

- The spare user task, initially waiting for the arrival of a wakeup message.
- Task TMR_PLUS_ONE_SPARE. This task is in charge of the:
  1. Transparent set up of the TIRAN Distributed Voting mechanism via the TIRAN_CreateTask function of the TIRAN Basic Services Library.
  2. Management of the replication of the input value.
  3. Declaration and insertion in the TIRAN TOM of the set of time-outs that represent an upper limit to the duration of the base tasks. This upper limit is set by the user and known by one of the hypotheses of the timed-asynchronous distributed system model introduced in Chapter 2 (all services are timed).
  4. De-multiplexing and delivering the output value.

In the absence of faults, task TMR_PLUS_ONE_SPARE would appear to the user as yet another version providing the same service supplied by tasks VERSION1,...,VERSION4. The only difference would be in terms of a higher reliability (see Chapter 10) and a larger execution time, mainly due to the voting overhead.
Location transparency in this case is supported by ARIEL, while replication transparency is supported by the TIRAN Distributed Voting mechanism. The degree of code intrusion in the application source code is reduced to the one instruction to spawn task TMR_PLUS_ONE_SPARE. No support for the automatic generation of makefiles is provided in the current version of the ARIEL translator, so the user needs to properly instruct the compilation of the source files written by the translator.

## 3.2 Recovery

The recovery section of the ARIEL script on p. 267 defines a recovery strategy for task TMR_PLUS_ ONE_SPARE. When any error is detected in the system and forwarded to the backbone, the backbone replies to this stimulus as follows:

1. It stores the error notification in its local copy of the DB.
2. It updates the $\alpha$-count variable related to the entity found in error.
3. If the notification is local, i.e., related to a local task, then the local component of the BB forwards the notification to the other components.
4. If the local BB entity plays the role of coordinator, it initiates error recovery by sending a “wakeup” message to the recovery interpreter.

This latter orderly reads and executes the r-codes. Table 2 shows a trace of the execution of some of the r-codes produced when translating a simplified version of the ARIEL script on p. 267. As already mentioned, RINT implements a virtual machine with a stack architecture. Line 5 starts the scanning of a guard. Line 6 stores on RINT’s run-time stack the current value of the phase of task 0 (the value of symbolic constant VERSION1). Line 7 compares the top of the run-time stack with integer

HAS_FAILED. The result, 0 (false) is stored on top of the stack. Line 8 checks the top of the stack for being 0. The condition is fulfilled, so a jump is made to line 18 of the r-code list. That line corresponds to the end of the current IF statement—the guard has been found as false, therefore the corresponding recovery actions have been skipped. Then, on line 19, some internal variables are reset. Line 20 starts the evaluation of the clause of the second IF of the recovery script. The scenario is similar to the one just described, though (on line 22) the phase of task 1 (that is, VERSION2) is found to be equal to HAS_FAILED. The following conditional jump is therefore skipped, and a stream of recovery actions is then executed: task VERSION2 is terminated on line 24, then value 10 (WAKEUP) is sent to task VERSION4 by means of r-code SEND (which sends a task the top of the run-time stack), and so forth, until line 33, which closes the current IF. A third guard is then evaluated and found as false. Clearing some internal structures closes the execution of RINT, which again "goes to sleep" waiting for a new "wakeup" message to arrive.

Figure 2, 3, and 4 show three different views to a TMR-and-one-spare system at work, as displayed by the TIRAN monitoring tool in a Netscape client (De Florio, Deconinck, Truyens, Rosseel, & Lauwereins, 1998). The first picture renders the current state and shape of the overall target application. Figure 3 summarizes the framework-level events related to processing node 0. In particular, a value domain failure is injected on the voter on node 2. This triggers the execution of a recovery script which reconfigures the TMR isolating the voter on node 2 and switching in the spare voter on node 3. The execution trace of the r-codes in this script is displayed in Fig. 4.

It is worth noting how, modifying the base recovery strategy of the ARIEL script on p. 267 does not require any modification in the application source code—not even recompiling the code, in case the r-codes are read from an external means (e.g., from a file). The two design concerns—that of the fault-tolerance software engineer and that of the application software engineer—are kept apart, with no repercussions on the maintainability of the service overall. To prove this, let us consider the ARIEL excerpt in Table 3. Such a strategy is again based on a TMR plus one spare, though now, before switching off the faulty version and in the spare, the current value of the $\alpha$-count filter related to the current version is compared with a threshold supplied by the user in the configuration section. If the value is below the threshold, it is not possible to assess that the fault affecting {VERSION1} is permanent or intermittent. In this case, the faulty task is restarted rather than substituted. In other words, another chance is given to that version, while its "black list" (its $\alpha$-counter) is updated. As mentioned in Chapter 6, research studies have shown that, whatever the threshold, the $\alpha$-count is bound to exceed it when the fault is permanent or intermittent. In the latter case, therefore, sooner or later the second strategy is to be executed, permanently removing the faulty version. It is worth noting how the adoption of such a scheme, which is no longer purely based on masking the occurrence of a fault, in general implies an execution overhead that may violate the expected real-time behavior—*during error-recovery*.

Further support towards graceful degradation when spares are exhausted could also be foreseen.

| | |
|---|---|
| 5 | IF statement. |
| 6 | STORE-PHASE: stored phase of task 0, i.e., 0. |
| 7 | COMPARING(9999 vs. 0): Storing 0. |
| 8 | Conditional GOTO, fulfilled, 18. |
| 18 | FI statement. |
| 19 | OA-RENEW. |
| 20 | IF statement. |
| 21 | STORE-PHASE: stored phase of task 1, i.e., 9999. |
| 22 | COMPARING(9999 vs. 9999): Storing 1. |
| 23 | Conditional GOTO, unfulfilled, 24. |
| 24 | KILLING TASK 1. |
| 25 | PUSH(10). |
| 26 | SEND MSG 10 to TASK 3. |
| 27 | PUSH(1). |
| 28 | SEND MSG 1 to TASK 3. |
| 29 | PUSH(3). |
| 30 | SEND MSG 3 to TASK 2. |
| 31 | PUSH(3). |
| 32 | SEND MSG 3 to TASK 0. |
| 33 | FI statement. |
| 34 | OA-RENEW. |
| 35 | IF statement. |
| 36 | STORE-PHASE: stored phase of task 2, i.e., 0. |
| 37 | COMPARING(9999 vs. 0): Storing 0. |
| 38 | Conditional GOTO, fulfilled, 48. |
| 48 | FI statement. |
| 49 | OA-RENEW. |

Table 2: Trace of execution of the r-codes corresponding to the recovery section of the ARIEL script on p. 267. Number 9999 is the value of constant `HAS_FAILED`.

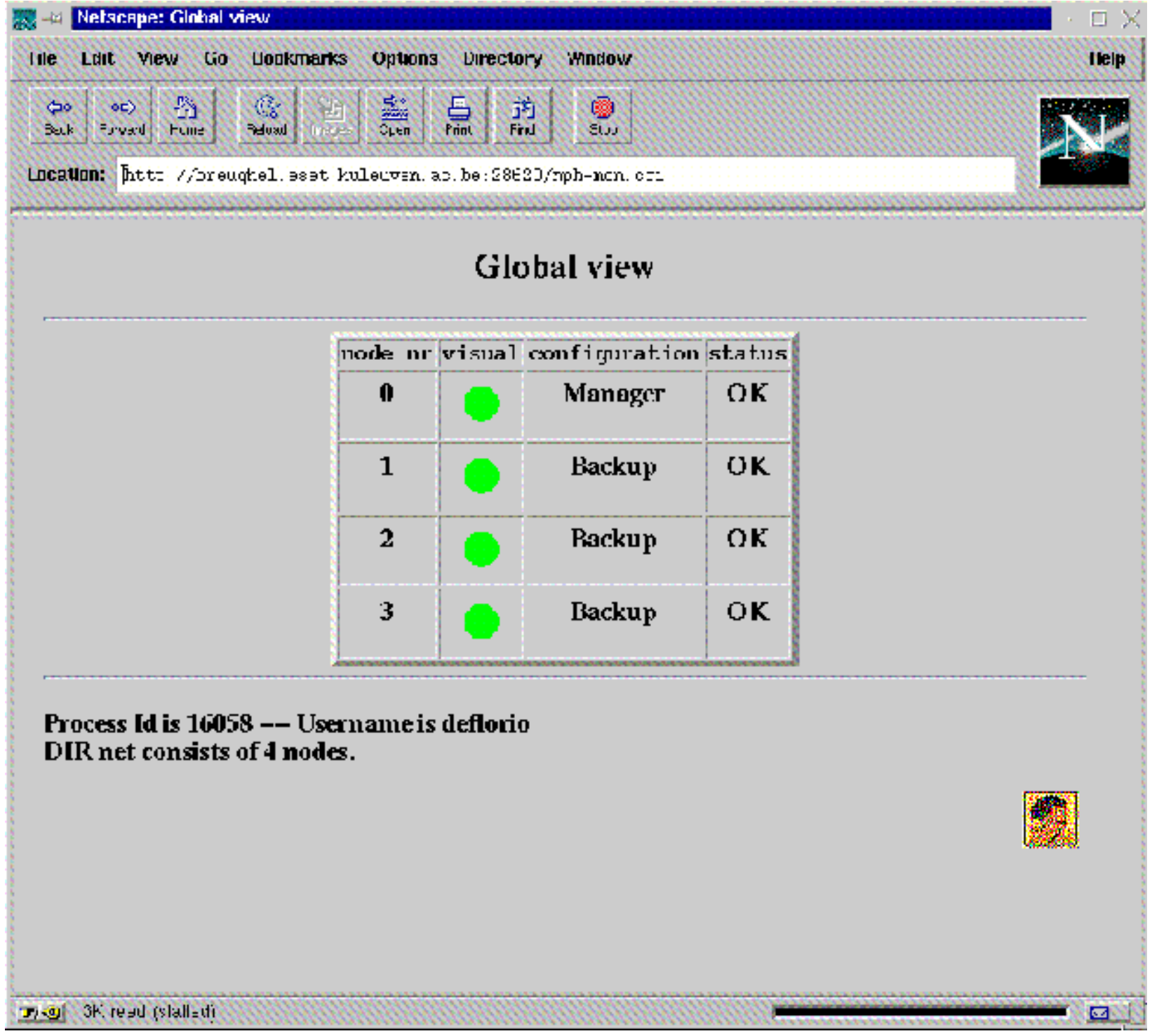

Figure 2: A global view of the state and shape of the target application, as rendered by a monitoring CGI script. In this case a four processing node system is used. Node 0 hosts the main component of the backbone. The circular icons are hypermedia links (see Fig. 5). The small icon on the bottom right links to the execution trace of the ARIEL interpreter.

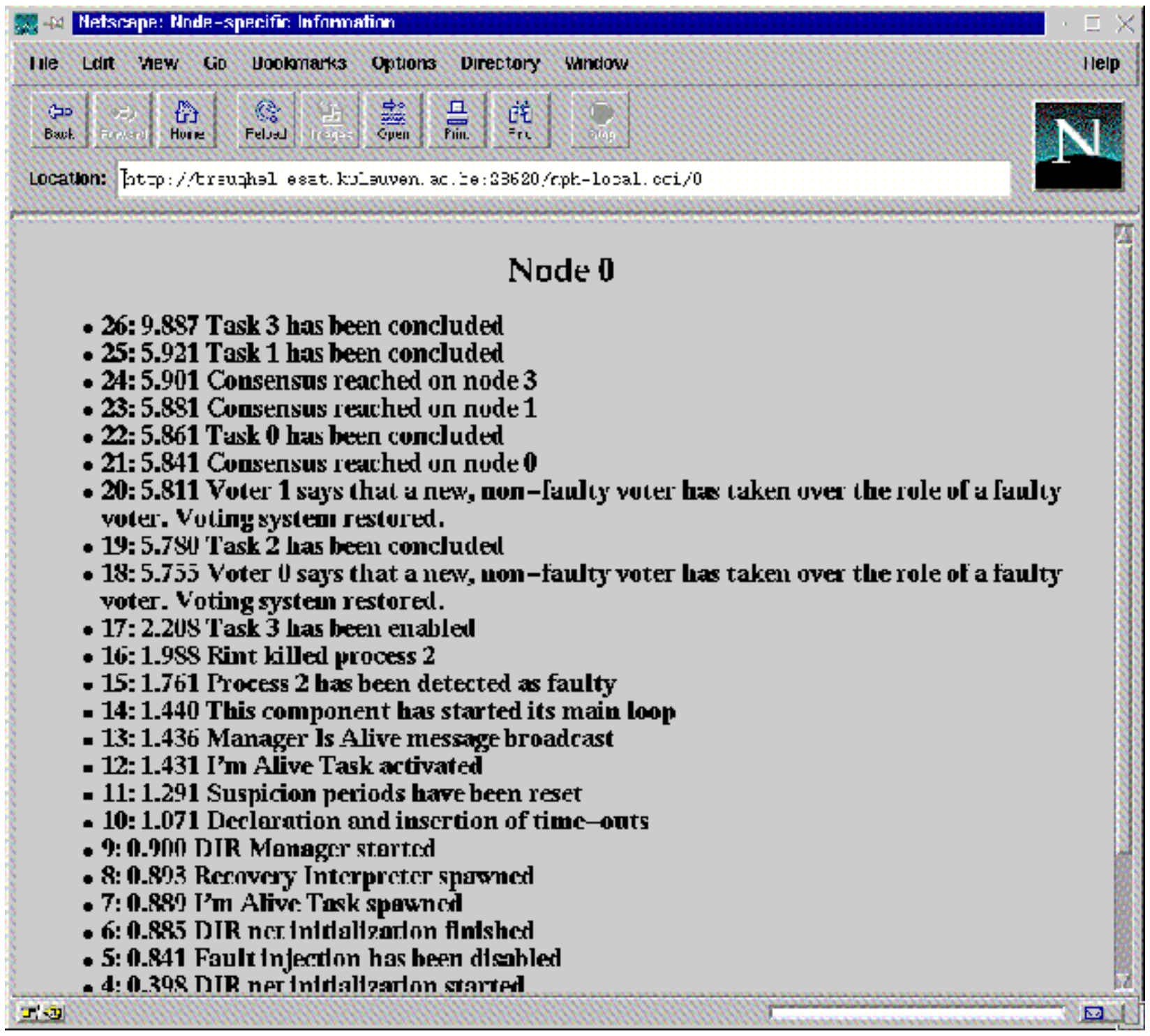

Figure 3: A view to the events tracked while monitoring processing node 0. This view can be obtained by selecting the top circular icon of Fig. 2. Events are ordered decreasingly with respect to their conclusion time. Times are expressed in seconds and measured through the local clock. Going from bottom to top: on line 15 process 2 is detected as faulty. On 16, that process is terminated. On 17, a new process (a spare voter) is awaken. On 18 and 20, respectively voter 0 and 1 acknowledge the reconfiguration. This allows retry a global consensus that succeeds (see lines 21, 23, and 24). Finally all voters acknowledge their termination.

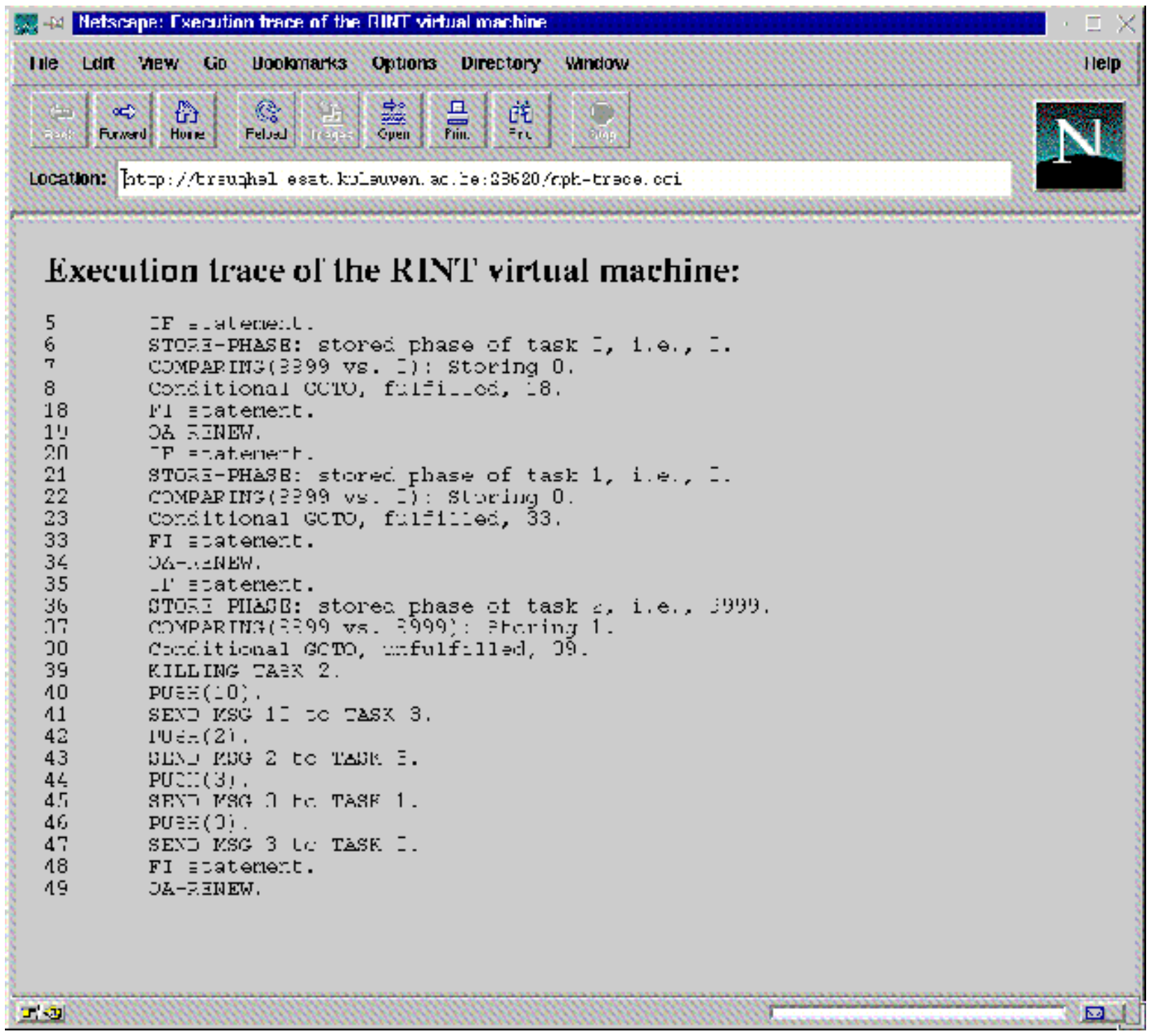

Figure 4: As soon as an error is detected—in this case, a voter has been found in minority—RINT starts executing the r-codes. This picture is the execution trace of the r-codes in Table 6 and Table 9 of Chapter 6. Numbers refer to the code lines in those tables. Note how, at line 37, the state information of task 2 is found to be equal to "9999" (HAS_FAILED). As a consequence, a number of actions are executed in lines 39–47. In particular, task 2 is terminated at line 39.

```
ALPHACOUNT {VERSION1} IS threshold = 3.0, factor = 0.4 END
IF [ PHASE (T{VERSION1}) == {HAS_FAILED} || FAULTY T{VERSION1} ]
THEN
      IF [ TRANSIENT T{VERSION1} ]
      THEN
          RESTART T{VERSION1}
          SEND {VERSION1} T{VERSION2}
          SEND {VERSION1} T{VERSION3}
      ELSE
          STOP T{VERSION1}

          SEND {WAKEUP} T{VERSION4}
          SEND {VERSION1} T{VERSION4}

          SEND {VERSION4} T{VERSION2}
          SEND {VERSION4} T{VERSION3}
      FI
FI
```

Table 3: An ARIEL script for a TMR-plus-one-spare system that exploits the $\alpha$-count technique to avoid unnecessary reconfigurations.

### 3.3 Lessons learned

Enhancing the TIRAN Distributed Voting mechanism allowed include two special services (spare support and fault-identification) into those offered by NVP systems. For those applications that best match with the MV application-level fault-tolerance approach, and with NVP in particular, those services allow further increase the dependability without being detrimental to SC—i.e., without intruding code. Eligible applications include coarse-grained, distributed or parallel applications, i.e., those addressed by $\mathcal{R}\varepsilon\mathcal{L}$. SA is clearly limited to the above class of provisions. A has not been considered here, but the code separation of ARIEL could be used to achieve some adaptability. Furthermore, safety-critical parallel or distributed airborne and spaceborne applications appear to be well addressed by those services. It must be remarked, though, that the safety-critical character of these applications calls for specific framework-level and system-level support, such as the one provided by the GUARDS architecture (Powell et al., 1999), a generic, parameterizable hardware and OS architecture for real-time systems, which straightforwardly supports the NVP scheme.

# 4 COMPOSING DEPENDABLE MECHANISMS: THE REDUNDANT WATCHDOG

This section describes how it is possible to compose dependable mechanisms addressing specific needs by means of the $\mathcal{RεL}$ architecture devised in TIRAN. The case reported in the rest of this section originated by a requirement of one of the TIRAN partners, ENEL SpA. In present day's fault-tolerance systems at ENEL, a hardware watchdog timer is available and integrated in several of their automation applications. Such device is a custom (non-COTS) hardware component that is capable of guaranteeing both high availability and high integrity. This allows its use in contexts where safety is the main concern.
As mentioned in Chapter 3, ENEL has started an internal plan that aims at renewing and enlarging their park of automation systems, also with the goal of improving their efficiency and quality of service. Within this plan, ENEL is investigating the possibility to substitute this hardware device with a software component, exploiting the redundancy of COTS parallel or distributed architectures, in such a way as to guarantee acceptable levels of safety and availability for their applications. While taking part in TIRAN, CESI posed the above problems to the consortium. The TIRAN consortium proposed the solution described in this section, based on the $\mathcal{RεL}$ approach.

## 4.1 An Industrial Problem: The Redundant Watchdog Requirements

Technicians of energy automation systems typically express requirements in textual form, capturing the main dependability needs of the applications. Considering the Primary Substation Automation System a list of dependability requirements has been collected and addressed in the TIRAN project (Botti et al., 1999). In this section the focus will be on two of those Application Requirements (referred as AR1 and AR2 below) that lead to the need of an enhanced watchdog mechanism. They are formulated as follows:

**AR1** : "If an erroneous situation can not be recovered according to required mode and within given time constraints, then a mechanism for the auto-exclusion of the system should be provided which, if not reset before the expiration of a pre-fixed time-out, disconnects the system from the plant, leaving the plant in an acceptable state, forcing the output to assume a pre-defined secure configuration, providing appropriate signalling to the operator and to the remote systems (as automation system failures should not affect the plant)."

**AR2** : "The auto-exclusion should guarantee a high availability, integrity and security — e.g. by a redundant and periodically tested mechanism, with auto-diagnostics."

The auto-exclusion functionality (as required by AR1 and AR2) has been traditionally supported by the so-called plant's watchdog (plantWD) mechanism, a dedicated hardware device with high integrity and availability degrees. In most cases the plantWD mechanism is used as an ultimate action of a fault tolerant strategy to detect un-recovered processing errors and to avoid their propagation. Errors are typically run-time violations occurring during the execution of an application process due to, e.g., a process that has crashed or is slowed down.

The watchdog mechanism (WD) cyclically sets a timer requiring an application process to explicitly reset it by sending an "I am alive" message before it reaches its deadline. If, for any reason, the application process is not able to send the message, the watchdog raises an error condition that has to be treated by some entity in some way. Depending on the global fault tolerance strategy adopted our plantWD is set to count either the double of or the same time of the basic application cycle.

In support of the migration to flexible software dependability services running on COTS platforms, the goal of developing a robust, software-based WD mechanism has been addressed by the TIRAN Project. A watchdog basic tool has been implemented characterized by the following Watchdog Requirements (WR) and Properties (WP):

**WR1** : "The WD has to survive at system reboot or reset, i.e. the memory it allocates for its counter is not to be cleared."

**WR2** : "In a distributed software architecture the application node's signals have to be put in a logic AND to actually signal the WD, i.e. the WD effectively stops to countdown only if on each node the execution has terminated correctly."

**WR3** : "The WD has to survive at node failures, i.e. whatever node faults the WD mechanism should be not compromised."

**WR4** : In order to guarantee correct operation of the WD mechanism, it is mandatory that the WD task is running at a higher priority than the tasks (that run on the same node) it supervises. WD tasks supervising tasks on other nodes must have appropriate priority to ensure proper operation. It is the responsibility of the application writer to ensure correct partitioning and priority allocation.

**WP1** : The watchdog task can be placed either on the same node where the application tasks run on or on a different node.

**WP2** : Placing the watchdog task on the same node where the application task runs on minimizes overhead and detection latency.

**WP3** : Placing the watchdog task on a different node with respect to the application node lowers the probability of a common failure for both application and watchdog task that would go undetected.

**WP4** : Detection latency is under the control of the application writer. The higher the frequency of sending "I'm alive" messages, the lower the detection latency.

**WP5** : Overhead is under the control of the application writer. The lower is the frequency of sending "I'm alive" messages, the lower is the overhead paid by the application task and the communication system.

**WP6** : WD is just one task which receives system clock ticks and application "I'm alive" messages. Both types of messages are received through interprocess communication and are asynchronous to WD task.

**WP7** : Being the WD in a distributed software architecture it is able to receive multiple signals and to apply a logical operation on them (i.e. in the case of the logical operation AND required by WR1 the WD will fire if at least one node does not produce its signal).

In Section 4 it will be shown how the requirement WR3 above may be fulfilled by instantiating more WD mechanisms and by applying different voting mechanisms to their firings. Such Redundant WatchDog (RWD) mechanism is characterized by the following design properties:

**RWP1** : Processing errors affecting WD replicas can be detected and recovered transparently by the RWD

**RWP2** : The number of WD replicas and the voting mechanism chosen determine a different improvement of the RWD dependability: e.g. Nreplica=3, allows a 2-out-of-3 voting (which can correct up to 1 fault); the selection of the suitable Nreplica and voting is a compromise among dependability and performance overhead, left to the application writer's experience.

**RWP3** : WD replicas can be placed all on the same node. This minimizes overhead and detection latency but does not increase the RWD dependability.

**RWP4** : WD replicas can be placed on different nodes. This minimizes the chance of a common failure affecting each WD replica.

## 4.2 The Strategy and Its Key Components

The strategy proposed within TIRAN exploits the following components:

- The Basic Services Library, and specifically its function `TIRAN_Send`, which multicasts a message to a logical, i.e., a group of tasks.
- The configuration support provided by ARIEL.
- The TIRAN BB and its DB.
- The recovery language ARIEL.
- The watchdog Basic Tool, i.e., a node-local software component in level 1.1 of the TIRAN architecture (see Chapter 6 for a description of the TIRAN architecture.)

The latter, which executes on a single processing node, can not guarantee the required degree of availability when used as a substitute of the ENEL hardware watchdog. This notwithstanding, the adoption of the above components allowed *composing*—rather than *programming*—a new prototypic DM, the so-called Redundant Watchdog (RW).

This composition is made in terms of the above elements of the TIRAN framework, with ARIEL playing the role of coordinator.
In order to introduce the strategy, the following scenario is assumed:

- A distributed system is available, consisting of three nodes, identified as $N_1, N_2$ and $N_3$.
- On each node of this system, a number of application tasks and an instance of the TIRAN watchdog are running.

The design goal to be reached is enhancing the dependability of this basic scheme by means of a technique that does not overly increase, at the same time, the complexity of the overall software tool.
The adopted strategy is now explained—each step has been tagged with a label describing the main $\mathcal{R}\mathcal{E}\mathcal{L}$ feature being exploited.

**Configuration:** Define and configure the three watchdogs by means of the provisions described in Chapter 6. In particular,

- Assign them the unique-ids $W_1, W_2$, and $W_3$.
- Specify that, on a missed deadline, a notification is to be sent to the TIRAN BB.
- Deploy the watchdogs on different nodes.

**Configuration:** Define logical $L$, consisting of tasks $W_1, W_2$, and $W_3$.

**Recovery:** Define in ARIEL an "*AND-strategy*", that triggers an alarm when each and every watchdog notifies the BB, an "*OR-strategy*", in which the alarm is executed when any of the three watchdog expires, and a "*2-out-of-3 strategy*", in which a majority of the watchdogs needs to notify the BB in order to trigger the alarm. In the current prototype, the alarm is a notification to the task the unique-id of which is $A$.

The configuration step is coded as in Table 4. Table 5 lists the recovery actions corresponding to the AND-strategy.
When a watched task sends "watchdog $L$" its heartbeats, the Basic Services Library relays these messages to the three watchdogs on the three nodes. In absence of faults, the three watchdogs[5] process these message in the same way—each of them in particular resets the internal timer corresponding to the client task that sent the heartbeat. When a heartbeat is missing, the three watchdogs expire and send a notification to the BB, one at the time. The reply of the BB to these notifications is the same: RINT is awoken and the r-codes are interpreted. The difference between the three strategies is then straightforward:

- The OR-strategy triggers the alarm as soon as any of the watchdog expires. This tolerates the case in which up to two watchdogs have crashed, or are faulty, or are unreachable. This intuitively reduces the probability that a missing heartbeat goes undetected, hence can be regarded as a "*safety-first*" strategy. At

```
INCLUDE "watchdogs.h"

TASK {W1} IS   NODE {N1}, TASKID {W1}
TASK {W2} IS   NODE {N2}, TASKID {W2}
TASK {W3} IS   NODE {N3}, TASKID {W3}

WATCHDOG {W1}
  HEARTBEATS EVERY {HEARTBEAT} MS
  ON ERROR WARN BACKBONE
END WATCHDOG

WATCHDOG {W2}
  HEARTBEATS EVERY {HEARTBEAT} MS
  ON ERROR WARN BACKBONE
END WATCHDOG

WATCHDOG {W3}
  HEARTBEATS EVERY {HEARTBEAT} MS
  ON ERROR WARN BACKBONE
END WATCHDOG

LOGICAL {L} IS  TASK {W1}, TASK {W2}, TASK {W3}  END LOGICAL
```

Table 4: Configuration of the Redundant Watchdog.

```
IF [ PHASE (TASK{W1}) == {EXPIRE} AND
     PHASE (TASK{W2}) == {EXPIRE} AND
     PHASE (TASK{W3}) == {EXPIRE} ]
THEN
     SEND {ALARM} TASK{A}
     REMOVE PHASE LOGICAL {L} FROM ERRORLIST
FI
```

Table 5: The AND-strategy of the Redundant Watchdog. Action `REMOVE` resets the phase corresponding to the tasks of logical $L$.

the same time, the probability of "false alarms" (mistakingly triggered alarms) is increased. Such alarms possibly lead to temporary pauses of the overall system service, and may imply costs.

- The AND-strategy, on the other hand, requires that *all* the watchdogs reach consensus before triggering the system alarm. It decreases the probability of false alarms but at the same time decreases the error detection coverage of the watchdog Basic Tool. It may be regarded as an "*availability-first*" strategy.
- Strategy 2-out-of-3 requires that a majority of watchdogs expire before the system alarm is executed. Intuitively, this corresponds to a trade-off between the two above strategies.

More sophisticated strategies, corresponding to other design requirements, may also be composed. Other schemes, such as **meta-watchdogs** (watchdogs watching other watchdogs) can also be straightforwardly set up.

# 5 Cactus

Cactus is a design framework and runtime system for implementing and executing configurable services based on an event-driven execution model. As mentioned in Chapter 2, configurable services such as configurable communication protocols, file systems, database systems, and middleware are a useful approach to explicitate the system and fault model requirements and a useful tool to reach truly portable *services*, that is systems characterized by dynamic and adaptive system and fault models. One such system is the Configurable Transport Protocol.
The Cactus programming model is a clever improvement on the classical layered system design principle: Systems have a macroscopic structure, that is, the traditional layers, called services in the Cactus terminology, and a microscopic structure, that is a web of modules representing the atoms constituting the services and connected non-hierarchically into a network of so-called "micro-protocols". Non-hierarchically means here that there is no communication restriction between modules, which can exchange information and cooperate regardless their macroscopic position (that is, even though the corresponding services are not contiguous to each other). This structuring technique is said to provide the designer with a model coupling enhanced flexibility with the required separation of design concerns.
In Cactus micro-protocols communicate through an even-driven execution model. In the above mentioned Configurable Transport Protocol, for instance, all service properties and functional components are implemented as micro-protocols, each of which is a list of event handlers in the familiar form of **guarded actions**, i.e., statements structured like in

$$g : a,$$

where $g$ is an event stating a significant change such as the arrival of a message and $a$ is one or more event handler assocuated to the guard $g$. When $g$ occurs, $a$ is executed. This is clearly the same principle used in the recovery language Ariel discussed in Chapter 6.

The Cactus designers did not choose to develop a custom event-based programming language such as Ariel to specify its guarded actions—instead, they preferred to come up with a set of conventions and a software library. This allowed produce Cactus implementations in several languages, such as C and Java.

As explained in (Hiltunen, Taïani, & Schlichting, 2006), Cactus has been used to build services ranging from low-level communication protocols to group membership and group remote procedure call. Application-level services such as a configurable distributed system monitoring service have also been developed with Cactus.

The clever design choices of Cactus place it in between a system like Ariel and library-based approaches; moreover, several similarities between Cactus and aspect-oriented programming have been pointed out in (Hiltunen et al., 2006).

One concludes by assessing SC as bad (the system requires the explicit coding and intertwining of the protocols), SA as limited (by Condor's model and structure), and A as potentially good (though dynamic trading of event handlers, e.g., appears not to be part of Condor.)

# 6 CONCLUSIONS

The reported experiences demonstrate how $\mathcal{RεL}$ allows fast-prototype complex strategies by composing a set of building blocks together out of those available in the TIRAN framework, and by building *system-wide, recovery-time coordination strategies* with ARIEL. This allows set up sophisticated fault-tolerance systems while keeping the management of their complexity outside the user application. The compact size of the ARIEL configuration scripts is one of the arguments that can be used as evidence to this claim. Transparency of replication and transparency of location are also reached in this case. No similar support is provided by the application-level fault-tolerance approaches reviewed in Chapters 4–6.

As a final remark let us observe how SC is guaranteed here by a custom, separate composition language (ARIEL), that SA is limited to the single class of addressed provisions, and that the possibility to trade off strategies or compose new ones at run-time brings to an excellent degree of A.

# Notes

[1]A read-only tuple is one that can only be read via either rd() or rdp().

[2]The rationale behind this decision is that, this way, the user cannot produce mock error notifications, which would increase the risk of Byzantine failures (Lamport, Shostak, & Pease, 1982).

[3]If available, error detection support at driver or kernel level may be also instrumented so to emit ENs.

[4]This way, also software design faults would be addressed.

[5]In this case study, three instances of the same software component have been used. Clearly this does not protect the system from *design faults in the watchdog component itself*. Using NVP when developing the watchdogs may possibly guarantee statistical independence between failures.

page

# MEASURING AND ASSESSING TOOLS

## 1 INTRODUCTION AND OBJECTIVES

As mentioned in Chapter 1, a service's dependability must be justified in a quantitative way and proved through extensive on-field testing and fault injection, verification and validation techniques, simulation, source-code instrumentation, monitoring, and debugging. An exhaustive treatment of all these techniques falls outside the scope of this book, nevertheless the author feels important to include in this text an analysis of the effect on dependability of some of the methods that have been introduced in previous chapters.

## 2 RELIABILITY ANALYSIS OF THE TIRAN DISTRIBUTED VOTING MECHANISM

As mentioned in Chapter 9, a number of applications are structured in such a way as to be straightforwardly embedded in a fault-tolerance architecture based on redundancy and consensus. Applications belonging to this class are, for instance, parallel airborne and spaceborne applications. The TIRAN Distributed Voting mechanism provides application-level support to these applications. This section analyses the effect on reliability of the enhancements to the TIRAN Distributed Voting mechanism described in the mentioned chapter, that is, management of spares, dealt with in Sect. 2.1, and fault-specific recovery strategies supported by the $\alpha$-count feature, analysed in Sect. 2.2.

### 2.1 Using ARIEL to Manage Spares

This section analyses the influence of one of the features offered by ARIEL—its ability to manage spare modules in $N$-modular redundant systems—that has been introduced and discussed in Chapter 6 and Chapter 9.
Reliability can be greatly improved by this technique. Let us first consider the following equation:

$$R^{(0)}(t) = 3R(t)^2 - 2R(t)^3, \qquad (1)$$

i.e., the equation expressing the reliability of a TMR system with no spares, $R(t)$ being the reliability of a single, non-replicated (simplex) component. Equation (1) can be derived for instance via Markovian reliability modeling under the assumption of independence between the occurrence of faults (Johnson, 1989).
With the same technique and under the same hypothesis it is possible to show that, even in the case of non-perfect error detection coverage, this equation can be considerably improved by adding one spare. This is the equation resulting from the Markov model in Fig. 1, expressed as a function of error recovery coverage ($C$, defined as the probability associated with the process of identifying the failed module out of those available and being able to switch in the spare (Johnson, 1989)) and time ($t$):

$$R^{(1)}(C,t) = (-3C^2 + 6C) \times [R(t)(1 - R(t))]^2 + R^{(0)}(t). \tag{2}$$

Appendix A gives some mathematical details on Eq. (2).

Adding more spares obviously implies further improving reliability. In general, for any $N \geq 3$, it is possible to consider a class of monotonically increasing reliability functions,

$$(R^{(M)}(C,t))_{M>0}, \tag{3}$$

corresponding to systems adopting $N + M$ replicas. Depending on both cost and reliability requirements, the user can choose the most-suited values for $M$ and $N$.
Note how quantity (2) is always greater than quantity (1) as $R^{(0)}(t)$ and $(-3C^2 + 6C)$ are always positive for $0 < C \leq 1$. Figure 2 compares Eq. (1) and (2) in the general case while Fig. 3 covers the case of perfect coverage. In the latter case, the reliability of a single, non-redundant (simplex) system is also portrayed. Note furthermore how the crosspoint between the three-and-one-spare system and the non-redundant system is considerably lower than the crosspoint between the latter and the TMR system—$R(t) \approx 0.2324$ vs. $R(t) = 0.5$.
The reliability of the system can therefore be increased from the one of a pure NMR system to that of $N$-and-$M$-spare systems (see Fig. 2 and Fig. 3).

## 2.2 Using the $\alpha$-count Feature

As it has been already mentioned, the `TRANSIENT` clause of ARIEL, exploiting the $\alpha$-count fault identification mechanism introduced in Chapter 6, can be adopted to associate different recovery or reconfiguration strategies:

- To the detection of a permanent or intermittent fault, in general requiring reconfiguration and redundancy exhaustion,
- and to the detection of a transient fault, that could be tolerated with a recovery technique such as resetting the faulty component.

This section describes a simple Markov reliability model of a TMR system discriminating between these two cases. Figure 4 shows the model. $T$ is the probability that the current fault is a transient one and $R$ is the probability of successful recovery for a component affected by a transient fault. To simplify the model, recovery is assumed to be instantaneous. The system degrades only when permanent faults occur. Solving the model brings to the following equation:

$$R^{\alpha}_{\text{TMR}}(t) = 3 \exp^{-2(1-RT)\lambda t} - 2 \exp^{-3(1-RT)\lambda t}. \tag{4}$$

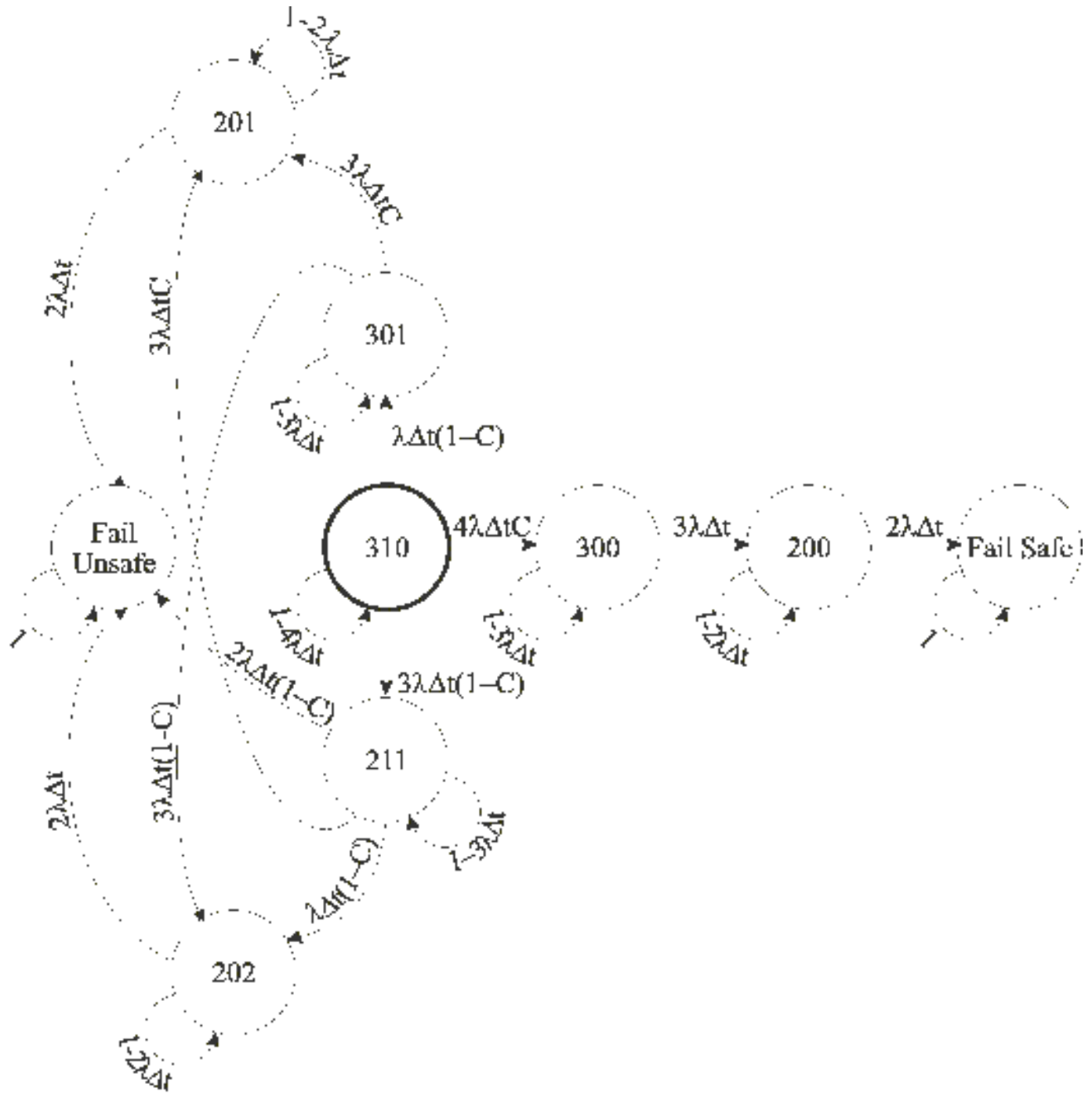

Figure 1: Markov reliability model for a TMR-and-1-spare system. $\lambda$ is the failure rate, $C$ is the error recovery coverage factor. A "fail safe" state is reached when the system is no more able to correctly perform its function, though the problem has been safely detected and handled properly. In 'Fail unsafe,' on the contrary, the system is incorrect, though the problem has not been handled or detected. Every other state is labeled with three digits, $d_1d_2d_3$, such that $d_1$ is the number of non-faulty modules in the TMR system, $d_2$ is the number of non-faulty spares (in this case, 0 or 1), and $d_3$ is the number of undetected, faulty modules. The initial state, 310, has been highlighted. This model is solved by Eq. (2).

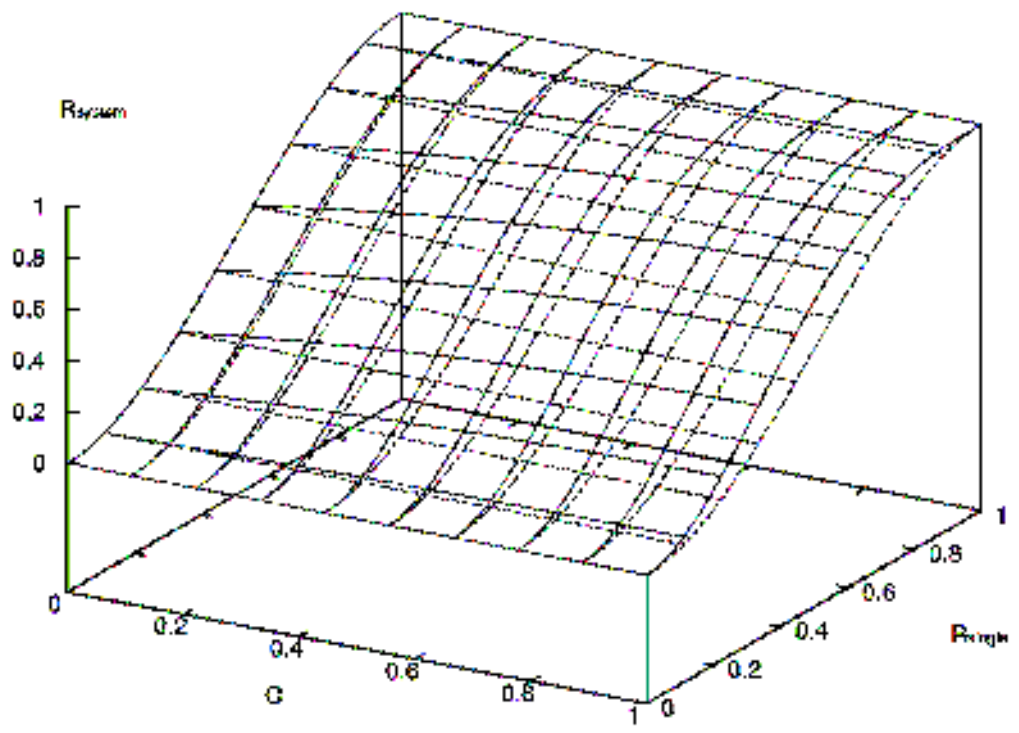

Figure 2: Graphs of Eq. (1) and (2) as functions of $C$ and of $R$. Note how the graph of (2) is strictly above the other.

Eq. 4 can be written as

$$R^{\alpha}_{\text{TMR}}(t) = 3R(t)^{2(1-RT)} - 2R(t)^{3(1-RT)}, \tag{5}$$

with $R(t)$ the reliability of the simplex system. Note how, in a sense, the introduction of the $\alpha$-count technique results in a modification of the exponents of Eq. 1 by a factor equal to $1 - RT$.

In general, the resulting system shows a reliability that is larger than the one of a TMR system. Figure 5 compares these reliabilities. As it is evident from that image, the crosspoint between the reliability graphs of the simplex and that of the TMR system extended with the $\alpha$-count technique is in general lower. Its exact value is given by the function

$$\text{crosspoint}(R, T) = 0.5^{1/(1-RT)},$$

which is drawn in Fig. 6.

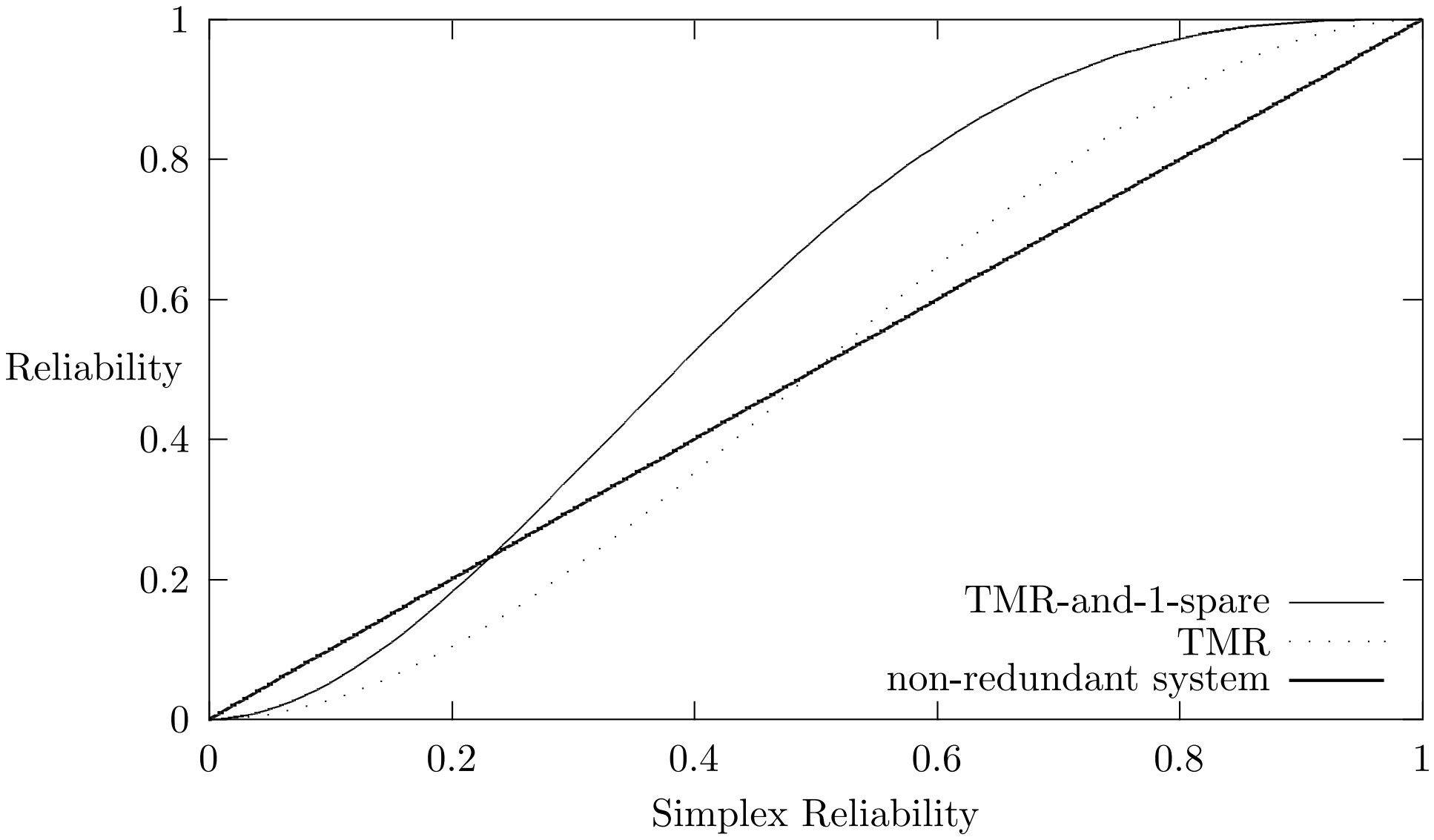

Figure 3: Graphs of Eq. (1) and (2) when $C = 1$ (perfect error detection coverage). The reliability of a single, non-redundant system is also portrayed.

# 3 PERFORMANCE ANALYSIS OF REDUNDANT VARIABLES

In Chapter 4 the concept of redundant variables was introduced: Data structures in which the degree of replication is not fixed once and for all at design time, but changes dynamically with respect to the disturbances experienced during the run time. Here it is shown how the performance of our approach was analyzed.

In order to analyze our system, a simulator, called "scrambler" was developed. Our scrambler tool allows to simulate a memory, to protect it with redundant variables, to inject memory faults (bit flips or "bursts" corrupting series of contiguous cells), and to measure the amount of redundancy actually used. Scrambler interprets a simple scripting language consisting of the following commands:

**SLEEP** $s$ , which suspends the execution of the scrambler for $s$ seconds,

**SCRAMBLE** $n, p$ , which repeats $n$ times action "scramble a pseudo-random memory cell with probability $p$",

**BURST** $n, p, l$ , which repeats $n$ times action "scramble $l$ contiguous cells with probability $p$",

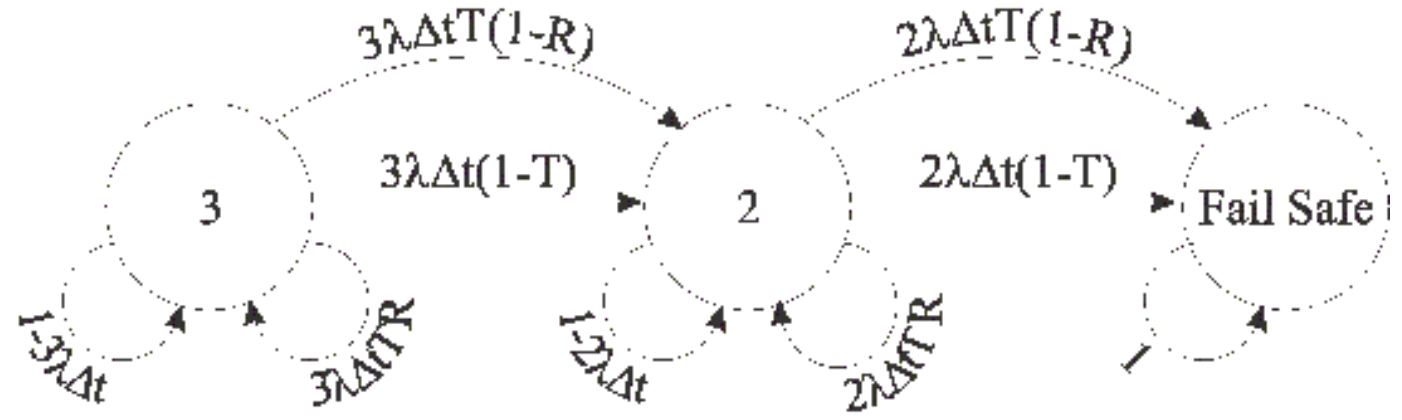

Figure 4: The Markov reliability model of a TMR system that gracefully degrades in the presence of a permanent or intermittent fault and "cleans up" (that is, rejuvenates) the components affected by transient faults.

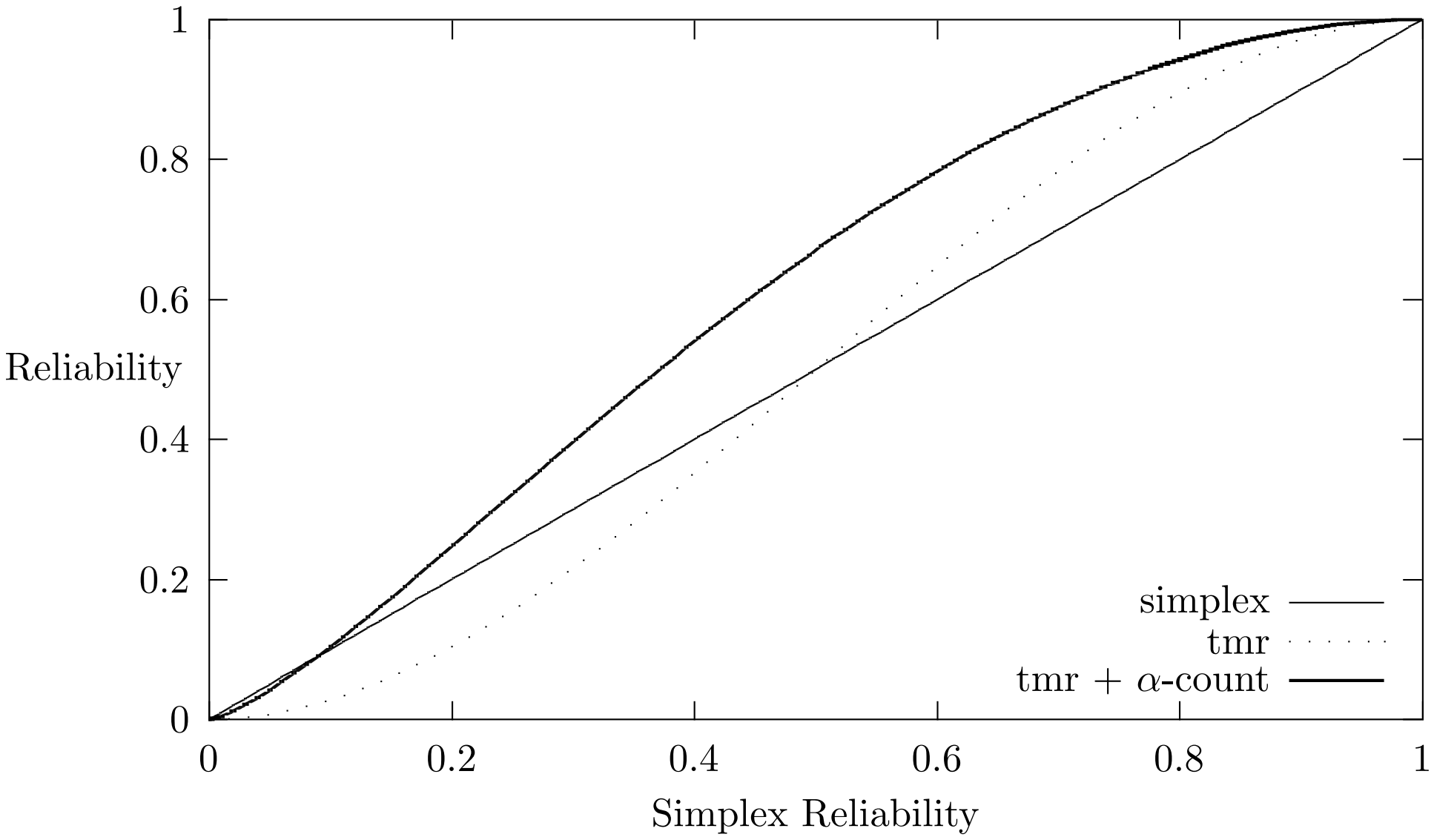

Figure 5: Reliability of a simplex system, of a TMR system, and of a modified TMR system exploiting the $\alpha$-count fault identification technique. In this case $T$ is 50% and $R = 0.6$.

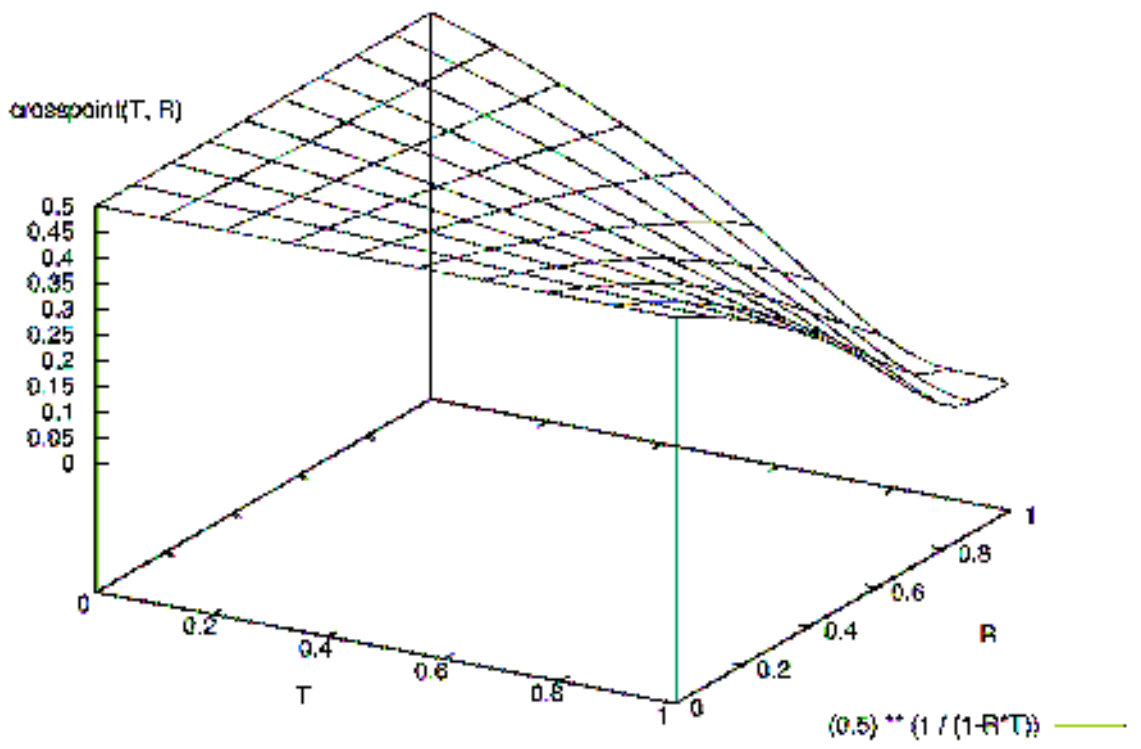

Figure 6: The crosspoint between the graph of the reliability of the simplex system and that of $R^{\alpha}_{\text{TMR}}$ as a function of $R$ and $T$.

**END** , which terminates the simulation.

The above commands can be used to compose a complex sequence of fault injections. As an example, the following script, corresponds to the following configuration: no faults for 1 second, then various disturbances occurring with Gaussian distribution[1], then no disturbances again for 5 seconds:

```
SLEEP 1                  // sleep 1 sec
SCRAMBLE 2000, 0.1053992 // scramble 2000 random cells
                         // with probability f(-3)
SCRAMBLE 2000, 0.3678794 // scramble 2000 random cells
                         // with probability f(-2)
BURST 2000, 0.7788008, 10// execute 2000 bursts of 10
                         // contiguous cells
                         // with probability f(-1)
SCRAMBLE 2000, 1         // scramble 2000 random cells
BURST 2000, 0.7788008, 10// execute 2000 bursts of 10
                         // contiguous cells
                         // with probability f(1)
SCRAMBLE 2000, 0.3678794 // scramble 2000 random cells
                         // with probability f(2)
SCRAMBLE 2000, 0.1053992 // scramble 2000 random cells
                         // with probability f(3)
SLEEP 5                  // sleep 5 secs
END                      // stop injecting faults
```

The idea behind these scripts is to be able to represent executions where a program is subjected to environmental conditions that vary with time and range from ideal to heavily perturbed. Scenarios like these ones are common, e.g., in applications servicing primary substation automation systems (Deconinck, De Florio, Dondossola, & Szanto, 2003) or spaceborne applications (Oey & Teitelbaum, 1981).

In the following a few experiments are described, together with the results obtained with scrambler.

All our experiments have been carried out with an array of 20000 redundant 4-byte cells and an allocation stride of 20 (that is, replicas of a same logical cell are spaced by 20 physical cells). In all the reported experiments the following script was run:

```
SLEEP 1
SCRAMBLE 10000, 0.9183156388887342
SCRAMBLE 10000, 0.9183156388887342
SLEEP 3
SCRAMBLE 10000, 0.9183156388887342
SCRAMBLE 10000, 0.9183156388887342
END
```

Concurrently with the execution of this script, 1500000 read accesses were performed in round robin across the array. The experiments record the number of scrambled cells and the number of read failures.

Scrambler makes use of standard C function “rand”, which depends on an initial seed to generate each pseudo-random sequence. In the reported experiments the same value has been kept for the seed, so as to produce exactly the same sequences in each experiment.

### 3.1 Experiment 1: Fixed, low redundancy

In this first experiment scrambler was executed with fixed (non adaptive) redundancy 3. Table 1 shows the setting of this experiment. The main conclusion one can draw from this run is that a statically redundant data structures provision in this case fails 193 times: In other words, for 193 times it was not possible to find a majority of replicas in agreement, and the system reported a read access failure. The total number of memory accesses is proportional to $3 \times 1500000 \times k$, where $k > 0$ depends on the complexity of the redundant read operation.

### 3.2 Experiment 2: Fixed, higher redundancy

Also experiment 2 has fixed redundancy, this time equal to 5. Table 2 shows the setting of this new experiment. Main conclusion is that the higher redundancy is enough to guarantee data integrity in this case: No read access failures are experienced. The total number of memory accesses is proportional to $5 \times 1500000 \times k$.

**$ scrambler faults.in 3 scrub noadaptive**

```
Scrambler::sleep(1)
run 1
run 50001
run 100001
run 150001
Scrambler::scramble(10000,0.918316)
Scrambler::scramble(10000,0.918316)
Scrambler::sleep(3)
run 200001
run 250001
```

*. . . lines omitted. . .*

```
run 650001
Scrambler::scramble(10000,0.918316)
Scrambler::scramble(10000,0.918316)
Scrambler::END
run 700001
run 750001
```

*. . . lines omitted. . .*

```
run 1500001
36734 scrambled cells, 193 failures, redundance == 3
redundance 3:  1500001 runs
redundance 5:  0 runs
redundance 7:  0 runs
redundance 9:  0 runs
redundance 11:  0 runs
```

Table 1: Experiment 1: Scrambler executes the script in file "faults.in". Parameters set redundancy to 3, select memory scrubbing to repair corrupted data when possible, and keep redundancy fixed.

```
$ scrambler faults.in 5 scrub noadaptive
Scrambler::sleep(1)
run 1
run 50001
run 100001
Scrambler::scramble(10000,0.918316)
Scrambler::scramble(10000,0.918316)
Scrambler::sleep(3)
run 150001
...lines omitted...
run 500001
Scrambler::scramble(10000,0.918316)
Scrambler::scramble(10000,0.918316)
Scrambler::END
run 550001
...lines omitted...
run 1500001
36734 scrambled cells, 0 failures, redundance == 5
redundance 3:  0 runs
redundance 5:  1500001 runs
redundance 7:  0 runs
redundance 9:  0 runs
redundance 11:  0 runs
```

Table 2: Experiment 2: Scrambler executes the same script as before; only, redundancy is now set to 5. No failures are observed.

```
$ scrambler faults.in 5 scrub adaptive
run 1
Scrambler::sleep(1)
run 50001
run 100001
run 150001
run 200001
Scrambler::scramble(10000,0.918316)
Scrambler::scramble(10000,0.918316)
Scrambler::sleep(3)
run 250001
...lines omitted...
run 600001
Scrambler::scramble(10000,0.918316)
Scrambler::scramble(10000,0.918316)
Scrambler::END
run 650001
...lines omitted...
run 1500001
36734 scrambled cells, 1 failures, redundance == 3
redundance 3:  1455404 runs
redundance 5:  6054 runs
redundance 7:  28967 runs
redundance 9:  9576 runs
redundance 11:  0 runs
```

Table 3: Experiment 3: Scrambler executes the same script as before; only, redundancy is now adaptive. No failures are observed, but the employed redundancy is mostly of degree 3.

### 3.3 Experiment 3: Adaptive Redundancy

In this last experiment adaptive redundancy was enabled and initially set to 5. Table 3 shows the resulting setting. Most worth noting is the fact that also in this case no read access failures show up, but the actual amount of redundancy required to reach this result is much lower. Consequently, also the total number of memory accesses, proportional to $(3 \times 1455404 + 5 \times 6054 + 7 \times 28967 + 9 \times 9576) \times k$, is considerably lower. Figure 7 shows how redundancy varies during the first 100000 read cycles. During this time frame no fault injection takes place. This is captured by our adaptation strategy, which decreases redundancy to 3 after 1000 cycles. Figure 8 depicts an interval where several fault injections do take place. These events are detected and trigger several adaptations.

### 3.4 Conclusions

The analysis of the performance of one of the application-level approaches introduced in Chapter 4, namely redundant variables, was described. The main focus here was on the methods and tools used to assess, in a quantitative way, the properties of our tool.

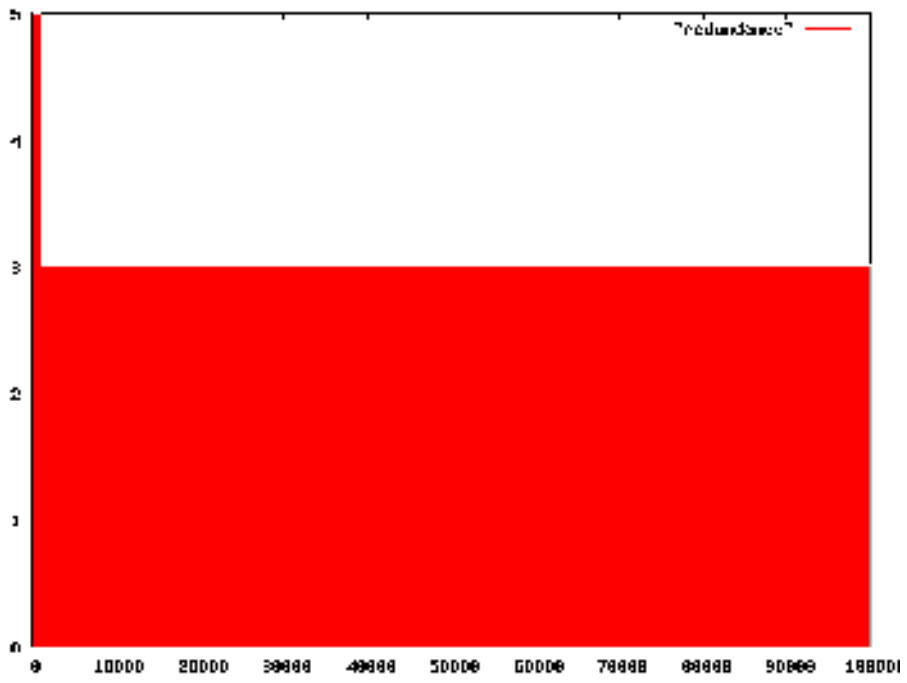

Figure 7: The first 100000 cycles of experiment 3 (Sect. 3.3). Note how redundancy drops to its minimum after the first 1000 cycles.

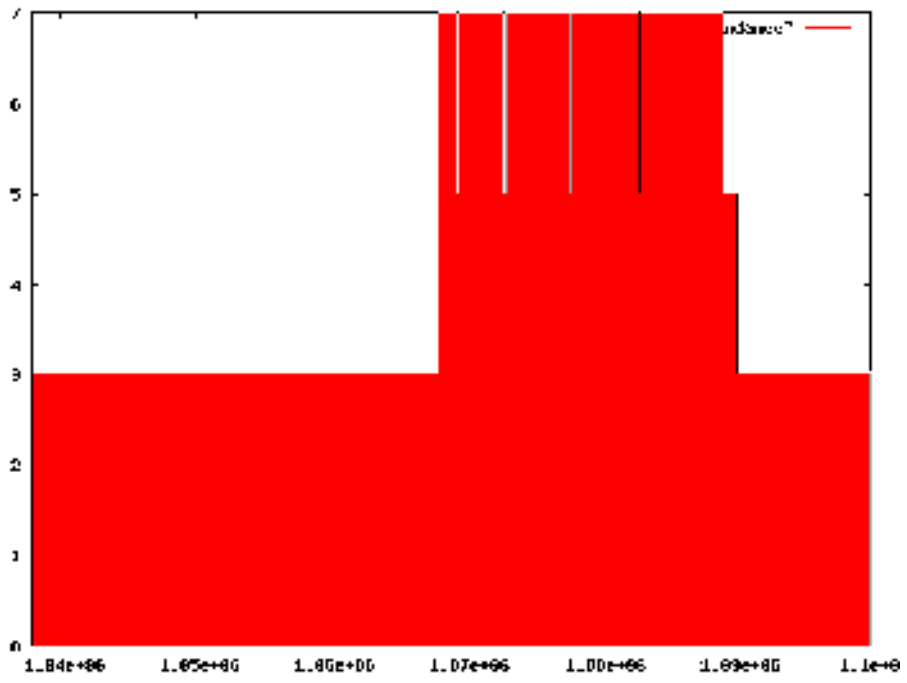

Figure 8: A section of experiment 3 in which fault injection takes place. Note how redundancy changes accordingly.

# 4 A TOOL FOR MONITORING AND FAULT INJECTION

To conclude this chapter a distributed, multimedia application for monitoring and fault injection is described. Such application was developed in the framework of Project EFTOS (Embedded Fault-Tolerant Supercomputing, already introduced in Chapter 3). It dynamically sets up a hierarchy of HTML pages reflecting the current status of an EFTOS-compliant dependable application running on a Parsytec CC system. These pages are fed to a WWW browser playing the role of hypermedia monitor. The adopted approach allows the user to concentrate on the high-level aspects of his/her application so as to quickly assess the validity of their dependability assumptions. This view of the system lends itself well for being coupled with a tool to interactively inject software faults in the user application.

## 4.1 Introduction

As systems get more and more complex, the need for a one-look snapshot of their activity is indeed ever increasing. This need has been strongly felt by the people involved in Project EFTOS (Deconinck, De Florio, Lauwereins, & Varvarigou, 1997), whose aim was to set up a software framework for integrating fault-tolerance into embedded distributed high-performance applications in a flexible and easy way.
Through the adoption of the EFTOS framework, the target application running on a parallel computer is plugged into a hierarchical, layered system whose structure and basic components have been described in detail in Chapter 3:

- at the lowest level, a set of parameterizable functions managing error detection (Dtools) and error recovery (Rtools). EFTOS supplies a number of these Dtools and Rtools, plus an API for incorporating user-defined EFTOS-compliant tools;
- at the middle level, the DIR net (detection, isolation, and recovery network) is available to coherently combine Dtools and Rtools, to ensure consistent strategies throughout the whole system, and to play the role of a backbone handling information to and from the fault-tolerance elements. To fulfill these requirements, the DIR net makes use of processes called Manager, Agents, and Backup Agents;
- at the highest level, these elements are combined into dependable mechanisms e.g., methods to guarantee fault-tolerant communication, voting methods and so on.

As explained in Chapter 3, during the lifetime of the application the EFTOS framework guards the service from a series of possible deviations from the expected behavior; this is done by executing detection, isolation, and reconfiguration tasks. For instance, a memory access exception caught in a thread by a trap handling Dtool may trigger a relocation of that thread elsewhere in the system. As another example, if an error is detected which affects a component of the DIR net itself, say a Manager, then the system will isolate that component and elect another one (actually, one of the Backup Agents) as the DIR Manager.
In order to allow the user to keep track of events such as those sketched above, the DIR net continuously prints on the system console short textual descriptions. Evidently such a linear, unstructured listing of events pertaining to different aspects of different actions taking place in different points of the user application, does not make up the best of mechanisms to gain insight in the overall state of the fault-tolerant system. On the contrary, a hierarchical, dynamic view of the structure and behavior of that system, including:

- Its current shape (on which node which components are running, and their topology),
- the current state of its components (for instance, whether they are regarded to be correct, faulty, or are being recovered),
- each component's running history,

appeared much more attractive.
Two main advantages from the adoption of such a strategy were expected to be:

- (At design and system validation time) the possibility to assist the user assessing and/or validating his/her EFTOS-based fault-tolerance design,
- (at run time) the possibility to shorten the latency between the occurrence of the event, its comprehension, and a proper reaction at user level[2].

This text describes the architectural solution that was successfully adopted within EFTOS to easily and quickly develop a tool fulfilling the above stated needs—a portable, highly customizable hypermedia monitor for the EFTOS applications making use of cheap, ready available off-the-shelf software components like e.g., an Internet Browser. It also shows how our monitor supplies the user with the needed structured information, and how it proved its usefulness within EFTOS. In particular an extension is described by means of which our monitor can be turned into a versatile tool for fault-injection.

## 4.2 Design Requirements

In order to quickly deliver human-comprehensible information from the gigantic data stream produced by an EFTOS application, two needs have been assessed:

- Deriving a hierarchical representation of the data.

  (Most of the produced data is available, but it has to be organized and made browsable in “layers”:

  – At the highest level, only the logical structure of the application should be displayed: Which nodes are used, the EFTOS components executed on each node, and their overall state;

  – at a medium level, a concise description of the events pertaining to each particular node should be made available;

  – at the lowest level, a deeper description of each particular event may also be supplied on user-demand),

- the use of multimedia.

  “An image is worth a thousand words”, they say, and maybe even more insight can be derived from the extensive use of colors, sounds, video-clips and so on. For instance, re-coloring a green image to red may lead the user into realizing that a previously fault-free state has turned into a problem. The use of colors traditionally associated to meanings, or whose meaning can be borrowed from well-known everyday situations (e.g., those of traffic lights) further speeds up the delivery of the information to the user.

Both aspects were already available in Internet browsers available at the time of EFTOS, such as Netscape, which were able to render hierarchies of dynamically produced HTML (Berners-Lee & Connolly, 1995) pages. Therefore it was decided to develop a distributed application piloting an Internet browser that in turn plays the role of a hypermedia renderer for the EFTOS system activity. This product, called the EFTOS Monitor, is described in the following sections.

## 4.3 The Architecture of the EFTOS Monitor

The EFTOS Monitor basically consists of three components (see Fig. 9):

1. A *client* module, to be run by the DIR net;
2. an “*intermediate*” module, to be run by a number of Common Gateway Interface (Kim, 1996) (CGI) scripts;
3. the “*renderer*” i.e., an Internet browser.

- The client part, together with the DIR net and the user application, runs on a Parsytec CC system (Parsytec, 1996b), a distributed-memory MIMD supercomputer consisting of powerful processing nodes based on PowerPC 604 at 133MHz, dedicated high-speed links, I/O modules and routers. The system adopts the thread processing model; as mentioned in Chapter 3, threads communicate via a proprietary message passing library called EPX (Parsytec, 1996a) (*Embedded Parallel extensions to uniX*). The main tasks of the client module are the set up and management of a database maintaining an up-to-date snapshot of the system activity, including the current mapping of the DIR net's components onto the processing nodes and the state and current activity of each component. This module also connects to the intermediate part via TCP sockets (as described for instance in (Comer, 1993)) and sends it signals at the very beginning and on each state transition.

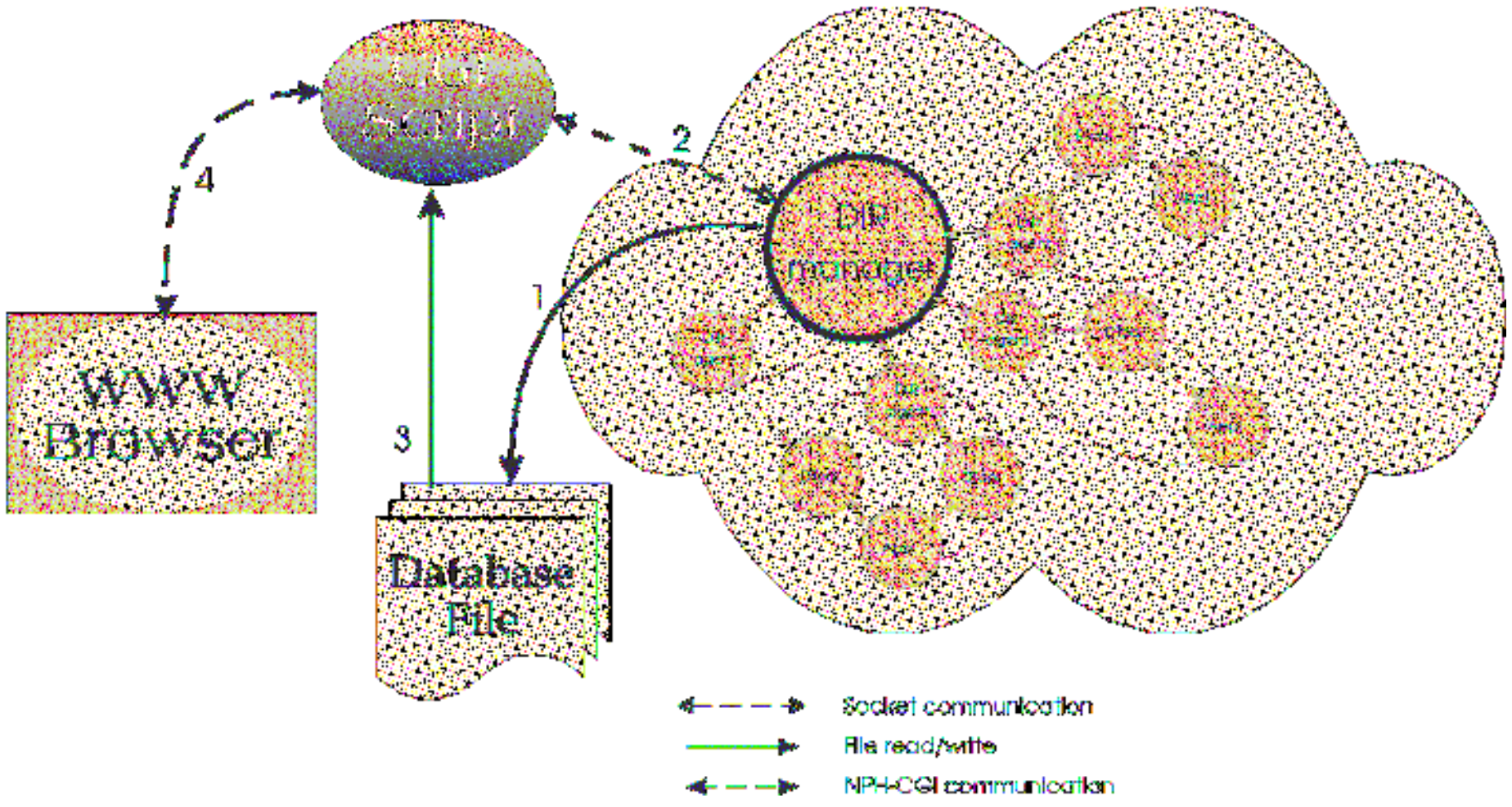

Figure 9: The architecture of the EFTOS Monitor: the CGI scripts and the EFTOS application share the same file system and communicate via a socket stream: Each time a new event takes place, the DIR net updates a special database (1) and sends a notification to the main CGI script (2). The latter reads the database (3) on the arrival of the notification and converts it into a HTML hypertext, which is then fed back (4) to a Netscape or another Internet browser for hypermedia rendering. The client part of the Monitor is integrated in the DIR net Manager process.

- The intermediate module consists of a hierarchy of CGI scripts spawned by an Apache HTTP (Berners-Lee, Fielding, & Frystyk, 1996; Fielding et al., 1997) server, all running on the workstation hosting the CC system. The root script of this hierarchy connects with the client module and acts as a TCP server: for each new stimulus, the snapshot file is read over and a HTML document is produced and fed to the renderer. A connection is also started up with this latter so as to be able to interact with it without the intervention of the HTTP daemon: Having done so, one CGI script may stay alive and produce multiple HTML requests. This special feature is known as "non-parsing header" (NPH) mode (Kim, 1996). Logically speaking, one may say that the intermediate module acts as a gateway between the CC system and the hypermedia renderer. Like mythical Janus (the god of gates, doors and doorways (Janus, n.d.); in Italian, *Giano*), one "face" of the system is turned to the client module and gathers its requests, while a second one is turned to the renderer and translates those requests in HTML. Because of this its main component has been called `cgiano`.

- The third component, the renderer, is a browser playing the role of a server able to display HTML data.

The application is started in two steps via a shell script whose first task is to run the renderer (or to reconnect to a previously run renderer: This latter is possible using e.g., remote control extensions (Zawinski, 1994) or an approach based on the Common Client Interface mechanism. The renderer is run with a uniform resource locator (URL) pointing to the root-level CGI script, which connects to Netscape in NPH-mode and starts listening for a TCP connection. As a second step, the shell script spawns the parallel application on the CC system. Then the application launches the DIR net and the Monitor's client module. The latter initializes the snapshot files, connects to the CGI script and sends it the first signal. The script reacts to that stimulus by translating the main snapshot file in HTML and requesting the renderer to display it. The top-left image in Fig. 10 shows a typical output of this phase: The EFTOS application appears to the user as an HTML table depicting the processing nodes in the user partition. The state of each module is illustrated by means of colors (green is "OK", red is "not OK", yellow means that the module is currently being recovered, and so on). In this way the user can immediately perceive whether a node is ready or not and which actions are carried out on it, as asked for in the requirements (Sect. 4.2.)
Information displayed in this HTML document only covers the logical structure of the application. If the user "clicks" any icon on this page, a high-level hypertextual description of the DIR net-events pertaining to that specific node is displayed in a separate browser (see Fig. 10, Window "Node-specific Information"). To keep this page up to date, an automatic reload is periodically performed. This technique is explained e.g., in (Kim, 1996).
This secondary document is in turn a hypertext whose links point to in-depth descriptions of each specific event (see Fig. 10, Window "Attached Information.")

### 4.4 Architecture Assessment

A number of observations may be drawn upon the above presented architecture; in particular:

- Our architecture is based on unmodified, low-cost, off-the-shelf hypermedia components.

- It is open, in the sense that the architecture is based on wide-spread standards e.g., the use of uniform resource locators within a World-Wide Web interconnection, the HTML language, TCP/IP sockets, the MIME classification, and so on;
- It is distributed, and in particular the renderer may run on any X11-compliant Display Server, including remote PCs running Cygwin/X11.

A possible alternative was to develop a custom application to play as a tailored monitoring tool. As an example, Scientific Computing Associates' TupleScope visual debugger is a custom X-based visualization and debugging tool for parallel programs using the LINDA approach introduced in Chapter 4. This may result in higher performance and possibly be more flexible but of course:

- It reasonably requires more time to develop even for a simple prototype;
- it requires custom design and development choices that may impact portability and supported features e.g., which software development environment and specifically which language and which libraries to use, or whether to restrict the hypermedia rendering to images or to use sounds as well—these choices may be simply skipped in our approach;
- distribution and hypermedia issues call for specific support which turn into higher costs and longer times.

For instance, TupleScope runs with the user application by adding a special linking option at compile time to the user application; this means it has been developed on purpose as a custom X11 application. Though it perfectly addresses its own goals, it has limited rendering capabilities (it only deals with static images) and would certainly require non-negligible efforts to adapt it towards other media. Moreover, TupleScope is available on a number of platform, though the costs of this portability and consequent support are certainly not negligible as well.

## 4.5 A Tool for Fault Injection

The same approach used to monitor the state of an EFTOS-compliant application is also effective in order to actively interact with it. Considering once again Fig. 9, a control path may be drawn starting from the user at his/her browser, then crossing a CGI script, and eventually reaching the user application. It is therefore fairly possible to add a layer to the hierarchy of HTML pages dynamically created by the intermediate modules so that the user may freely choose among a certain set of malicious actions to bring against an EFTOS application, including for instance:

- An integer division-by-zero,
- a segmentation violation,

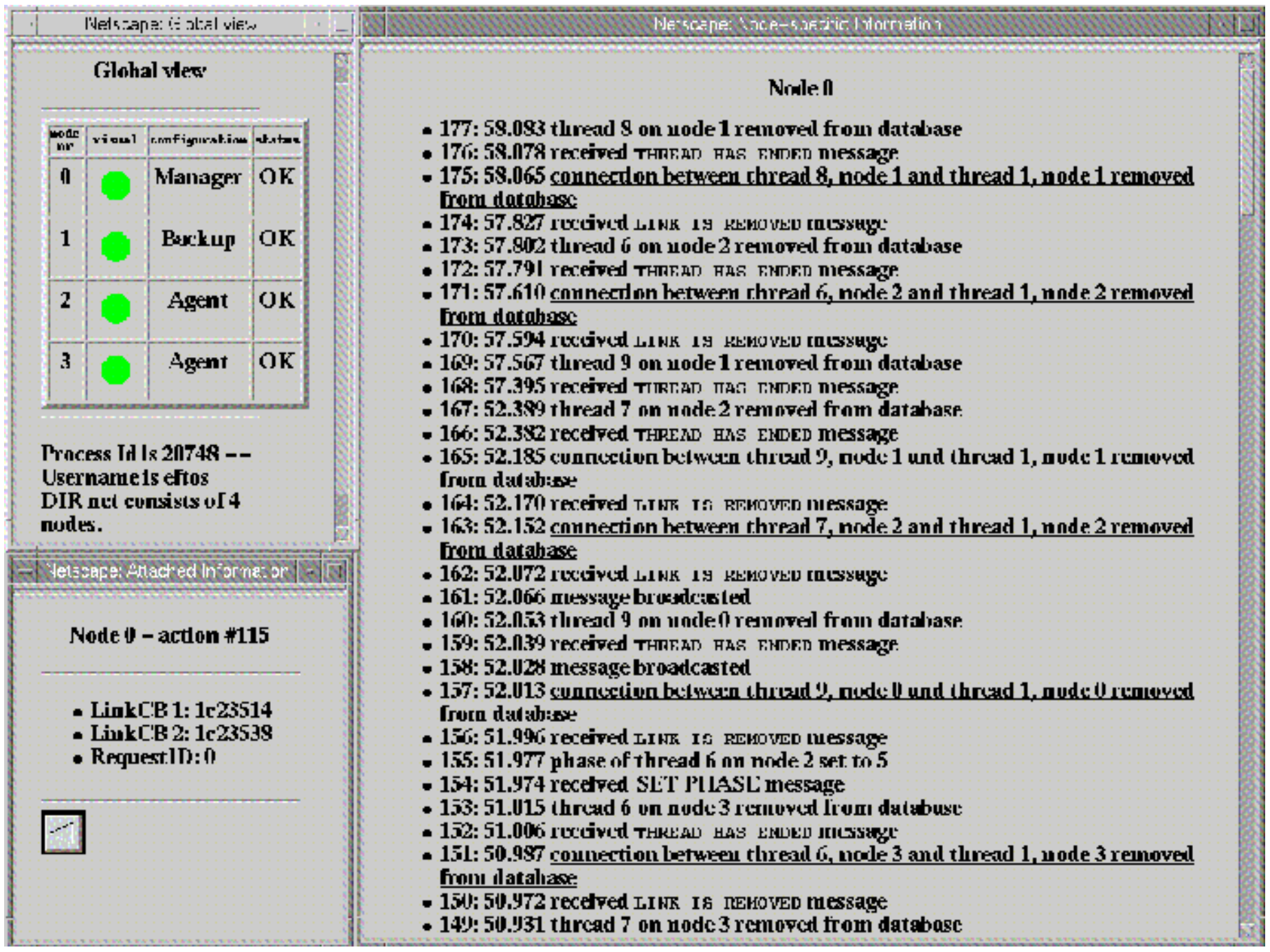

Figure 10: The three windows of the EFTOS Monitor. In window 1 ("Global view"), the `visual` column contains graphical hyperlinks pointing to second-level information about the corresponding processing node at the same row. `Configuration` is the DIR net-role. `Status` may be one of the following values: `OK`, `Faulty`, `Isolated`, `Recovering`, and `Killed`. Some minor information is also displayed at the bottom of the page. The right-hand hypertext (window "Node-specific information") is the result of "clicking" on the top circular icon and enumerates the actions that have just taken place on node 0, fresher-to-older. The elapsed time (in seconds) corresponding to each event is displayed. Underlined sentences may be further expanded by clicking on them e.g., the bottom-left image reports about action number 115 of the hypertext.

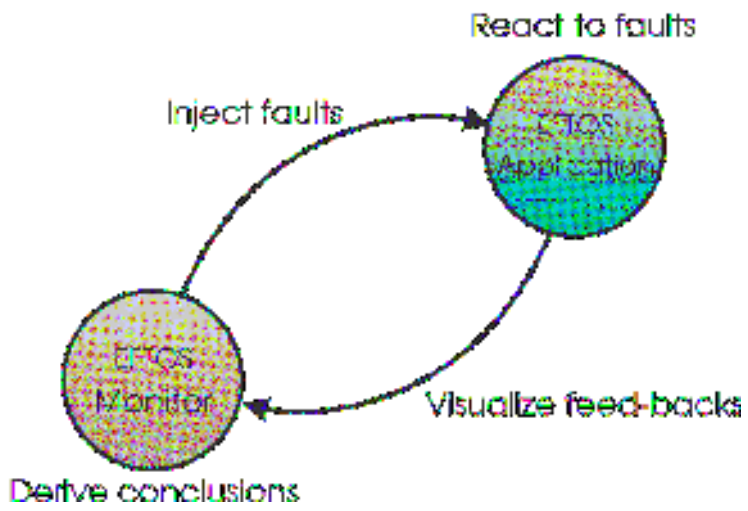

Figure 11: The recursive loop of fault injection and monitoring.

- a link failure,
- rebooting a processing node,
- killing a thread.

These requests would then reach a CGI script, be translated in appropriate system- or application-level actions, which would then be executed or turned into fault-injection requests to be fulfilled by the DIR Manager. As an example of system-level action, the CGI script may directly execute a system command to reboot one node in the CC system. As of application-level actions, the Manager may for instance ask the trap handler tool (see Chapter 3) to trigger a specific signal like `SIGSEGV` (segmentation violation) on a certain thread; or it may request a watchdog timer tool (again, described in Chapter 3) on a particular node to behave like if it had detected a time-out. As a direct consequence of the injection of these faults, a number of detection, isolation, and recovery actions take place on the system according to the EFTOS-based fault-tolerance strategies adopted by the designer in his/her application. These actions will then be reported in the snapshot files and displayed by the Monitor. This process, summarized in Fig. 11, may be modeled after a recursive loop like follows:

---

```
do {
    Inject fault;
    Observe feed-back;
    Derive conclusions;
    Correct the fault tolerance model;
} while (model is unsatisfying).
```

---

This procedure should result in a useful tool for rapidly assessing a design, trying alternative fault-tolerance strategies, and overloading the system with malicious attacks aiming at verifying its resilience, with a quick and meaningful feed-back from the system.

## 4.6 Conclusions

A distributed application has been presented, for monitoring the fault tolerance aspects of an embedded parallel application and for interactively injecting faults into it. The overall system makes up an integrated environment in which

- the application,
- a graphical rendering of the results, and
- real-time interactions

cyclically evolve. By doing so the researcher is made able to verify the hypothesis he/she is formulating about the system.

The design choice to adopt low-cost, off-the-shelf components for hypermedia rendering revealed to be cost-effective, speed up the development process, match the design requirements, and pave the way towards more ambitious capabilities and features. In particular, the use of an Internet browser as hypermedia renderer allows inherit the benefits of the volcanic evolutions of web services, HTML languages, HTTP protocol, multimedia capabilities of the browsers, and so on.

The high degree of openness proven by this heterogeneous application basing itself on uniform communication mechanisms and standardized access interfaces guarantees portability and makes it also a good starting point towards the development of similarly structured applications ranging from remote equipment control to hypermedia multi-user environments.

The deeper insight gained from the EFTOS Monitor on the run-time aspects of our dependable applications has turned it into an invaluable tool to speed up the development of our application-level fault-tolerance provisions.

# 5 CONCLUSION

We have described three examples of approaches used to assess the dependability of application-level provisions: Reliability analysis is used to quantify the benefits of coupling an approach such as recovery languages to a distributed voting mechanism (De Florio, Deconinck, & Lauwereins, 1998); a custom tool is used to systematically inject faults onto the adaptively redundant data structure discussed in Chapter 4 (De Florio & Blondia, 2008); a hypermedia application to watch and control a dependable service is then used for monitoring and fault-injection (De Florio, Deconinck, Truyens, Rosseel, & Lauwereins, 1998).

# A MATHEMATICAL DETAILS RELATED TO EQ. 2

The basic steps leading to Eq. 2, i.e.,

$$R^{(1)}(C,t) = (-3C^2 + 6C) \times [R(t)(1-R(t))]^2 + R^{(0)}(t)$$

are described in what follows.

The Markov reliability model of Fig. 1 brings to the following set of equations:

$$\left\{ \begin{array}{lcl} p_{310}(t+\Delta t) & = & p_{310}(t)(1-4\lambda\Delta t) \\ p_{300}(t+\Delta t) & = & p_{300}(t)(1-3\lambda\Delta t) + p_{310}(t)4\lambda\Delta t C \\ p_{200}(t+\Delta t) & = & p_{200}(t)(1-2\lambda\Delta t) + p_{300}(t)3\lambda\Delta t \\ p_{\mathrm{FS}}(t+\Delta t) & = & p_{\mathrm{FS}}(t) + p_{200}(t)2\lambda\Delta t \\ p_{211}(t+\Delta t) & = & p_{211}(t)(1-3\lambda\Delta t) + p_{310}(t)3\lambda\Delta t(1-C) \\ p_{301}(t+\Delta t) & = & p_{301}(t)(1-3\lambda\Delta t) + p_{310}(t)\lambda\Delta t(1-C) \\ p_{201}(t+\Delta t) & = & p_{201}(t)(1-2\lambda\Delta t) + p_{301}(t)3\lambda\Delta t C + \\ & & p_{211}(t)3\lambda\Delta t C \\ p_{202}(t+\Delta t) & = & p_{202}(t)(1-2\lambda\Delta t) + p_{301}(t)3\lambda\Delta t(1-C) + \\ & & p_{211}(t)\lambda\Delta t(1-C) \\ p_{\mathrm{FU}}(t+\Delta t) & = & p_{\mathrm{FU}}(t) + p_{201}(t)2\lambda\Delta t + \\ & & p_{211}(t)2\lambda\Delta t(1-C) + p_{202}(t)2\lambda\Delta t. \end{array} \right.$$

The above equations can be written as follows:

$$\left\{ \begin{array}{lcl} \frac{dp_{310}(t+\Delta t)}{dt} & = & -4\lambda p_{310}(t) \\ \frac{dp_{300}(t+\Delta t)}{dt} & = & -3\lambda p_{300}(t) + 4\lambda C p_{310} \\ \frac{dp_{200}(t+\Delta t)}{dt} & = & -2\lambda p_{200}(t) + 3\lambda p_{300}(t) \\ \frac{dp_{\mathrm{FS}}(t+\Delta t)}{dt} & = & 2\lambda p_{200}(t) \\ \frac{dp_{211}(t+\Delta t)}{dt} & = & -3\lambda p_{211}(t) + 3\lambda(1-C)p_{310}(t) \\ \frac{dp_{301}(t+\Delta t)}{dt} & = & -3\lambda p_{301}(t) + \lambda(1-C)p_{310}(t) \\ \frac{dp_{201}(t+\Delta t)}{dt} & = & -2\lambda p_{201}(t) + 3\lambda C p_{301}(t) + 3\lambda C p_{211}(t) \\ \frac{dp_{202}(t+\Delta t)}{dt} & = & -2\lambda p_{202}(t) + \lambda(1-C)p_{211}(t) + 3\lambda(1-C)p_{301}(t) \\ \frac{dp_{\mathrm{FU}}(t+\Delta t)}{dt} & = & 2\lambda p_{202}(t) + 2\lambda p_{201}(t) + 2\lambda(1-C)p_{211}(t). \end{array} \right.$$

For any state $s$, let us now call $L_s = L(p_s(t))$, where $L$ is the Laplace transform. Furthermore, as (310) is the initial state, it is reasonable to assume that $p_{310}(0) = 1$ and $\forall s \neq (310) : p_s(0) = 0$. Then taking the limit of the above equations as $\Delta t$ goes to zero and taking the Laplace transform brings to

$$\left\{ \begin{array}{lcl} L_{310} & = & \frac{1}{s+4\lambda} \\ L_{300} & = & \frac{4C}{s+3\lambda} - \frac{4C}{s+4\lambda} \\ L_{200} & = & \frac{6C}{s+4\lambda} - \frac{12C}{s+3\lambda} + \frac{6C}{s+2\lambda} \\ L_{\mathrm{FS}} & = & \frac{C}{s} - \frac{3C}{s+4\lambda} + \frac{8C}{s+3\lambda} - \frac{6C}{s+2\lambda} \\ L_{211} & = & \frac{3(1-C)}{s+3\lambda} - \frac{(3(1-C)}{s+4\lambda} \\ L_{301} & = & \frac{1-C}{s+3\lambda} - \frac{1-C}{s+4\lambda} \\ L_{201} & = & 6C(1-C)(\frac{1}{s+4\lambda} - \frac{2}{s+3\lambda} + \frac{1}{s+2\lambda}) \\ L_{202} & = & 3(1-C)^2(\frac{1}{s+4\lambda} - \frac{2}{s+3\lambda} + \frac{1}{s+2\lambda}). \end{array} \right.$$

Inverting the Laplace transform, the following probabilities can be found:

$$\left\{ \begin{array}{lcl} p_{310}(t) & = & \exp^{-4\lambda t} \\ p_{300}(t) & = & 4C\exp^{-3\lambda t} - 4C\exp^{-4\lambda t} \\ p_{200}(t) & = & 6C\exp^{-4\lambda t} - 12C\exp^{-3\lambda t} + 6C\exp^{-2\lambda t} \\ p_{211}(t) & = & 3(1-C)\exp^{-3\lambda t} - 3(1-C)\exp^{-4\lambda t} \\ p_{301}(t) & = & (1-C)\exp^{-3\lambda t} - (1-C)\exp^{-4\lambda t} \\ p_{201}(t) & = & 6C(1-C)(\exp^{-4\lambda t} - 2\exp^{-3\lambda t} + \exp^{-2\lambda t}) \\ p_{202}(t) & = & 3(1-C)^2(\exp^{-4\lambda t} - 2\exp^{-3\lambda t} + \exp^{-2\lambda t}) \end{array} \right.$$

(only useful states have been computed).

Let us denote with $R$ the reliability of the basic component of the system, i.e., $\exp^{-\lambda t}$, and $R_{\mathrm{TMR}}$ as the reliability of the TMR system based on the same component. The reliability of the three and one spare system, $R^{(1)}(C,t)$, is given by the sum of the above probabilities:

$$\begin{aligned} R^{(1)}(C,t) &= R^4(-3C^2+6C)+R^3(6c^2-12C-2)+R^2(-3C^2+6C+3) \\ &= (-3C^2+6C)(R(1-R))^2+(3R^2-2R^3) \\ &= (-3C^2+6C)(R(1-R))^2+R_{\text{TMR}}, \end{aligned}$$

which proves Eq. (2).

# Notes

[1]The chosen probabilities correspond to Gaussian $f(x) = \exp(-x^2/4)$ for $x = \pm 3, \pm 2, \pm 1$, and 0.

[2]Indeed, the high volume of data coming out of such a complex system is very likely to at least delay the appearance of the failure in the so-called *user's universe* i.e., "where the user of a system ultimately sees the effect of faults and errors" (Johnson, 1989); in some cases it may also make it transparent to the user altogether.

page

# CONCLUSIONS AND APPENDICES

## 1 AN INTRODUCTION AND SOME CONCLUSIONS

We have reached the end of our discussion about application-level fault-tolerance protocols, which were defined as the methods, architectures, and tools that allow the expression of fault-tolerance in the application software of our computers. Several "messages" have been given:

- First of all, fault-tolerance is a "pervasive" concern, spanning the whole of the system layers. Neglecting one layer, for instance the application, means leaving a backdoor open for problems.
- Next, fault-tolerance is not abstract: It is a function of the target platform, the target environment, and the target quality of service. The tools to deal with this are the system model and the fault model, plus the awareness that 1) all assumptions have a coverage and 2) a coverage means that, sooner or later, maybe quite later but "as sure as eggs is eggs," cases will show up where each coverage will fail.
- This means that there is a (even ethical) need to design our systems thinking of the consequences of coverage failures at mission time, especially considering safety critical missions. I coined a word for those supposed fault-tolerant software engineers that do not take this need into account: Endangeneers. Three well-known accidents have been presented and interpreted in view of coverage failures in the fault and system models.
- Next, the critical role of the system structure for the expression of fault-tolerance in computer applications was put forth: From this stemmed the three properties characterizing any application-level fault-tolerance protocol: Separation of concerns, adequacy to host different solutions, and support for adaptability. Those properties address the following question: Given a certain fault-tolerance provision, is it able to guarantee an adequate separation of the functional and non-functional design concerns? Does it tolerate a fixed set of faulty scenarios, or does it dynamically change that set? And, is it flexible enough as to host a large number of different strategies?

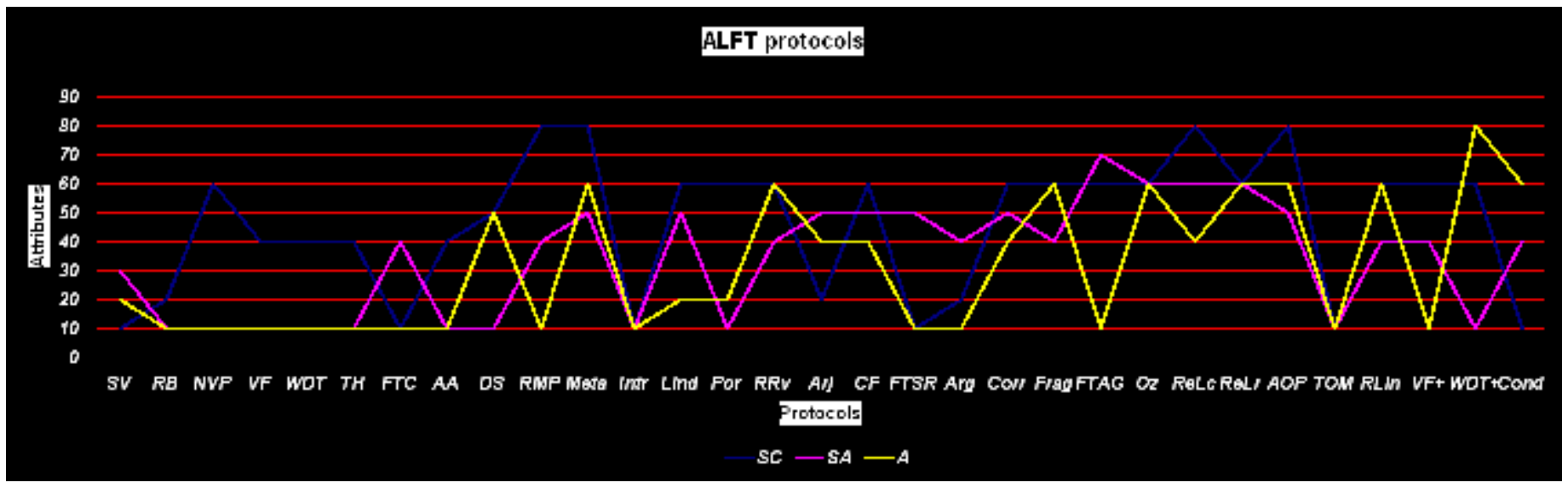

Figure 1: Application-level software fault-tolerance protocols according to the three structural attributes SC, SA, and A.

- Then it has been shown that there exist a large number of techniques and hence of system structures able to enhance the fault-tolerance of the application. Each of these techniques has its pros and cons, which we tried to point out as best as we could. We also attempted to qualify each technique with respect to the above mentioned properties[1]. A summary of the results of this process is depicted in Fig. 1.
- Another key message is that complexity is a threat to dependability, and we must make sure that the extra complexity to manage fault-tolerance does not become another source of potential failures. In other words, simplicity must be a key ingredient of our fault-tolerance protocols, and a faulty fault-tolerant software may produce the same consequence of a faulty non fault-tolerant software—or maybe direr.
- Finally, we showed with some examples that adaptive behaviour is the only way to match the ever mutating and unstable environments characterizing mobile systems. As an example, static designs would make bad use of the available redundancy.

As the reader will have noticed, this book also has different levels of deepenings: Some approaches are sketched, some others are explained in deep detail. As mentioned already this is due to the fact that this book represents the author's current vision and *summa* of personal experiences. This is particularly true for ARIEL, the system described in Chapter 7. ARIEL is described in large detail there, then used in Chapter 10 in two of the hybrid cases described there. One of its component, the DIR net, is described in detail in Chapter 9. Finally, in what follows, a view to the internals of the ARIEL system is provided so as to give the reader an idea of how a fault-tolerance architecture for the application layer was actually implemented. I trust this to be particularly useful to young researchers in the initial phase of a master or doctoral program in resilient computing.
Enjoy!

## Notes

[1]Clearly it is very difficult to consider a quantitative metrics for using the above properties; we limited ourselves to a qualitative, if not subjective, assessment, based on the information in our possessions and in some cases on first-hand experiences.

page

# CONCLUSIONS AND APPENDICES

## 1 AN INTRODUCTION AND SOME CONCLUSIONS

We have reached the end of our discussion about application-level fault-tolerance protocols, which we defined as the methods, architectures, and tools that allow the expression of fault-tolerance in the application software of our computers. Several "messages" have been given:

- First of all, fault-tolerance is a "pervasive" concern, spanning the whole of the system layers. Neglecting one layer, for instance the application, means leaving a backdoor open for problems.
- Next, fault-tolerance is not abstract: It is a function of the target platform, the target environment, and the target quality of service. The tools to deal with this are the system model and the fault model, plus the awareness that 1) all assumptions have a coverage and 2) a coverage means that, sooner or later, maybe quite later but "as sure as eggs is eggs," cases will show up where each coverage will fail.
- This means that there is a (even ethical) need to design our systems thinking of the consequences of coverage failures at mission time, especially considering safety critical missions. I coined a word for those supposed fault-tolerant software engineers that do not take this need into account: Endangeneers. We also showed three well-known accidents and we interpreted them in view of coverage failures in the fault and system models.
- Next, we put forth the critical role of the system structure for the expression of fault-tolerance in computer applications: From this stemmed the three properties characterizing any application-level fault-tolerance protocol: Separation of concerns, adequacy to host different solutions, and support for adaptability. Those properties address the following question: Given a certain fault-tolerance provision, is it able to guarantee an adequate separation of the functional and non-functional design concerns? Does it tolerate a fixed set of faulty scenarios, or does it dynamically change that set? And, is it flexible enough as to host a large number of different strategies?

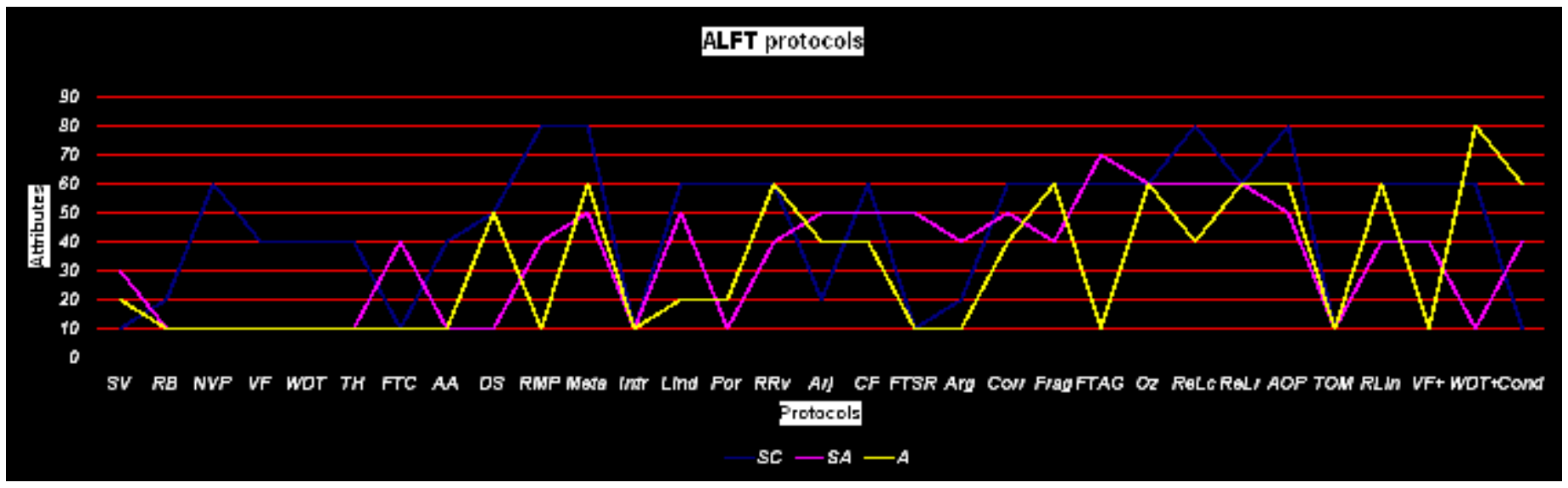

Figure 1: Application-level software fault-tolerance protocols according to the three structural attributes SC, SA, and A.

- Then we have shown that there exist a large number of techniques and hence of system structures able to enhance the fault-tolerance of the application. Each of these techniques has his pros and cons, which we tried to point out as best as we could. We also attempted to qualify each technique with respect to the above mentioned properties[1]. A summary of the results of this process is depicted in Fig. 1.

- Another key message is that complexity is a threat to dependability, and we must make sure that the extra complexity to manage fault-tolerance does not become another source of potential failures. In other words, simplicity must be a key ingredient of our fault-tolerance protocols, and a faulty fault-tolerant software may produce the same consequence of a faulty non fault-tolerant software—or maybe direr.

- Finally, we showed with some examples that adaptive behaviour is the only way to match the ever mutating and unstable environments characterizing mobile systems. As an example, static designs would make bad use of the available redundancy.

As the reader will have noticed, this book also has different levels of deepenings: Some approaches are sketched, some others are explained in deep detail. As mentioned already this is due to the fact that this book represents the author's current vision and *summa* of personal experiences. This is particularly true for ARIEL, the system described in Chapter 7. ARIEL is described in large detail there, then used in Chapter 10 in two of the hybrid cases described there. One of its component, the DIR net, is described in detail in Chapter 9. Finally, in what follows, a view to the internals of the ARIEL system is provided so as to give the reader an idea of how a fault-tolerance architecture for the application layer was actually implemented. I trust this to be particularly useful to young researchers in the initial phase of a master or doctoral program in resilient computing.
We provide in what follows an introduction to Lex and YACC, two tools that are often useful when dealing with linguistic structures for computers. The provided information is not a full manual, which would be out of the scope of this book, but more a summary of the minimal information required to fully comprehend the next section, devoted to the ARIEL internals.
Enjoy!

# A TOOLS FOR TRANSLATORS AND COMPILERS: LEX AND YACC

This section describes two well-known tools for crafting efficient translators and compilers: The LEX lexical analyzer and the YACC syntactical analyzer.

## A.1 Introduction

The capability to compose valid sentences in a given language, as well as to verify that a given string represents a valid sentence in a given language builds upon two lower-level capabilities:

1. Classification, which is the ability to turn a stream of characters into a stream of lexical entities (words, punctuation, delimiters), and
2. Verification, that is, the ability to recognize the syntactical correctness of a sentence, starting from a stream of lexical entities.

The first capability is also known as lexical analysis, the second as syntactical analysis.
Given, for instance, the mathematical expression:

$$sin(a + \sqrt{}(0.4)),$$

the first capability means translating the stream of characters representing the above expression, i.e.
`('s', 'i', 'n', ' ','(', 'a',` ...),
into a stream of tokens, or syntactical atoms:
`("sin", '(', "a", '+', "sqrt",` ...);
Syntactical analysis is indeed the tool that allows us to verify the syntactical correctness of a sentence, given a certain set of rules, called "grammar"—in the above example, the grammar of well-formed mathematical formulae.
The above mentioned capabilities are experienced by human beings as inherent and natural abilities, of which one has not even full awareness. When one has to set up, e.g., an interpreter of a computer language, or any other software module that needs recognize a given structure in its input stream, then it is useful to set up a hierarchical structure at the base of which there are tools for lexical and syntactical analysis. Such tools are software systems that ease the development of lexical and syntactical analyzers. Two well known standard utilities for this are LEX and YACC (or their GNU counterpart, Flex and Bison). LEX and YACC allow to speed up considerably the development of parsers, translators, compilers, interpreter, conversion tools and have been used for more than 20 years by the author of this book to craft various tools, including those described in this chapter and in chapters 7, 10, and 11. LEX and YACC have been especially designed for combined use and for hosting user-defined C routines. Custom versions supporting C++ and Java also exist. In what follows we strictly follow the following two sources: (Levine, Mason, & Brown, 1992) and (Johnson, 1975).

## A.2 The LEX Lexical Analyzer

LEX may be defined as a "tokenizator": Given a stream of chars, LEX performs a classification of groups of contiguous characters. These groups are called tokens, i.e., words and symbols that are *atomic* from the viewpoint of syntactical analysis.
For instance, LEX can translate string $sin(a + sqrt(0.4))$ in a set of couples "(token, token number)", e.g., as follows:

- "sin", `FUNCTION`
- "(", `'('`
- and so forth.

The token number identifies the class the token belongs to.
LEX can be used

- either as a stand-alone tool, so to perform simple translations or compute statistics on the lexical atoms,
- or in conjunction with a parser generator. YACC is the natural choice, but it is always possible to choose another parser.

LEX writes a deterministic finite state automaton from a list of **regular expressions**. Regardless the number of rules supplied by the user, and regardless their complexity, the LEX finite state automaton breaks the input stream into tokens in a time that is proportional to the length of the input stream. The number of rules and their complexity only influence *the size* of the output source code.
The general structure of a LEX program is as follows:

> [ *Definitions* ]
> %%
> [ *Rules* ]
> [ %%
> *User functions* ]

*Definitions* and *User functions* can be missing. Hence, the minimum size LEX program is the following one:

%%

LEX performs its classification via a list of regular expressions (regex) that the user needs to supply through a standard language.
Regex describe *patterns of characters* to be located in the text. LEX reads these regular expressions and produces a finite state machine that recognizes those patterns. Finite state machines are indeed the simplest conceptual tool with which to recognize words expressed by regex. LEX uses the same regular expressions recognizer used by most of those UNIX tools that offer pattern matching services: `vi`, `sed`, `awk`, `find`, `grep`, for instance, adopt a similar set of agreement based on a similar set of "meta-characters", including
`" \ [ ] ^ - ? . * + | { } $ / ( ) % < >`.
(Perl, archie, and others, adopt slightly different sets). In what follows we provide a brief description of the most important among the meta-characters:

`"` the quotation mark operator is the simplest meta-character: all the characters of a string between quotation marks are interpreted as plain (non-meta) characters.

`[ ...]` Squared parentheses (pair []) specify classes of characters. For instance, `[xyz]` means: "*a single* `x`*,* `y` *or* `z` *char*"

The hyphen sign between any two chars $a$ and $b$ means that all the chars between ord($a$) and ord($b$) are specified. For instance, `[A-Z]` means "*any uppercase letter*", while `[A-Za-z]` means: "*any letter*".

Furthermore, `[\40-\176]` for instance selects a range of characters, that is, the one between *octal*(40) and *octal*(176).

`[^ ...]` Character "`^`" , within the squared parentheses, means "complementary set". For instance, `[^0-9]` means "*any character but the digits*".

`\` (Backslash) has the same meaning it has, e.g., in the C language function `printf`: It turns a meta-character into a plain character.

`.` (Dot) means "*any character but* `'\n'`".

`?` The question mark goes after optional strings of characters. For instance, `ab?c` means: "*either* `'ac'` *or* `'abc'`".

`*` Postfix operator "star" means *zero* or more instances of a given class. As an example, `[^a-zA-Z]*` means "*zero or more instances of non-alphabetic chars*".

`+` Postfix operator "plus" means *one* or more instances of a given class. For instance, `[xyz]+` means "*any non-empty string, of any size, consisting of any of the characters* `'x'`*,* `'y'` *and* `'z'`", such as e.g. `xyyyyyyzz`.

Operators `()` and `|`. Parentheses group a set of characters into one object. For instance, in `(xyz)+`, operator `+` is applied to string `xyz`. Within a group, the OR between entities is specified via meta-character `|`. For instance,

`(ab|cd+)?(ef)*`

means "*zero or more instances of string* `"ef"`*, possibly preceded either by string* `ab` *or by* `cd+` *(*`c` *followed by one or more instances of* `d`)".

`^`: This char, if not within square parentheses, means "at begin-of-file or right after a newline."

`$`: This means "at the end of a line" or "at end-of-file", i.e., if the following char is either `'\n'` or `EOF`. For instance, `(riga|row)$` means "string `riga` or string `row` followed either by `\n` or by `EOF`.

`/`: Infix operator slash checks whether an entity is followed by another one. For instance, `a/b` means "character `a`, only when followed by character `b`". Note that `ab/\n` is equivalent to `ab$`.

{}: Curly brackets have two meanings:

- When grouping two comma-separated numbers, as in `(xyz){1,5}`, they represent a *multiple instance*. The above example means "*from one to five instances of string* `xyz`".
- When grouping letters, they represent the value of a regex alias (see further on).

`%` Character `%` is reserved character, not a meta-character.

A LEX source file may include up to three sections; the first one is the one including the LEX definitions. Definitions include a list of regular expressions:

```
letter          [a-zA-Z]
letters         {letter}+
```

This is, so to say, the grammar of the LEX definitions:

1. At column 1, an identifier is supplied,
2. then some blank or tab chars,
3. and finally a regular expression.

After its definition, the identifier becomes an alias for its regular expression.
To dereference an alias one has to put curly brackets around it.
The Rules section is mainly a list of *associations* in the form

$$r \Rightarrow a$$

where *r* is a regular expression and *a* is a list of *actions*, i.e., user defined C language statements that are executed when the corresponding regular expression is recognized. An example follows:

```
%%
begin       printf("{");
end         {
                putchar('}');
            }
```

When no rule is verified, a default rule is executed: `ECHO`, meaning "print the current character". For readers accustomed to thinking in C, this means that the finite state automaton produced by LEX has a `switch` statement ending with "`default:  ECHO;`."
Practical consequences of this include, e.g., that there is no need to supply rules for the so called "literal tokens," i.e., single characters whose token number is equal to their ASCII code. Another consequence is that in order to "sift out" some portion of text from the input stream, one needs recognize it explicitly and to associate a null action to it. As an example, to remove newline characters, one needs write this simple program:

```
%%
\n        ;
```

Some simple transformations can be useful in order to facilitate the import of a file. For instance some word processors, such as TEX, regard paragraphs as multiple lines of text separated by two or more line feed characters, while other programs such as Word regard paragraphs as a single line and separate paragraphs with one or more line feed character. When one wants to import some TEX text into Word the following simple script may be useful:

```
%%
\n\n        ECHO;
\n          putchar(' ');
```

It converts every single `\n` into a character space.
When a regular expression is recognized, the corresponding string (the token) is copied in a string pointed by a `char*` called `yytext`. This is true also for literal tokens.
The corresponding script is similar to the previous one:

```
%%
[^\n]\n[^\n]  { putchar(yytext[0]);
                putchar(' ');
                putchar(yytext[2]);
              }
```

It is sometimes interesting to take a look at the source code of the program produced by LEX; for instance, on my system running cygwin and flex, action `ECHO` turns out to actually be this simple macro:

```
#define ECHO puts(yytext)
```

Variable `int yyleng` is set by LEX to the number of characters of the string that verifies the current rule; in other words,

`yyleng == strlen(yytext).`

The following excerpt shows an example of how to use `yyleng`:

```
%%
[0-9]+          dig += yyleng;
[a-zA-Z]+       alp += yyleng;
(.|\n)          oth++;
```

Clearly the above excerpt is not a completely meaningful LEX program:

1. Variable `dig` etc. have not been declared.
2. No output message is provided at the end.

Nevertheless, this is not a buggy LEX program: The objective of LEX is to produce a C program, not necessarily an *error free* one: No checks are done on the syntactic correctness of the output program. Typical situations are that some syntax errors slip in the actions (indeed, actions are simply copied as strings into the output program). Later on we will describe how to introduce C text into the output program so as to declare variables and produce output messages at end of processing time.
A number of functions are available to the LEX user:

`yymore()` instructs LEX so that the next matched string is attached to the current value of `yytext`. An example follows:

```
%%
\"[^"]*   {
           if (yytext[yyleng-1] == '\\')
               yymore();
           else
               do_that(yytext);
          }
```

`yyless()` “sends back” a given number of characters. This can be very useful in some cases. For instance, in the early days of the C programming language, `a =- b` had the same meaning of `a -= b` (that is, subtract `b` from `a`). This is no more the case, so the first form could result in a hidden bug when current compilers deal with “old dusty deck” software. To detect possible problems one may use the following filter:

```
%%
=-[a-zA-Z] {
           printf("Operator =- is ambiguous: ");
           printf("not recognized.\n");
           yyless(yyleng-2);
           manage_assignment();
           }
```

The form `yyless(`$x$`)` pushes back onto the input `yyleng` $- x$ characters.

`int input()` reads the next input character. Character `NULL` (bitwise equivalent to `(int)0`) is interpreted as the end-of-file condition).

`void output(char c)` writes `c` onto the output stream.

`void unput(char c)` "pushes back" `c` into the input stream.

The user can choose between a standard version of these functions or make use of his or her own versions of those functions, which must have the same name and prototype.

int yywrap(void)

This system (or user-) function is called when an `EOF` is encountered. The system version of this function returns `1`, which means "end of processing." The user can substitute this function with a new version which, if it returns `0`, lets the execution continue until a new `EOF` is encountered. By doing so it is possible, e.g., to process more than one input file during the same run. Another use of `yywrap()` is to allow the user to specify end-of-job functions (for instance, printing the final output), which is one of the requirement we highlighted for the code fragment at page 319.

LEX adopts two steps to select which user rule to apply:

1. The rule that recognizes the largest string is always preferred.
2. If more than one rule recognizes largest strings, it is chosen the rule the user has specified first in the LEX script.

Within a same rule, LEX returns the largest possible string:

```
%%
\'.*\'   { yytext[0] = '[';
           yytext[yyleng-2] = ']';
           printf("%s",yytext);
         }
```

produces a program that, when reading an input string such as `'hi' -he said- 'how are you?'`, writes the following string on the output: `[hi' -he said- 'how are you?]`.
When LEX selects which rule to execute, it creates an ordered list of possible candidates. The one to be executed is the one at the top of the list. When that action includes macro

```
REJECT;
```

the following two actions take place:

1. The input string is sent back onto the input stream.
2. The rule is removed from the list. The rule that is selected is therefore the new top one.

`REJECT` is useful, e.g., to count all the "digrams" (that is, all the couples of contiguous alphabetic characters) in a given text:

```
%%
[A-Z][a-z] {   digram[yytext[0]][yytext[1]]++;
               REJECT;
           }
(.|\n)     ;
```

Each digram in the text is located by the first rule, as it returns a string of *two* characters while the second one returns a string of just one character. `REJECT` writes back the two characters of the digram onto the standard input stream and rejects the first rule. The second one is then executed. As a result, a character is removed from the input stream.
LEX allows to include in the output C source code any useful information (header files, declaration of global variables and so forth). An example follows:

```
%%
[a-z][a-z] {  extern int dig[26][26];
              dig[yytext[0]-'a'][yytext[1]-'a']++;
              REJECT; }
(.|\n)     ;
%%
int dig[26][26];
int yywrap() { int i, j;
        for (i=0; i<26; i++)
          for (j=0; j<26; j++)
             if (dig[i][j])
                printf("digram [%c%c] = %d\n",
                            'a'+i,'a'+j, dig[i][j]);
        return 1;
}
```

Inclusion can be done in three "zones" of the output source file:

1. At the end of the file (as it is done in the above example).

2. At the beginning of the file, that is, before any of the functions.

3. At the beginning of function `yylex()`.

The three zones in the output source code correspond to the following zones of the LEX script:

1. In *User Functions.*

2. In *Definitions*,

3. On top of *Rules*, i.e., right after the first %%;

Case **1** is trivial—one has just to write the required C code, as shown in the above example. For **2** and **3**, we need to distinguish the text to be processed by LEX from the text that needs be copied verbatim in the output file. To do so, one can follow any of these ways:

- `[`
  `t]+.*` (at least a blank space or tab character at column zero, then the data to be flushed onto the output file.)

- Anything between `%{` and `%}`.

In most UNIX environments, in order to process a LEX source code available in file *source.l*, one needs execute the following commands:

1. `lex` *`source.l`* (that is we use LEX to translate *source.l* into the C program *lex.yy.c*; the same syntax is used with GNU flex.)

2. `cc -o output lex.yy.c -ll` (that is we use the C compiler to read the C program produced by LEX and compile it with the assistance of the LEX library, producing an executable file called *output* .)

3. `./output` (that is, we execute the output file.)

File `lex.yy.c` includes function `yylex()` i.e., the actual scanner. Compiling `lex.yy.c` with the system library `libl.a` (or `libfl.a` in the case of FLEX) a `main()` function is automatically supplied, which calls function `yylex()`.
The user can substitute this default `main()` with one of his or her own design. Doing this, one can choose between either automatically generating an executable or "pipelining" LEX output to other programs—for instance, a syntactical analyzer such as YACC.

## A.3 Syntactical Analysis with YACC

YACC, whose name stands for "Yet Another Compiler-Compiler," has been defined by its authors as a system for describing the input structure of a program. Indeed, the YACC programmer is required to supply:

1. The syntactical structure of the input, and
2. C code to be executed when the syntax rules are recognized.

On the basis of the above input data, YACC produces a C program with a parsing routine. The parsing routine calls a lower level routine, called `yylex()`, in order to get the next lexical atoms in the input stream. It goes without saying that LEX produces exactly one such routine.
YACC works with grammars of type **LALR(1)**, plus rules to solve ambiguities. As its names tell, YACC is just one of many "compiler compilers", and many new powerful syntactical analyzers have been designed, a noteworthy example being ANTLR (Parr, 2007). Some of them are open software and can be easily downloaded and compiled from the Internet. YACC (or its GNU sibling, Bison) is likely to be available on any UNIX distribution, so we decided to focus on its syntax. Another reason to go for YACC with respect to other tools is that its syntax has been designed so as to make it very easy to master its functions for someone who knows LEX already.

### A.3.1 Structure of a YACC Script

The general structure of a YACC script strictly follows the one of a LEX script:

> [ *Definitions* ]
> %%
> *Rules & Actions*
> [ %%
> *User functions* ]

In particular, the structure of *Rules & Actions* is similar to the corresponding section of a LEX script: It includes a set of *grammar rules*, plus *actions* that are associated to each rule. Each time a rule is recognized, the corresponding actions are executed. Actions may return values and use the values returned by other actions.
YACC rules have the following structure:

*lhs* : *rhs* ;

where *lhs* is a non-terminal symbol and *rhs* is a sequence of zero or more terminal or non-terminal symbols, "literals" (see below), and actions. Identifiers for terminal and non-terminal symbols follow the rules of the C language, with the addition that character `'.'` is considered as a letter.
A literal is a constant character defined as follows:

```
literal : QUOTE char QUOTE
        | QUOTE BACKSLASH char QUOTE
        | QUOTE BACKSLASH od od od QUOTE
         ;
```

where in this case `QUOTE` is character `"` and `od` is an octal digit. The pipe character "`|`" is the YACC way to represent alternative definitions—it reads out as "or". It is used when more than one rule has the same *lhs*.
As with LEX, the parentheses `%{` and `%}` allow to include in the output of YACC any C source code. This code is global with respect to the parser function and to the user functions.
YACC uses a number of identifiers starting with "`yy`" for internal purposes. As a consequence it is a good idea to avoid that prefix: While the YACC syntax is more or less standard, clearly its implementations may be completely different from each other or from future ones. A variable starting with "`yy`" may translate in a hidden design fault.
Lexical atoms (the tokens) must be explicitly declared in *Definitions.* This is done, for instance, by writing one or more lines such as the following one:

$$\texttt{\%token}\ name_1\ name_2\ \ldots$$

All the symbols that have not declared as tokens are implicitly declared as non-terminals (non-terminals). An important property that must be guaranteed for any YACC source program is that each non-terminal must be the *lhs* of at least one rule.
The declaration of the start symbol of the grammar may be done as follows:

`%start` *name*

in *Declarations.* If this specification is missing, it is assumed that the start symbol is the *lhs* of the first grammar rule specified by the user.
A special token marks the end-of-input. This is called end-marker in YACC lingo. If the tokens encountered between the start of processing and the end-marker (not including the latter) *verify* the start symbol, then the parsers successfully stops processing after having read the end-marker. Reading the end-marker before the start symbol is verified leads to an error.
Within each rule, the programmer can specify some *actions* to be executed each time that rule is recognized while analysing the input stream. Actions may return values and use the values returned by other actions. Also the tokens returned by `yylex()` may have values. Actions are a group of C statements between curly brackets. Each action can return a value by setting variable `$$`. For instance:

```
{ action(); $$=1; }
```

returns 1. Also the rules may return values. This value is either the value of the first component or the value of variable `$$`. For instance:

```
A : B;
```

is equivalent to

```
A : B { $$ = $1; } ;
```

The following example shows how it is possible to use the values returned by previous rules:

```
expr    :   '('   expr   ')'
            {    $$ = $2;   }
        ;
```

In other words, $$i$ is the value returned by RHS[$i$]. An example follows: Let us suppose we wanted to build a syntax analysis tree and for that we are using a function `node()` that allocates a new syntax object and returns its address. Then such functions could be called in rules such as the following one:

```
expr :   expr    infix_op     expr
         {  $$ = node( $2, $1, $3);  }
     ;
```

Values returned by rules and actions are integers by default.
As we mentioned in previous section, function `yylex()` returns an integer—the token number. This number is either a literal (when in $[0, 255]$) or a symbolic constant $s > 256$ that describes the lexical "class" the recognized string belongs to. One such class could be, for instance, `NUMBER`, identifying all numerical symbolic constants. Function `yylex()` also returns the actual string that was found in the input. That string is kept in variable

`extern` **X** `yylval;`

where **X** is either `int` or can be defined by the user. An example follows:

```
%%
[0-9]+  { yylval=atoi(yytext); return NUMBER; }
```

The choice of which integers to use with tokens can be done

*automatically* by YACC, which associates the integers from 256 one by one to the tokens that have been declared with the `%token` keyword; or

*implicitly* for literals, to which it is associated their ASCII code; or

*explicitly* by the YACC programmer, who can associate an integer greater than 0 after the name of a token or a literal in section *Declarations*.

Token numbers must be different. When executing `yacc` with the `-d` option, a header file is created, called `y.tab.h`, which contains all the token numbers. This file can be included, e.g., in the LEX script, through a statement such as the following one:

```
%{
#include "y.tab.h"
%}
```

Note that the C program produced by LEX can be either compiled separately or even included in the YACC output program by specifying in *User functions* the following statement:

```
#include "lex.yy.c"
```

### A.3.2 Associativity rules

Some arithmetical operators have their own associativity rules, and by agreement there are priorities between them. Therefore a method is required in order to set a priority among operators and to choose beforehand the type of associativity that is required.
The kind of associativity of an operator can be defined in YACC by the three directives:
`%left %right %nonassoc`
Such directives also represent an alternative way to declare tokens and literals with respect to `%token`. For instance,

```
%right  '='
%left   '-'  '+'
```

selects right association for the assignment operator (that is, $a = b = c$ is interpreted as $a = (b = c)$) and left association for `'+'` and `'-'`.
Each row defines a priority level. The earlier the specification appears in the source file, the lower its priority:
`'='` $\prec$ (`'+'`, `'-'`) $\prec \cdots$
For instance,

$$a = b = c * d - e - f/g;$$

is interpreted as

$$a = (b = (((c * d) - e) - (f/g)));$$

Keyword `%nonassoc` specifies that a certain operator must *not* be applied more than once. For instance, in Fortran the following expression
`A .LT. B .LT. C`
is not valid. The `%nonassoc` tokens catch such conditions.
There are cases in which a same sign, for instance `'-'`, has two different meanings and priorities:

```
expr : expr '=' expr
     | expr '*' expr
     | expr '-' expr
     | '-' expr
```

Unary "minus" has greater priority than that of diadic "minus". In such cases one can make use of a fictitious token and operator `%prec`:

```
%left '+' '-'
%left '*' '/'
%left UMINUS
%%
expr : expr '-' expr
     |   ....
     | '-' expr   %prec   UMINUS /* same priority of UMINUS */
```

As we have seen already, the value stack of YACC is by default based on short integers. The user can override that choice and choose any other type. In this case the value stack is organized as a vector of `union`'s. The programmer can declare such `union` and associate the name of its members with the tokens and non-terminals that return a value. When the user does declare the `union`, the following string is attached to any reference like `$$` or `$`$i$:

.*field-name*

This is a possible example

```
%union {
   char  *String;
   double Real;
   int    Integer;
}
```

An equivalent way to shape this union is by defining explicitly type `YYSTYPE`:

```
typedef union {
      char  *String;
      double Real;
      int    Integer;
   } YYSTYPE;
```

After that definition one can associate a field-name to a token like follows:

```
%left   <Integer>   '+'  '-'
%right  <Real>      '='
```

It is even possible to associate a field-name to a non-terminal:

```
%type   <String>   expr
%type   <Real>     number
```

Another possibility is to associate a field-name to an action:

```
 expr : '(' strexp ')'
         { $<Real>$ = atof( $<String>2 );
         }
      ;
```

that is, "`$<`", followed by a field-name, followed by "`>$`"

## Notes

[1]Clearly it is very difficult to consider a quantitative metrics for using the above properties; we limited ourselves to a qualitative, if not subjective, assessment, based on the information in our possessions and in some cases on first-hand experiences.

page

# B The Ariel internals

In this section we would like to provide the reader with an (abridged) description of the internals of the Ariel translator, art, so as to give a real-life example of the way tools such as Lex and YACC can be used to craft application-level fault-tolerance provisions.

Art is built by compiling the Lex source code “Ariel.l” and the YACC source code “Ariel.y”, together with several ancillary C source codes. Art takes as input an Ariel script and produces as output:

1. A number of configuration files describing a distributed application and an execution platform.
2. An error recovery specification, expressed in the form of a pseudo-code for the Ariel virtual machine.

The configuration files are to be compiled with the sources of the distributed application, while the error recovery specifications are compiled with the Ariel virtual machine, which then executes concurrently with the distributed application. Each time a fault is detected in the system, the Ariel virtual machine executes the error recovery specification interpreting the pseudo-code. This pseudo-code is called “r-code” (recovery code) and represents the “virtual machine language” for a stack machine called RINT. More details on this can be found in Chapter 7.

We begin with the Ariel’s scanner, “ariel.l”, then we provide an excerpt of some ancillary source code used by Ariel’s parser, and we conclude with an abridged view to the latter.

## Ariel’s scanner

Source code Ariel.l starts with a “%{” … “%}” section, within which a few comment lines introduce the source code, its version and so forth. C header files are included and some global variables are declared. An example is “lines”, an integer starting at 1 and counting the number of lines being processed:

```
%{
/*************************
**
**     File: ariel.l
**
**     Description: scanner
**
**     Language: lex
**
**     History:
**     Version 2.0d   08-Feb-2006
**      - added "Watchdog"
```

*… lines omitted …*

```
**************************

#include <string.h>
int lines=1;

%}
```

Ariel.l continues with its definitions. These include:

- Ariel’s keywords, such as “IF” or “THEN”;
- generic lexical atoms such as “NUMBER” or “REAL”;
- atoms such as “THREAD” or “NODE”, defining the Ariel identifiers for Ariel’s “entities”, in this case respectively processes or computers;

* atoms such as “AGENT” or “MANAGER”, defining the roles that are to be assigned to certain processes:

```
IF       ([Ii][Ff])
FI       ([Ff][Ii])
ELSE     ([Ee][Ll][Ss][Ee])
THEN     ([Tt][Hh][Ee][Nn])
```

***… lines omitted …***

```
THREAD  ([Tt]([0-9]+)
GROUP   ([GgLl]([0-9]+)
NODE    ([Nn]([0-9]+)
DIGIT   [0-9]
NUMBER  ("-")?[0-9]+
REAL    ((({NUMBER})?{DOT})?{NUMBER})
AGENT           ([Aa][Gg][Ee][Nn][Tt][Ss]?)
BACKUPAGENT
(([Bb][Aa][Cc][Kk][Uu][Pp][Ss]?){AGENT}?)|([Aa][Ss][Ss][Ii][Ss][Tt][Aa][Nn][Tt][Ss]?)
MANAGER         ([Mm][Aa][Nn][Aa][Gg][Ee][Rr])
ROLE            ({AGENT}|{BACKUPAGENT}|{MANAGER})
```

Other definitions specify Ariel’s guards or pre-conditions. Such guards are to check whether a given entity has been stopped or restarted, or detected as faulty, or isolated, and so forth:

```
KILLED          "-"?(([Kk]([Ii][Ll][Ll][Ee][Dd])?)|([Ss][Tt][Oo][Pp][Pp][Ee][Dd]))
RESTARTED       "-"?[Rr]([Ee][Ss][Tt][Aa][Rr][Tt][Ee][Dd])?
PRESENT         "-"?[Pp]([Rr][Ee][Ss][Ee][Nn][Tt])?
RUNNING         "-"?([Rr][Uu][Nn][Nn][Ii][Nn][Gg])
ISOLATED        "-"?[Ii]([Ss][Oo][Ll][Aa][Tt][Ee][Dd])?
FAULTY          "-"?[Ff]([Aa][Uu][Ll][Tt][Yy])?
```

There follow some definitions for the configuration of single version and multiple version fault-tolerance provisions. As an example, these are the Lex definitions corresponding to a watchdog timer:

```
WATCHDOG        [Ww][Aa][Tt][Cc][Hh][Dd][Oo][Gg]
KEYW_AT         ([Aa][Tt])
WATCHES         [Ww][Aa][Tt][Cc][Hh][Ee][Ss]
HEARTBEATS      [Hh][Ee][Aa][Rr][Tt][Bb][Ee][Aa][Tt][Ss]
EVERY           [eE][vV][eE][Rr][Yy]
MS              [Mm][Ss]
US              [Uu][Ss]
REBOOT          [Rr][Ee][Bb][Oo][Oo][Tt]
EndWatchdog     ([Ee][Nn][Dd](([ ]+)?({WATCHDOG})))
```

Then the Ariel’s rules are listed. As an example, when art encounters the token “GID” (in Ariel, the identifier of a group of tasks, e.g. “G42”), the corresponding actions set the return value, yylval, with the group number (42) and with the role of the given entity (in this case, `ID_NORMAL | ID_GROUP`, which means “a user-defined group of tasks”). Note how yylval is in this case is not a scalar value but a complex data type defined by the YACC “%union” statement (see Section 3):

```
%%
{GID}           {
                yylval.id.role = ID_NORMAL | ID_GROUP;
                sscanf(yytext+1, "%d", &yylval.id.id);
                return GID;
                }
```

Another few examples of Ariel rules. Some of them set yylval (e.g. “NUMBER”, corresponding to an integer input, or “REAL”, namely a real number); some others simply return a token number to the calling parser (e.g., definition “WATCHDOG” returns the token number represented by the symbolic constant having the same name). Token numbers are created automatically by YACC in this case. The last rule catches unrecognized characters, which are printed on the output and ignored:

```
{NUMBER}        {
                        sscanf(yytext, "%d", &yylval.integer);
                        return NUMBER;
                }
{REAL}          {
                        sscanf(yytext, "%f", &yylval.real);
                        return REAL;
                }
{WATCHDOG}      return WATCHDOG;
{WATCHES}       return WATCHES;
{HEARTBEATS}    return HEARTBEATS;
{EVERY}         return EVERY;
{MS}            return MILLISEC;
{US}            return MICROSEC;
{REBOOT}        return REBOOT;
{EndWatchdog}   return KEYW_ENDWATCHDOG;
{THRESHOLD}     return THRESHOLD;
{ALPHACOUNT}    return ALPHACOUNT;
{FACTOR}        return FACTOR;
{EndAlpha}      return  KEYW_ENDALPHA;
.               fprintf(stderr, "Lex: unrecognized char: %s\n", yytext);
```

## Ariel's r-code

As we mentioned already, the output of the art translator include the translation of the Ariel script into r-code. To explain in more detail how this is done, we provide here an excerpt of the header files and source codes that create the r-code. Finally, we briefly describe the main loop of RINT, the run-time executive of the r-code.

### 1. R-code.h

Header file "rcode.h" defines the opcode of the RINT virtual machine. These are organized as a series of integer numbers between 0 and LAST_RCODE (in the current implementation, 56). The opcodes are organized into classes, consisting of one or more opcode. The header file also defines a number of predicates, such as isstoprcode or issetrcode, returning a non-zero value when the argument satisfies the predicate. Rcode.h consists of a compile-time section, used when compiling the Ariel translator "art", and a run-time section, used when compiling  the r-code virtual machine RINT.

```
/* The R-opcodes */

#define R_STOP          000000
#define isstoprcode(x)     (! x)



#define R_SET_ROLE      1
#define R_SET_DEF_ACT   2

#define issetrcode(x)   (x>=R_SET_ROLE && x<=R_SET_DEF_ACT)


#define R_AND           3
#define R_OR            4
#define R_NOT           5
#define R_INC_NEST      6
#define R_DEC_NEST      7
#define R_STRVAL        8
#define R_STRPHASE      9

#define is1operandrcode(x)   (x>=R_AND && x<=R_STRPHASE)
```

```
#define R_FALSE         10
#define R_GOTO          11
#define R_PUSH          12
#define R_FUNCTION_CALL 13
#define R_OANEW         14
#define R_CLEAR         15
#define R_PAUSE         16

#define is2operandrcode(x)   (x>=R_FALSE && x<=R_PAUSE)


#define R_KILLED       17
#define R_RESTARTED    18
#define R_PRESENT      19
#define R_ISOLATED     20
#define R_FAULTY       21
#define R_FIRED        22
#define R_PAUSED       23
#define R_REINTEGRATED 24

#define R_STRERRN      25
#define R_STRERRT      26
#define R_COMPARE      27

#define isclause(x)   (x>=R_KILLED && x<=R_REINTEGRATED)
#define istestrcode(x)   (x>=R_KILLED && x<=R_COMPARE)


#define R_KILL         28
#define R_WARN         29
#define R_ANDWARN      30
#define R_START        31
#define R_RESTART      32
#define R_REBOOT       33
#define R_CHKERR       34
#define R_ENABLE       35
#define R_REMOVE       36
#define R_REMOVE_ALL   37
#define R_SEND         38
#define R_GET          39
#define R_SET          40
#define R_CONST        41
#define R_ADD          42
#define R_SUBTRACT     43
#define R_MULTIPLY     44
#define R_DIVIDE       45
#define R_COMPLEMENT   46
#define R_LOGS         47
#define R_LOGI         48
#define R_LOGC         49 // VdF, Oct 2002
#define R_LOGV         50 // VdF, Oct 2002
#define R_EVENT        51 // VdF, 30 Oct 2002
#define R_ISOLATET     52 // VdF, 30 Oct 2002
#define R_ISOLATEG     53 // VdF, 30 Oct 2002
#define R_ISOLATEN     54 // VdF, 30 Oct 2002
#define R_SETPHASE     55 // VdF, 13 Oct 2002

#define isactionrcode(x)   (x>=R_KILL && x<=R_REMOVE_ALL)

#define R_DEADLOCKED   56
#define LAST_RCODE     R_DEADLOCKED
```

There follow the definition of type rcode_t, which is a triplet of integers consisting of an opcode and two arguments. The rcode translation of the Ariel script is stored into an array of rcode_t triplets, called compile_time_rcode:

```
typedef int rcode_t[3];
```

```
#ifdef COMPILETIME
static rcode_t compile_time_rcode[RCODE_MAX_CARD];
#endif
```

The rcode.h header file then defines the data structures for the management of Ariel nested IF statements, which need to be translated into their rcode equivalent by means of conditional and unconditional branch instructions (R_FALSE and R_GOTO). Their meaning is explained in Sect. 2.

```
struct goto_t {
   int pc;              /* pc is the value of the RINT program counter */
   struct goto_t *next;
};

typedef struct {
        int pc;
        struct goto_t *gotos;
        } if_t;

static if_t ifs[RCODE_MAX_NEST], *iftop;
static int  ifp;
```

This concludes the compile-time section of header file rcode.h. There follow the declarations for the r-code functions, i.e., the functions that are executed by RINT when interpreting the opcodes. All these functions read the three integers corresponding to an r-code and return an integer exit value:

```
#ifndef COMPILETIME

int   R_Stop(int,int,int);
int   R_And(int,int,int);
int   R_Or(int,int,int);
int   R_Not(int,int,int);
int   R_OArenew(int,int,int);
int   R_Clear(int,int,int);
```

*... lines omitted ...*

```
int   R_IsolateGroup(int,int,int);
int   R_IsolateNode(int,int,int);
int   R_Deadlocked(int,int,int);
```

An extra function called R_Nop (for “no operation”) is also declared. Such function corresponds to outdated opcodes still to be reorganized. Finally, rfunc, an array of function pointers is defined statically (that is, at compile time). Note that entry rfunc[*i*] corresponds to the *i*-th opcode:

```
int   R_Nop(int a,int b,int c); // { return 0; }

static int (*rfunc[])(int,int,int) =
 { R_Stop,              // 0
   R_Nop,               // 1
   R_Nop,               // 2
   R_And,               // 3
   R_Or,                // 4
   R_Not,               // 5
   R_NestIn,            // 6
   R_NestOut,           // 7
   R_StoreVal,          // 8
```

*... lines omitted ...*

```
   R_IsolateGroup,      // 54
   R_IsolateNode,       // 55
   R_Deadlocked,        // 56
 };
```

## 2. R-code.c

Source file rcode.c defines, among others, the functions to generate the rcode-equivalent of an Ariel recovery script. Some of these functions are very simple, such as rcode_stop:

```
/***************************************************************************/
/*      generates the R_STOP R-code, which closes an R-code object file      */
/***************************************************************************/
int rcode_stop()
{
  if (pc < RCODE_MAX_CARD)
  {
     compile_time_rcode[pc][0] = R_STOP;
     compile_time_rcode[pc][1] = -1;
     compile_time_rcode[pc][2] = -1;
     pc++;
     return pc;
  }
  return 0;
}
```

Some others are complex and intertwined, such as those corresponding to the IF THEN … ELSE / ELIF … FI statements:

```
/***************************************************************************/
/*      generates the R_FALSE R-code, corresponding to the beginning of      */
/*      `if/then/elif/else/fi' statements.                                   */
/***************************************************************************/
int rcode_if()
{

  if (pc < RCODE_MAX_CARD)
  {
     if (iftop == NULL)        /* top-level if  */
     {
         iftop = &ifs[0];
         ifp = 0;
     }
     else                      /* non top-level if */
     {
         ifp++;
         iftop = &ifs[ifp];
     }

     iftop->gotos = NULL;
     iftop->pc = pc;

     compile_time_rcode[pc][0]     = R_FALSE;   /* new R_FALSE statement */
     compile_time_rcode[pc][DEST]  = 0;         /* goto value initially set to 0 */
     compile_time_rcode[pc][2]     = -1;        /* unused */

     pc++;
     return pc;
  }
  return 0;
}

/***************************************************************************/
/*      generates the R_GOTO R-code (unconditioned branch). It is used to    */
/*      jump to the next `elif' or the `else' part of an `if' statement.     */
/***************************************************************************/
int rcode_goto(int where)
{
  if (pc < RCODE_MAX_CARD)
```

```
  {
     compile_time_rcode[pc][0] = R_GOTO;
     compile_time_rcode[pc][1] = where;
     compile_time_rcode[pc][2] = -1;

     pc++;
     return pc;
  }
  return 0;
}

/**************************************************************************/
/*   generates the R_GOTO R-code statement which closes the current section  */
/*   of an `if' statement, together with the R_FALSE R-code which starts a   */
/*   new `elif' section.                                                     */
/**************************************************************************/
int rcode_elif(int goto_address)
{
/* when we encounter an `elif' statement, three things have to be managed:

   1. first, a goto_t block has to be allocated and linked to the list, and
      an incomplete goto statement has to be added to rcode list;
   2. secondly, compile_time_rcode[iftop->pc][3] i.e., the running value, has to
      be updated with the current value of the pc register;
   3. third, a new `false' statement has to added, and iftop->pc should be
      updated with the current value of the pc register.
 */
```

*… lines omitted …*

```
}

/**************************************************************************/
/*      generates the R_GOTO R-code, corresponding to the beginning of       */
/*      an `else' section of an `if' statement. Ends the R_FALSE R-code      */
/*      R-code corresponding to previous section.                            */
/**************************************************************************/
int rcode_else(int goto_address)
{
/* when we encounter an `else' statement, two things have to be managed:

   1. first, a goto_t block has to be allocated and linked to the list, and
      an incomplete goto statement has to be added to rcode list;
   2. secondly, compile_time_rcode[iftop->pc][3] i.e., the running value, has to
      be updated with the current value of the pc register;

   No new `false' statement has to be added, nor iftop->pc should be
   updated with the current value of the pc register, as it was the case
   with rcode_elif().
 */
```

*… lines omitted …*

```
}


/**************************************************************************/
/*      ends the pending R_GOTO's, corresponding to all sections of the      */
/*      current `if'. Ends also the top R_FALSE.                             */
/**************************************************************************/
int rcode_fi()
{
/* Once we encounter a `fi' it obviously means that a whole `if' statement
   is over; when this happens we still have to adjust the whole list of goto's
   so as to point to this statement.

   When everything is done for each and every goto, then it's time to pop
   the `if' statement off the stack. When we reach the bottom we also have
   to reset iftop to NULL.
 */
```

*… lines omitted …*

```
}
```

Rcode.c also defines rflush, the function that, at the end of processing time, flushes the contents of compile_time_rcode into the output files. One such file is “trl.h”, an example of which is available in Table 6 of Chapter 7. Function rflush is called by the main function of the Ariel.y parser.

```
/***********************************************************************/
/*    flushes the R-codes to the object file, resets counters accordingly    */
/***********************************************************************/
int rflush()
{
   extern FILE *f;

   n = fwrite( (void*) &compile_time_rcode[0][0], sizeof(rcode_t), pc, f);
   fflush(f);
```

## 3. Rint.c

This section briefly sketches the source code of the RINT virtual machine, the interpreter of the r-code generated by the art translator. Its main function is rather simple:

```
int main(void) /* RINT main module */
{
  rdump = fopen("rdump.txt", "w");       /* rdump logs all the executed rcodes */

  if( BSL_InitLibrary() == BSL_ERROR)  { /* connects to the BSL (re: Chap.7, Sect.3.2) */
                fprintf(stderr, "RINT: failed to initialise BSL\n");
                BSL_CloseLibrary();
                return -1;
        }

  dowhile();                             /* executes a loop in function “dowhile” */

  if( BSL_CloseLibrary() == BSL_ERROR) { /* and disconnects */
                fprintf(stderr,
                "RINT: failed to close BSL. Clean up the system manually!\n");
                return -1;
        }

  return 0;
}
```

Clearly the greater part of the complexity of RINT lies in its dowhile function, which just waits continuously for a waking message after which it executes the rcodes one by one:

```
dowhile ()
{
        int line;
        extern int rcode_card;

        size = sizeof(message_t);
        while (1)
        {
                RINTGetMessage();                 /* “Wake up!” */

                printf("Recovery starts (%d r-codes).\n", rcode_card);
                BSL_ResetTimerBase();             /* Zero time counters */

                line = 0;                         /* line (program counter) is zeroed */
                while ( line < rcode_card ) {     /* while there are rcodes… */

                    … lines omitted …

                                                  /* execute the function
```

```
                                        corresponding to the current rcode
                                       */
                rfunc [ rcodes[line][0] ] (rcodes[line][1], rcodes[line][2], line);
                line++;
        }
```

The execution of the rcodes implies the use of a run-time stack for the evaluation of arithmetical and Boolean expressions. Such stack is accessed through function R_Push and R_Pop, defined elsewhere. After the execution of any arithmetical or Boolen operation, the result is pushed onto the stack. This is for instance the execution of R_False,_the function corresponding to recovery opcode R_FALSE:

```
                  if (R_Pop() == FALSE)          // check the top of the stack
                  {
                        fprintf(rdump, "%d \tConditional GOTO, fulfilled, %d.\n",
                                        line, line + rcodes[line][1]);

                        line += rcodes[line][1]; // if FALSE, execute jmp
                        continue;
                  }
```

Another simple example is given by function R_Or (the run-time executive for opcode R_OR):

```
int R_Or(int op1, int dummy, int rline)
{
        /* RCODE: R_OR
           ARIEL:  part of the expressions within IF statements */
        int t = R_Pop();
        int s = R_Pop();
        R_Push( t || s );
        fprintf(rdump, "%d \tOR statement (popped %d and %d, pushed %d).\n",
                        rline, t, s, t || s);
        return  t || s;
}
```

After having described the Ariel scanner and Ariel's compile-time and run-time components, we can now focus our attention on the Ariel parser.

## Ariel's parser

The Ariel parser is the program that actually instructs the translation from the high-level Ariel source into the lower-level rcodes and configuration header files. A thorough description of this section would be inappropriate, so we concentrate on a few particularly meaningful fragments.

As already mentioned, the structure of a YACC parser closely resembles that of LEX scanners. For instance, like Ariel.l, also Ariel.y begins with a "%{" … "%}" section. Among other things, in that section we include header file "rcode.h", described in previous section, and define the type of yylval through the "%union" statement:

```
%{

... lines omitted ...

#include "rcode.h"

typedef struct {
          unsigned
          int role;   /* a bit pattern with one or more of these
                         bit positions turned on:
                           ID_GROUP, ID_THREAD, ID_NODE, ID_ENTITY,
                           ID_NORMAL, ID_STAR, ID_DOLLAR */
```

```
          int id;     /* an integer identifying the entity */
        }             ident_t;

%union {
        float real;                // value of the just scanned real number
        int   integer;             // value of the just scanned integer number
        ident_t  id;               // role and id of an entity
        struct { char name[64];    // symbolic name of an rcode opcode
                 int  rcode;
        } string;
        char  quoted_string[64];   // value of a quoted string
        int   status;              // status info
}
```

Ariel also defines 26 integer variables, each for each letter of the English alphabet. The Definitions section of YACC also makes room for them:

```
static int var[26];  /* Ariel's 26 pre-defined and pre-initialised variables */
```

Several sections follow, which describe the data structures for the configuration of some of the "Basic Tools" (see Chapter 7, Sect. 2.3 for more detail on this). One such tool is the watchdog timer, a Single-version software fault-tolerance provision that would normally be configured in the functional source code of the user application. Ariel allows configuring such tools in the Ariel (non-functional) source code. Here is the YACC section of the watchdog:

***… lines omitted …***

```
/************************* watchdogs *****************************/
typedef struct {
                int watching, /* id of the watching task */
                    watched,  /* id of the watched task */
                    rate,
                    unit,
                    action,
                    target, running;
        }       watchdog_t;

#define    MAX_WDOGS    32
watchdog_t watchdog[MAX_WDOGS];
int        w_sp = 0;
#define    INCWATCH     ++w_sp;
#define    WATCHTOP     watchdog[w_sp]
#define    BADWATCH     (w_sp > MAX_WDOGS)
#define    MAX_WD_FNAME 80
#define    MAX_WD_NAME  40
int watchdog_flush(watchdog_t*, int); // creates the source codes for the watchdogs
/********************** end watchdogs ****************************/
```

Another section defines a restoring organ (see Chapter 4, Section 3 on the EFTOS Voting Farm, the source of which was used here). Note that this is a software fault-tolerance provision that would normally require a non-negligible amount of code intrusion, which is considerably reduced in this case:

```
/************************ nversion *****************************/
typedef struct { int version,           // version id
                     task,              // task id
                     timeout,           // max duration of task
                     unit;              // millisecs or microsecs
        } version_t;
#define    MAX_VERS     7               // up to 7 versions (7-MR system)
typedef struct {
                int running,            // parsing flag
                    task;               // task id of the restoring organ
                version_t va[MAX_VERS]; // the versions
                int va_num;             // running number of encountered versions
                int voting;             // voting algorithm
```

```
                char *metric;            // name of the metric function
                int success, error;
                int versmin, versmax;    // versions can be given in any order
        }       nversion_t;
#define    MAX_NVERS    16               // up to 16 restoring organs
nversion_t nversion[MAX_NVERS];
int        nv_sp = 0;                    // stack pointer for nversion[]
#define    INCNVERS     ++nv_sp;
#define    NVERSTOP     nversion[nv_sp]
#define    BADNVERS     (nv_sp > MAX_NVERS)
#define    MAX_NV_FNAME 80
#define    MAX_NV_NAME  40
int nversion_flush(nversion_t*, int);   // creates the sources of the restoring organs
/*********************** end nversion ****************************/
```

Another section defines the tokens, (optionally) their type, and precedence/associativity rules:

```
%token <string> ROLE
%token <integer> NUMBER VAR
%token <real> REAL
%token <id> GID NID TID

%token IF ELSE ELIF FI THEN
%token KILL RESTART START ISOLATE
%token KEYW_TASK KEYW_TASKID KEYW_NODE KEYW_IS KEYW_ALIAS KEY
%token KEYW_LOGICAL KEYW_ENDLOGICAL
%token WATCHDOG WATCHES HEARTBEATS EVERY
%token REBOOT KEYW_ENDWATCHDOG
%token NVERSION KEYW_VERSION TIMEOUT VOTING
%token MAJORITY ALGORITHM
%token METRIC SUCCESS KEYW_ENDNVERSION
```

*... lines omitted ...*

```
%token <integer> LET VAL
%token <integer> EQ NEQ GT GE LT LE
%token <integer> MILLISEC MICROSEC
%left AND OR
%left NOT

%token <status> KILLED RESTARTED PRESENT
%token <status> ISOLATED FAULTY DEADLOCKED
%token <status> REINTEGRATED FIRED PAUSED

%type <status> status
%type <id> id
%type <integer> compare
%type <integer> seconds
%type <integer> expression linexp

/* precedences/associativity */
%left '+' '-'
%left '*' '/'
%left UMINUS
```

We get then to the main YACC section, that of Rules & Actions. The start symbol is “rlstats”, which states that an Ariel script consists of one or more “rlstat” or erroneous lines:

```
%%
rlstats:
        | rlstats rlstat
        | error '\n'
                {
                fprintf(stderr, "\tLine %d: syntax error.\n", lines);
                errors++;
                }
        ;
```

It is the definition of "rlstat" that lists all possible cases for a valid Ariel statement. Here it follows its (abridged) specification:

```
rlstat:   '\n'
        | definition '\n'
                {
                        if (rec)
                        {
                        fprintf(stderr, "\tLine %d: semantical error: ", lines);
                        yyerror("Can't define roles in strategy section");
                        errors++;
                        }
                }
        | section '\n'
        | include '\n'
        | timeouts '\n'
        | definitions '\n'
        | identifiers '\n'
        | alphacounts '\n'
        | aliases '\n'
        | logicals '\n'
        | entrypoints '\n'
        | stacksizes '\n'
        | watchdog '\n'
        | nversiontask '\n'
        | injection '\n'
        ;
```

An exhaustive description not being appropriate here, we decided to focus our attention on some of the above mentioned statements. We begin with an example of a configuration statement: this is the syntax of statements for the configuration of watchdogs.

```
watchdog:       watchdog_start watchdog_args watchdog_end
        ;

watchdog_args:
        |       on_error watchdog_args
        |       heartbeats watchdog_args
        |       w_alphacount watchdog_args
        ;

watchdog_start:      WATCHDOG KEYW_TASK NUMBER WATCHES KEYW_TASK NUMBER '\n'
                {
                        WATCHTOP.running = 1; /* flag: we're within a watchdog */
                        WATCHTOP.watching = $3;
                        WATCHTOP.watched = $6;
                        taskram[$3].actor = RE_WD;
                }
        ;

w_alphacount:ALPHACOUNT KEYW_IS THRESHOLD '=' REAL ',' FACTOR '=' REAL KEYW_ENDALPHA '\n'
                {
                        alphas[WATCHTOP.watching].used = 1;
                        alphas[WATCHTOP.watching].threshold = $5;
                        alphas[WATCHTOP.watching].k = $9;

                }
        ;

seconds :       MILLISEC | MICROSEC
        ;

heartbeats:     HEARTBEATS EVERY NUMBER seconds '\n'
                {
                        if (WATCHTOP.running)
                        {
                                WATCHTOP.rate = $3;
```

```
                                WATCHTOP.unit = ($4 == MILLISEC)? 0:1;
                        }
                        else
                        {
                                fprintf(stderr, "\tLine %d: semantical error: ", lines);
                                yyerror("bad use of ON ERROR.");
                                errors++;
                        }
                }
        ;

on_error:       ON ERROR WARN KEYW_TASK NUMBER '\n'
                {
                        if (WATCHTOP.running)
                        {
                                WATCHTOP.action = 0;
                                WATCHTOP.target = $5;
                        }
                        else
                        {
                                fprintf(stderr, "\tLine %d: semantical error: ", lines);
                                yyerror("bad use of ON ERROR.");
                                errors++;
                        }
                }
        |       ON ERROR REBOOT '\n'
                {
                        if (WATCHTOP.running)
                                WATCHTOP.action = 1;
                        else
                        {
                                fprintf(stderr, "\tLine %d: semantical error: ", lines);
                                yyerror("bad use of ON ERROR.");
                                errors++;
                        }
                }
        |       ON ERROR RESTART '\n'
                {
                        if (WATCHTOP.running)
                                WATCHTOP.action = 2;
                        else
                        {
                                fprintf(stderr, "\tLine %d: semantical error: ", lines);
                                yyerror("bad use of ON ERROR.");
                                errors++;
                        }
                }
        ;


watchdog_end:   KEYW_ENDWATCHDOG '\n'
                {
                        WATCHTOP.running = 0;
                        INCWATCH;
                        if (BADWATCH)
                        {
                                fprintf(stderr, "\tLine %d: semantical error: ", lines);
                                yyerror("Too many watchdogs have been defined.");
                                errors++;
                        }
                }
        ;
```

Apart from converting configuration statements into configured header files and source code fragments, the other important task of the Ariel translator is producing the rcode equivalent of the Ariel recovery script. As an example, herein we describe how the IF THEN … ELSE … FI statements are dealt with:

```
section:        if elif else fi
        ;
```

```
if:     {
                /* each IF increments the IF nesting counter */
                rcode_single_op(R_INC_NEST);

              … lines omitted …

        }
        IF '[' expr  ']'     Sepp  /* Sepp is whitespaces */
        {
                 /* at first we issue an incomplete R_FALSE
                    statement that is meant to jump to the ELSE
                    clause or to the FI statement if the
                    expression will be evaluated as false
                  */

                 rcode_if();

               … lines omitted …
        }
        THEN Sepp actions
                        {
                                /* No maintenance is required by THEN */
                        }
        ;
else:   |        /* else may be missing */
                        {
                                /* at first, before ELSE, we issue an
                                   incomplete R_GOTO corresponding to the
                                   Beginning of the ELSE clause
                                   and we record where we are
                                 */
                                goto_pc = rcode_goto(0);
                        }
                ELSE
                        {
                                /* after ELSE, we complete the R_FALSE R-code
                                   issued at the closest IF statement.
                                 */
                                rcode_else(goto_pc);
                        }
                Sepp actions
                        {
                                /* there go the else actions… */
                        }
        ;

fi:             FI
                        {
                                rcode_fi();

                                /* each FI implies decrementing the nesting counter */
                                rcode_single_op(R_DEC_NEST);

                              … lines omitted …
                        }
        ;
```

The possibility to have nested IF is expressed very simply:

```
actions:
        | actions action
        ;

action: '\n'
        |
        section            Sepp
        |
        recovery_action    Sepp
        ;
```

The actual error recovery actions are defined as follows:

```
recovery_action:  KILL    id
                {
                        /* This means "KILL *", which is not likely to be
                           a sensible recovery action, hence is not allowed
                         */
                        if ($2.role & ID_STAR)
                        {
                                fprintf(stderr, "\tLine %d: semantical error: ", lines);
                                yyerror("Can't use `*' with KILL");
                                errors++;
                        }

                        /* GENERATE_RACTION is, in its simplest form, a
                           rcode_raction($1, $2.role, $2.id)
                         */
                        GENERATE_RACTION(R_KILL, $2);

                }
        |         RESTART TID
                {
                        GENERATE_RACTION(R_RESTART, $2);
                }
        |         RESTART GID
                {
                        GENERATE_RACTION(R_RESTART, $2);
                }
        |         RESTART NID
                {
                        GENERATE_RACTION(R_REBOOT, $2);
                }
        | CALL NUMBER
                {
                        /* CALL takes its argument from the stack;
                           first argument is arg counter, 0 in this case
                         */
                        rcode_twoargs(R_PUSH, 0);

                        /* the available function calls are
                           identified by integers */
                        rcode_twoargs(R_FUNCTION_CALL, $2);
                }
        | CALL NUMBER '(' list ')'
                {
                        for (i=card_list-1; i>=0; i--)
                        {
                                rcode_twoargs(R_PUSH, list[i]);
                        }
                        rcode_twoargs(R_PUSH, card_list);

                        rcode_twoargs(R_FUNCTION_CALL, $2);
                }

        | LET VAR '=' linexp
                {
                        rcode_twoargs(R_SET, $2);
                        /* linexp pushes its result on top of the evaluation stack; then
                           The code for R_Set is very simple:
                                      int R_Set(int index, int dummy1, int dummy2)
                                      { int t = R_Pop();
                                        return var[index] = t;
                                      }
                         */
                }
        | LOG NUMBER
                {
                        rcode_twoargs(R_LOGI, $2);
```

```
            }
        | LOG CLOCK
            {
                    rcode_single_op(R_LOGC);
            }
        | LOG VAR
            {
                    rcode_twoargs(R_LOGV, $2);
            }
        ;
```

*…* **many** *lines omitted …*

The last section of the YACC source code Ariel.y includes the whole scanner and defines the main function:

```
%%
#include "lex.yy.c"

main(int argc, char *argv[])
```

After a long list of definitions, the management of the art's command arguments, and the opening of the input files, the main function executes yyparse, i.e., the actual parser. As a result – provided that the parsing concluded successfully – several data structures are filled with configuration data and the actual output r-code. The processing ends with a series of "flush" calls:

```
yyparse();

rflush();

if (w_sp > 0)
    {
            watchdog_flush(watchdog, w_sp);
            fprintf(stderr, "Watchdogs configured.\n");
    }

if (nv_sp > 0)
    {
            nversion_flush(nversion, nv_sp);
            fprintf(stderr, "N-version tasks configured.\n");
    }
```

*… lines omitted …*

```
}
```